\documentclass[10pt, english, onehalfspacing, liststotoc, parskip, headsepline]{HKUThesis} 
\usepackage{aas_macros}
\usepackage[utf8]{inputenc} 
\usepackage[T1]{fontenc} 
\usepackage{fontawesome} 
\usepackage{wrapfig} 
\usepackage{tikz} 

\usepackage{mathpazo} 
\usepackage[natbib=true,maxbibnames=5,firstinits=true,style=authoryear,backend=biber]{biblatex}%

\usepackage[autostyle=true]{csquotes} 

\AtBeginEnvironment{enquote}{\itshape} 

\usepackage{rotating} 

\usepackage[normalem]{ulem} 

\thesistitle{Reliable Tests of Faint-end UV Luminosity Functions in Strong Lensing Fields}

\supervisor{Prof. Jeremy \textsc{Lim}}

\cosupervisor{Prof. Tom \textsc{Broadhurst}}

\examiner{}

\degree{Doctor of Philosophy}

\author{Jiashuo \textsc{Zhang}}

\addresses{}

\subject{Computer Vision}


\university{University of Hong Kong}

\bsuniversity{HKU}
\msuniversity{HKU}

\department{Department of Physics}

\group{Laboratory of The Student}

\faculty{Faculty of Science}

\AtBeginDocument{
\hypersetup{pdftitle=\ttitle} 
\hypersetup{pdfauthor=\authorname} 
\hypersetup{pdfkeywords=\keywordnames} 
}
\begin{document}

\frontmatter 
\pagestyle{plain} 


\begin{titlepage}
\addtocounter{page}{-1}
\begin{center}

\begin{tabular}{cc}
    \includegraphics[align=c, width=0.18\textwidth]{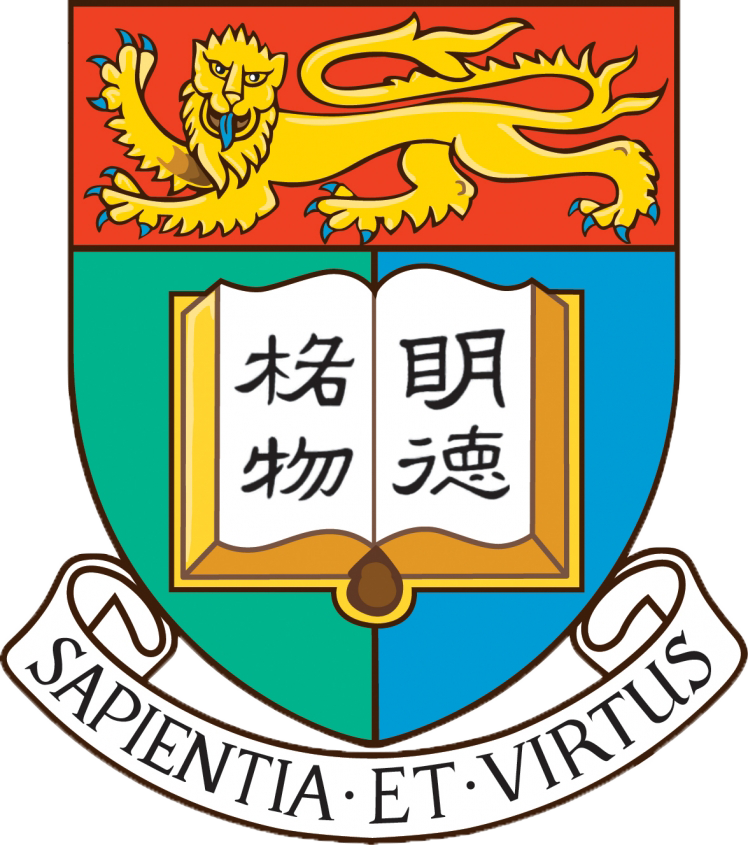} &  
    {\scshape \huge \black{\univname}} 
\end{tabular}

\vspace{0.5cm}
\textsc{\Large Doctoral Thesis}\\[0.5cm] 

\rule[0.4cm]{13cm}{0.1pt}\\
{\huge \bfseries \ttitle\par}\vspace{0.4cm} 
\rule{13cm}{0.1pt}\\ \vspace{1.5cm}
 
\begin{minipage}[t]{0.4\textwidth}
\begin{flushleft} \large
\emph{Author:}\\
\href{http://#}{\authorname} 
\end{flushleft}
\end{minipage}
\begin{minipage}[t]{0.4\textwidth}
\begin{flushright} \large
\emph{Supervisor:} \\
\href{https://www.eee.hku.hk/~elam/}{\supname} \\ 
\emph{Co-Supervisor:} \\
\href{https://www.eee.hku.hk/~hso/}{\cosupname} 
\end{flushright}
\end{minipage}\\[1.6cm]
 
\vfill

\large \textit{A thesis submitted in fulfillment of the requirements\\ for the degree of \degreename}\\[0.3cm] 
\textit{in the}\\[0.4cm]
\deptname\\\facname\\[1.6cm] 
 
\vfill

{\large \usdate\today}\\[4cm] 

\vfill
\end{center}

\end{titlepage}

\blankpage
\addtocounter{page}{-1}


\begin{abstract}
\addchaptertocentry{\abstractname} 
Dark matter comprises $\sim 85\%$ of the entire mass of the Universe, but the fundamental nature of its constituent particles remains elusive. In this thesis, I test for two competitive dark matter models: the conventional heavy particle paradigm ($p$CDM), and dark matter being ultralight bosons of mass $\sim 10^{-22}$eV ($\psi$DM). Owing to the macroscopic wave-like nature of the ultralight particles, $\psi$DM models predict the abundance of intrinsically faint galaxies to be suppressed relative to $p$CDM models, leading to a faint-end turnover in galaxy luminosity functions (LF). 

In this thesis, I will test for such a faint-end turnover induced by $\psi$DM models, exploiting the strong lensing power by massive galaxy clusters to probe intrinsically fainter magnitudes. 
A key challenge for such an analysis, however, is contamination by low-z galaxies sharing similar observed spectral energy distributions (SEDs) as high-z counterparts. As I discuss in this thesis, such a contamination issue is generally severe and may wash out the faint-end turnover signatures anticipated from $\psi$DM models. More worryingly, this contamination issue seems to be overlooked, or not properly addressed, by the existing LF faint-end constructions in lensing fields, casting doubts on the corresponding mass constraints derived. In this thesis, I thus plan to perform a more reliable test of the faint-end LF (and hence the $\psi$DM paradigm) in strong lensing fields, after properly mitigating low-z interlopers. 

The key messages of this thesis will be as follows. First, I demonstrate that $\sim 50\%$ of the purported $3.5\leq z\leq 5.5$ galaxies within existing photometric redshift catalogs constructed for Hubble Frontier Fields (HFF) are low-z interlopers. Secondly, I demonstrate that individual mitigation of interlopers can be achieved with the combination of deep observations from \textit{Hubble Space Telescope} (HST) and \textit{James Webb Space Telescope} (JWST), by simultaneously observing the rest-frame optical and Balmer wavelengths of high-z galaxies. For HFF fields without supplementary data, machine learning methods will also be shown useful in preserving such a mitigating power, in the sense that an "interloper identifier" with excellent performance can be constructed. Finally, following the individual mitigating power on low-z contaminants achieved, I then derive clean $3.5\leq z\leq 5.5$ and $6\leq z\leq 10$ samples for a more reliable test on the faint-end turnovers in LFs. I will show that I found no evidence for such a turnover, leading to a constraint on the $\psi$DM mass of $>2.97\times10^{-22}$eV at 95\% confidence. 

Following recent observational motivation from dwarf galaxies, I will also interpret my derived mass bound in a scenario where dark matter is composed of multiple particle copies, each with distinct mass. With linear perturbation analysis, I argue the derived mass bound is likely on an effective de Broglie scale governing the collective behavior of the entire $\psi$DM budget, owing to the gravitational equilibrium reached on macroscopic scales. I also emphasize that the multi-copy $\psi$DM paradigm is strongly motivated from the String Axiverse perspective. As such, my analysis contributes to better understanding (and also simulating, as I also provide short-cuts for determining the initial matter power spectra) structure formation in String Theory motivated contexts. 

\hfill

(WORD COUNT : 490)

\vspace{2cm} 
\begin{flushright}
\hfill Signed: \underline{\hspace{5cm}}\\ 
\hfill Author : \authorname \\ 
\end{flushright}

\vspace{0.5cm} 
\begin{flushright}
\hfill Signed: \underline{\hspace{5cm}}\\ 
\hfill Supervisor : \supname \\ 
\end{flushright}

\end{abstract}

\pagestyle{empty}
\newpage
\addtocounter{page}{-1}
\begin{center}
\vspace*{2cm}
\huge{ \bf \ttitle}
\end{center}

\vspace{20mm}
\begin{center}
by

\vspace{10mm}
{\bf \authorname}\\
B.Sc. \& M.Sc. \textit{Imperial College London}
\end{center}

\vspace{30mm}
\begin{center}
A Thesis Submitted in Partial Fulfilment \\
of the Requirements for the Degree of \\
Doctor of Philosophy \\
\vspace{10mm}
at \\
\vspace{10mm}
\univname\\
\monthyeardate\today
\end{center}




\begin{declaration}
\setcounter{page}{1}
\addchaptertocentry{\authorshipname} 

\vspace{0.6cm}
I, \authorname, declare that this thesis titled, \enquote{\ttitle}, which is submitted in fulfillment of the requirements for the Degree of Doctor of Philosophy, represents my own work except where due acknowledgement have been made. I further declared that it has not been previously included in a thesis, dissertation, or report submitted to this University or to any other institution for a degree, diploma or other qualifications.

\vspace{2cm} 
\begin{flushright}
\hfill Signed: \underline{\hspace{5cm}}\\[2em] 
\hfill Date: \underline{\hspace{1.5cm} \usdate\today \hspace{1.5cm}}\\ 
\end{flushright}

\end{declaration}



\begin{acknowledgements}
\setcounter{page}{2}
\addchaptertocentry{\acknowledgementname} 
\vspace{1cm}

\noindent I would like to express my heartfelt gratitude to Prof. Jeremy Lim and Prof. Tom Broadhurst for their support, supervision, encouragements and many insightful discussions throughout my PhD studies. I am also particularly grateful for Prof. Jeremy Lim for having the best research group I have ever encountered, and for having me in it. I would like to give special thank to Juno, Keith, Arsen, and James for helping me to catch up on astronomy knowledge, and for providing many productive discussions on my projects. I am also very grateful for my collaborators from the PEARLS team and Giorgio and Leo from HKUST for their help and encouraging support on my projects. I would like to thank Prof. Lixin Dai, Prof. Stephen Ng, and Prof. Patrick Kelly for being my examiners. I thank Haichen, Leif, and Shaozhuo for keeping the theorist within me. I thank Rudrani and Zoe for helping me co-organizing the Journal Club. I thank my friends at office: Alex, Amruth, Haoyang, Janet, Jin-hong, Stanley, Tom, Zhihong, Zijian for having many fun times together. Last but not least, I thank all my family, close friends and mentors throughout my life for shaping me into a better person and a determined truth seeker.

\begin{flushright}
    \authorname \\
    University of Hong Kong \\
    \usdate\today
\end{flushright}

\end{acknowledgements}



\tableofcontents 






\mainmatter 

\pagestyle{thesis} 


\chapter{Introduction}
\label{chap:intro}

The galaxy luminosity functions (LFs) - the number density of galaxies as a function of their luminosity - at different redshifts are one of the most fundamental observables for testing cosmological models of structure formation and growth. In particular, the faint end behavior of the LF (for galaxies with $M\gtrsim-15$) is crucial for understanding how galaxies form hierarchically and for learning about the nature of dark matter. If the abundance of intrinsically faint galaxies is found to be suppressed, it could suggest that dark matter is made of ultralight particles whose wave-like behavior on large scales counters gravitational collapse. To better detect these low-luminosity galaxies at high redshifts, observations have increasingly turned to massive galaxy clusters for use as gravitational lenses. A key challenge in lensing field LF tests, however, is contamination by low-z galaxies that appear similar in their spectral energy distributions (SEDs) as high-z counterparts within the same observing wavelength range (e.g., of \textit{Hubble Space Telescope}, HST). As I demonstrate in this thesis, such contamination is generally severe and may wash out the faint end turnover anticipated from ultralight wave dark matter ($\psi$DM) models. But I also demonstrate that contaminants could be mitigated with additional observations that extend the wavelength coverage (e.g., from the \textit{James Webb Space Telescope}, JWST) and with machine learning methods. My goal in this thesis is to perform a more reliable test on the faint end of UV LF in lensing fields after mitigating low-z interlopers, with an ultimate aim of assessing the viability of the $\psi$DM scenario.

\section{Galaxy Luminosity Function}


Our current understanding of structure formation suggests that all galaxies are formed within gravitationally collapsed density inhomogeneities known as halos, the bulk gravitational potential of which is contributed by dark matter. These inhomogeneities originate from primordial quantum fluctuations in the density of dark matter, which are initially tiny as reflected by CMB anisotropy measurements \citep{Smoot1992ApJ}, but would grow (linearly) over time as denser regions gradually attract more matter from their surroundings. Unlike ordinary matter, dark matter inhomogeneities are not influenced by the pressure gradient sourced by cosmic radiation, and hence can clump efficiently. Dark matter inhomogeneities would eventually grow massive enough (leading to order unity density perturbations) and collapse under gravity to form dark matter halos. As the Universe expands, baryons decouple from cosmic radiation (in particular photons, after recombination) and gradually fall into the gravitational potential of dark matter halos. Over time, baryons (at that stage, gas) trapped would dissipate heat, eventually become cool enough to form stars and give rise to visible (proto-)galaxies \citep{1804.03097}. It is also contemplated that structure formation proceeds hierarchically, that is, smaller halos or galaxies formed would merge to form bigger halos or galaxies, creating cosmic structures that span a wide range in galaxy sizes. 

\begin{figure}
\centering
\begin{subfigure}{.47\textwidth}
  \centering
  \includegraphics[width=\linewidth]{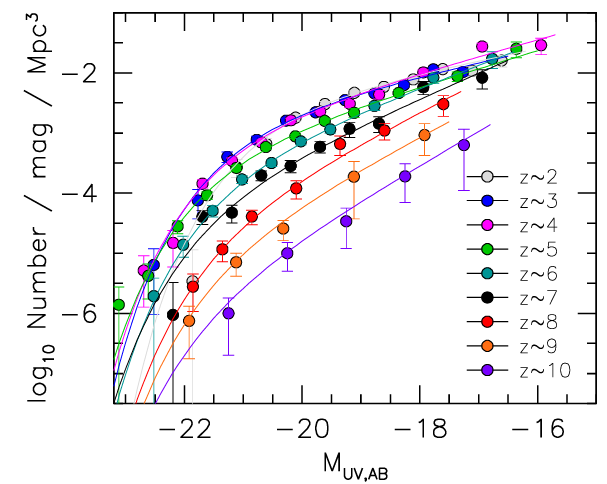}
\end{subfigure}%
\hfill
\begin{subfigure}{.49\textwidth}
  \centering
  \includegraphics[width=\linewidth]{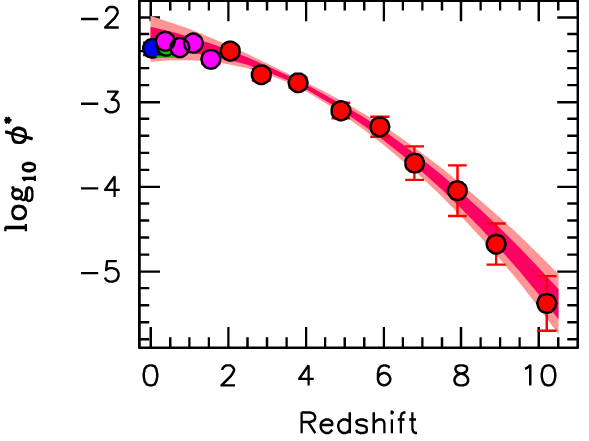}
\end{subfigure}
\caption{\textit{Left}: Galaxy UV Luminosity Functions as measured by \cite{Bouwens2021} at different redshifts, exhibiting an exponential suppression at bright magnitudes and steadily increasing to fainter plotted magnitudes following a power-law. Solid lines are the best fit Schechter functional form to the measurements. \textit{Right}: Overall amplitude $\phi^*$ of UV LFs at different redshifts based on the fitting shown in the left panel. It is seen that $\phi^*$ increases faster at higher redshifts than at lower redshifts, suggesting an "accelerated" evolution. }
\label{Accelerated_Evolution}
\end{figure}

A good tracer of hierarchical structure formation and evolution is galaxy LFs. An example is the 'accelerated' evolution discovered in the UV LFs, i.e., LF measured in rest-frame UV wavelength (see Sec.\ref{sec_counts_construct}) of galaxies to better probe recent star formation, prior to the launch of JWST \citep{Oesch2012, Bouwens2014B, Oesch2018, Ishigaki2018, Bouwens2021}. This is seen as a faster increase in the overall amplitude of the UV LF ($\phi^*$) at higher redshifts than at lower redshifts, as reproduced in the right panel of Fig.\ref{Accelerated_Evolution}. Such an "accelerated" evolution suggests the galactic evolution is dominated by the growth of dark matter halos themselves, with a roughly unchanging star formation efficiency\citep{Bouwens2021}. 

The launch of JWST pushed the redshift frontier for the UV LF construction to as high as $z\sim 25$ \citep{2503.15594, 2504.05893}, revealing a surprising over-abundance of bright $M_{uv}<-20$ galaxies than anticipated from a constant star formation efficiency model \citep{2208.01612, 2311.04279, 2501.00984}. Several explanations were offered for this discovery, including very bursty or stochastic star formation in the early Universe \citep{2307.15305,2405.13108}, reduced dust attenuation \citep{2208.00720}, or star formation efficiency is instead higher than $z<10$ \citep{2208.01612, 2303.04827}. Some also argue that constant star formation efficiency remains viable, provided stellar populations at $z>10$ are younger than typically anticipated and hence there are more intrinsically bright galaxies \citep{2501.03217}.


\section{LF Faint End and Nature of Dark Matter}

In this thesis, I will focus in particular on the faint end of UV LF, which is also of critical importance from many perspectives, such as reionization, baryonic physics, and, most importantly for this thesis, the nature of dark matter. For instance, if the steep slope observed in blank field \citep{Finkelstein2015, Bouwens2014B, Bouwens2021} continues to much fainter magnitudes ($M_{UV}\gtrsim-13$), this would suggest reionization before z$\sim$6 could have been dominantly driven by an abundant amount of low mass star-forming galaxies \citep{Bunker2004, Robertson2013}. More recently, JWST allowed for better investigations of the ionizing photon production rate of dwarf galaxies at high redshifts, and the production rate was found to exceed the commonly anticipated level by a factor of 4 \citep{2308.08540}, and increasing with decreasing luminosity \citep{2211.12548}. These suggest dwarf galaxies can explain/dominantly drive reionization with a more modest escape fraction than needed by luminous galaxies. 

\begin{figure}
\centering
\begin{subfigure}{.4\textwidth}
  \centering
  \includegraphics[width=\linewidth]{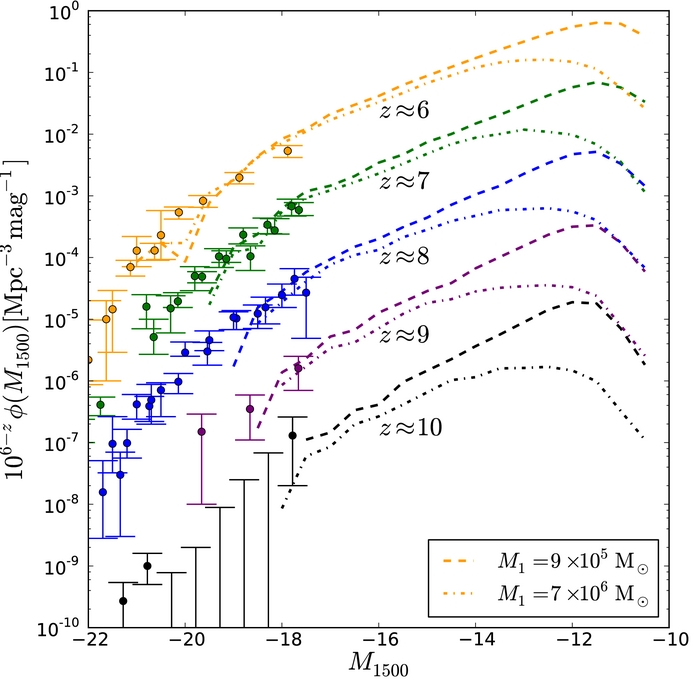}
\end{subfigure}%
\hfill
\begin{subfigure}{.53\textwidth}
  \centering
  \includegraphics[width=\linewidth]{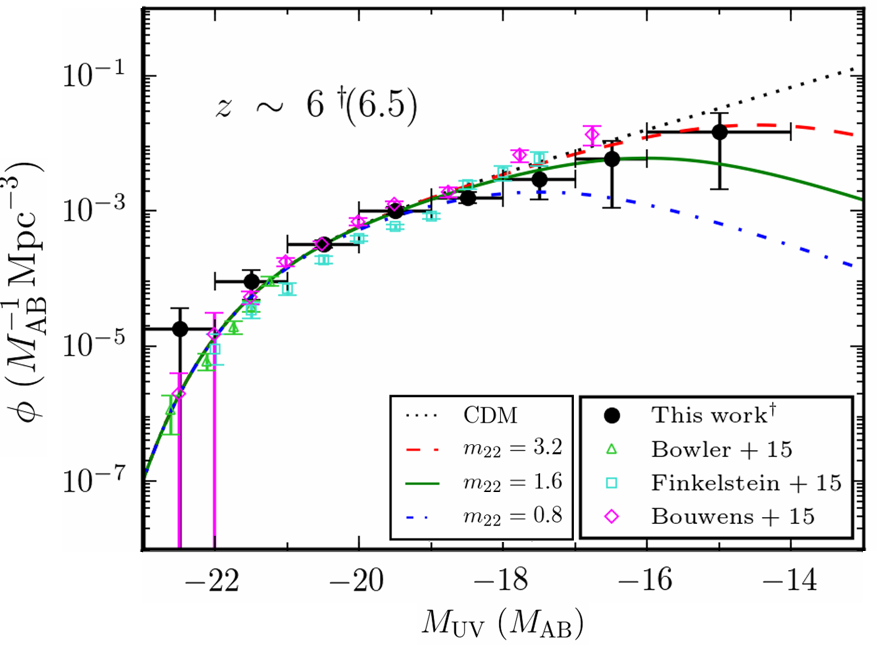}
\end{subfigure}
\caption{\textit{Left}: UV LFs with baryon physics induced faint end turnover at $M_{UV}\sim -12$ as obtained by cosmological simulation of \cite{Gnedin2014}. Dashed and dotted curves correspond to two different mass resolutions, and data points are a compilation of observations available at the time. \textit{Right}: UV LFs with $\psi$DM induced faint end turnover in comparison with UV LFs observed by \cite{Leung2018} (black data point) and others available at the time. The black dotted curve corresponds to the heavy particle CDM paradigm with UV LF steadily increasing to faint magnitudes, and colored curves correspond to UV LFs anticipated in the $\psi$DM paradigm ($m_{22}$ is $\psi$DM mass in unit of $10^{-22}$eV) as parametrized by \cite{Schive2016}. }
\label{UVLF_turnover}
\end{figure}

Conventionally, Cold Dark Matter (CDM) is thought to be composed of weakly\footnote{By weakly interacting, it actually means for WIMPs to have electro-weak scale self-interaction. Such a particle with mass $\sim100$GeV majestically produces just the right amount of thermal relic abundance to explain all of the dark matter; this coincidence has ever since been known as the "WIMP miracle". } interacting massive particles (WIMPs) of mass $\sim100$GeV. Such particles have strong theoretical motivations from (supersymmetric) extension of the Standard Model of particle physics, and cosmological simulations adopting the particle nature for CDM have been successful at explaining the large-scale structure. In the heavy particle CDM paradigm (hereafter, $p$CDM), there exists an ever-growing population of lower mass halos, leading to a steady increase in UV LFs towards fainter magnitudes until $M_{UV}\sim -12$, where a flattening or turnover is also anticipated from baryonic physics. The corresponding turnover is the result of insufficient gas cooling in lower mass halos halting star formation \citep{Munoz2011, Jaacks2013, Gnedin2014}, with examples UV LFs could be found in the left panel of Fig.\ref{UVLF_turnover} (taken from \cite{Gnedin2014}).  


Despite its success on the large scale, $p$CDM fails to reproduce certain astronomical observations on smaller scales. For instance, cosmological simulations significantly over-predict the number of dwarf satellite galaxies around the Milky Way, leading to the famous "missing satellite" problem \citep{Moore1999ApJ, Klypin1999ApJ}. One proposed solution for the "missing satellite" problem was that many of the dwarf satellite galaxies could be invisible if their stars are stripped out of the gravitational potential through tidal stripping. Subsequent analysis, however, indicates that $p$CDM also over-predicts the abundance of massive and dense sub-halos that ought to be luminous, leading to the "too-big-to-fail" problem \citep{BK2011MNRAS}. Furthermore, decades of experimental searches have continuously reported null detection of theoretically motivated heavy particle candidates \citep{2024arXiv241017036A, 2024arXiv241212972T, 2025arXiv250113665Y}, adding to the challenges faced by the traditional $p$CDM paradigm. To address these small-scale crises faced by $p$CDM, alternative dark matter models have been proposed. One notable CDM alternative is ultralight bosons with mass $\sim 10^{-22}$eV, first proposed by \cite{hu2000} to address the "core-cusp problem". The anticipated core profile was confirmed to form through cosmological simulations \citep{Schive2014}, owing to the presence of solitons at the center of dark matter halos. 

Because of their extremely light masses, ultralight bosons have astronomical-scale de Broglie wavelengths (a few hundred pc in massive galaxies, tens of pc in massive galaxy clusters) and lead to a natural suppression on the small-scale structure formation with their wave-like nature (hence the name wave dark matter, $\psi$DM). This suppression also manifests itself as a faint end turnover on UV LFs\footnote{Alternative dark matter models such as Warm Dark Matter or Self-interacting Dark Matter gives different suppression of low-mass halos and hence different faint-end suppression signature on UV LFs.} (independent of baryonic physics) at a position dependent on the mass of ultralight bosons. This is illustrated by colored curves in the right panel of Fig.\ref{UVLF_turnover} (taken from \cite{Leung2018}), where a lighter $\psi$DM mass leads to UV LF turning over at a brighter magnitude. These $\psi$DM-induced turnovers could also be seen to occur at magnitudes much brighter than those induced by $p$CDM+baryonic physics\footnote{A faint end turnover at the same position of $M_{UV} \sim -12$ would correspond to a $\psi$DM mass of $8.9\times 10^{-22}$eV at $z\sim 6$. But $\psi$DM-induced faint end turnover is also redshift dependent. As a consequence, the $\psi$DM model could still be tested against the $p$CDM model at different redshifts.} ($M_{UV} \sim -12$), allowing for the tests of these two different dark matter scenarios. 

\section{Gravitational Lensing and Magnification Bias}

Since the launch of the HST, significant progress has been made in constructing UV LF in deep "blank" fields \citep{Bouwens2014B, Bouwens2021, Finkelstein2015}, reaching as faint as $M_{UV} \lesssim -16$. JWST enabled the construction of UV LF at higher redshifts ($z>9$) as previously summarized, but reaching magnitudes $M_{uv} < -17$ \citep{2302.02429,2306.06244, 2501.00984} not as faint as accessed at lower redshifts. To better probe the faint end of the UV LF relevant to testing $\psi$DM versus $p$CDM models, one promising approach is to leverage the power of strong gravitational lensing.

\begin{figure}
     \centering
     \includegraphics[width=0.9\textwidth]{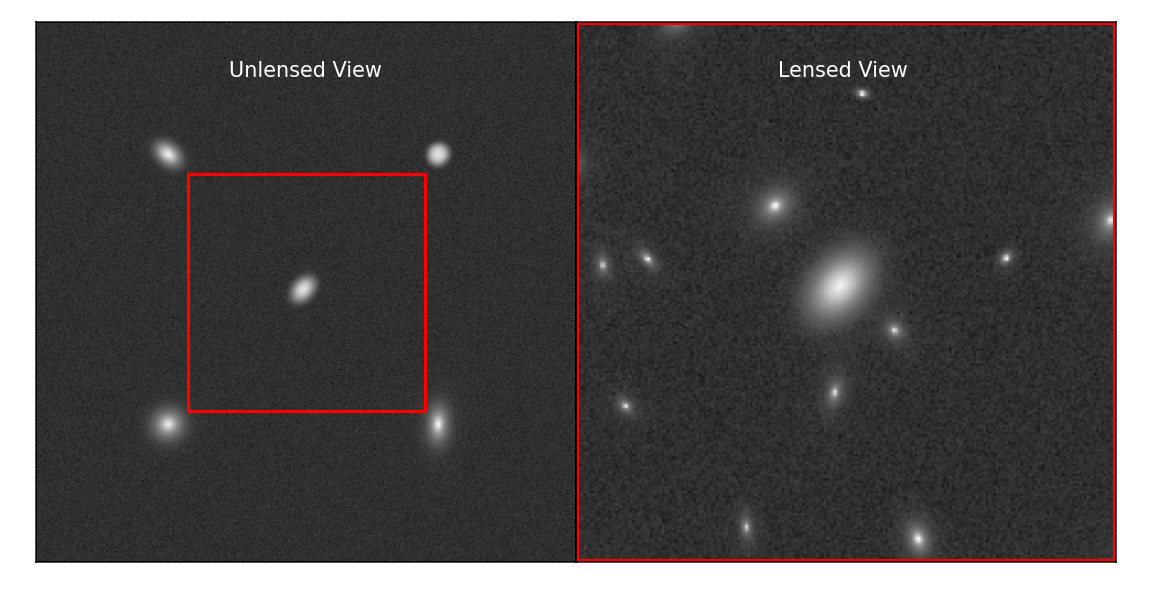}
     \caption{Demonstration of effects brought by lensing magnification with simulated images. On the left, we present an unlensed view hosting five bright galaxies. The lensed view toward the same observing direction is shown on the right, in which many fainter galaxies are newly magnified to be observable. But the sky area probed by the lensed view is also seen reduced, as indicated by the red contour in the left panel.}
    \label{mag_bias_demo}
\end{figure}

Owing to its magnifying effect, gravitational lensing has been demonstrated to be powerful for detecting faint, distant galaxies \citep{Bradley2023, Atek2023, Atek2023B} and even stars at redshifts as high as $z=6.2$ \citep{Welch2022Nature}. And over the past decades, observations have increasingly turned to massive galaxy clusters for use as strong gravitational lenses to better detect galaxies at high redshifts. For example, the search for high-redshift galaxies was a primary goal of the Cluster Lensing And Supernova Survey with Hubble (CLASH) program, comprising imaging of twenty-five galaxy clusters with the HST in sixteen filters spanning near-UV to near-IR. More recently, the newly launched JWST also showcased the ability of gravitational lensing to unveil galaxies in the early universe from imaging at near- to mid-infrared wavelengths. Many of General Observer (GO) and Guaranteed Time Observations (GTO) time are also allocated to programs targeting massive galaxy cluster fields, including but not limited to Prime Extragalactic Areas for Reionization and Lensing Science (PEARLS), Ultradeep NIRSpec and NIRCam ObserVations before the Epoch of Reionization (UNCOVER), and Strong LensIng and Cluster Evolution (SLICE).

While lensing magnification enables the detection of individual fainter high-$z$ galaxies, it does not necessarily increase the overall surface number density of high-$z$ galaxies observed. Instead, the observed surface number density is governed by two competing effects brought by lensing magnification. Fig.\ref{mag_bias_demo} illustrates this by comparing the unlensed (left) and lensed (right) view toward the same observing direction. It could be seen that gravitational lensing magnifies fainter galaxies to be detectable, but at the expense of reducing surveyed sky area within the same telescope field of view (indicated by the red contour). This interplay between lowering the effective detection threshold and decreasing the probed cosmic volume has been formalized mathematically, and is known as magnification bias \citep{Tom1995}.

For the shown illustration, the lensed view has a higher surface number density of galaxies than the unlensed view. This is the result of sufficiently abundant fainter galaxies newly lensed to be accessible, and when this happens, magnification bias is termed \textit{positive}. Conversely, if newly detected faint galaxies are not abundant enough to compensate for the sky area reduction effect, gravitational lensing would instead lower the surface number density observed, resulting in \textit{negative} magnification bias. Finally, when the newly detected faint galaxies are just abundant enough to balance the volume reduction effect, lensed and unlensed views have the same observed surface number density, and magnification bias is called \textit{null/critical}. 

Following the above discussion, it could be concluded that magnification bias, specifically the observed surface number density of galaxies, sensitively depends on the abundance of fainter galaxies (i.e., the faint end of the UV LF). Owing to this sensitivity, magnification bias could be used to test different cosmological models, in particular, $\psi$DM versus $p$CDM models. As commented by \cite{Leung2018}, magnification bias is a superior way for such a test for the following reason. In the context of strong gravitational lensing, a given patch of the background sky can be lensed to appear at multiple positions on the image plane, giving rise to multiply lensed images. Any attempt to directly measure the number density of background galaxies (i.e., to construct the UV LFs) must address this multi-occurrence. On the other hand, magnification bias holds separately on each image pixel as a local observational effect, allowing for independent treatments of the multiply lensed images of the same background region. As a consequence, magnification bias avoids the need to identify multiply lensed images and correction of cosmic volume, facilitating a considerably simpler and direct analysis. 

\section{Discrepancies on Faint End UV LF in Lensing Fields}

\begin{figure}[h]
\centering	
\includegraphics[width=0.95\textwidth]{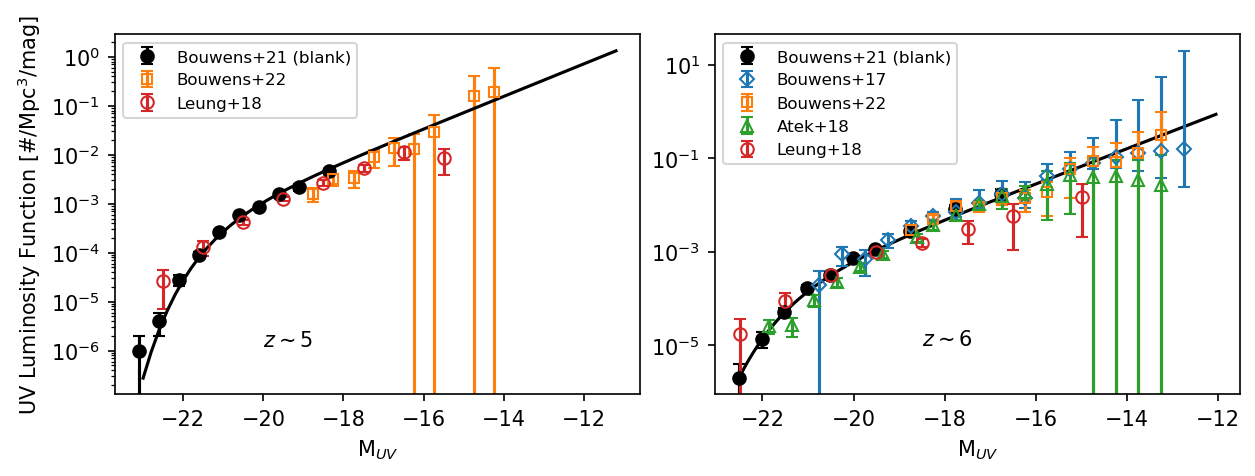}
\caption{Comparison of UV LF at $z\sim5$(left) and $z\sim6$ (right) constructed in HFF lensing fields by various studies and also that from deep blank fields by \cite{Bouwens2021} (black solid line indicating faint end extrapolation). Measurements by \cite{Leung2018, Atek2018} and \cite{Bouwens2017} all hint at a faint end turnover, despite at different magnitudes, whereas measurements by \cite{Bouwens202205} suggest the opposite.}
\label{UVLF_Faint_END_discrepancy}
\end{figure}

By targeting six massive galaxy cluster lenses, the HFF program has produced the deepest lensed images of HST and lowered the luminosity threshold for UV LF construction by 2-4 magnitudes \citep{Livermore2017, Atek2018, Ishigaki2018, Yue2018, Bouwens202205}. However, contrary results have been reported on the faint end of UV LF as reproduced in Fig.\ref{UVLF_Faint_END_discrepancy}. In particular, some studies claimed hints of faint end turnover \citep{Bouwens2017, Atek2018, Leung2018}, whereas others argued otherwise at the same magnitudes \citep{Ishigaki2018, Bouwens202205}.

One potential explanation for this discrepancy could be the significant uncertainties associated with high magnification factors \citep{Atek2018}. Notably, substantial systematic errors have been observed due to variations in lens models when the lensing magnification factor exceeds $\mu > 30$ \citep{Bouwens2017}. When magnification factors are underestimated in highly magnified regions, cosmic volumes would be overestimated after lensing corrections, consequently leading to an apparent turnover at the faint end of the UV LF. Conversely, an overestimation in lensing magnification factors would instead lead to a faint-end turn-up or wash out faint-end turnovers induced by baryonic physics (in $p$CDM paradigm) or $\psi$DM. However, as more strong lensing constraints become available (particularly with the ability of JWST to probe higher redshifts), improvements in lens models may address this issue. Narrowing to a smaller range of lensing magnification factors for analysis may also help in this respect. 



\begin{figure}[h]
\centering	
\includegraphics[width=0.6\textwidth]{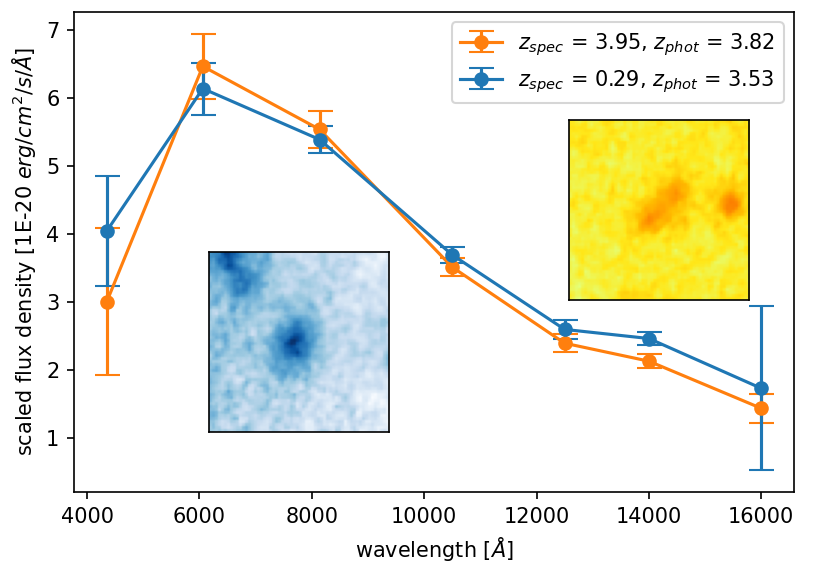}
\caption{Comparison of SED in Hubble Frontier Fields filters of a $z_{spec}$=0.29 galaxy (blue curve) with a $z_{spec}$=3.95 galaxy (orange curve). The values are taken from the S18 catalog, and the SED for the low spec-z source is scaled up by a factor of 2.75 for a better comparison of shapes. The scaled SEDs are seen to be very similar in shape, which leads to a misidentification problem of the low-z sources, as is demonstrated here by their photo-z estimates. In the inserted figures, we also show a 1.8 arcsec$^2$ F160W band cutout image (using similar colors as their SEDs) for both galaxies. Both galaxies are of similar apparent size, but the high-z galaxies are seen to be slightly more elongated. This slight visual difference, however, is not enough to refute the wrongly estimated photo-z for the low spec-z galaxy: the blue cutout could also be interpreted as the lensed image of a compact high-z galaxy affected less by lensing shear. }
\label{clusz_vs_highz_SED}
\end{figure}

\begin{figure}
\centering
\begin{subfigure}{.49\textwidth}
  \centering
  \includegraphics[width=\linewidth]{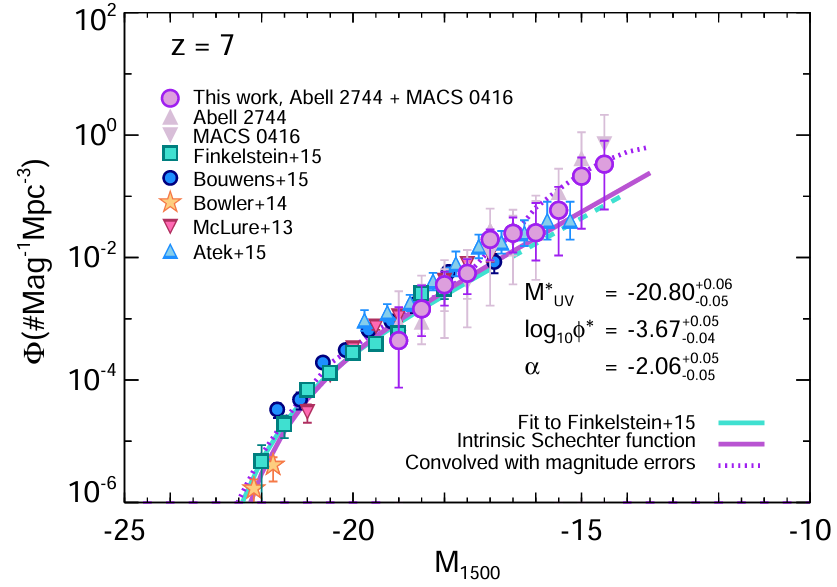}
\end{subfigure}%
\hfill
\begin{subfigure}{.495\textwidth}
  \centering
  \includegraphics[width=\linewidth]{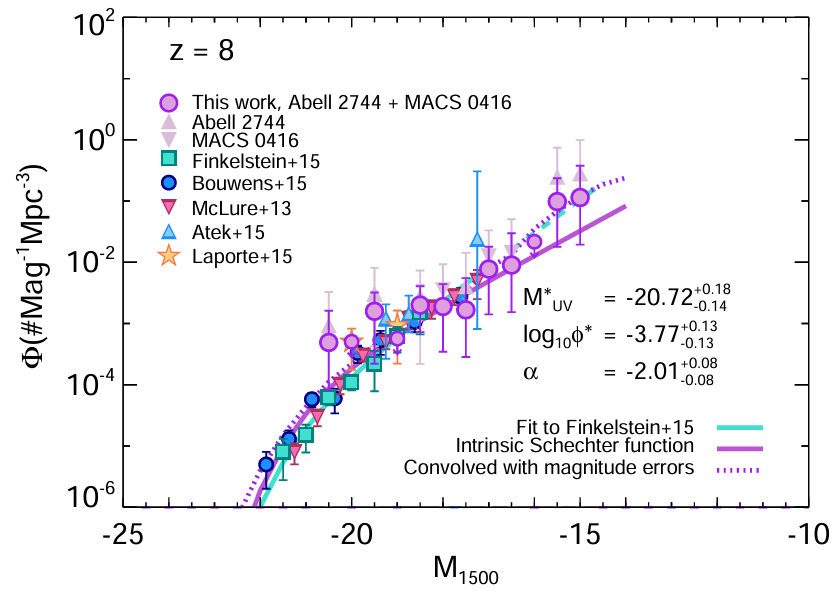}
\end{subfigure}
\caption{UV LFs at $z\sim 7$ (left) and $z\sim 8$ (right) measured by \cite{Livermore2017} (purple circles) in strong lensing fields Abell 2744 and MACS J0416. A clear elevation could be seen to occur relative to the intrinsic Schechter function (purple solid line, obtained from fits to combined HFF and CANDELS data) at magnitudes $M\sim -15$ of relevance for the $\psi$DM test. This elevation could be the result of contamination by misidentified lower redshift galaxies.}
\label{Livermore_excess}
\end{figure}

A more critical issue, in search of faint end turnover signatures, is that lower redshift galaxies may be confused as high-z galaxies owing to the similarity in the observed SEDs, thereby contaminating the sample\footnote{In this thesis, these misidentified lower-$z$ galaxies will be inferred to as contaminants/interlopers.}. Specifically, as depicted by Figure \ref{clusz_vs_highz_SED}, the redshifted Balmer/4000\AA breaks of quiescent galaxies at low redshift (z = 0.29) can resemble the redshifted Lyman break of star-forming galaxies at high redshifts (z = 3.95) over the wavelength range spanned by the HST. Although both the CLASH and HFF fields benefit from observations with the Multi-Unit Spectroscopic Explorer (MUSE) at the Very Large Telescope (VLT), these observations are not sufficiently deep for measuring the redshifts of relatively dim galaxies. 

The possibility that HFF photometric redshift catalogs may be contaminated has been previously mentioned by \cite{Leung2018}. In particular, they noticed a significant overlap in the color-color and color-magnitude diagrams between the cluster members and the $z\sim 4$ candidates cataloged by \cite{Coe2015} (C15), suggesting a potential cross-contamination. Having contamination may also explain why some UV LF measurements in HFF cluster fields are elevated at the very faint end, for instance in \cite{Livermore2017} (reproduced in Fig.\ref{Livermore_excess} for $z\sim7,8$) and \cite{Bouwens202205}. Such an elevation could be understood as dimmer (lower signal-to-noise) galaxies are easier to misidentify, and hence lead to a larger elevation\footnote{With the same logic, contamination may easily wash out faint-end turnovers and hence impede the assessment of viability of, e.g., $\psi$DM paradigm. } at fainter magnitudes. For JWST, operating at longer wavelengths than HST, similar contamination/confusion may also occur, but at a higher redshift range. For instance, dusty lower redshift galaxies at $z\sim5$ could be mistaken as galaxies at $z\sim16$, as has been demonstrated by \cite{2023arXiv230315431A}. 






\begin{figure}
    \centering
    \includegraphics[width=0.95\linewidth]{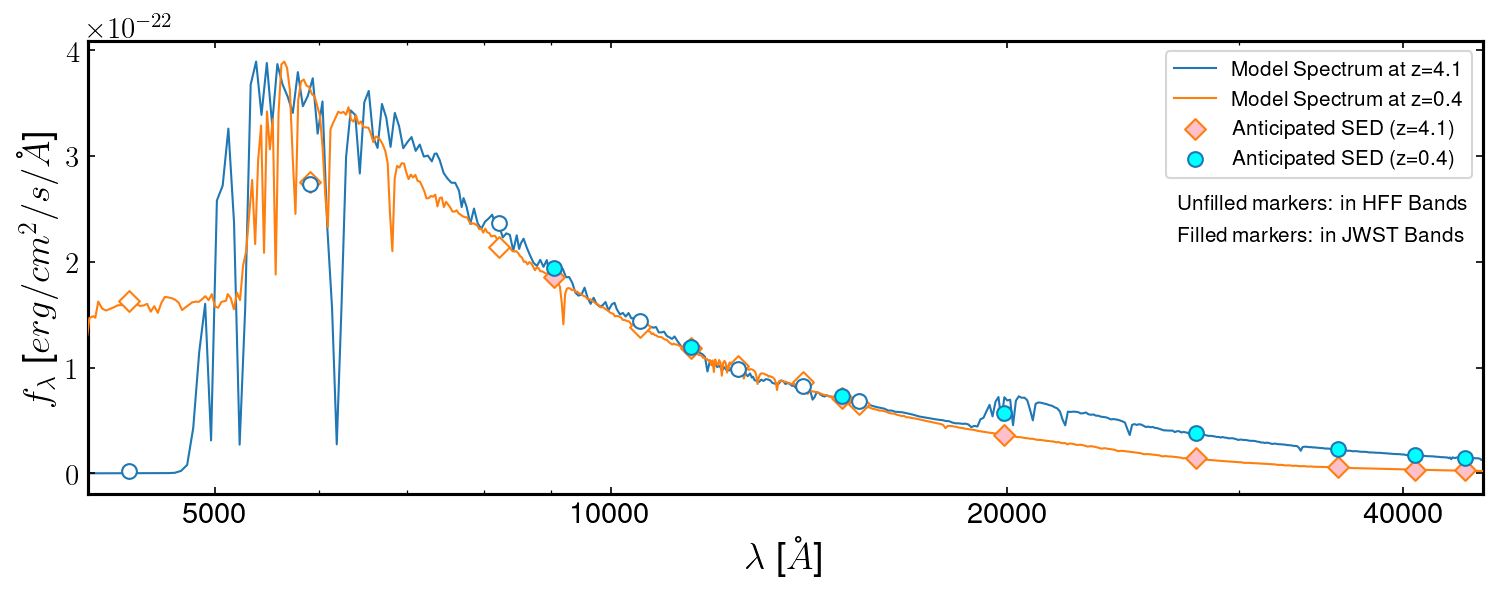}
    \caption{Model spectrum (orange) of an early-type galaxy at $z=0.4$ leads to similar observed SEDs as the model spectrum (blue) of a star-forming galaxy at $z=4.1$ in HFF bands (unfilled markers), and hence may be misidentified to $z\sim 4$. On the other hand, JWST observations (filled markers) cover the rest-frame Balmer break of the $z=4.1$ galaxy. This allows for more accurate identifications of their true redshifts, subsequently enabling individual mitigation of interlopers. }
    \label{JWST_VS_HFF_SED}
\end{figure}

The combination of deep HST and JWST data, however, holds promise for effective individual mitigation of interlopers. I illustrate this with Fig.\ref{JWST_VS_HFF_SED}, presenting an example model spectrum\footnote{These spectra were extracted from \href{https://massivestars.stsci.edu/starburst99/docs/default.htm}{Starburst99} database.} of a star-forming galaxy at $z=4.1$ in blue, and an early-type galaxy at $z=0.4$ in orange. Cyan circles and orange diamonds represent anticipated flux measurements given these spectra in HFF (unfilled markers) and JWST (filled markers) bands, from which it can be seen that JWST covers the rest frame Balmer break of the $z=4.1$ galaxy, lifting the degeneracy seen in the anticipated SEDs across HFF wavelengths. As a consequence, high-z galaxies and low-z interlopers can be effectively separated\footnote{For the other HFF fields without supplementary JWST observations, machine learning methods will be shown helpful in mitigating interlopers in those fields, avoiding the need for separate time-intensive JWST observations.}, enabling more reliable tests of faint-end UV LF in strong lensing fields. The last argument also applies to higher redshifts, motivating us to test for (stronger) signatures of $\psi$DM above $z>6$. 

\section{Thesis Outline}

This thesis is structured as follows. In Chap.\ref{chap:gnuastro}, I will first introduce a new and novel source detection algorithm, \texttt{GNU Astronomy Utility}, that I used to construct a photo-z catalog (combining HST and JWST observations) for the strong lensing field of MACS J0416. This constructed photo-z catalog will subsequently be used for individually identifying interlopers in Chap.\ref{chap:contamination_jwst}, and testing for faint-end turnovers induced by $\psi$DM above $z>6$ in Chap.\ref{chap:UVLFtesting}. For the benefit of future users, the introduction to \texttt{GNU Astronomy Utility} will be as thorough as possible. I will also comment on the key advantages and disadvantages of cataloging with \texttt{GNU Astronomy Utility} in comparison with using traditional cataloging methods (e.g., with \texttt{Source Extractor}).

In Chap.\ref{chap:contamination}, I expand on the potential contamination issue (by misidentified low-z interlopers) discussed for existing HFF photo-z catalogs, and provide quantitative estimates on the contamination levels. This analysis heavily relies on the simultaneous observation of the HFF program towards nearby blank fields, 6 arcmin away from respective galaxy clusters. These blank fields are expected to suffer less contamination (as cluster members are non-present), and were observed to similar depth as in the cluster fields, allowing for a reliable construction of UV LF over the same observed magnitude range. By extrapolating UV LF as measured from parallel fields using the magnification bias method, I then demonstrate that the contamination level is generically high over $3.5\leq z_{phot}\leq 5.5$ and about $\sim 50\%$ for HFF photometric redshift catalogs provided by \cite{Shipley2018, Coe2015} and the ASTRODEEP team.

With the availability of deep JWST data (by the PEARLS team, \cite{Windhorst2022}) on M0416, I then construct the JWST+HST combined photo-z catalog using \texttt{GNUastro} in Chap.\ref{chap:contamination_jwst}. I will demonstrate that individual mitigation of interlopers is made possible as HST and JWST simultaneously observe the rest frame Lyman and Balmer breaks of $z\sim4$ galaxies. The galaxy properties of the identified interlopers will also be investigated, revealing they are mostly dwarf galaxies and consist of both quiescent cluster members and (potentially) star-forming field galaxies. In Chap.\ref{chap:contamination_jwst} I also demonstrate that machine learning methods can extract subtle differences between genuine high-z galaxies and low-z interlopers already encoded in existing HFF-based measurements. This allowed for individual mitigation of interlopers in the remaining five HFF cluster fields, while avoiding the need for time-intensive supplementary JWST observations.  

Established with interloper mitigated samples (through machine learning and also the combination of JWST+HST observations), I perform a more reliable lensing field test on the faint-end of UV LFs in Chap.\ref{chap:UVLFtesting}. More specifically, I will test for any evidence of faint end turnover as predicted by the $\psi$DM model and previously hinted by \cite{Leung2018}. By examining the observed surface number density of $3.5\leq z\leq 5.5$ galaxies (after machine learning mitigation) with magnification bias, however, no such evidence was found. This led to a bound on the $\psi$DM mass of $2.97\times 10^{-22}$eV at $95\%$ confidence level. Similarly, by examining the observed surface number density of carefully selected $6\leq z\leq 10$ galaxies in the M0416 lensing field, a less constraining bound of $>2.53\times 10^{-22}$eV at $95\%$ confidence level was derived. I will comment on the contradictory results between my and those of \cite{Leung2018} (also relying on magnification bias), demonstrating that data incompleteness might have accidentally led them to claim a faint-end turnover (favoring the mass of $1.6\times 10^{-22}$eV).

In Chap.\ref{chap:multi_axion}, I will also discuss how the above mass bounds could be interpreted in a scenario where $\psi$DM is not comprised of a single particle copy (as commonly assumed), but instead composed of multiple copies, each with distinct mass. Such a multi-copy $\psi$DM scenario is well motivated from the perspective of String Axiverse, and will increasingly gain attention owing to recent astronomical evidence from dwarf galaxies \citep{Pozo2024PhRvD}. Using linear perturbation analysis, I will show that different $\psi$DM copies reach a gravitational equilibrium on large scales. Correspondingly, the suppression on small-scale structures is likely dominated by an effective mass determined by their relative contribution to the full dark matter budget. I also provide an analytical formula for this effective mass, as well as a simplified method for determining the initial conditions (more specifically, suppressions to matter power spectrum relative to the $p$CDM paradigm) relevant for future multi-$\psi$DM cosmological simulations. 

Finally, in Chap.\ref{chap:summary}, I provide a comprehensive summary of the results presented in this thesis, and mention several future prospects.

My paper presenting the contamination level estimates for HFF photo-z catalogs in Chap.\ref{chap:contamination} has also been published at \cite{Zhang2025}. Results in the other chapters are being prepared for publications.


\chapter{Methodology: Cataloging with GNU Astronomy Utility}
\label{chap:gnuastro}

In this chapter, I introduce a new and novel source detection algorithm that I later used (see Chap.\ref{chap:contamination_jwst}) to construct my own photo-z catalog combining deep JWST and HST observations. The constructed catalog will subsequently be used to (1) test if interlopers in existing HFF catalogs (in particular, catalogs provided by \cite{Shipley2018}, S18) over redshift $3.5\leq z\leq 5.5$ could be individually identified, and (2) over redshift $6\leq z\leq 10$, test for any suppressed abundance of intrinsically faint galaxies as anticipated by $\psi$DM model (see Chap.\ref{chap:intro}). As such, my primary targets (for detection and analysis) are faint galaxies, which are generally difficult to detect. 

Conventionally, source detection algorithms such as \texttt{Source Extractor}(\cite{Bertin1996}, hereafter \texttt{SExtractor}) detect galaxies as a group of pixels above a certain brightness threshold. By lowering the detection threshold, \texttt{SExtractor} could be tailored to detect faint galaxies. On the other hand, lowering the detection threshold also increases the risk of detecting strong random noise peaks as fake galaxies. As a consequence, the faintest galaxies that one could detect with confidence are limited by the random sky noise when using \texttt{SExtractor}. 

In addition, a lot of faint galaxies (especially gravitationally lensed high-z galaxies) may sit beside a very bright neighbor in strong lensing fields. \texttt{SExtractor} attempts to deblend such sources from their bright neighbors by effectively allowing for a 'local' detection threshold (determined internally), with additional conditions that the faint galaxy needs to have a large enough local brightness peak, and to contribute a sufficient fraction of the total observed brightness. Such a deblending strategy, however, faces the same issue of either detecting strong noise peaks as fake galaxies (for too low a local threshold), or, conversely, being unable to deblend the faint galaxies from their much brighter neighbors. One attempted resolution in literature is to model and subtract lights from bright cluster members (see, for instance, \cite{Shipley2018, Molino2017}) such that faint galaxies may be detected in relative isolation. Nevertheless, real galaxies have more complicated light profiles than analytic models (e.g., Sersic/Moffat profiles). Correspondingly, the modeling of light could be imperfect, subsequently leading to the generation of artificial sources near the center of subtracted galaxies. 

For my analysis, to detect faint galaxies as completely as possible from deep and crowded cluster field images, I opt to perform source detection using a newly invented source detection algorithm, \texttt{GNU Astronomy Utilities} (hereafter \texttt{GNUastro}, v0.19) \citep{Akhlaghi2015, Akhlaghi2019}. \texttt{GNUastro} detects astronomical signals by looking for correlations among neighboring pixels, a property not shared by random noise. With this detection philosophy, \texttt{GNUastro} was shown able to detect very dim sources, even pixel-wise below sky level \citep{Akhlaghi2015}. In crowded/bright regions, \texttt{GNUastro} could also deblend neighboring galaxies effectively as individual local brightness peaks. And the majority of fake detections (from strong noise peaks) could be avoided, by conditioning the real galaxies to have signal-to-noise ratios higher than most (e.g., 99\%) noise fluctuations in sky background regions without pixel-wise correlated signals. 

In the remainder of this chapter, I further expand on how \texttt{GNUastro} detects galaxies, by detailing three separate subprograms involved\footnote{More detailed instructions could be found in the \texttt{GNUastro} manual \href{https://www.gnu.org/software/gnuastro/manual/}{here}}: \texttt{NoiseChisel} (Sec.\ref{sec_noiseChisel}), \texttt{Segment} (Sec.\ref{subsec_segment}), and \texttt{mkatalog} (Sec.\ref{sec_mkcatalog}). For the benefit of future users, I will also provide a short discussion on the advantages and disadvantages when cataloging with GNUastro in Sec.\ref{sec_pro_cons_gnuastro}. In particular, I will demonstrate that \texttt{GNUastro} outperforms \texttt{SExtractor} at detecting faint galaxies (also separating them from bright neighbors) in general, and is capable of generating photometries consistent with traditional methods. The drawbacks, however, are that \texttt{GNUastro} may split up galaxies if galaxies host multiple visible features, and may be effort-intensive in obtaining the best input parameters. The photo-z catalog newly constructed for the field of M0416 using \texttt{GNUastro} will be introduced in Chap.\ref{chap:contamination_jwst}.

\section{NoiseChisel}
\label{sec_noiseChisel} 

\begin{figure}
     \centering
     \includegraphics[width=0.8\textwidth]{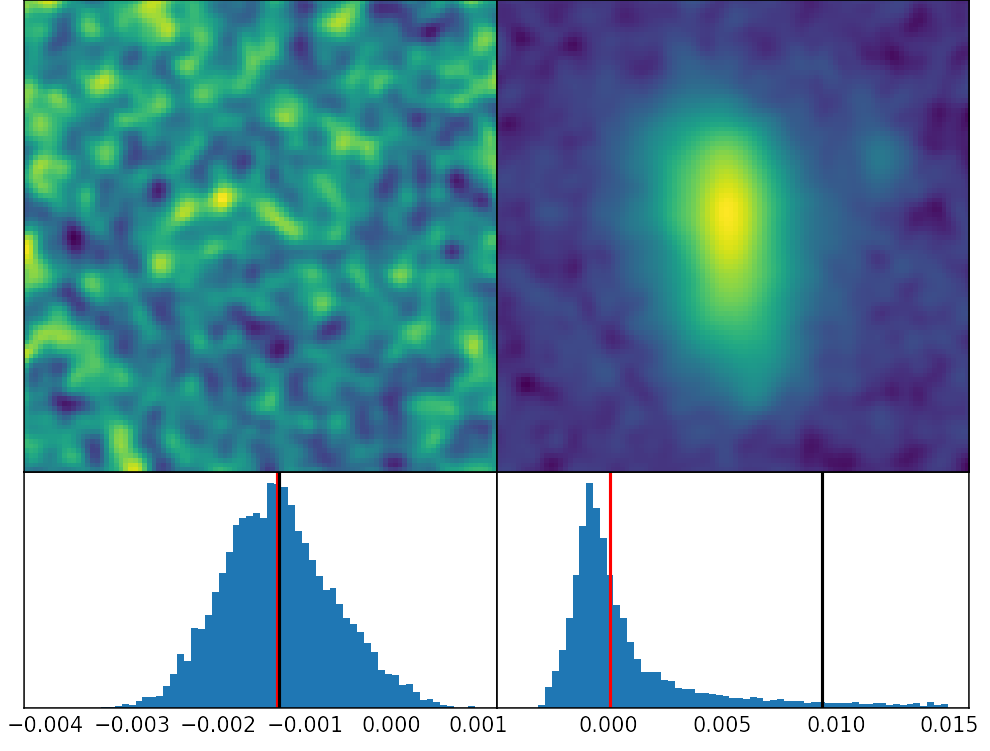}
     \caption{Presence of galaxies (in the right panel) skews the pixel value distribution to be more asymmetric and less Gaussian-like than random sky regions (left panel). \texttt{NoiseChisel} makes use of this fact to separate blank regions from regions containing astronomical signals, and this is done by measuring the deviation between median (vertical red line) and mean (vertical black line) of pixel value distributions shown in the bottom panels.}
    \label{gnuastro_skewness}
\end{figure}

Given an input image, \texttt{NoiseChisel} detects neighborly correlated signals from astronomical sources and separates them from the sky region. To achieve this signal-sky separation, \texttt{NoiseChisel} first breaks the field of view into smaller tiles (with user-definable size) and measures the distribution of pixel values in each tile. In Fig.\ref{gnuastro_skewness}, I present the pixel value distributions for two example tiles, where one is relatively blank (left) and the other contains a bright galaxy (right). It could be seen that the pixel value distribution in the blank region is more Gaussian-like and more symmetric, whereas the presence of a galaxy skews the pixel value distribution to be asymmetric. A good measure for this asymmetry in the pixel value distribution is the difference between the median and mean values. In Fig.\ref{gnuastro_skewness}, I mark the median of each distribution as a vertical red line and the mean as a vertical black line. Clearly, median agrees well with mean for the distribution of blank region in the left panel, whereas they are strongly deviated owing to the presence of a galaxy in the right panel. By requiring the difference between the mean and median of the pixel value distribution to be small enough (by default, 0.01 quantile), \texttt{NoiseChisel} is then able to separate blank (i.e., sky background) tiles from tiles containing astronomical signals. 

\begin{figure}
     \centering
     \includegraphics[width=0.9\textwidth]{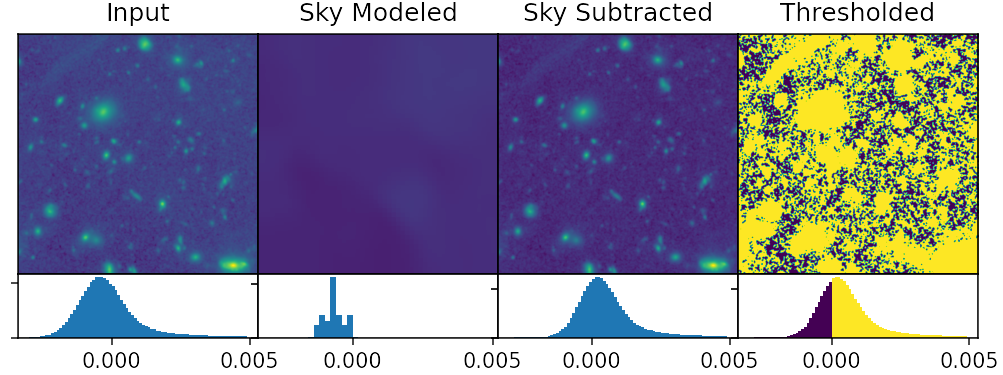}
     \caption{Given an input image (first panel from left), \texttt{NoiseChisel} models the smooth background sky (second panel) by interpolating the distribution of the blank tiles identified. This modeled sky will be removed from the input image to produce the sky-subtracted image (third panel), which will subsequently be used for source detection. Removal of smooth background could be seen to bring pixel value distribution (bottom panels) to peak closer to zero, signaling the recovery of over-subtracted signals in the default data reduction pipelines. In claiming primary detection regions, \texttt{NoiseChisel} notices that astronomical signals are still neighborly correlated even pixel-wise below sky level (often well reflected by the median in pixel values). Consequently, the thresholding applied in the rightmost panel could be seen to fall below the median of the pixel value distribution.}
    \label{check_qthresh}
\end{figure}

The distribution of the identified blank tiles will be further interpolated to cover the entire image, leading to a modeled smooth background/sky map\footnote{Key parameters that control the modeling of smooth background include (1) \texttt{--tilesize}, tile size for measuring pixel value distribution and hence determine the splitting between sky and detection regions; (2)\texttt{--meanmedqdiff}, the maximum quantile difference between mean and median of pixel value distribution to claim a tile as blank; and (3)\texttt{--interpnumngb}, number of neighboring blank tiles to compare with for each sky tile identified (tile significantly deviating from its neighbors will be rejected prior to interpolation), this reflects the anticipated smoothness of underlying background. For more details and how to optimize these parameters, please refer to \texttt{GNUastro} manual.  }. This modeled sky map will subsequently be subtracted from the input images, the effect of which could be seen in Fig.\ref{check_qthresh} for an example input image shown in the leftmost panel. For each panel of Fig.\ref{check_qthresh}, I present on the bottom the respective pixel value distribution. Interestingly, the pixel value distribution was found to peak below zero for the presented input image, reflecting that the smooth background was over-subtracted by the default data reduction pipeline. The modeled sky map from \texttt{NoiseChisel} is provided in the second panel from left, and it could be seen to capture well the underlying intensity variation of the input image. Correspondingly, after removing the modeled sky background, the residual image shown in the third panel is more uniform in regions without visible galaxies. For the sky-subtracted image, I also notice the pixel value distribution now peaks closer to zero, signaling the 'recovery' of some over-subtracted lights.




For the sky-subtracted image, the peak of distribution (often close to the median) is instead slightly above zero. This observation suggests that the field is rather crowded, i.e., there exist diffuse astronomical signals contributing dominantly to the 'sky level'. Or, in other words, neighborly correlated signals exist below the "sky level" (often reflected by the median of pixel value distributions). To avoid removing these diffuse lights from detection, \texttt{NoiseChisel} then defines the preliminary detection regions with a thresholding level (by default, at 0.3 quantile) below the median of the pixel value distribution. For the shown example in Fig.\ref{check_qthresh}, regions above the quantile threshold of 0.3 (indicated in the bottom panel as the boundary between yellow and purple histograms) are colored in yellow in the Thresholded image. By a direct comparison with the sky-subtracted image, the thresholded yellow regions could be seen to successfully capture diffuse features around astronomical objects. As a matter of fact, with this detection philosophy, \texttt{NoiseChisel} was shown able to detect faint galaxies visually non-existent (i.e., buried under sky random noise) in shallow observations, where their existence could be justified in deeper observations \citep{Akhlaghi2015}!



\begin{figure}
     \centering
     \includegraphics[width=0.9\textwidth]{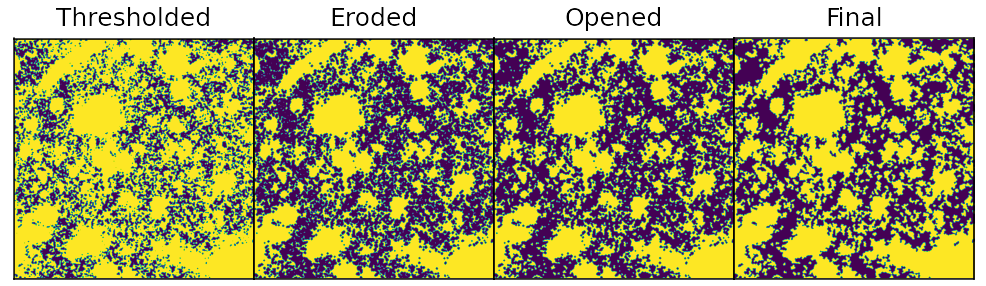}
     \caption{Given a thresholded image (leftmost panel, same as in Fig.\ref{check_qthresh}), \texttt{NoiseChisel} consequently applies \texttt{erosion} and \texttt{opening} to effectively carve out sky random noises and isolate individual detection regions, giving rise to the second and third panels from the left, respectively. However, some of the detected regions may still be the result of strong random noise fluctuations. To remove these regions, \texttt{NoiseChisel} repeats the step of \texttt{erosion} and \texttt{opening} on non-thresholded sky regions to measure the SNR of sky random peaks. Real galaxies are then defined to be the detected regions with a signal-to-noise ratio higher than most of the sky random peaks, a condition seen to remove most of the small isolated detection 'islands', giving rise to the final detection map in the last panel.}
    \label{check_qthresh_det}
\end{figure}

In the Thresholded image of Fig.\ref{check_qthresh}, the preliminary detection regions are seen as all connected up owing to random noise fluctuations. To mitigate this issue and 'isolate' individual detection regions, \texttt{NoiseChisel} invokes two operations to effectively \textit{chisel out} the noise connecting (also at the outskirts of) the detected regions. The first operation is called \texttt{erosion}, which removes pixels immediately neighboring non-thresholded sky regions. The effect of applying twice the erosion could be seen from the eroded image of Fig.\ref{check_qthresh_det}, where individual detected regions could be seen better separated. In the meantime, some small yellow "islands" persist. These small detected regions could either be compact isolated galaxies or be from strong local random noise, which are generally smaller than the former. 

To remove the smaller 'islands' corresponding to random sky noise, a simple approach is to apply additional erosion, but at a price of simultaneously removing signal pixels at the outskirts of detected regions. As a resolution to this dilemma, \texttt{NoiseChisel} designed a second operation, \texttt{opening}, to first remove pixels neighboring sky regions and subsequently recover the signal pixels by dilating the detected region. With this operation, small isolated "islands" will be removed before dilating the detected regions, and hence will not be recovered. This fact is illustrated in the opened image of Fig.\ref{check_qthresh_det}, in which small yellow 'islands' are seen to be less abundant. 

Among the remaining 'islands', however, some may still be false detections from strong random noise peaks. To further remove those, \texttt{NoiseChisel} runs erosion and opening operations separately in the non-thresholded sky region to determine the typical strength (i.e., signal-to-noise ratio, SNR) of strong noise fluctuations. With reference to these noise peaks, a detected region is considered 'real' only if its SNR is sufficiently above the SNRs of most (by default, 99\%) noise peaks. In the final image of Fig.\ref{check_qthresh_det}, such a selection strategy could be seen to nicely remove most of the very small 'islands' that were persistent in the opened image. From the shown example, it could also be seen that detected regions that passed the SNR selection are also (slightly) inflated prior to \texttt{NoiseChisel} output. This applied dilation is to recover some of the previously removed signal pixels, and also, to fill up any small blank regions inside the detection region. 

The final output from \texttt{NoiseChisel} then includes the inflated detection regions, the modeled smooth background, and the smooth sky-subtracted input image. For my future catalog construction, there are correspondingly two applications of  \texttt{NoiseChisel}. First, it will be applied to detect astronomical signals\footnote{Besides the parameters for smooth background modeling, key parameters for astronomical signal detection include (1) \texttt{--qthresh}, quantile level for thresholding in defining preliminary detection region; (2) \texttt{--erode} and \texttt{--opening}, number of erosion and opening to be applied respectively; (3) \texttt{--snquant}, minimum SNR (in quantile of SNR distribution for random noise in non-detection/sky region) needed to claim a detected region as real; (4) \texttt{--detgrowquant}, faintest brightness level till which the detected regions could be inflated (in quantile of full pixel value distribution).} on the detection image, which is often an weighted stack of different filters. In addition, \texttt{NoiseChisel} will also be applied on individual filters to model and remove the smooth background, so that I have accurate brightness measurements\footnote{Technically, this may not be necessary for most galaxies when performing local aperture photometries, so long as galaxies are sufficiently small compared to the variation scale of the smooth background. For studies focusing on large and extended features (or diffuse galaxies that are relevant for my analysis), however, properly removing the smooth background would be crucial. }. In both cases, I would fine-tune the input parameters of \texttt{NoiseChisel} to give a sensible background modeled (as demonstrated by Fig.\ref{check_qthresh}). For the first application, I would also check (with the aid of residual image) whether the final detected regions include all astronomical sources visible. 


\section{Segment}
\label{subsec_segment}

While \texttt{NoiseChisel} was designed to (hence be able to) detect correlated signals and thus uncover faint galaxies/features (as shown by \cite{Akhlaghi2015}), there is one unwanted inconvenience: i.e., the final detected regions are not the segmentation for individual galaxies, and hence cannot be used for photometry measurements directly. This fact could be seen by comparing the Final image of Fig.\ref{check_qthresh_det} with the Input image in Fig.\ref{check_qthresh}, where neighboring bright galaxies are seen to fall on the same patch of detected region. To separate these individual galaxies, \texttt{GNUastro} introduces \texttt{Segment} to detect galaxies as individual local maxima.

\texttt{Segment} operates as follows. Given a detected region, \texttt{Segment} would start from the highest pixel value (the corresponding brightest pixel will be given a label) and iterate down towards the lowest value. For each pixel value it iterates to, if the pixel neighbors a single labeled object, this pixel is also assigned to the same labeled object. Or, if the iterated pixel does not neighbor any of the previously labeled objects, it corresponds to a new local maximum and would be labeled as a new object. Finally, if the iterated pixel neighbors multiple claimed objects, it is considered the boundary between those objects. In \texttt{Segment}, labeled regions corresponding to individual local maxima are called \textit{clumps}, and boundaries separating different clumps are called \textit{rivers}. 

\begin{figure}
     \centering
     \includegraphics[width=0.9\textwidth]{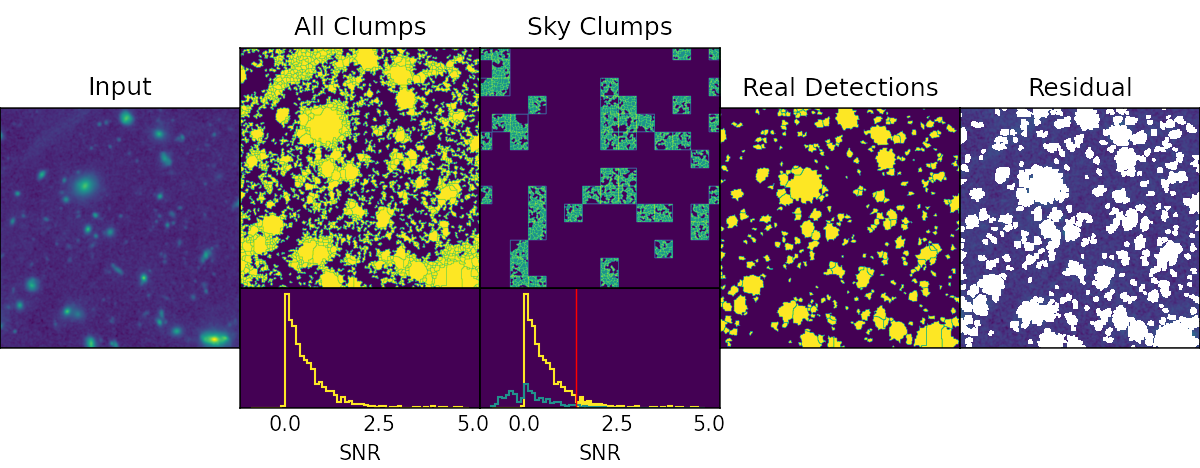}
     \caption{After obtaining the final detected regions with \texttt{NoiseChisel} given an input image (shown in the leftmost panel), \texttt{Segment} separates individual galaxies within the detected region with a watershed method. The result of which is shown in the second panel from left, with distribution of SNRs of these smaller regions (i.e., \textit{clumps}) shown in the bottom panel. Akin to removing fake detections from strong sky random peaks in \texttt{NoiseChisel}, \texttt{Segment} separately measures the SNRs of sky clumps within non-detected sky regions, and conditions real galaxies to be clumps in the detected regions with higher SNRs than most of the sky clumps. This is reflected in the bottom figure of the third panel from left, where the red vertical line marks the SNR threshold for claiming real detections (corresponding to shaded yellow histograms). The result of this SNR cut is shown in the 'Real Detections' image, and the remaining clumps (subsequently dilated) could be seen to nicely capture all visible galaxies in the input image by visually inspecting the residual image shown in the last panel. }
    \label{check_segment}
\end{figure}

In Fig.\ref{check_segment}, I show the clumps thus obtained in the All Clumps panel given the final detected regions from Fig.\ref{check_qthresh}. Here, each clump is shown as a yellow region, with rivers separating different clumps shown as green contours. By comparison with the Input image, visible galaxies are now well separated, and they stand out as clumps of large sizes. Note that there are also many smaller clumps filling up the detected regions between different galaxies. These clumps come from random fluctuations and hence need to be removed.

To remove these random fluctuations, \texttt{Segment} invokes a similar philosophy as in \texttt{NoiseChisel} and compares the SNR\footnote{The definition of SNR for clumps could be found in \cite{Akhlaghi2019}, and SNRs are measured relative to the rivers enclosing each clump. } of clumps with strong noise peaks in the non-detected sky region. In the Sky Clumps image, I present clumps identified in the sky regions. Notice the images are again split into different tiles, and only sky tiles (i.e., those with median and mean of pixel value distribution close enough) are used to define sky clumps. In the bottom panel of the All Clumps and Sky Clumps image, I present the SNR distributions in yellow for clumps in the detection region, and in green for clumps in the sky region. With reference to the green histogram from sky clumps, \texttt{Segment} considers yellow clumps in the detected regions with SNR above 95$\%$ (by default, user specifiable with parameter \texttt{--snquant}) of the green histogram as real detection, and rejects the lower SNR ones. In the Real Detection image, high SNR clumps thus selected are preserved as yellow regions. As with \texttt{NoiseChisel}, these real detections are also slightly inflated by a small amount prior to outputting (as the final segmentation of individual galaxies). In the Residual image, I masked out the final galaxy segmentations thus obtained from the Input image, and it could be seen that there are no visible galaxies left undetected.

Later in Sec.\ref{sec_pro_cons_gnuastro}, the watershed local maximum finding strategy of \texttt{Segment} will be shown effective in separating faint galaxies from their bright neighbors. There is also one intrinsic issue with this strategy, however, that is galaxies with multiple visible features (introducing multiple local maxima) may be broken into smaller parts. An example of this issue would be the galaxy sitting at the lower right corner of the Input image in Fig.\ref{check_segment}, where the specified galaxy is seen split into three separate parts. Similarly, strong random fluctuations may also result in unwanted splitting up of galaxies, a problem more pronounced for intrinsically faint and diffuse galaxies. The unwanted splitting of galaxies, however, could be mitigated to some extent if the image is smoothed (removing finer brightness variations) prior to clump identification. For this reason, Gaussian smoothing is applied by \texttt{Segment} by default, often with an over-sampled kernel of size 1.5 pixels. Naturally, to avoid splitting up bright galaxies (e.g., the one shown in Fig.\ref{check_segment}), a wider kernel is needed. But with the same kernel, \texttt{Segment} may also lose power in separating small faint galaxies from their bright neighbors. As I further discuss in Sec.\ref{catalog_for_M0416}, for my catalog construction, I therefore fine-tuned the smoothing kernel size for \texttt{Segment} to best avoid galaxies being split into smaller pieces. In the meantime, my adopted smoothing kernel was not too wide, such that I would lose the power of separating faint galaxies from their brighter neighbors.


\section{mkcatalog}
\label{sec_mkcatalog}
Just like other source detection algorithms, GNUastro allows for dual-mode aperture photometry to be extracted for each galaxy, and this is accomplished using the third subprogram \texttt{mkcatalog}. Instead of relying on traditional elliptical or circular apertures, however, \texttt{mkcatalog} utilizes the pixels (i.e., rivers) that enclose the non-regularly shaped segmentation (i.e., clumps) of each galaxy to model the local background. Specifically, the local background is modelled as the mean level of the river pixels, and hence, the brightness of each galaxy is measured to be the sum of respective pixel values, further subtracting off the product of the clump area with the mean sky level from the river. The photometries obtained through this method will be compared with example measurements using conventional methods in Section \ref{sec_pro_cons_gnuastro}, and results will be shown as largely consistent with existing measurements.

Besides the segmentation map for clumps, \texttt{Segment} actually also outputs the segmentation map for \textit{objects}, where \textit{objects}\footnote{The concept of \textit{object} was introduced to specifically mitigate the issue that galaxies may still be split up into smaller pieces after smoothing. } are obtained by merging neighboring clumps together if clumps are not strongly separated (i.e., sufficiently brighter than the river separating them). And by default, \texttt{mkcatalog} only constructs a catalog for \textit{objects}, hence the command \texttt{--clumpscat} will be needed to output a catalog separately based on clump segmentations. For my future analysis, I will not consider the labeled \textit{objects} as, first, there are no rivers separating neighboring \textit{objects} on the \textit{objects} segmentation map, and hence local backgrounds could not be removed for brightness measurements. Moreover, some faint galaxies successfully identified as individual clumps may also be merged back with their brighter neighbors.

To construct the photometry catalog for clumps, the following commands are useful. Brightness measurements could be initiated with command \texttt{--brightness}, and \texttt{--brightnesserr} for uncertainties. For dual-mode photometry, the value files for measuring the brightnesses of clumps also need to be separately specified with \texttt{--valuesfile}. In practice, these value files are just the smooth background-subtracted image of individual filters. The location of the detected galaxies could either be reported as the position of the brightest pixel (initiated with \texttt{--xpeak} and \texttt{--ypeak}), or as the flux-weighted center (initiated with \texttt{--x} and \texttt{--y}). For my catalog construction, I will adopt the former to be better in line with the detection philosophy of \texttt{Segment}. \texttt{mkcatalog} also incorporates commands to measure shape parameters of elliptical morphology, for instance \texttt{--semimajor}, \texttt{--semiminor}, and \texttt{--positionangle}. Half-light radii of detected galaxies may also be estimated with the command \texttt{--halfsumradius}, corresponding to the circularized radius of the area containing half the total brightness. For more details on other measurements allowed, and examples of invoking \texttt{mkcatalog}, please refer to the \texttt{GNUastro} manual. 


\section{Pros and Cons in using GNUastro}
\label{sec_pro_cons_gnuastro}

\begin{figure}
     \centering
     \includegraphics[width=0.9\textwidth]{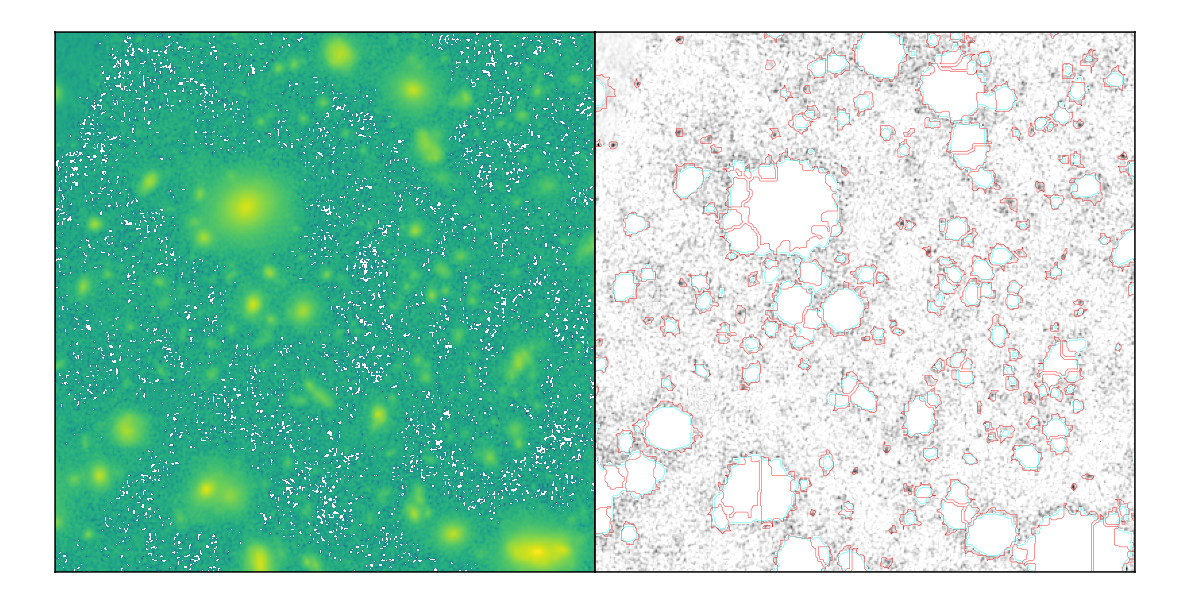}
     \caption{\textit{Left}: Detection image used by \cite{Shipley2018} for generating S18 catalogs. \textit{Right}: Comparison of segmentation outputs from \texttt{GNUastro} (red contours) and \texttt{SExtractor} (cyan contours enclosed blank regions) on the residual image given the detection image shown on the left. \texttt{GNUastro} is seen able to detect many small isolated galaxies (likely high-z galaxies) missed by \texttt{SExtractor}. But \texttt{GNUastro} also misses a few regions detected by \texttt{SExtractor} owing to low SNRs. For brighter galaxies with individual visible features and faint galaxies suffering from strong random noise, the galaxy may be split into smaller pieces, with the river separating neighboring clumps running across the relatively bright region. In such a case, the resulting low SNRs of the corresponding clumps may lead to unwanted rejection of the galaxies (this may also explain some of the \texttt{GNUastro} undetected regions), but as previously commented in Sec.\ref{subsec_segment}, such an issue can be mitigated by convolving the input image prior to clump-finding. }
    \label{compare_SExtractor}
\end{figure}

To better realize the strengths and disadvantages of cataloging with \texttt{GNUastro}, in this section, I compare its galaxy detection method with \texttt{Source Extractor}, the conventional tool for galaxy detection. For consistency, I will use the same detection image, which was provided by \cite{Shipley2018} for the massive galaxy cluster M0416. This detection image was obtained by stacking F814W to F160W band images from HFF. To reveal the fainter galaxies, \cite{Shipley2018} also removed bright cluster galaxies carefully. In the left panel of Fig.\ref{compare_SExtractor}, I present this detection image over the same FOV as previously considered in Fig.\ref{check_qthresh}, and I will restrict to this shown cutout for further comparison.

In the panel of Fig.\ref{compare_SExtractor}, I then compare the segmentation output from \texttt{SExtractor} provided by \cite{Shipley2018} with the segmentation output from \texttt{GNUastro}. Here, I masked out the detected region by \cite{Shipley2018} as regions enclosed by cyan contours, and indicate \texttt{GNUastro} detected objects (clumps) as red contour-enclosed regions. First of all, I notice there are a few masked-out regions without \texttt{GNUastro} detection. These tend to be faint features/galaxies, hence the corresponding clump regions were removed owing to their low SNR. As previously commented, galaxies with sub-features and faint galaxies suffering from strong local random fluctuations might be split into smaller pieces. In such cases, the river separating the neighboring clumps may run across the (relatively) bright regions of the galaxy and lead to a low SNR (or unreliable photometries) for the clumps. This may also explain some of the \texttt{GNUastro} undetected regions in comparison with \texttt{SExtractor} in Fig.\ref{compare_SExtractor}.

\begin{figure}
     \centering
     \includegraphics[width=0.9\textwidth]{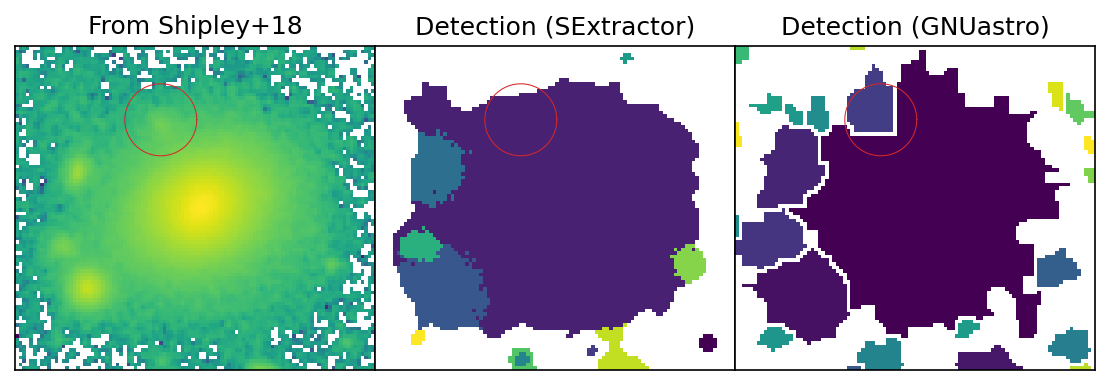}
     \caption{Illustration of \texttt{GNUastro}'s ability in deblending neighboring sources. In the shown example here, the faint source indicated by red circle is nicely captured by the program whereas it was non-detected by \texttt{SExtractor}.}
    \label{compare_deblending}
\end{figure}

On the other hand, \texttt{GNUastro} was able to detect many small galaxies missed by \texttt{SExtractor}. These are isolated red-contour enclosed objects visible in the residual image, and their compactness suggests they may very well be high-$z$ galaxies of particular relevance to my future analysis. Furthermore, by zooming in to a region around a very bright galaxy in Fig.\ref{compare_deblending}, it could be seen that \texttt{GNUastro} has successfully deblended the red circle enclosed faint galaxy from its bright neighbor, whereas \texttt{SExtractor} fails to do so\footnote{Another recently developed software \texttt{ProFound} works similarly to \texttt{GNUastro} at deblending neighboring sources. But this program sacrifices speed in comparison to \texttt{SExtractor} (also \texttt{GNUastro}) as commented by \cite{1802.00937}, and hence is not ideal for large FOV. }. These observations then justify that \texttt{GNUastro} is better suited for my future analysis, in particular, it can detect faint galaxies more completely than \texttt{SExtractor}. 

\begin{figure}
     \centering
     \includegraphics[width=0.9\textwidth]{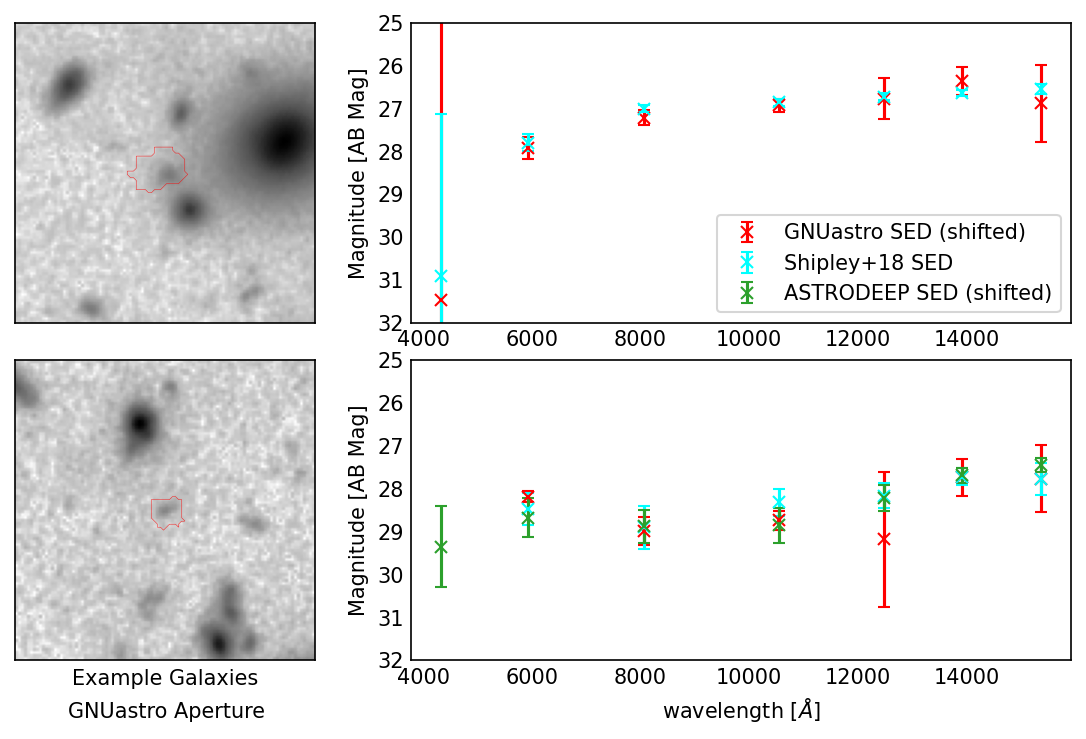}
     \caption{SEDs as measured by \texttt{GNUastro} (red) in comparison with S18 (cyan) and ASTRODEEP (green) catalogs. The top panel presents a galaxy located next to a bright neighbor, while the bottom panel features a galaxy in relative isolation. To enable a better comparison of shapes, the measurements from \texttt{GNUastro} and ASTRODEEP have been adjusted to align with the level of the S18 data. Overall, the SEDs from \texttt{GNUastro} appear to be largely consistent with existing measurements. }
    \label{compare_SED}
\end{figure}

In Fig.\ref{compare_SED}, I also compare the SEDs derived from \texttt{GNUastro} with several existing measurements based on the standard method (with circular aperture) from S18 and ASTRODEEP catalogs. As could be seen from the left panel, I consider both a galaxy in relative isolation and a galaxy with a bright neighbor. To compare only the shape of SEDs\footnote{This is to account for the fact that different segmentation sizes lead to different total brightness. Also, the shape of SEDs is often more dominating for photo-z determinations.}, I shifted the \texttt{GNUastro} and ASTRODEEP measurements to match the level as reported by S18 catalogs\footnote{Note the object in the top panel was not detected in the ASTRODEEP catalog, which also used \texttt{SExtractor} for source detection}. It could be seen that the SED measured by \texttt{GNUastro} agrees well with existing measurements for both of the considered galaxies. As the last comment, here I have deliberately selected two relatively faint galaxies; hence, the agreement in SEDs with existing measurements could naturally be expected to be better for brighter galaxies.

Overall, I conclude that \texttt{GNUastro} outperforms \texttt{SExtractor} at (1) detecting isolated faint/compact galaxies likely of relevance to high-$z$ studies, (2) separating faint galaxies from their bright neighbors. \texttt{GNUastro} is also capable of extracting sensible SEDs, even with non-regularly shaped segmentation and local aperture of width 1 (for river) in pixel units. On the other hand, galaxies could be split into smaller pieces, leading to unwanted rejection of the corresponding sources or unreliable photometries. This issue then demands the fine-tuning of \texttt{GNUastro} parameters, in particular the size of the smoothing kernel. In practice, \texttt{GNUastro} also has the disadvantage of having to separately fine-tune the smooth background modelling parameters\footnote{I have, in fact, attempted to achieve auto-optimization of the program parameters, asserting that the smooth-background subtracted images have a blank region following a Gaussian distribution. The auto-optimization of \texttt{NoiseChisel} parameters is currently achieved with a gradient descent approach, but the corresponding program is still being developed and needs further testing. } for each filter, and this is in general a time-consuming task. It should nevertheless be noted that \texttt{GNUastro} is a software still being developed, and the aforementioned inconveniences/issues may be improved in later releases.

\chapter{Large Contamination in HFF Photo-z Catalogs}
\label{chap:contamination}


\section{Spurious $z\sim4$ Excess} 
\label{HFFcontamin_Sec_Intro}


\begin{figure}
\includegraphics[width=0.97\textwidth]{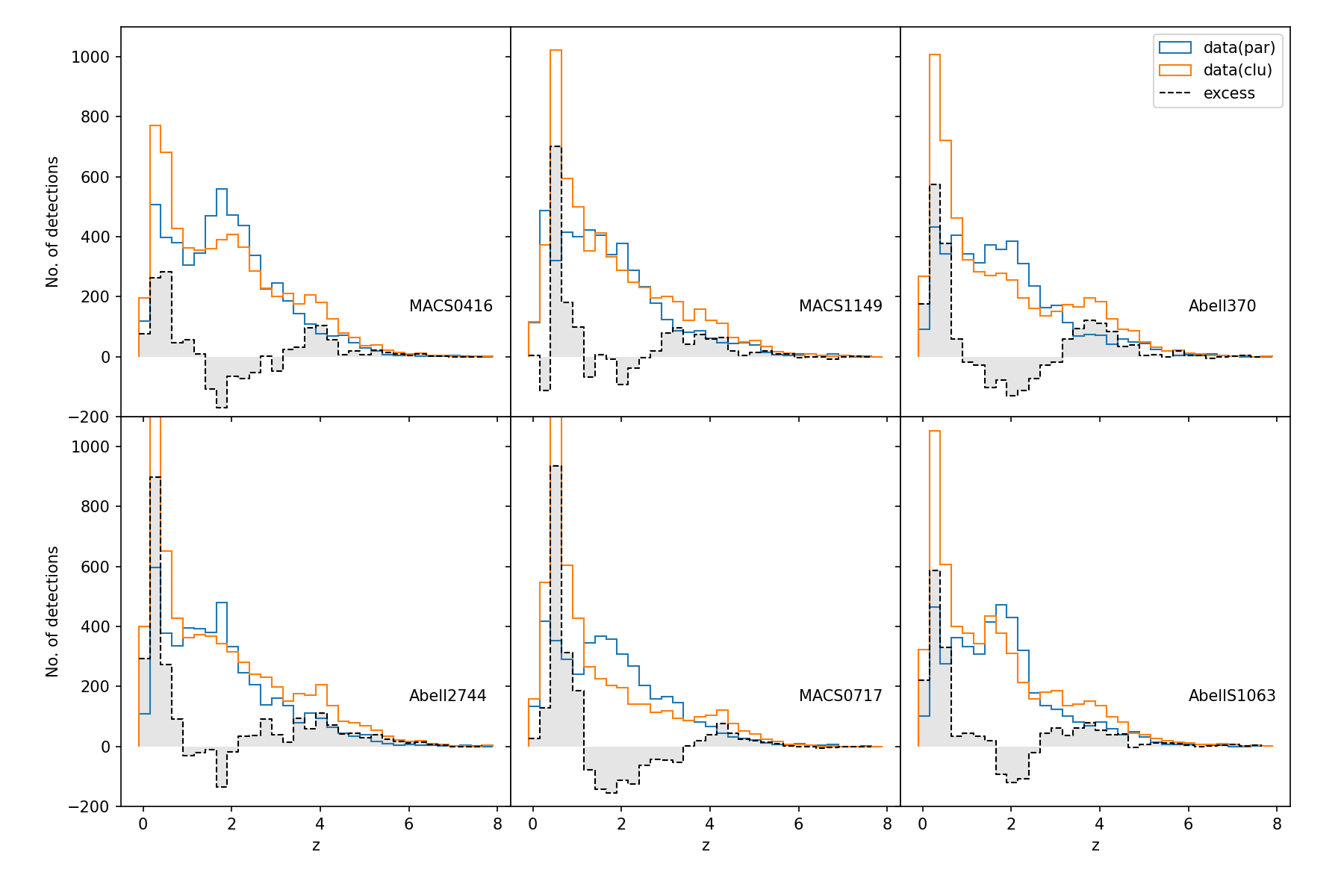}
\caption{Histogram of photometric redshifts in HFF cluster fields (orange) compared with respective parallel fields (blue) from S18 catalog. Following \cite{Shipley2018}, here I have only used sources that are non-stellar objects, with reasonable photometric redshift fitting, i.e. sources flagged with \texttt{use$\_$phot}=1. For a better visual comparison, I plot also the difference between histograms as shaded black steps. The shaded black regions for each pair of fields could be seen as composed of three parts. A part where I have excess in the cluster field at the cluster redshift, this comes from cluster members. A part with a deficit in the cluster around redshift $\sim2$, which I will show is the result of negative magnification bias over this redshift range in Sec.\ref{bias}. Finally, a second excess part above redshift $\sim$3.5, which I will argue in this paper as likely coming from the misidentification problem of low-z passive galaxies. I will also provide a quantitative estimate on the contamination fraction in Sec.\ref{bias}. }
\label{excess}
\end{figure}

\begin{figure}
\includegraphics[width=0.97\textwidth]{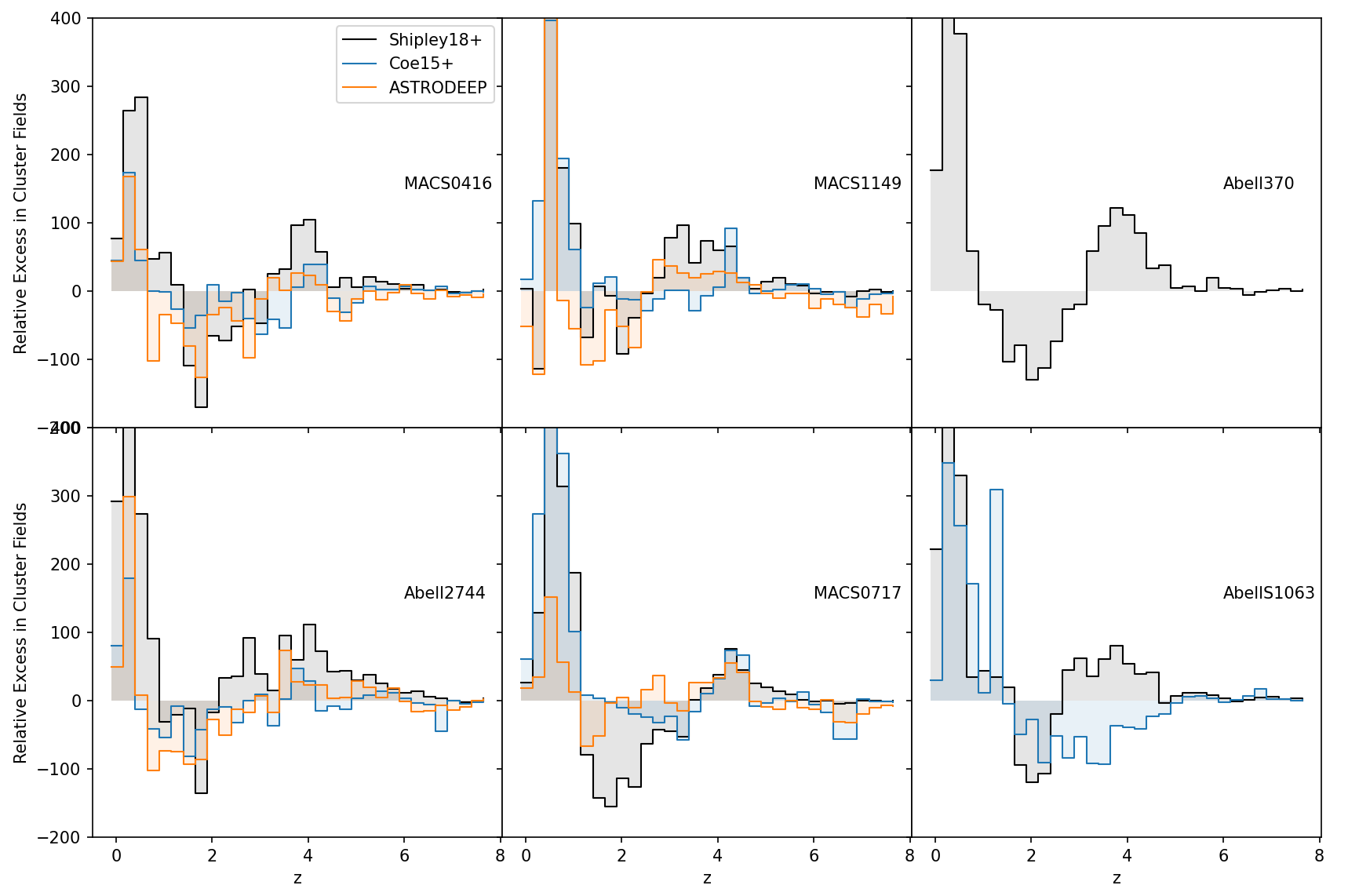}
\caption{Relative excess in the number of galaxies detected in cluster fields than in parallel fields for three separate catalogs: S18 (black histograms), ASTRODDEP (orange histograms), and C15 (blue histograms) catalogs. The excess for S18 catalogs is as obtained from Fig.\ref{excess}. As with S18 catalogs, here I have focused only on galaxies considered to have a reliable photo-z determination. This amounts to selecting \texttt{RELFLAG}=1 galaxies from the ASTRODEEP catalogs, and \textit{odds}$>$0.5 galaxies from the C15 catalogs (see Sec.\ref{sec_contamination_othercata} for more detail). The resulting distributions are largely consistent with the distribution from S18 catalogs: that there is first an excess in the cluster field at the cluster redshift, followed by a deficit in the cluster field around $z\sim 2$, and finally, another excess in the cluster field at $z\sim 4$.  }
\label{excess_allcata}
\end{figure}



In Chap.\ref{chap:intro}, I mentioned that low-z galaxies with similar observed spectral energy distributions (SEDs) as high-z galaxies may be misidentified, with some tentative evidence from the analysis of \cite{Leung2018} and also the excess seen at the faint-end of UV LF constructed by \cite{Livermore2017}. In this chapter, I further discuss this potential contamination issue. In particular, I aim to quantify how severely it contaminates the existing photometric redshift catalogs. 

In relevance to this discussion, \cite{Shipley2018} (S18) recently reported an excess in the number of $z\sim 4$ galaxies cataloged in HFF cluster fields than in accompanying parallel fields - nearby blank regions located 6 arcmin away from each cluster field. The reported excess is reproduced in Fig.\ref{excess}, which shows the number of galaxies\footnote{Here, and through out this thesis, I use only those flagged with \texttt{use$\_$phot}=1, which are non-star-like galaxies considered to have reasonable photometries and photo-z estimations.} against redshift for each cluster (orange histogram) and parallel field (blue histogram). \cite{Shipley2018} attributed this enhancement to intrinsically low-luminosity galaxies at high redshifts made detectable by the magnification power of gravitational lensing, and claimed that similar enhancements are seen in the ASTRODEEP catalogs compiled by \cite{Merlin2016}, \cite{Castellano2016}, \cite{DiCriscienzo2017} for the HFF fields. 

Following this claim, I investigate in Fig.\ref{excess_allcata} the number of galaxies detected in both cluster fields and parallel fields for ASTRODEEP catalogs (available only for four of the HFF clusters), and also for C15 catalogs (available for five of the HFF clusters). Indeed, while small, the ASTRODEEP catalog reports a relative excess in the cluster field as commented by \cite{Shipley2018}. Some small excess could also be argued for C15 catalogs, with a particularly strong excess for MACS 0717. The C15 catalog for the field of AbellS 1063 instead shows a deficit at $z\sim 4$, which I will further discuss in later sections. 

In what follows, I investigate the alternative possibility (to gravitational lensing as attributed by S18) of whether the excess seen in Fig.\ref{excess} and Fig.\ref{excess_allcata} is the result of severe contamination, e.g., by misidentified cluster members. In particular, I will examine, using magnification bias, whether the number density of galaxies observed in cluster fields is compatible with the faint end extrapolation of UV LF derived from accompanying parallel fields (free, of course, from cluster member contamination). Using this method, I will demonstrate that about half the galaxies purported to be at $z\sim4$ are in fact low-z interlopers.

The rest of this chapter will be structured as follows. In Sec.\ref{section_implications}, I first present evidence based on various diagnostics for severe contamination among galaxies purported at $z\sim 4$ in S18 catalogs. I will focus mainly on the redshift range of $3.5\leq z \leq 5.5$, chosen to best capture the excess observed in the cluster fields. To more quantitatively assess the degree of contamination, I apply the concept of magnification bias. The detailed methodology will be described in Sec.\ref{Method}. In particular, I will introduce \textit{exclusion regions} and outline how UV LF could be measured from the accompanying parallel fields. Then in Sec.\ref{bias}, I present the main results of my magnification bias analysis. I would first test my methodology over the redshift range of $1.2\leq z\leq 2.4$, which is expected to be free of contamination. I then extend to the redshift range of $3.5\leq z \leq 5.5$ and report the contamination level, with potential consequences discussed in Sec.\ref{contaminated_LF_sec}. For the interest of the broader community, I comment on prospects of mitigating interlopers in Sec.\ref{sec_mitigating}, highlighting the need for additional deep observation from JWST. Finally, for completeness, I apply the same magnification bias computation to ASTRODEEP and C15 catalogs in Sec.\ref{sec_contamination_othercata} and report the contamination level to be similarly high $\sim 50\%$. I then conclude in Sec.\ref{conclusion_section}. Detailed stellar properties of these contaminants will be investigated in more detail in Chap.\ref{chap:contamination_jwst} once JWST observations are included. Throughout my analysis, I assume a simple two-component flat $\Lambda$CDM cosmology with $H_0 = 70$ km/s/Mpc and $\Omega_\Lambda = 0.7$.

\section{Statistical Evidence}
\label{section_implications}
\subsection{color versus surface brightness}

\begin{figure}
\centering
\begin{subfigure}{.5\textwidth}
  \centering
  \includegraphics[width=.85\linewidth]{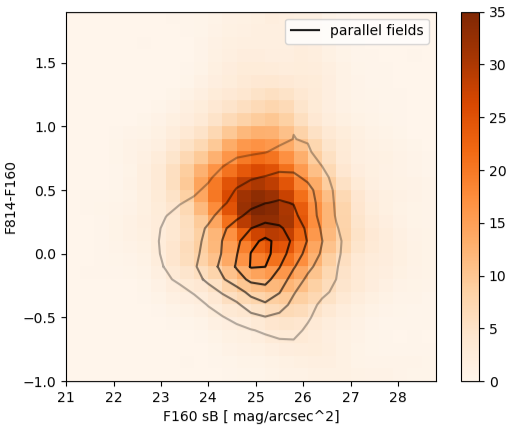}
  \caption{F814-F160 vs F160 band surface brightness.}
\end{subfigure}%
\hfill
\begin{subfigure}{.45\textwidth}
  \centering
  \includegraphics[width=.9\linewidth]{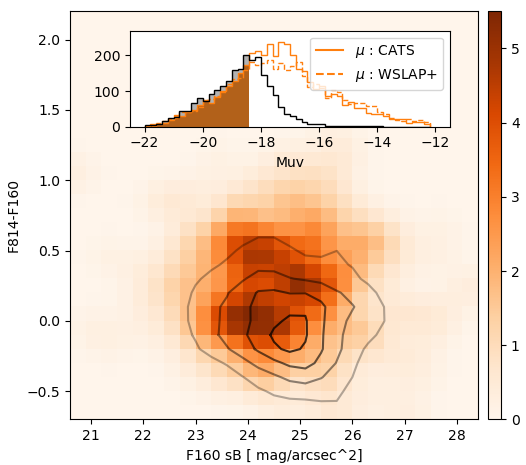}
  \caption{color vs surface brightness with $M_{uv}$ cut.}
\end{subfigure}
\caption{Comparison of intrinsic properties between $3.5\leq z_{phit}\leq5.5$ high-z candidates in the cluster and parallel fields, using only \texttt{use\_phot}=1 galaxies from S18 catalogs. Here I focus dominantly on lensing invariant observables: color F814-F160 capturing the SED redwards of the Lyman break, and F160W band surface brightness capturing their intrinsic brightness. In both figures (a) and (b), the color-surface brightness distribution is shown as a background 2d histogram with a mild degree of Gaussian smoothing for the cluster fields, and as black contours of different transparency for the parallel fields. Figure (a) used all \texttt{use\_phot}=1 galaxies, and indicates $3.5\leq z_{phit}\leq5.5$ candidates in cluster fields to be statistically redder than in parallel fields. But, as revealed by the insert of figure (b), in cluster fields I have a lot intrinsically fainter galaxies lensed to be observable: after lensing correction (dashed with WSLAP+ and solid with CATS model), the histograms of $M_{uv}$ for cluster fields (orange) peak at a lower magnitude than that for parallel fields (black). To remove those intrinsically fainter galaxies, which may also be statistically redder, I impose $M_{uv}<-18.4$ (shaded region under histograms) for figure (b). The redder population relative to parallel fields persists after the $M_{uv}$ cut, demonstrating contamination even among the UV bright sample.}
\label{color_sB_con}
\end{figure}

Color and surface brightness are two well-established lensing invariant observables, enabling a sensible comparison between the intrinsic properties of high-z galaxy candidates in lensing cluster fields and non-lensing parallel fields. In what follows, I compare the color and surface brightness of $3.5\leq z\leq 5.5$ galaxy candidates in cluster fields with more robust high-z candidates in parallel fields. Any statistical discrepancy would indicate contamination. I adopt F814W-F160W as color to compare SED redwards of the redshifted Lyman breaks (or Balmer breaks for interlopers). This reflects the stellar age of galaxies and is dependent on the amount of dust extinction. For surface brightness, as a proxy for the stellar mass of galaxies, I adopt the longest wavelength band - F160W - as passive cluster members would be most easily detected. 

Fig.\ref{color_sB_con}(a) shows the corresponding color versus surface brightness diagram. The distribution of $3.5\leq z\leq 5.5$ candidates in cluster fields is shown as a background 2d histogram (after a mild degree of Gaussian smoothing), with the distribution for parallel fields shown by black contours of different transparency. As is apparent, $3.5\leq z\leq 5.5$ candidates in the cluster fields are statistically redder than those in the parallel field at similar surface brightness. Contamination by passive cluster members can explain this, as they tend to be redder than high-z star-forming galaxies owing to older stellar populations. Intrinsically fainter High-z galaxies, on the other hand, may also be redder, either being less star-forming or more dust-extincted. Moreover, gravitational lensing could magnify these intrinsically fainter and redder high-z galaxies to be abundantly observed in the cluster fields, providing an alternative explanation for the difference observed. 


To remove the intrinsically fainter galaxies from consideration, I impose a cut on absolute magnitude. The insert of Fig.\ref{color_sB_con}(b) shows the distribution of absolute UV magnitudes ($M_{uv}$) at rest frame 1600$\AA$ for high-z candidates in cluster fields as orange histograms and in parallel fields as black histograms. The cluster field histograms used either CATS models (solid, see Sec.\ref{sec_mag_bias_equation} for more detail) or WSLAP+ models (dashed) to correct for lensing magnifications, and the resulting distributions are similar. As can be seen, over the selected redshift interval, galaxies in the cluster field peak at lower luminosities than those in the parallel field. The intrinsically fainter galaxies could be removed by imposing a cut at where the parallel field histogram peaks. The main panel of Fig.\ref{color_sB_con}(b) then shows the color-surface brightness comparison repeated for the remaining sample (shaded region under histograms). Evidently, the difference seen in Fig.\ref{color_sB_con}(a) between cluster and parallel fields persists, providing solid evidence for contamination even among this intrinsically bright sample.

\subsection{Color-magnitude diagrams}
\label{sec_cm_diagram}

\begin{figure*}
\centering
\includegraphics[width=0.97\textwidth]{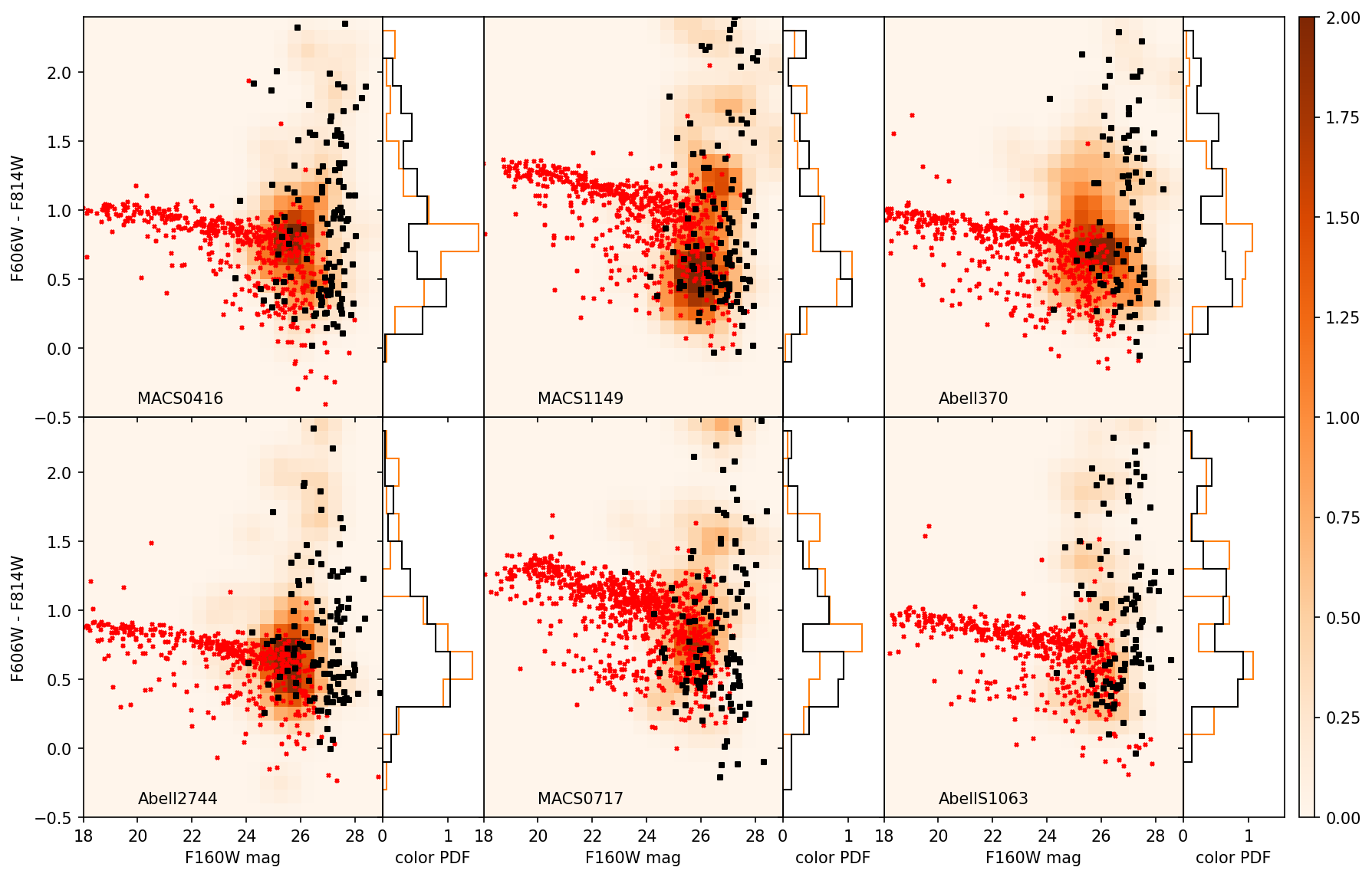}
\caption{Color-magnitude diagram for the UV-bright sample with $M_{uv}<-18.4$ to test the plausibility of misidentified cluster members as a source of contamination. Red dots are galaxies in the cluster fields with photo-z falling in $z_{clu} \pm 0.15$, and are plotted to indicate the position of the cluster Red Sequence. Black points are UV bright $3.5\leq z \leq 5.5$ galaxies in the parallel fields, and the smoothed 2d histogram background marks the distribution of 3.5-5.5 galaxies in the cluster fields. To the right of each figure, I also compare the color distribution in the cluster fields (orange histogram) with that in the parallel fields (black histogram). I note there is a strong concentration in the cluster field at the faint end extension of the cluster Red Sequence, demonstrating the contaminants could indeed come from misidentified cluster members. Galaxies in the parallel fields are seen to have a similar distribution in F606-F814. This simply reflects the cause of the misidentification problem, that redshifted Lyman breaks of high-z galaxies fall in the same bands as Balmer breaks of the cluster members and are of similar strength. }
\label{cm_high_z}
\end{figure*}

If contaminants hinted by Fig.\ref{color_sB_con}(b) are misidentified cluster members, I anticipate them to strongly concentrate towards the cluster's red sequence. Fig.\ref{cm_high_z} tests for this concentration by plotting F160W band magnitudes against the color straddling the Balmer break of cluster members - F606W-F814W. For each galaxy cluster, the red sequence and its faint end extension are indicated by $z_{clu} - 0.15 \leq z_{phot} \leq z_{clu} + 0.15$ galaxies\footnote{This range is chosen to capture the complete red sequence while avoiding features away from the red sequence.} shown as red dots. The distribution of $M_{uv}<-18.4$ $3.5 \leq z_{phot} \leq 5.5$ candidates in cluster fields is presented as the orange background, with their F606W-F814W color distributions also shown on the right as orange histograms. 

As is apparent, high-z candidates in cluster fields display prominent concentration towards the respective red sequence at the faint end, suggesting they may indeed be misidentified cluster members. High-z candidates in parallel fields, shown as black data points, are seen to span a similar range in color (see also the black histograms on the right). This reflects the cause of the misidentification problem - that the Balmer breaks of cluster members could be easily confused as the Lyman breaks from $3.5\leq z\leq 5.5$ with a comparable strength. Nevertheless, high-z candidates in parallel fields are seen offset from the cluster red sequence concentration, as well as high-z candidates in cluster fields, magnitude-wise. This further motivates the possibility of contamination in cluster fields.

\subsection{Radial distribution}
\label{sec_radial_distribution}

\begin{figure*}
\centering
\includegraphics[width=0.97\textwidth]{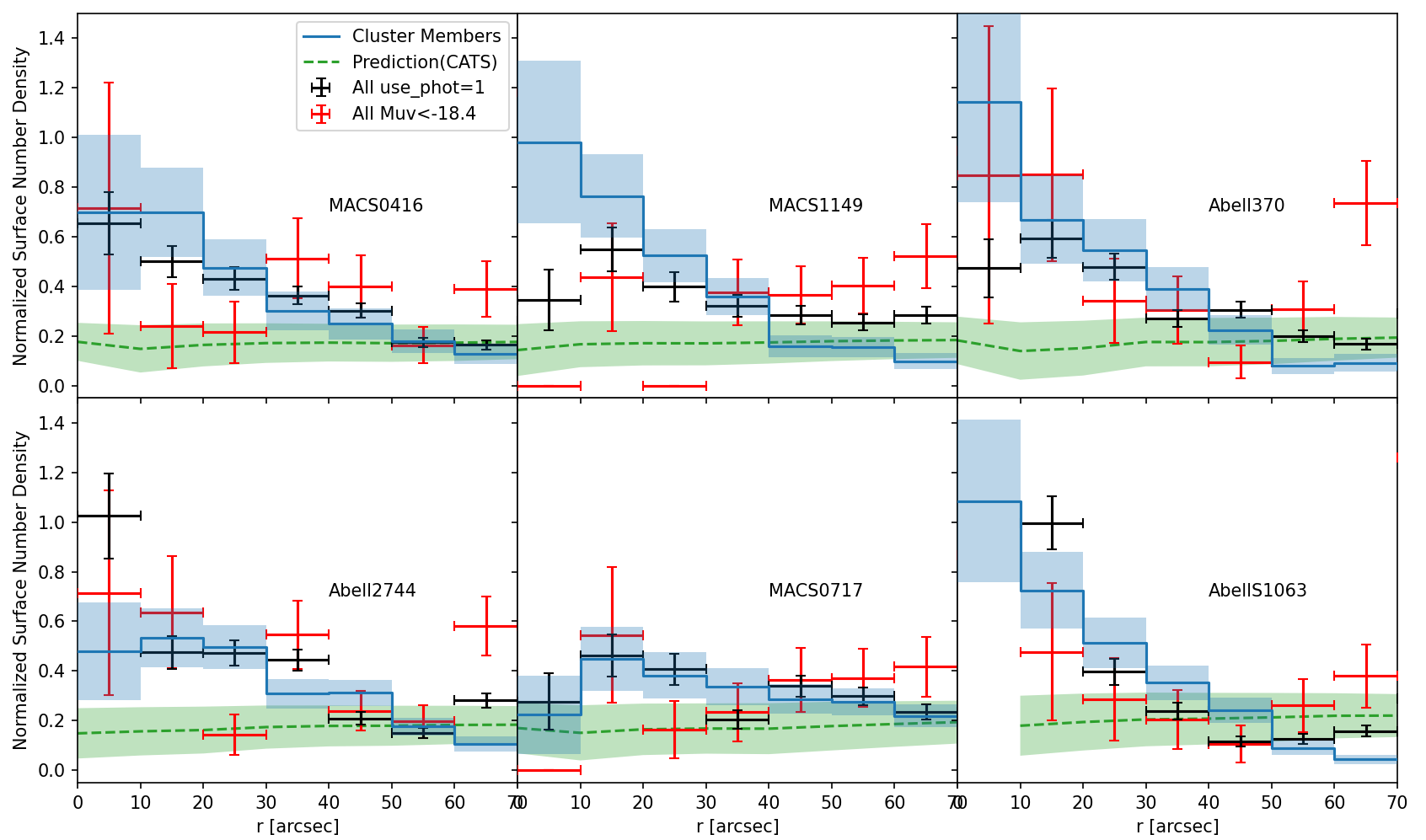}
\caption{Normalized surface Number density in different radial bins over $3.5\leq z\leq 5.5$ (black data points for all \texttt{use\_phot}=1, red data points for only UV-bright galaxies with $M_{uv}<-18.4$) compared with that for cluster members (blue step). Here, by cluster members, I refer to spectroscopically confirmed galaxies with $z_{clus} - 5\sigma_z \leq z_{spec} \leq z_{clus} + 5\sigma_z$, where $\sigma_z$ is the standard deviation of the cluster redshift peak in the spectroscopic redshift histograms. $r$ is the radial distance in arc-second to the FOV center, which I adopted from \cite{Lotz2017}. The radial distributions for full \texttt{use\_phot}=1 sample are seen to follow cluster members, whereas the same clustering trend is less obvious for $M_{uv}<-18.4$ sample owing to poorer statistics. By contrast, the predicted magnification bias, as shown in the green shaded region based on the CATS model, is seen decreasing with radius. Moreover, surface number density seems to increase for the $M_{uv}<-18.4$ sample beyond 40 arcsec. I demonstrate in later sections that this reflects the negative bias expected over this redshift range, suggesting the $M_{uv}<-18.4$ sample suffers less from contamination by misidentified cluster members. The latter point is also consistent with my observation from Fig.\ref{color_sB_con}(b) that the contamination peak is seen as less dominant after the $M_{uv}$ cut. } 
\label{cluster_member_distribution}
\end{figure*}

If contaminants are misidentified cluster members, I also anticipate them to display a radially concentrated distribution about the respective cluster center. To make this check, I computed the surface number density of galaxies in 10-arcsecond wide concentric annulus bins, with the field of view (FOV) centers adopted from the \citep{Lotz2017} paper. 

Fig.\ref{cluster_member_distribution} plots the radial distribution (in surface number density) of all $3.5 \leq z \leq 5.5$ candidates in the cluster fields as black data points, and for $M_{uv} < -18.4$ galaxies only as red data points. For an easier visual comparison, the data points are normalized by the respective cumulative sum over different radial bins. The radial distributions for cluster members are presented as blue steps, where I define cluster members as galaxies with spectroscopic redshifts in the range $[z_{clus} - 5\sigma_z, z_{clus} + 5\sigma_z]$, with $\sigma_z$ being the standard deviation of the cluster redshift peak in the spectroscopic redshift histograms.

Notably, the measured densities using all $3.5 \leq z \leq 5.5$ candidates follow a similar radial distribution as the cluster members, adding to the evidence (Sec.\ref{sec_cm_diagram}) that most high-z candidates are likely misidentified cluster members. The same agreement with cluster members is also suggested by $M_{uv} < -18.4$ galaxies, although with a large scatter due to poorer statistics. 

By contrast, gravitationally lensed $3.5\leq z\leq 5.5$ galaxies are expected to be less populated in the inner region than in the outer region, as demonstrated by my magnification bias prediction using the CATS model (green dashed line, more detail in Sec.\ref{sec_mag_bias_equation}). As I will show in Sec.\ref{bias}, this is the result of a slight negative bias observed over this redshift range.

\section{How to Estimate Contamination Level}
\label{Method}

To quantitatively assess the degree of contamination, I would employ magnification bias and predict the surface number density of gravitationally lensed high-z galaxies. By doing so, I identify any excess population observed as contamination from lower redshifts.

In the remainder of this section, I address how I determine the detection thresholds and galactic luminosity functions needed for the magnification bias analysis. I start in Sec.\ref{sec_data_complete} by imposing spatial uniformity in detection thresholds, a desired property for the best implementation of magnification bias. Subsequently, I determined the data completeness thresholds common to both cluster and parallel fields for the S18 catalogs. Then in \ref{sec_counts_construct}, I measure the UV LF from parallel fields, which naturally suffer less from contamination by misidentified cluster members. The detailed methodology is presented in Sec.\ref{sec_mag_bias_equation}. Finally, in Sec.\ref{subsec_uv_bright_sample}, I comment on the data-complete sample of UV-bright galaxies data (spanning a wide redshift range) needed for analysis. 

\begin{figure}
\centering
\includegraphics[width=0.8\textwidth]{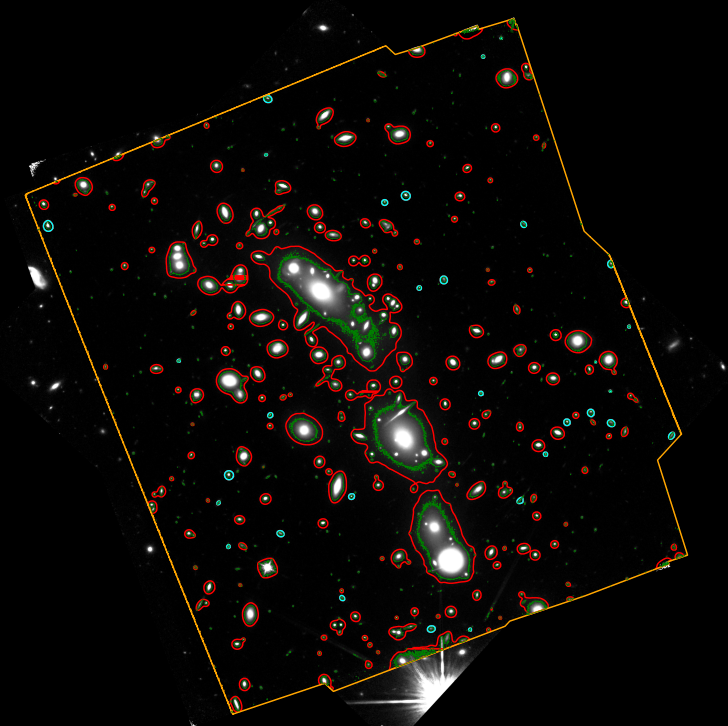}
\caption{Exclusion and common region for cluster field of M0416. To attain a more uniform detection threshold, I follow \cite{Leung2018} and mask out bright regions above $0.005$e/s on smoothed F160 images from analysis. These regions are enclosed by red contours. Cyan contours enclose bright regions above $0.005$e/s but contain a single galaxy candidate with redshift above 1.2 (as per S18 catalogs); such regions are preserved for accurately measuring galaxy number counts at the bright end. To best exploit the HFF data, I further restrict to the common HFF coverage region indicated by the orange contours. The background image is from the HFF F160W band. }
\label{exclusion_region}
\end{figure}

\begin{figure*}
\centering
\includegraphics[width=0.95\textwidth]{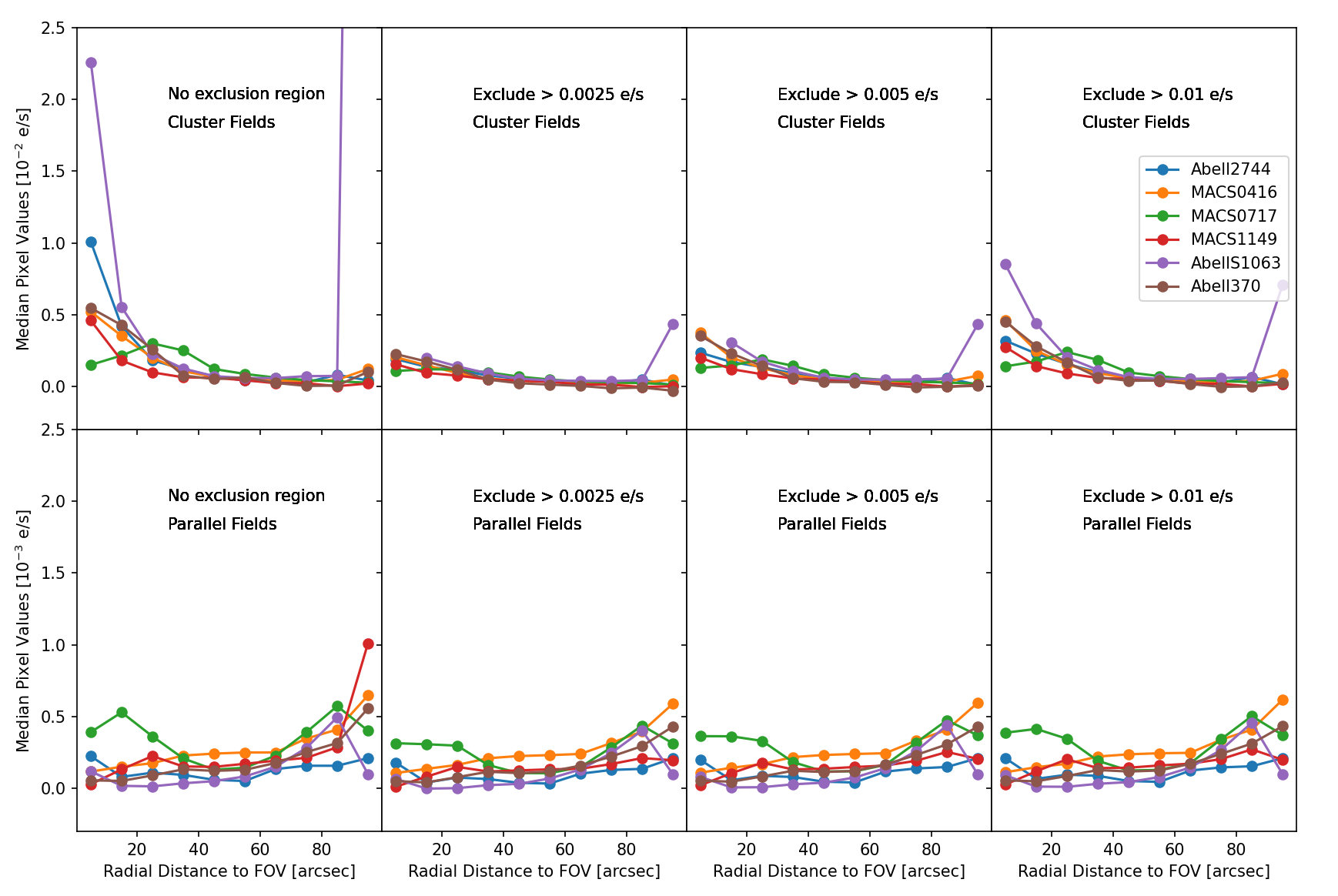}
\caption{Comparing different choices to define bright regions.}
\label{compare_exclu_definition}
\end{figure*}

\begin{figure}
\centering
\includegraphics[width=0.8\textwidth]{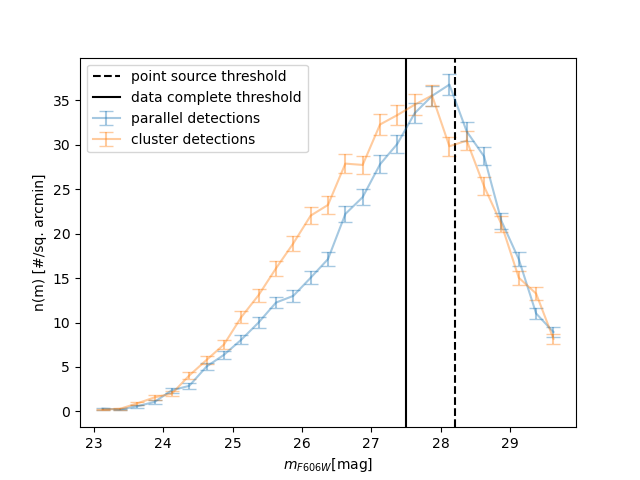}
\caption{Surface number densities of galaxies (cluster field as orange, parallel field as blue) as a function of F606 band apparent magnitudes. $5\sigma$ threshold for a point source over a 0.7" circular aperture is shown as a black dashed line, below which parallel densities drop drastically. Cluster field densities drop at a brighter magnitude, which I attribute to galaxies of the same brightness being larger than in the parallel field.  To impose the same completeness threshold in both the cluster and parallel fields, I define a conservative threshold at 0.25 mag brighter than where the cluster field number density drops dramatically, as indicated by the solid black vertical line. }
\label{neg_data_compl}
\end{figure}

\subsection{Uniform Completeness Thresholds}
\label{sec_data_complete}

In cluster fields, the observable surface number density of galaxies could be predicted by extrapolating the underlying galaxy luminosity function to magnitudes fainter than the detection threshold. The presence of bright galaxies, however, hinders the detection of dimmer galaxies and results in a non-uniform detection threshold across the cluster fields (likewise for parallel fields but to a lesser degree). To mitigate this spatial variation, I mask out bright regions in both cluster and parallel fields as described below. 

Fig.\ref{exclusion_region} shows one of the cluster field - MACSJ0416 - as imaged by the HFF F160W band\footnote{HFF images are available on \href{https://archive.stsci.edu/prepds/frontier/}{MAST archive}, among which I adopted their 30mas v1 release. WSLAP+ and CATS models adopted in this paper could also be found on the lens model page. }. In this band, bright cluster cluster members are most easily detected, allowing for the best identification of the bright regions. In Fig.\ref{exclusion_region}, I identify bright regions above a pixel unit of 0.01$e/s$  (corresponding to a brightness threshold of 23.33 mag/arcsec$^2$) as regions enclosed by green contours. These contours are seen to mostly enclose bright cluster members and display fine-scale structure tracing non-physical noise peaks around bright objects. To smooth over the noise peaks, I apply smoothing to F160W band images with a Gaussian kernel of size (standard deviation) 16 pixels. Examples of smoother contours are shown in red and cyan, for which I used a different brightness threshold (0.005$e/s$, or 24.08 mag/arcsec$^2$) for a clearer visual comparison. Cyan contours enclose bright regions containing a single $z_{phot} > 1$ galaxy; such regions will be preserved for analysis.  

To reflect upon a more uniform detection threshold, I test whether the median brightness level is similar across the remaining region. 
Fig.\ref{compare_exclu_definition} shows the median brightness level measured in different radial annulus bins for individual clusters and parallel fields. Here I compare the resulting distribution for three different brightness thresholds as well as for the case without any brightness threshold. As could be seen, the measured distributions in parallel fields are all relatively flat in all four panels. In the cluster fields, the most stringent cut of 0.0025$e/s$ (24.83 mag/arcsec$^2$) results in the flattest distribution. Motivated by this flatness, the choice of 0.0025$e/s$ was adopted by \cite{Leung2018} to remove most of the galaxy cluster lights for their UV LF faint end test above redshift $z>4$. For my investigation, however, I was interested in the misidentified cluster members that are more populated towards the inner region. Hence, to preserve a larger sample and better test for contamination, I choose a less stringent cut of 0.005$e/s$. In Fig.\ref{exclusion_region}, the final exclusion regions thus defined for M0416 are shown as regions enclosed by red contours. Note that with the choice of 0.005$e/s$, there remains a noticeable increase in brightness level for each cluster field in the innermost region. The implication of this rise will be discussed in the result section. 

To exploit the full availability of HFF photometric measurements, I further restrict my analysis to the common area covered by all HFF bands. This is shown as the region enclosed by the orange contour in Fig.\ref{exclusion_region}.

After imposing exclusion regions and common regions, I determine the data completeness threshold for each band by plotting the measured surface number density of galaxies as a function of apparent magnitudes. Fig.\ref{neg_data_compl} shows the example result for the F606W band, where the orange histogram combined all available cluster fields and the blue combined all parallel fields. The number density of galaxies in the cluster fields is seen elevated to that in the parallel fields over apparent magnitudes $\sim 25-27$. This could be the result of either magnified distant galaxies, or abundant dim cluster members, or both. The 5$\sigma$ detection threshold for point sources in both fields is indicated by a dashed vertical line. As can be seen, whereas the number density of galaxies drops dramatically below this threshold in the parallel fields, the number density in the cluster fields begins to drop at about 0.3 mag brighter than this threshold. This difference is caused presumably by differences in the distribution of galaxy sizes in the two fields, such that at the same apparent magnitude, there are more spatially extended galaxies in the cluster fields (e.g., extended cluster members, or lensed distant galaxies with large magnification factor) than in the parallel fields. To define the same completeness level for both fields, I conservatively adopt a magnitude of 27.5
as indicated by the vertical solid line in Fig.\ref{neg_data_compl}. Data complete thresholds in other wavelength bands are determined similarly for S18 catalogs, and are 28.0 for the F435W band, 27.25 for the F814W to F160W bands.  


\subsection{Luminosity Function}
\label{sec_counts_construct}

\begin{figure}
\centering
\includegraphics[width=0.8\textwidth]{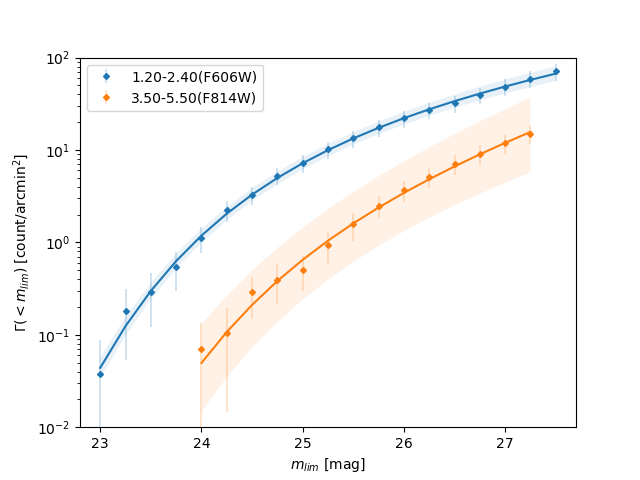}
\caption{Galaxy number counts over redshift 1.2-2.4 (blue) and 3.5-5.5 (orange). Each data point denotes the unlensed density of all sources brighter than the corresponding magnitude down to the data completeness threshold, where magnitudes are from the bands (indicated in legend) immediately longwards the Lyman break. Data points are averaged over measurements from six parallel fields, and error bars reflect both Poisson statistics as well as dispersion about the plotted average. Solid curves of the same color are the fitted Gamma functions with shaded region corresponding to fitting uncertainties.}
\label{low_LF}
\end{figure}

\begin{figure}
\centering
\includegraphics[width=0.8\textwidth]{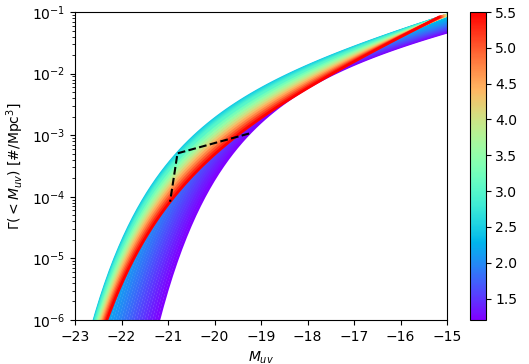}
\caption{Integrated UV LFs at different redshifts extracted from my fitted LF-redshift relations. The color bar reflects the redshift, which ranges from z=1.2 to z=5.5. The black dashed line joins the position of $M_*$ at different redshifts.}
\label{LF_zs}
\end{figure}


The primary focus of this chapter is to perform magnification bias analysis (in particular, contamination level estimates) over $3.5 \leq z \leq 5.5$. Prior to that, however, I first test my methodology against a lower redshift range, $1.2 \leq z \leq 2.4$, over which galaxies have distinct SEDs to cluster members and hence less likely to suffer from the misidentification problem. To derive UV luminosity functions needed for both redshift ranges, in Fig.\ref{low_LF} I plot the surface number density of galaxies in the parallel fields as a function of limiting magnitudes in blue for $1.2 \leq z_{phot} \leq 2.4$ and in orange for $3.5\leq z_{phot} \leq 5.5$. Here, to cope with the convention of UV LF measurements, apparent magnitudes in the higher redshift range are measured in the band (F814W) that captures the rest frame $1600\AA$ at the average redshift of 4.5. At average redshifts $z<2$, however, redshifted Lyman breaks may fall in the same band as $1600\AA$ or be very close. Hence, I adopted the band capturing rest frame 2000$\AA$ (i.e., F606W) instead for redshift range $1.2 \leq z_{phot} \leq 2.4$. 


The measured functions in Fig.\ref{low_LF} are essentially the integrated form of UV LFs, and UV LFs are known to be well-parameterized by Schechter functions. For subsequent magnification bias analysis, I therefore fitted the corresponding integral form (Gamma functions $\Gamma(<m_{lim})$) to achieve future extrapolation toward fainter magnitudes. Conversion to absolute magnitudes at rest frame 1600$\AA$ is done via the following equation derived from \cite{Leung2018}:

\begin{equation}
\begin{aligned}
   & M_{UV}(m_{obs}, \lambda_{eff}, z) =  m_{obs}  \\
    & + 2.5 \log_{10} \bigg{[} \frac{\mu}{(1+z)^{\beta + 1}} \bigg{(} \frac{\lambda_{eff}}{1600} \bigg{)}^{\beta + 2} \bigg{(} \frac{10 \text{pc}}{D_L(z)} \bigg{)}^{2} \bigg{]}
\end{aligned}
\end{equation}

\noindent where $\lambda_{eff}$ is the pivotal wavelength of the band used for apparent magnitudes, $D_L$ is the luminosity distance, and $\beta$ is the typical UV continuum slope of galaxy SEDs. The factor of $(\lambda_{eff}/1600)^{\beta+2}$ is needed to correctly convert absolute magnitude from rest frame $\lambda_{eff}/(1+\bar{z})$ to 1600$\AA$. For $\beta$, \cite{Bouwens2014B}, \cite{Bouwens2009}, \cite{Hathi2016} have measured its relation with $M_{UV}$ at different redshifts. By collecting these relations and fitting against redshift, I obtained $\beta(M_{UV},z)= -0.118 z -1.288 - (0.029 z + 0.003)(M_{UV}+19.5)$ for my magnitude conversion\footnote{Over redshift $3.5\leq z\leq5.5$, typical uncertainty in $\beta$ is of order of 0.1, see e.g. \cite{Bouwens2014B}, giving a typical uncertainty in $M_{UV}$ of order $<$0.3. I consider such a small uncertainty to not result in significant changes to the presented analysis.}. 

In Fig.\ref{low_LF}, the best-fit Gamma functions over $1.2 \leq z \leq 2.4$ and $3.5\leq z \leq 5.5$ are plotted as solid lines, with fitting uncertainties indicated by shaded regions. Obtained Schechter parameters are also provided in the first two rows of Table.\ref{LF_parameters}, where $\phi$ is the LF amplitude (in cosmic volume density), $M_*$ is the characteristic luminosity controlling exponential suppression at the bright end, and $\alpha$ is the faint end logarithmic slope. 

Having introduced how UV LF parameters could be determined at given averaged redshifts, I next consider redshift variation in UV LFs. This is relevant for wide redshift intervals such as $3.5\leq z\leq 5.5$, as redshift variation in UV LF could affect the accuracy of the magnification bias predicted. In order to perform magnification bias over an arbitrarily narrow redshift interval, I investigate how LF parameters vary with redshift by breaking the full redshift range of interest - $1.2 \leq z \leq 5.5$ - into smaller redshift bins. 

In determining the best binning to achieve a more robust UV LF measurement, I first tried the division of the full range of interest into ten smaller redshift bins with a redshift bin size of 0.3 at $z<3$ and 0.5 at $z\geq 3$. However, further investigation showed that a change in the corresponding binning choice could lead to large $\sim \pm 15\%$ fluctuation in the estimated contamination level owing to the insufficient number of galaxies in each redshift bin. To resolve this issue, I was inspired by the fact that a galaxy at redshift $z=2$ could be assigned to either the $1.5-2.5$ sample or the $1-2$ sample for statistical measurements. Hence, I decide to adopt a non-conventional approach and allow for different smaller redshift bins to overlap. This way I lose 'statistical independence' between adjacent bins, but gain more LF sampling points and also enough number of detections in all bins. Such an approach allowed us to more robustly determine the UV LF-redshift relation, and it was verified that different binning choices lead only to a controlled variation ($\sim\pm3\%$) in the final contamination level I later report. I present in Table.\ref{LF_parameters} my adopted redshift bins, and it can be seen that the chosen bin size is 0.6 at $z<2$, 0.8 between $2\leq z<3$, and 1 at $z\geq 3$.

\begin{figure}
    \centering
    \includegraphics[width=0.85\linewidth]{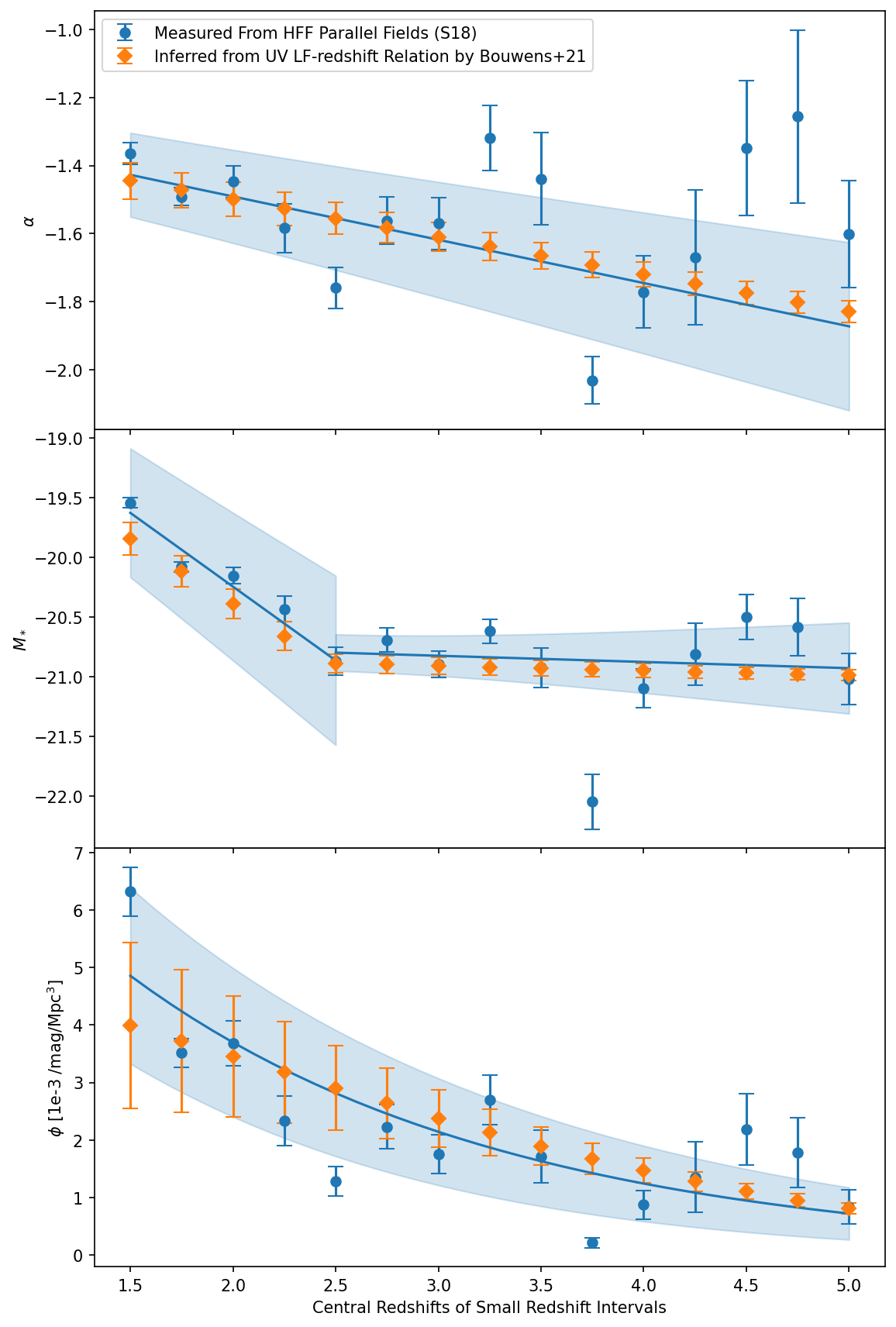}
    \caption{Blue data points present the Schechter parameters I measured in small redshift bins. These measured values are seen to fluctuate about the overall trend indicated by the blue solid curves, corresponding to the UV LF-redshift relations I reported in the main text. I also indicated the UV Schechter parameters (orange diamonds) at these sampling redshifts inferred from the UV LF-redshift relation provided by \cite{Bouwens2021}. It could be seen that \cite{Bouwens2021} measurements have much smaller error bars than my determination, but my reported UV LF-redshift relation is largely consistent with theirs within error bars. }
    \label{UVLF_cf_Bouwens}
\end{figure}

In each bin chosen, I separately determined the Schechter parameters that are presented in Table.\ref{LF_parameters}, also presented as blue data points in Fig.\ref{UVLF_cf_Bouwens}. These measurements were found to fluctuate about an overall trend that could be well parameterized by the following relation with redshift:

\begin{equation}
    \phi(z) = 10^{(-0.237\pm 0.050) z + (1.042 \pm 0.115)} \text{  [1e-3/mag/Mpc}^3],
\end{equation}

\begin{equation}
    M_*(z) = \begin{cases}
         (-1.237 \pm 0.230) z + (-17.772 \pm 0.415) \\
         \text{ if } z \leq 2.47\\
         (-0.052 \pm 0.139)(z-2.48) + (-20.798 \pm 0.154) \\ 
         \text{ if } z > 2.47
        \end{cases}
\end{equation},
\begin{equation}
    \alpha(z) =  (-0.127 \pm 0.045) z + (-1.237 \pm 0.103)
\end{equation}

\noindent, which are also explicitly presented as the blue solid curves in Fig.\ref{UVLF_cf_Bouwens}. Following this parametrization, in Fig.\ref{LF_zs} I plot UV LFs at all redshifts over $1.2\leq z\leq 5.5$ in their integrated form (to cope with Fig.\ref{low_LF}), with redshift indicated by the color bar. I also present the "trajectory" of $M_*$ with redshift as a black dashed line. Overall, I can see that at higher redshifts, the faint end slope is steeper, the UV LF amplitude is lower, and that exponential suppression occurs at a brighter magnitude. These observed trends agree with existing UV LF measurements from literature, and I comment that my determined LF-redshift relations are (within measurement uncertainties) akin to what \cite{Bouwens2021} obtained. The latter point could be better seen from Fig.\ref{UVLF_cf_Bouwens}, where I presented the inferred Schechter parameters based on \cite{Bouwens2021} UV LF-redshift relation at the same redshifts (orange diamonds). It could be seen \cite{Bouwens2021} measurements have much smaller fitting uncertainties, but my reported UV LF-redshift relation is largely consistent with theirs within error bars. 

\begin{table}[h]
\centering
\begin{tabular}{cccc}
z range & $\phi$ & $\alpha$ & $M_*$\\
 & (1e-3/mag/Mpc$^3$) &  &  (AB mag)\\
\hline
\hline
 1.20-2.40 & 4.715$\pm$0.388 & -1.374$\pm$0.042 & -19.909 $\pm$0.052 \\ 
3.50-5.50 & 0.459$\pm$0.198 & -1.886$\pm$0.142 & -21.468 $\pm$0.232 \\ 
\hline
1.20-1.80 & 6.322 $\pm$ 0.427 & -1.365 $\pm$ 0.032 & -19.542 $\pm$ 0.044 \\
1.45-2.05 & 3.517 $\pm$ 0.246 & -1.491 $\pm$ 0.027 & -20.081 $\pm$ 0.044 \\
1.70-2.30 & 3.687 $\pm$ 0.392 & -1.447 $\pm$ 0.046 & -20.153 $\pm$ 0.066 \\
1.85-2.65 & 2.337 $\pm$ 0.433 & -1.584 $\pm$ 0.072 & -20.433 $\pm$ 0.109 \\
2.10-2.90 & 1.281 $\pm$ 0.261 & -1.759 $\pm$ 0.060 & -20.870 $\pm$ 0.119 \\
2.35-3.15 & 2.228 $\pm$ 0.387 & -1.562 $\pm$ 0.070 & -20.692 $\pm$ 0.103 \\
2.50-3.50 & 1.754 $\pm$ 0.332 & -1.571 $\pm$ 0.076 & -20.898 $\pm$ 0.112 \\
2.75-3.75 & 2.693 $\pm$ 0.434 & -1.319 $\pm$ 0.095 & -20.621 $\pm$ 0.103 \\
3.00-4.00 & 1.717 $\pm$ 0.459 & -1.439 $\pm$ 0.136 & -20.924 $\pm$ 0.167 \\
3.25-4.25 & 0.209 $\pm$ 0.082 & -2.031 $\pm$ 0.070 & -22.048 $\pm$ 0.230 \\
3.50-4.50 & 0.870 $\pm$ 0.252 & -1.772 $\pm$ 0.105 & -21.096 $\pm$ 0.161 \\
3.75-4.75 & 1.353 $\pm$ 0.611 & -1.670 $\pm$ 0.198 & -20.814 $\pm$ 0.260 \\
4.00-5.00 & 2.186 $\pm$ 0.616 & -1.349 $\pm$ 0.197 & -20.501 $\pm$ 0.188 \\
4.25-5.25 & 1.780 $\pm$ 0.602 & -1.256 $\pm$ 0.254 & -20.586 $\pm$ 0.239 \\
4.50-5.50 & 0.832 $\pm$ 0.299 & -1.601 $\pm$ 0.156 & -21.019 $\pm$ 0.216
\end{tabular}
\caption{Fitted Schechter parameters over various redshift intervals.}
\label{LF_parameters}
\end{table}

\subsection{Implementing Magnification bias}
\label{sec_mag_bias_equation}

As previously discussed, magnification bias encompasses two competing effects brought by lensing magnifications on the observed surface number density of galaxies (per unit solid angle of sky). First, lensing makes detectable faint galaxies by raising their brightness to above the detection threshold, or, equivalently, by effectively lowering the detection threshold. The anticipated increase in the observed surface number density of galaxies, however, is countered by the decrease in the physical area of sky probed, as the latter also is magnified by lensing. Depending on how steeply the LF continues to fainter magnitudes, lensing magnification may result in a net increased (positive magnification bias), decreased (negative magnification bias), or leave no effect (null/critical magnification bias) on the overall surface number density of galaxies in cluster fields relative to in parallel fields. 

Taking this more mathematically, I shall denote $\Gamma(<m_{lim},z)$ as the surface number density of \textit{unlensed} galaxies at redshift $z$ observed above a detection threshold of $m_{lim}$:
\begin{equation}
\Gamma(<m_{lim},z) = \frac{dV(z)}{A_{obs}} \int^{m_{lim}}_{-\infty} \phi(m',z)dm'.
\label{surface_number_density}   
\end{equation}
\noindent In the above expression, $dV(z)$ is the infinitesimal cosmic volume element about redshift $z$, $A_{obs}$ is the area of observing field of view (FOV), and $\phi(m,z)$ is the \textit{underlying} LF measured in apparent magnitudes. According to magnification bias, the net effect of lensing magnification factor $\mu_z$ is then to result in an observed surface number density of \textit{lensed} galaxies $n_{len}$ given by
\begin{equation}
n_{len} (z,\mu_z,m_{lim}) = \frac{\Gamma(<(m_{lim}+2.5\log_{10}\mu_z),z)}{\mu_z},
\label{bias_equation}   
\end{equation}
\noindent where term $2.5\log_{10}\mu_z$ after $m_{lim}$ corresponds to effective lowering of detection threshold, and the $1/\mu_z$ term reflects the reduction effect on probed cosmic volume. 

In the remainder of this chapter, I apply magnification bias to predict the surface number density of galaxies over $3.5\leq z\leq 5.5$ at different levels of lensing magnifications. I then test for contaminants as any excess population observed in the cluster fields. To measure the observed surface number density at different levels of lensing magnification, I divide cluster field FOVs into ten magnification bins according to the following lens models: (1) v4 CATS \textit{lenstool} models \citep{Caminha2016, Caminha2017, Jauzac2016, Lagattuta2017, Limousin2016, Mahler2018}, adopted as a representative of parametric\footnote{The comprising components of mass distribution are described by analytical functions, with component parameters constrained by lensing data via MCMC.} lens modeling techniques; (2) v4.1 WSLAP+ models \citep{Lam2014, Diego2015a, Diego2015b, Diego2016a, Diego2016b, Diego2018}, representative of free-form\footnote{Projected mass density is modeled as a grid of Gaussian functions.} lens modeling techniques; and (3) additional two state-of-the-art parametric glafic models\footnote{These glafic models are not to be confused with existing ones on HFF archive. My adopted models are with the most updated lensing constraints at the time of writing (for the paper), and were carefully examined for their predictive power.} available only for MACS0416 (provided by collaborator) and Abell370 \citep{Keith2022}. To ensure comparable statistics, magnification bins are equally spaced in logarithmic scale with a logarithmic width of 0.2 up to a magnification factor of 100.\footnote{Galaxies with predicted magnification factor above 100 have large uncertainties owing to sitting very close to critical curves, such highly magnified sources are all put into the highest magnification bin available.} 

\begin{figure}
\centering
\includegraphics[width=0.8\textwidth]{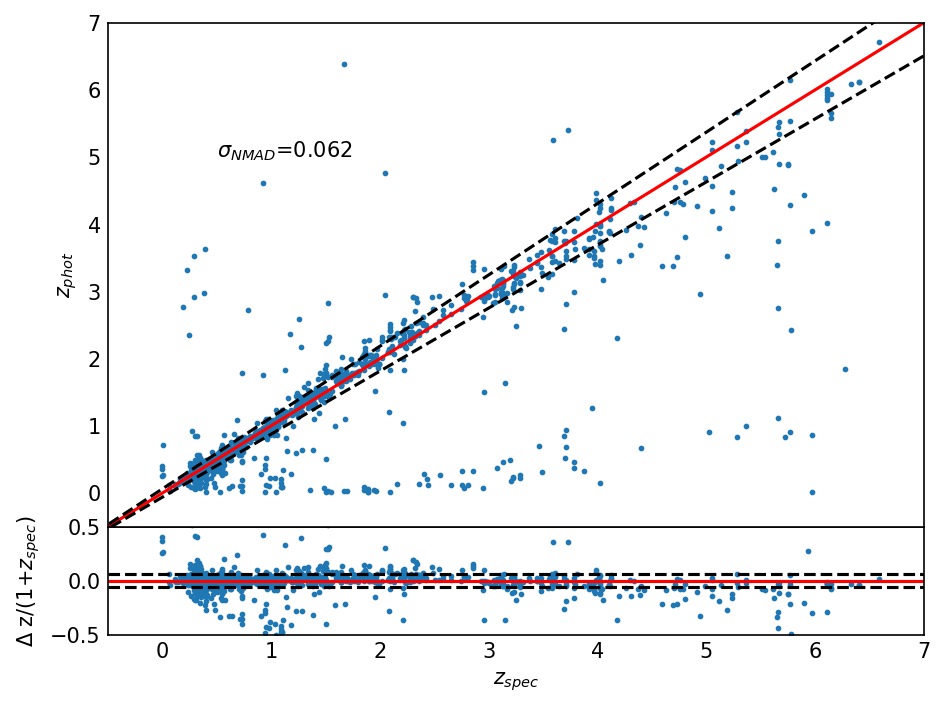}
\caption{Spec-z plotted against reported photo-z for galaxies from S18 catalogs (combining all HFF cluster fields) to determine $z_{NMAD}$. The normalized deviations $(z_{phot}-z_{spec})/(1+z_{spec})$ are provided in the bottom panel, and the scatter range is seen well captured by the black dash lines (corresponding to $\pm 0.062$ and $z_{NMAD} \sim 0.062(1+z)$).}
\label{specz_vs_photz_shipley}
\end{figure}

At any particular redshift $z$, I predict the surface number density of galaxies in each magnification bin by collecting the number of galaxies predicted on each image plane pixel, and subsequently dividing this number by the total area. Similarly, for the same magnification bin, I measure the observed surface number density of galaxies by counting the number of UV bright galaxies (see Sec.\ref{subsec_uv_bright_sample}) over redshifts $z \pm z_{NMAD}$. Here, $z_{NMAD}$ is the typical scatter\footnote{This scatter is often measured through the normalized median absolute deviation (NMAD) $\sigma_{NMAD}$, which is defined as $\sigma_{NMAD} = 1.48 \times \text{median}(|\Delta z - \text{median}(\Delta z)|)$, where $\Delta z \equiv z_{phot}- z_{spec}$.} (reflective of uncertainties) of reported photo-z about spec-z for S18 catalogs, and I measured it to be $z_{NMAD} \sim 0.062(1+z)$ from Fig.\ref{specz_vs_photz_shipley}. Correspondingly, when calculating both the observed and predicted surface number density over a wide redshift range (e.g. $1.2\leq z\leq 2.4$ and $3.5\leq z\leq 5.5$), I use redshifts separated by $\geq 2 z_{NMAD}$ (given the sampling size is $z \pm z_{NMAD}$) to individually implement magnification bias. The obtained surface number densities at different redshifts are later added up to yield the full density for the wide redshift range. 

With the same philosophy and the same set of magnification bias maps at different redshifts, I can also predict surface number density in different radial bins. This is how the shown prediction in Fig.\ref{cluster_member_distribution} was obtained.

\subsection{UV Bright Sample}
\label{subsec_uv_bright_sample}

\begin{figure*}
\includegraphics[width=0.9\textwidth]{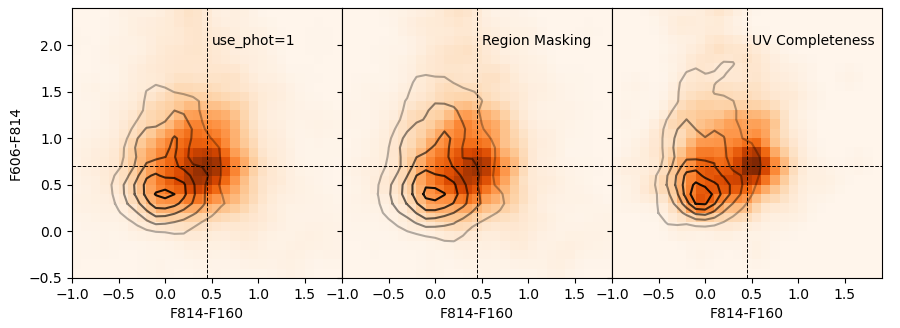}
\caption{Color-color distribution of the remaining S18 detections after different stages of selection.}
\label{color_check_selection_bias}
\end{figure*}


In Table.\ref{sample_remaining}, the number of \texttt{use$\_$phot}=1 galaxies from S18 catalogs over redshift ranges $1.2\leq z\leq 2.4$ and $3.5\leq z\leq 5.5$ are listed under \textit{use$\_$phot=1} column. The size of the sub-sample after applying exclusion regions is listed under the \textit{Region Masking} column, and it could be seen that this selection leaves about $30\%$ of the sample in the cluster fields and $40\%$ in the parallel fields. I demonstrate in Fig.\ref{color_check_selection_bias} that this sub-sample is reflective of the full \texttt{use$\_$phot}=1 sample. Here, background histograms and black contours are the F606-F814 vs F814-F160 distribution of $3.5\leq z\leq 5.5$ galaxies in the cluster and parallel fields, respectively. I also marked the peak of cluster field distribution by the dashed cross, which is not seen to vary as I restrict to \textit{Region Masking} sample. 

As magnification bias analysis relies on UV luminosity functions, I in fact only need UV bright galaxies in the band covering their rest frame UV wavelengths. As discussed in Sec.\ref{sec_counts_construct}, this rest frame wavelength is chosen to be $1600\AA$ at $z\leq2$ but $2000\AA$ at $z<2$. The final analysis sample thus comprises galaxies brighter than data completeness thresholds in the respective band. The corresponding sample sizes are listed in the \textit{Completeness} column of Table.\ref{sample_remaining} for both redshift ranges. Again, in the last panel of Fig.\ref{color_check_selection_bias}, I demonstrate that restricting to this sub-sample does not alter significantly the color-color distribution. I.e., overall, I am left with $\sim 17-18\%$ of \texttt{use$\_$phot}=1 galaxies for analysis. But as I showed in Fig.\ref{color_check_selection_bias}, the UV bright sample is representative of the whole \texttt{use$\_$phot}=1 sample. In other words, the contamination inferred using only the UV bright sample is expected to be reflective of the contamination level for the full galaxy sample.

\begin{table}[h]
    \centering
    \begin{tabular}{ccccc}
    \hline
       Field  & z range & \texttt{use$\_$phot}=1 & Region Masking & Completeness \\
    \hline
    \hline 
        A2744clu & 1.2-2.4 & 1593 & 491 & 238 \\
 & 3.5-5.5 & 917 & 382 & 148 \\
A2744par & 1.2-2.4 & 1751 & 765 & 326 \\
 & 3.5-5.5 & 454 & 211 & 69 \\
M0416clu & 1.2-2.4 & 1822 & 720 & 319 \\
 & 3.5-5.5 & 839 & 507 & 120 \\
M0416par & 1.2-2.4 & 2224 & 925 & 344 \\
 & 3.5-5.5 & 506 & 317 & 75 \\
M0717clu & 1.2-2.4 & 966 & 361 & 159 \\
 & 3.5-5.5 & 582 & 261 & 140 \\
M0717par & 1.2-2.4 & 1593 & 745 & 330 \\
 & 3.5-5.5 & 329 & 197 & 80 \\
M1149clu & 1.2-2.4 & 1561 & 539 & 266 \\
 & 3.5-5.5 & 680 & 320 & 101 \\
M1149par & 1.2-2.4 & 1761 & 847 & 376 \\
 & 3.5-5.5 & 392 & 228 & 79 \\
AS1063clu & 1.2-2.4 & 1614 & 482 & 201 \\
 & 3.5-5.5 & 673 & 259 & 87 \\
AS1063par & 1.2-2.4 & 1895 & 952 & 427 \\
 & 3.5-5.5 & 398 & 191 & 70 \\
A370clu & 1.2-2.4 & 1222 & 395 & 165 \\
 & 3.5-5.5 & 873 & 358 & 169 \\
A370par & 1.2-2.4 & 1682 & 850 & 381 \\
 & 3.5-5.5 & 409 & 224 & 66 \\
 \hline
Cluster total & 1.2-2.4 & 8778 & 2988 & 1348 \\
 & 3.5-5.5 & 4564 & 2087 & 765 \\
Parallel total & 1.2-2.4 & 10906 & 5084 & 2184 \\
 & 3.5-5.5 & 2488 & 1368 & 439 \\
 \hline
    \end{tabular}
    \caption{Number of $1.2\leq z\leq 2.4$ and $3.5\leq z\leq 5.5$ sources left from S18 catalogs after each selection step. }
    \label{sample_remaining}
\end{table}

\begin{figure*}
    \includegraphics[width=0.97\textwidth]{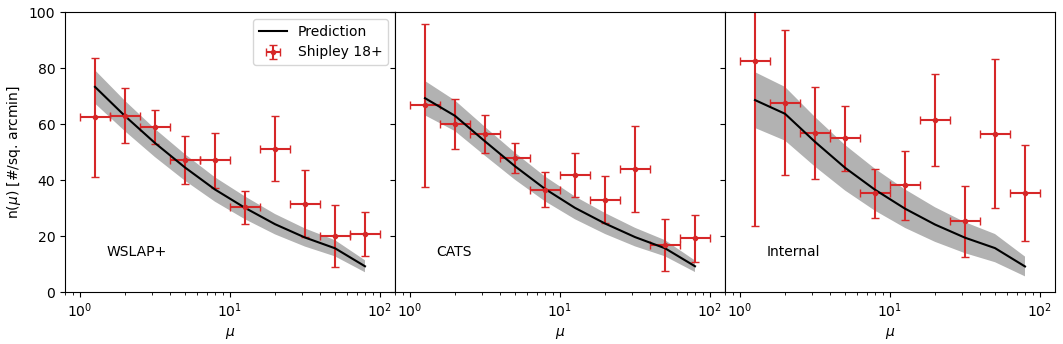}
    \caption{Magnification bias analysis over $1.2<z<2.4$ for S18 catalogs. Predicted magnification bias (black line) as a function of magnification factor $\mu$ compared with derived densities (red data points) when adopting glafic, WSLAP+, and cATS models, respectively. Data points are after averaging over available cluster fields for each catalog. The shaded region for prediction reflects measurement uncertainty in CLF-redshift relations. Predicted surface density is seen to decrease as magnification increases, indicating a negative magnification bias. The derived densities are seen to follow a similar trend as prediction, but may with a systematic offset. I also notice that data in some magnification bins are severely deviating from the predicted level, suggesting I have contamination in those bins.}
    \label{neg_bias_magbin_diego_collective}
\end{figure*}

\begin{figure*}
     \centering
     \includegraphics[width=0.97\textwidth]{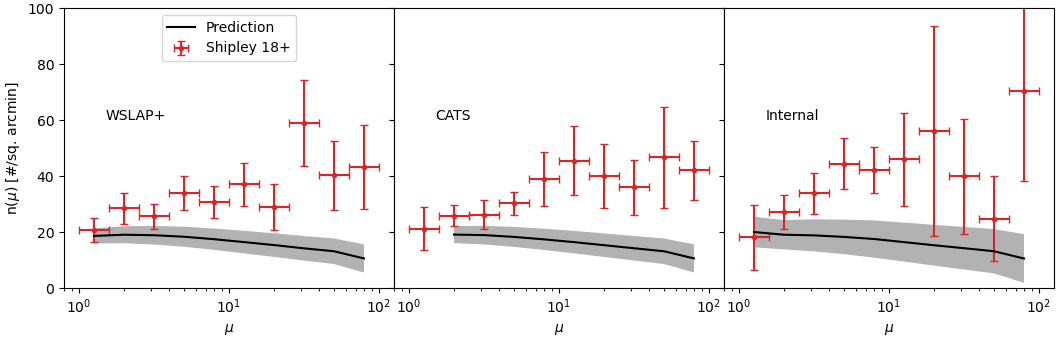}
     \caption{Magnification bias analysis over $3.5\leq z \leq 5.5$ for S18 catalogs. Predicted surface density is seen to decrease as magnification increases, indicating a negative magnification bias. On the other hand, the cataloged sources are seen to be more populated in the higher magnification regions, demonstrating strong contamination. }
        \label{pos_bias_magbin_diego_collective}
\end{figure*}

\section{Predicted vs Observed galaxy density}
\label{bias}

I start by testing my methodology over the redshift range of $1.2<z<2.4$, which I expect to suffer no contamination by early-type cluster members (Sec.\ref{sec_counts_construct}). Fig.\ref{neg_bias_magbin_diego_collective} plots the predicted surface number densities in different magnification bins as black solid curves for each lens model. The densities are averaged over available HFF cluster fields to mitigate cosmic variance effects. For the WSLAP+ and CATS lens models, I had all HFF fields available, but for the Internal models, only the M0416 and A370 fields were used. This results in larger uncertainties (shaded region) for the Internal model predictions. As a reminder, the predicted densities are based on UV luminosity function redshift relations derived from parallel fields. The shaded regions reflect the uncertainty in the luminosity function propagated into the magnification bias estimates. While the different lens models predict slightly varying density distributions, they all decrease toward higher magnification regions, indicating a negative magnification bias.

I next compare my predictions to the observed surface number densities from the UV bright sample, which are plotted as red data points in Fig.\ref{neg_bias_magbin_diego_collective}. The measured densities also average over available cluster fields, with vertical error bars incorporating Poisson statistics and variance between fields. The observed densities show the same trend of decreasing toward higher magnification regions, consistent with the predicted negative magnification bias. Referring back to Fig.\ref{excess}, negative bias thus provides a natural explanation for the relative deficit of galaxies in the cluster fields observed around $z\sim2$. 

While the predicted and observed trends agree regardless of lens model, a strong excess is observed in high magnification regions $\mu >10$. This may indicate contamination. For instance, these could be faint cluster members with poorly-estimated redshifts, and are naturally more populated towards inner, higher magnification regions.

Fig.\ref{pos_bias_magbin_diego_collective} shows a similar magnification bias analysis over the redshift range $3.5\leq z\leq 5.5$. In contrast to the strong negative bias shown in Fig.\ref{neg_bias_magbin_diego_collective}, the predicted magnification bias (black curve) is only mildly negative. The measured galaxy surface number densities (red points), on the other hand, display a strong increasing trend towards higher magnification regions. This deviation supports my previous hypothesis (Sec.\ref{section_implications}) that contaminants may be misidentified cluster members. 

The excess above predictions could also be converted to a measure of contamination level. I calculate this by collectively counting the total measured and predicted number of galaxies over $3.5\leq z \leq 5.5$. This way, the contamination levels are estimated to be $59.38 \pm 11.12 \%$ based on CATS models, $53.48 \pm 10.33 \%$ based on WSLAP+ models, and $59.53 \pm 18.52 \%$ based on Internal models. I note that regardless of lens model, more than half of $3.5\leq z \leq 5.5$ galaxies in the S18 catalogs are contaminants from misidentified cluster members! Furthermore, I remind readers that the inner regions of clusters may have a non-uniform, shallower detection threshold (see Sec.\ref{sec_data_complete}). This would instead imply the galaxy sample in higher magnification bins is incomplete, hence the number of lensed galaxies is over-predicted. As such, the actual contamination levels may also be higher than what is reported above. While more precise determination of the local detection thresholds and hence the contamination level is hard, my above analysis already demonstrated severe contamination ($>50\%$) in the S18 UV bright sample over $3.5\leq z\leq 5.5$. As the UV bright sample reflects the full \texttt{use\_phot}=1 sample (see Sec.\ref{subsec_uv_bright_sample}), I expect also a similar contamination level for the entire S18 catalogs.

\subsection{Using less/more stringent exclusion regions} 

For completeness, I further investigated the contamination level when adopting a different brightness threshold (0.01e/s and 0.0025e/s) for defining the exclusion region. The contamination level was found to be $54.77\% \pm  11.68\%$ with a less stringent brightness threshold (0.01 e/s), consistent with my estimate using 0.005e/s for defining exclusion regions. On the other hand, the contamination level was found to be $32.61\% \pm 13.68\%$ when using a more stringent brightness threshold (0.0025e/s). This lower contamination level is the result of rejecting more cluster members, which could be interlopers, sitting in the inner regions.

\section{Potential consequences: unphysical turn-ups in LF}
\label{contaminated_LF_sec}


Thus far, I have demonstrated that photo-z catalogs suffer from severe contamination. To my great surprise and worry, the estimated contamination level across the redshift range of $3.5\leq z\leq 5.5$ is as high as $\sim 60\%$! And on average (over estimates based on different lens models, using the number of cluster fields available as weights), the contamination level is found to be $56.87 \pm 11.84 \%$. Such a severe contamination level, as I demonstrate below, has profound implications regarding the test of cosmological models.

\begin{figure}
\centering
\begin{subfigure}{.45\textwidth}
  \centering
  \includegraphics[width=\linewidth]{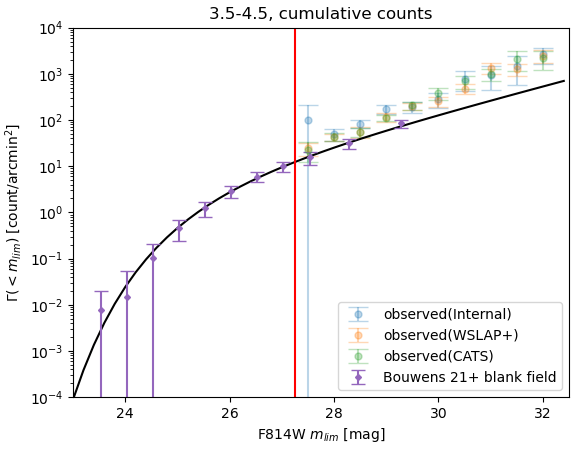}
\end{subfigure}%
\hfill
\begin{subfigure}{.45\textwidth}
  \centering
  \includegraphics[width=\linewidth]{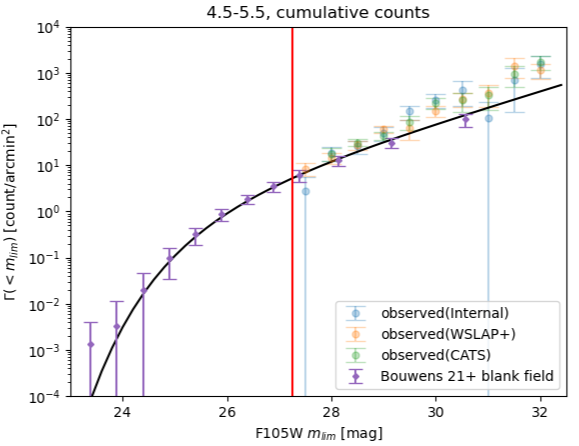}
\end{subfigure}
\caption{Cumulative UV LFs implied by the observed surface number densities of galaxies in cluster fields in comparison with deeper blank field measurements from \cite{Bouwens2021} over redshifts (a) 3.5-4.5 and (b) 4.5-5.5. The implied values are obtained by multiplying the measured surface number density in each magnification bin with the respective mean magnification factor, and are shown as colored data points depending on the lens model involved. Owing to lensing magnification, these implied values are all fainter than the data completeness threshold (red vertical line) in the band capturing rest frame 1600$\AA$. For comparison, I also plot my derived LF (integrated and extrapolated to fainter magnitudes) at median redshifts 4 and 5 from parallel fields as black solid lines. As could be seen, implied values (colored data points) rise above my derived LF (black curves) immediately after the data completeness limit, suggesting faint end up-turns in UV LFs. Such upturns, however, are not consistent with the deeper blank field measurements from \cite{Bouwens2021}, which are shown as purple data points. }
\label{reproducing_data}
\end{figure}

Recall from Fig.\ref{cm_high_z}, the clustering (dominated by contaminants) observed in the analysis sample about the cluster red sequence occurs mainly at faint magnitudes. In terms of UV LF construction, this may then lead to heavily over-estimated values at the faint end if contaminants are not properly dealt with. Such an impact is demonstrated by Fig.\ref{reproducing_data}, in which I reverse engineered\footnote{Multiplying measured surface number density of galaxies with mean magnification factor gives LF integrated up to the respective lowered threshold.} magnification bias equation Equ.\ref{bias_equation} to obtain implied cumulative LF values (colored data points) fainter than the unlensed data completeness threshold (red vertical line). As could be seen, the implied value is significantly higher than both (1) faint end extrapolation of my derived UV LF from parallel fields, shown as black solid lines; and (2) more recent blank field measurements\footnote{What \cite{Bouwens2021} measured was LFs. Cumulative LFs were calculated from LF measurements using the Simpson's rule.} from \cite{Bouwens2021}, shown as purple data points, over the same redshift range and covering magnitudes fainter than S18's data completeness threshold. This (having contamination) may also explain why some UV LF measurements from cluster fields (e.g. \cite{Livermore2017, Bouwens202205}) are elevated at the very faint end, as previously discussed in Chap.\ref{chap:intro}. Similarly, having contamination may wash out faint end turnovers in UV LFs, such as those predicted by baryonic physics or by alternative models for dark matter (e.g., $\psi$DM), and hinder the test of corresponding models. Effective ways to mitigate interlopers are desperately needed for more reliable faint-end UV LF constructions/tests.

\section{Prospect of Mitigation}
\label{sec_mitigating}


In this section, I mention some of my attempts to identify and mitigate contaminants (by better understanding their properties) when only HFF images are available. Detailed implementation of individual identification by incorporating deep JWST measurements is reserved for later chapters. 

\subsection{Cross-check with MUSE catalogs}

I begin with a straightforward and simple attempt to individually identify the contaminants by cross-matching $3.5\leq z\leq 5.5$ candidates in the S18 analysis sample with more recent spectroscopic redshift catalogs provided by \cite{Richard2021} in the fields of A2744, A370, and M0416. Any sources matched to spec-z source with $z_{spec}$ much lower than their claimed photo-z are then contaminant candidates.

By requiring the source separation to be smaller than 0.3 arcsec for a match, I identify 4 potential contaminant candidates (all in the M0416 field) listed in Table.\ref{matched_interloper_speczs}. The four matched candidates are of apparent magnitude $\sim 23-25$ in the F160W band, and as MUSE catalogs have a much brighter data complete threshold\footnote{The data complete threshold in the F814 band is $\sim24.5$ for the MUSE catalog but $\sim27.25$ for Shipley catalogs}, there were no other fainter candidates matched this way. 

\begin{table}
    \centering
    \begin{tabular}{c|c|c|c|c}
        RA & DEC & photo-z & spec-z & zconf $^a$ \\
        \hline
        64.0255364 & -24.0766491 & 3.8068 & 0.4519 & 1\\
        64.0480544 & -24.0749681 & 3.9578 & 0.4023 & 1\\
        64.0446720 & -24.0725194 & 4.0006 & 0.4004 & 2\\
        64.0239677 & -24.0729296 & 3.9586 & 1.3700 & 1
        \end{tabular}
    \caption{Matched contaminant candidates among relevant photo-z 3.5-5.5 candidates.\\   
    $^a$ zonf = 1 means spec-z is based on a single low signal-to-noise or ambiguous emission line, or several very low signal-to-noise absorption features. zonf = 2 means spec-z is based on a single emission line without additional information, or several low signal-to-noise absorption features, or zconf=1 galaxies whose confidence is increased by the identification of multiply-imaged systems. }
    \label{matched_interloper_speczs}
\end{table}

\begin{figure*}
\centering	
\includegraphics[width=0.97\textwidth]{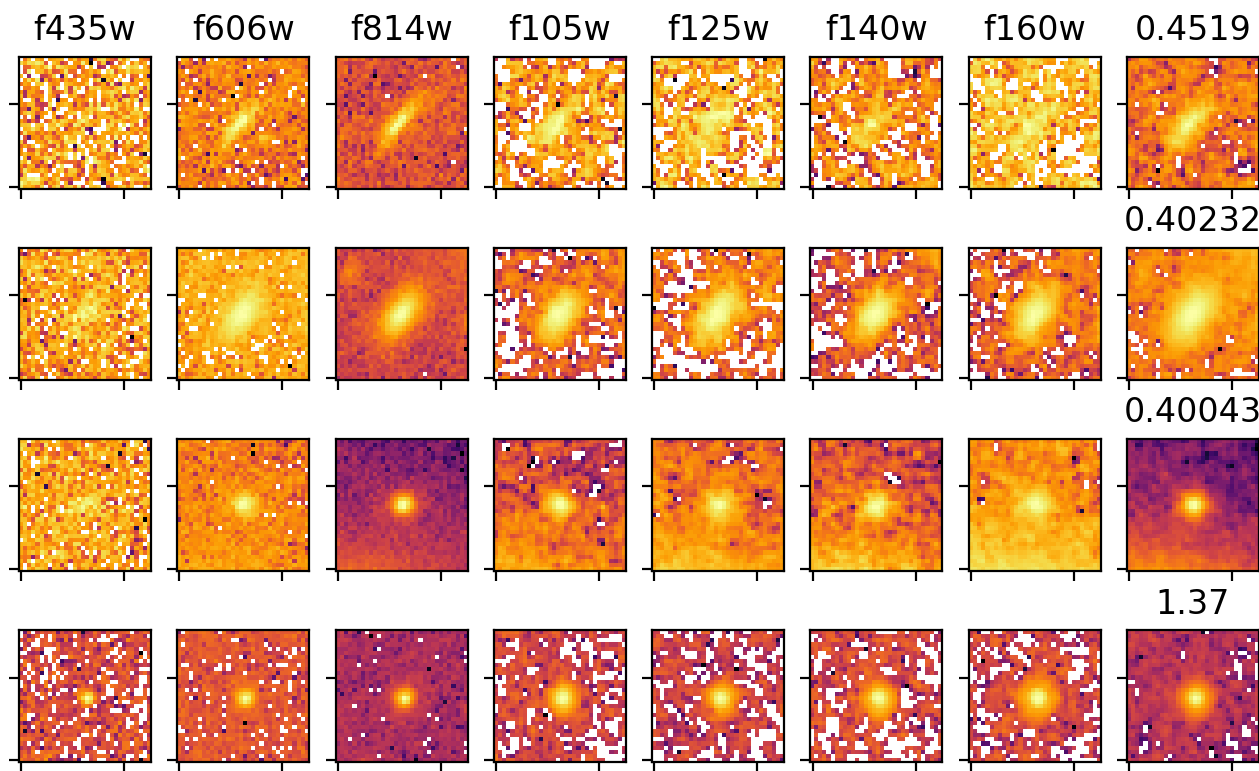}
\caption{ Cutouts of four spectroscopic-wise interloper candidates within $3.5<z_{phot}<5.5$ sample. The cutouts are 4 square arcseconds in size and are provided in all HFF bands as well as stacked images (last column). Their matched spectroscopic redshift is provided on top of the stacked image. Except for the galaxy in the first row, the other galaxies are clearly detected in the F435W band, indicating they cannot be high-z dropout galaxies. }
\label{confirmed_interloper_cutouts_example}
\end{figure*}

It is seen, however, that the spec-zs of the four matched contaminant candidates all have low confidence flags, motivating further visual inspection. In Fig.\ref{confirmed_interloper_cutouts_example}, I present the corresponding cutouts of these four candidates in all HFF bands as well as on a stacked image (weighted sum from F814 to F160 bands) following the same order as they are listed in Table.\ref{matched_interloper_speczs}. Their matched spec-z is also provided on top of the rightmost (stacked) image. The $z_spec=0.4519$ candidate is seen as very elongated and with an apparent curvature not commonly seen for disk galaxies. As such, I consider this galaxy as more likely to be a lensed arc from the high-z Universe. For the other three galaxies, there are some features visible in the F435W band, justifying their interloper nature as they cannot have redshifts higher than 3.5. Two of these three galaxies are seen to have a galaxy cluster spec-z, reflecting that they are likely misidentified cluster members. For the record, the inferred high-z and low-z identities for these four galaxies were later confirmed with the combination of JWST observations. 

I also attempted to confirm more contaminants by cross-matching the S18 analysis sample with MUSE Pilot-WING (\cite{Lagattuta2022} for A370) catalog, which extended previous MUSE coverage to 14 arcmin$^2$, covering for the first time the outskirts of A370. But \cite{Lagattuta2022} catalog also had a much brighter effective detection threshold (e.g., 25.5 in F814W band) than HFF, and hence no more spec-z confirmed contaminants could be claimed. 

\subsection{Visual inspection}
\label{sec_contamin_visualinspect}
The three contaminants successfully identified in the previous subsection suggest that contaminants have at least two morphological types, diffuse elongated and roundish compact. Here to check whether these two morphologies are more commonly seen in cluster fields than in parallel fields, I perform a thorough visual inspection on the analysis sample of an example cluster field of A2744 and its accompanying parallel field.

\begin{figure*}
\centering	
\includegraphics[width=0.97\textwidth]{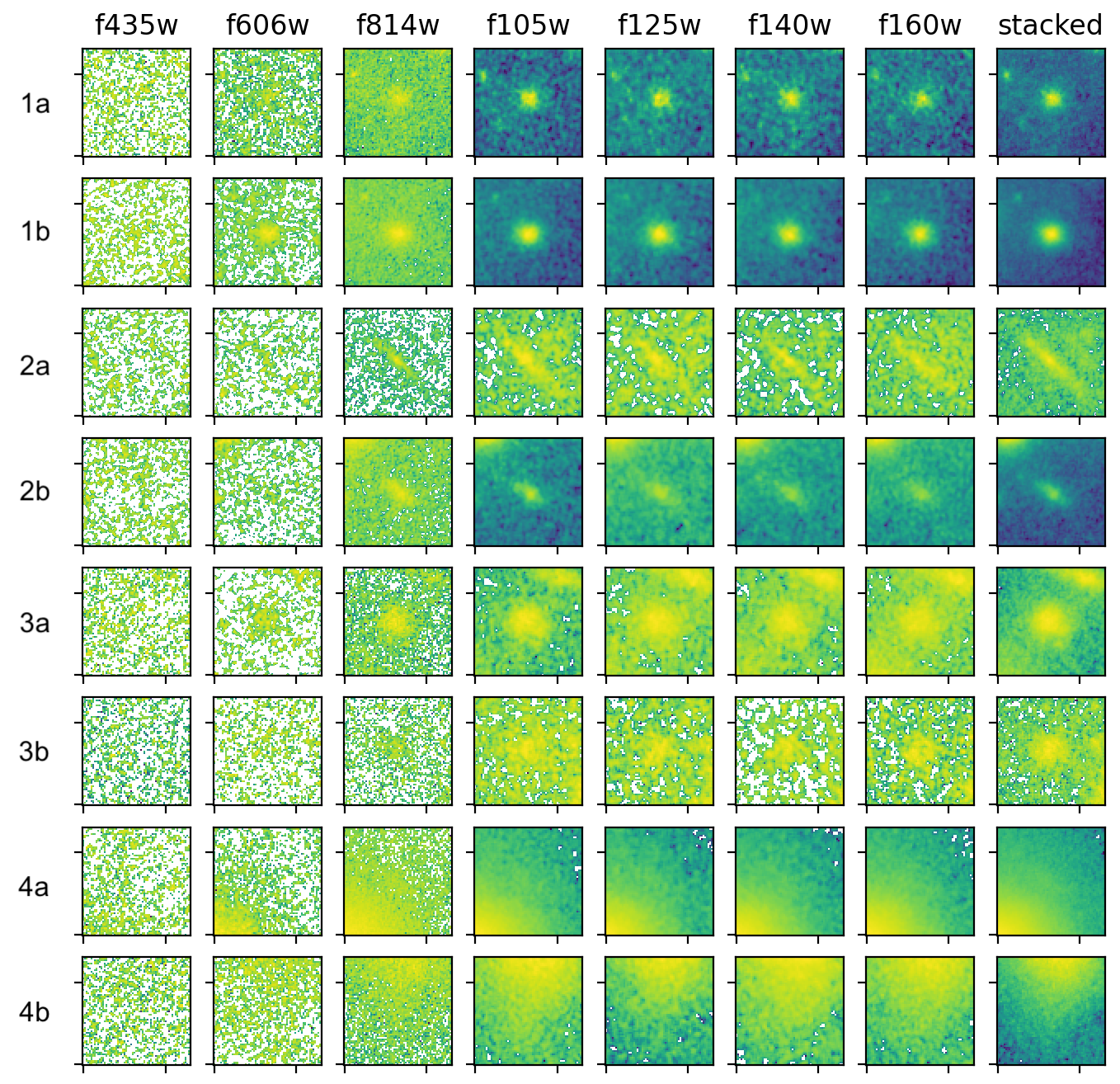}
\caption{ Example cutouts for four types of galaxies seen in the A2744 cluster field in different HFF bands, as well as in a stacked image.}
\label{LBG_cutouts_example}
\end{figure*}

With reference to the galaxies in the parallel field, which I assume to be dominated by genuine high-z galaxies, it was found that high-z candidates in the cluster fields generally fall into four categories: 

\begin{enumerate}
    \item $\sim 32\%$ are compact sources. I provide cutouts for two examples of such galaxies in (1a) and (1b) of Fig.\ref{LBG_cutouts_example} in all HFF bands and stacked images. It can be seen that these resemble the two roundish compact contaminants identified in the previous section, suggesting they could be interlopers from low-z. For instance, some of the compact sources could be globular clusters as suggested by \cite{Shipley2018}, or dwarf galaxies. It is worth mentioning, however, that galaxies of this morphological type also dominate the parallel field sample. Hence, some of such compact galaxies may also be genuine high-z galaxies, leaving their real identities ambiguous. 
    
    \item $\sim 18\%$ of the sample are galaxies that are elongated, including gravitationally lensed arcs. Two examples are provided in (2a) and (2b) of Fig.\ref{LBG_cutouts_example}. For the very elongated lensed arcs, I consider them as safely coming from the high-z Universe. Likewise, for elongated galaxies with a small observed size, these could be disky star-forming galaxies at high-z. But from the big diffuse elongated contaminant identified at $z_{spec}=0.40232$, the larger galaxies in this sample could be bright cluster members with a strong D4000\AA break, possibly as a result of strong dust attenuation. 
    
    \item $\sim 33\%$ of the sample are \textit{big} and \textit{roundish} galaxies visually with a diffuse light profile. Two example cutouts are provided in (3a) and (3b) of Fig.\ref{LBG_cutouts_example}. These galaxies are rare in the parallel field, and given their unexpectedly large size, I suspect these are mostly dim early-type galaxies (likely cluster members) and are the primary source of contamination. 
    
    \item The remaining $\sim 17\%$ of the sample is spurious detection in that no features were visible. Two examples are provided in (4a) and (4b) of Fig.\ref{LBG_cutouts_example}. These may either be very faint galaxies only observable after removing bright cluster galaxies, or fake detections caused by noise. 
    
\end{enumerate}

These results suggest the majority of contaminants are likely to be big, roundish galaxies visually with a diffuse light profile, rather than big elongated or compact galaxies that I identified in the spec-z cross-matching exercise. In the subsequent sections, I will also use a parameter reflecting the concentration of the light profile to support this argument. 

\subsection{color, surface brightness, concentration}
\label{sec_contamination_csbcon}

Previously, I have explicitly used color and surface brightness -- two lensing invariants -- to argue for the existence of contaminants. Here, I take a better look at the color-surface brightness distribution of my analysis sample (previously with only a brutal force $M_{UV}$ cut) and also define a third lensing invariant -- concentration parameter $S$.

\begin{figure}
\centering
\begin{subfigure}{.48\textwidth}
  \centering
  \includegraphics[width=\linewidth]{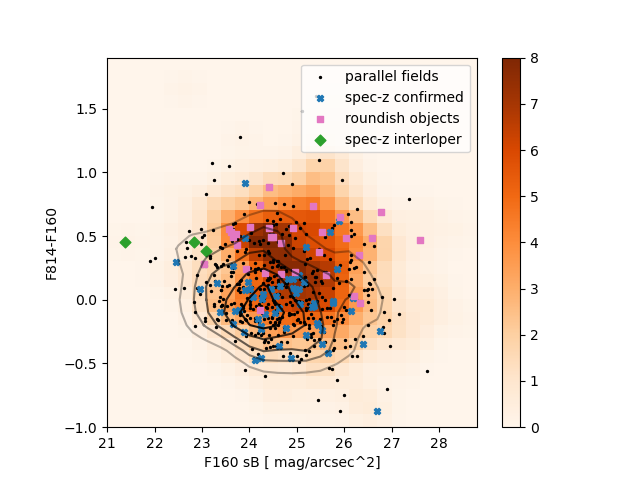}
  \caption{F814-F160 vs F160 band surface brightness.}
\end{subfigure}%
\hfill
\begin{subfigure}{.48\textwidth}
  \centering
  \includegraphics[width=\linewidth]{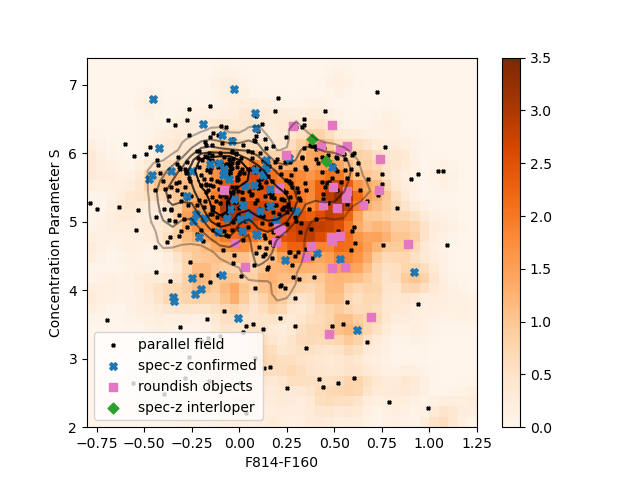}
  \caption{F814-F160 vs concentration parameter S.}
\end{subfigure}
\caption{Diagnostics with color F814-F160, surface brightness in F160W band, concentration parameter $S$. (a) color-surface brightness plots. (b) color-concentration plots. In both plots, black dots are high-z objects in parallel fields dominated by genuine Lyman break galaxies. I also indicate isocontours of their phase space population by different transparency: the less transparent a contour is, the higher the occupation value it corresponds to. I see most cluster field objects are redder in color at the same surface brightness and dimmer in surface brightness at the same color.}
\label{color_sB_con_detailed}
\end{figure}

In Fig.\ref{color_sB_con_detailed}(a), I compare the color-surface brightness distribution (2d histogram background) of $3.5\leq z\leq 5.5$ analysis sample in the cluster fields (known to be dominated by contaminants) with a more robust high-z sample composed of (1) high-z candidates in the parallel fields (black dots), and (2) spec-z confirmed high-z galaxies in the cluster fields (blue crosses). As in Fig.\ref{color_sB_con}, I marked the concentration of the analysis sample in parallel fields by contours of different transparency. Roundish objects previously identified during the visual inspection stage are also shown as pink squares, and they are seen to occupy the same region as most $3.5\leq z\leq 5.5$ candidates in cluster fields. Albeit with a much brighter surface brightness, the three spec-z-wise confirmed contaminants (green diamonds) also have a similar color to most contaminants. Combining these observations, it could be argued that most contaminants have redder colors (possibly owing to an older stellar population) than generic high-z galaxies at the same surface brightness, and are likely to have big, roundish profiles.

To better justify the interlopers are likely with diffuse light profiles, I define a measure of light profile concentration using the ratios between Kron-radius\footnote{Kron-radius is defined as the weighted average radius with weighting being radial surface brightness profile.} and half-light radius. More specifically, I translate this ratio to an estimate of the Sersic index following conversion provided by \cite{Graham2005}, and I denote this translated parameter as the concentration parameter $S$. Given $S$ is defined from ratios, this parameter is by nature a lensing invariant, except for a few galaxies in regions with very strong spatial variation in magnification factors (e.g., near critical curves). 

In Fig.\ref{color_sB_con_detailed}(b), I plot the distribution of $S$ against color F814W-F160W. It could be seen that there is a strong concentration around $ S \sim 5$ among the high-z candidates in the cluster fields, and also among roundish objects that I previously picked out. Parallel field galaxies and spec-z confirmed high-z galaxies in the cluster fields, on the other hand, are seen to concentrate at a slightly higher concentration parameter $S\sim5.5$. This supports my previous statement that contaminants are likely with more diffuse light profiles than genuine high-z galaxies. The three spec-z confirmed contaminants were not seen with a visually diffuse profile. In Fig.\ref{color_sB_con_detailed}(b), this fact is reflected by two galaxies having a concentration parameter $S\sim 6$, and the third one beyond the shown range with $S\sim 9$.

\subsection{Lyman Break Galaxy-like criteria}
\label{sec_LGB_failure}

In literature, UV LFs are often constructed with the aid of Lyman Break Galaxy-like (LBG-like) selection criteria, which are color-based and independent of photo-z measurements. To investigate whether such selection rules could statistically mitigate contaminants in the lensing fields, I repeat the same magnification bias analysis over $3.5\leq z\leq 5.5$ using only LBG-like galaxies selected by the criteria\footnote{We apply the corresponding LBG-like criteria separately to $3.5\leq z \leq 4.5$ and $4.5 \leq z \leq 5.5$ samples, and to both cluster fields and parallel fields. } provided by \cite{Bouwens202203}. 

These LBG-like criteria were found very stringent and preserved only $\sim 34\%$ of the UV bright sample, rendering it hard to re-measure the UV LFs from the parallel fields. In such a scenario, I adopted the blank field UV LFs derived by \cite{Bouwens2021} (not by \cite{Bouwens202203} as they considered lensing fields), which were also based on LBG-like galaxies. For the LBG-like criteria selected sample, the new (average) contamination level was found to be $11.04 \pm 11.79\%$. This is much lower than previously reported and compatible with zero, signaling a potential success. 

\begin{figure}
\centering
\begin{subfigure}{.45\textwidth}
  \centering
  \includegraphics[width=\linewidth]{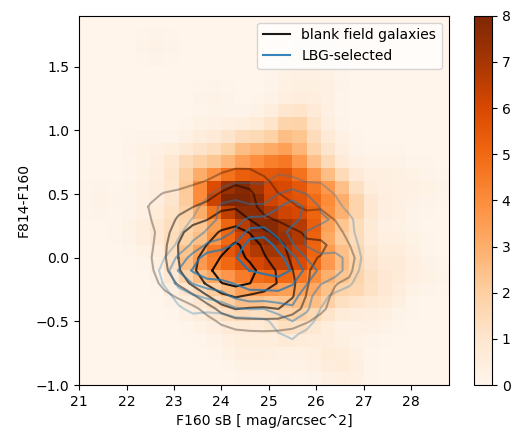}
  \caption{F814-F160 vs F160 band surface brightness}
\end{subfigure}%
\hfill
\begin{subfigure}{.45\textwidth}
  \centering
  \includegraphics[width=\linewidth]{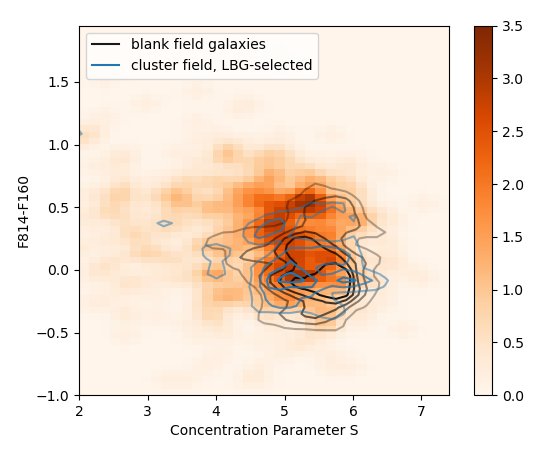}
  \caption{F814-F160 vs concentration parameter S}
\end{subfigure}
\caption{LBG-selected galaxies on (a) color-surface brightness plots and (b) color-concentration plots, with their concentration indicated by contours of different transparency. It could be seen that LBG-criteria selected galaxies are widespread and sit in-between the "safe" (indicated by black contours from parallel fields) and "contaminated" (strong concentration on the background 2d histogram) regions. As such, LBG-like galaxies may still be contaminated by low redshift interlopers. }
\label{color_sB_con_on_LBG}
\end{figure}

To learn better if contaminants are really mitigated, I further examined where the LBG-like criteria selected galaxies sit on color-surface brightness, Fig.\ref{color_sB_con_on_LBG}(a), and on color-concentration plots, Fig.\ref{color_sB_con_on_LBG}(b). It could be seen that the selected galaxies (distributions marked by blue contours) occupy a region in-between the analysis sample in parallel fields (black contours) and the 'contaminants region' from cluster fields (strong concentration in the background 2d histogram). A persistent concentration near $S\sim5$ could also be seen on the color-$S$ plot, and this will later be understood as coming from field interlopers after I individually identified these contaminants (Chap.\ref{chap:contamination_jwst}). These observations then cast doubts on whether LBG-like criteria are truly effective at statistically mitigating interlopers, despite their magnification bias-wise success. Moreover, I caution readers that LBG criteria not only reject interlopers but also some high-z galaxies as well. Correspondingly, while the sample is arguably 'cleaner', it is less complete. In particular, LBG-like conditions select dominantly blue galaxies; hence, UV LF derived based on LBG-like galaxies might not be optimal for tests involving intrinsically faint/red high-z galaxies. The same problem also applies to other statistical selection rules. As such, while it seems well-motivated to define color-surface brightness-concentration parameter selection rules for mitigating interlopers, I refrain from making such an attempt. 

Finally, for the analysis of \cite{Leung2018}, they 'mitigated' the potential contamination issue by restricting to the redshift range of $z\leq 4.75$, as this sample was found to have a lesser degree of overlap with cluster members. As a quick cross-check, I examined the contamination level over $4.8\leq z \leq5.5$, and found it to be $12.50\pm 28.41\%$ (i.e., consistent with zero) when using 0.0025 e/s for defining exclusion regions as \cite{Leung2018} did. When using 0.005 e/s for defining exclusion regions as I previously did, the contamination level was instead found to be $46.81 \pm 21.86\%$. The large difference between the two reported values then suggests that contaminants (especially cluster members) are dominantly removed by \cite{Leung2018} with more stringent exclusion regions. Nonetheless, as will become clear in a later chapter (Chap.\ref{chap:contamination_jwst}), field galaxies also contribute significantly to the contamination. These field galaxies have a more uniform distribution across the FOV, and hence cannot be removed with more stringent exclusion regions. Finally, the large contamination level reported when defining exclusion regions with 0.005 e/s demonstrates that a redshift cut cannot be used to mitigate contaminants. Similarly, other redshift-motivated selection rules (for instance, \cite{Bouwens202205} excluded the anticipated confusing redshift range of each galaxy cluster for analysis) for lensing field high-$z$ analysis could also be problematic, and I caution readers in applying such selection rules.

\subsection{JWST comes to rescue?}

The best hope to individually mitigate interlopers thus lies in obtaining deeper observations near rest frame $\sim 4000\AA$ of the high-z galaxies, such that their Balmer breaks are observed. This could be achieved, for instance, with the aid of the newly launched JWST. Examples of such helpful NIRCam observations on HFF clusters include those from PEARLS (GTO 1176, \cite{Windhorst2022} targets include M0416, M1149, A2744), UNCOVER (GO 2561, \cite{2022arXiv221204026B}, targets A2744) and CANUCS (GTO 1208, \cite{2022PASP134b5002W}, targets include M0416, M1149) team. These NIRCam observations could further be complemented with spectroscopic observations on these clusters, such as those by GLASS (ERS 1324, \cite{Treu2022}) and the CANUCS team for better lens model construction and spectroscopic mitigation of the contaminants. In Chap.\ref{chap:contamination_jwst}, I will demonstrate that the combination of deep HST with deep JWST images (from the PEARLS team) indeed allows for individual mitigation of interlopers, as the observed Balmer breaks of high-z galaxies are strong enough to be distinguished from low-z interlopers. 


\section{For Other photometric catalogs} 
\label{sec_contamination_othercata}

\begin{figure}
     \centering
     \includegraphics[width=0.8\textwidth]{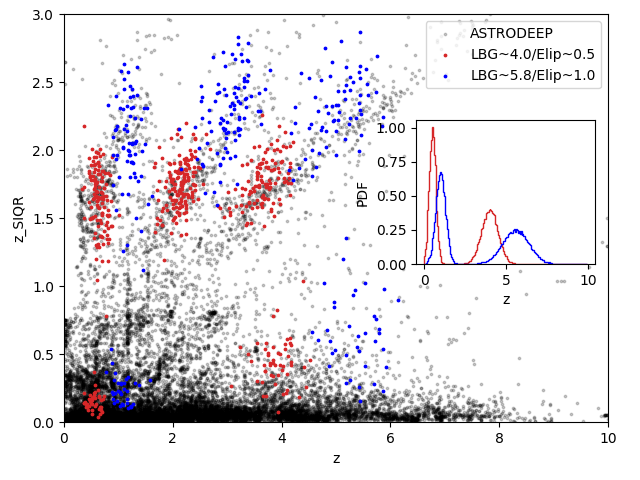}
     \caption{Distribution of ASTRODEEP sources (black) with simulated galaxies on median photo-z versus SIQR plane. Red (Blue) corresponds to the population with SED that could either be interpreted as from a Lyman break galaxy at $z\sim$4.0(5.8) or from dim non-star-forming elliptical galaxies at $ z\sim$0.5 (1.0), respectively. The PDF from which photometric redshifts are drawn for each population is shown in the insert with the same color. I observe three distinct stripes in the high SIQR regions, where the middle stripe corresponds to interlopers in the ASTRODEEP catalog misidentified not to z$\sim$4 but to z$\sim$2 owing to their way of reporting photo-z.}
        \label{contaminating_median}
\end{figure}

The analysis in previous sections has primarily focused on S18 catalogs. I now return to the excess identified in Fig.\ref{excess_allcata} for ASTRODEEP catalogs and C15 catalogs, and investigate the corresponding contamination levels.

ASTRODEEP catalogs are available for four of the HFF clusters: \cite{Merlin2016, Castellano2016} for A2744 and M0416, and \cite{DiCriscienzo2017} for M0717 and M1149. C15 catalogs were provided by \cite{Coe2015}, and are available for five of the HFF clusters except for the A370 field (whose near-IR band observations were only partially complete at the time). 

As with Shipley catalogs, for analysis, I consider only sources with a reliable photo-z estimation. This was achieved by selecting \texttt{RELFLAG} = 1 source for ASTRODEEP catalogs, where this condition refers to having reliable photometry in at least five of the HST bands so that the galaxy's photometric redshift is well-constrained. For C15 catalogs, I selected sources with photo-z \texttt{odds} parameter $>$ 0.5; these correspond to galaxies with a fairly strong peak in their posterior redshift distribution about the reported redshift. 

\begin{figure}
    \centering
    \includegraphics[width=0.9\linewidth]{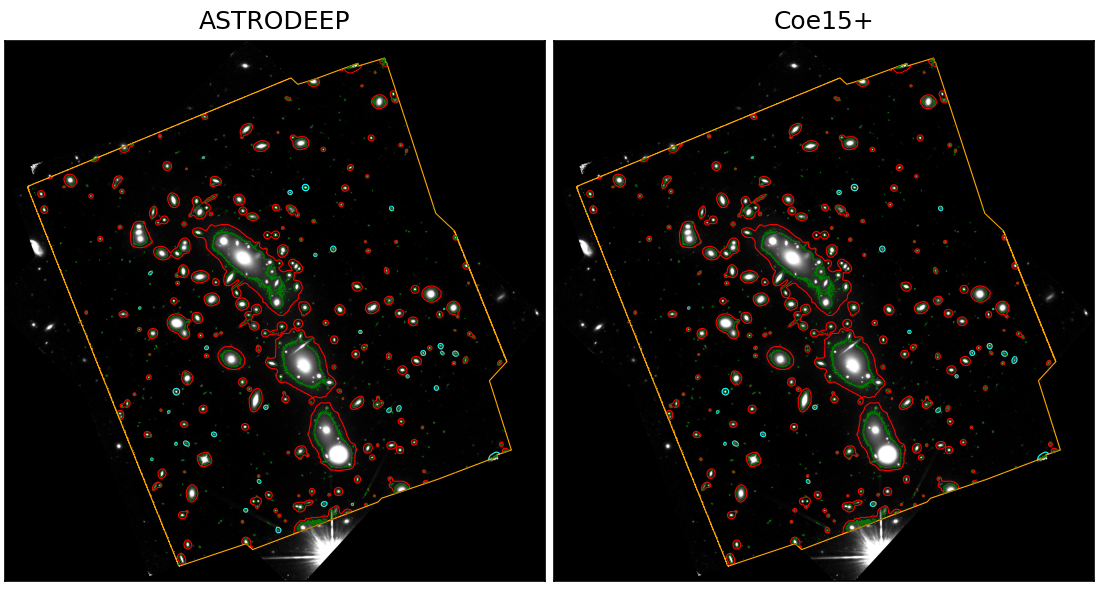}
    \caption{Re-selection of bright regions containing a single $z_{phot}>1$ galaxy for ASTRODEEP and C15 catalogs when defining exclusion regions.}
    \label{exclusion_ASTRODEEP_C15}
\end{figure}

\begin{figure}
\centering
\begin{subfigure}{.48\textwidth}
  \centering
  \includegraphics[width=\linewidth]{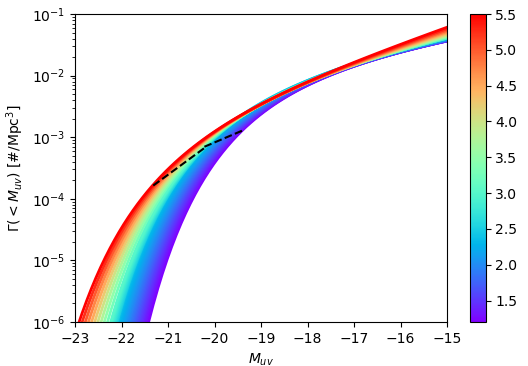}
  \caption{For ASTRODEEP catalog}
\end{subfigure}%
\hfill
\begin{subfigure}{.48\textwidth}
  \centering
  \includegraphics[width=\linewidth]{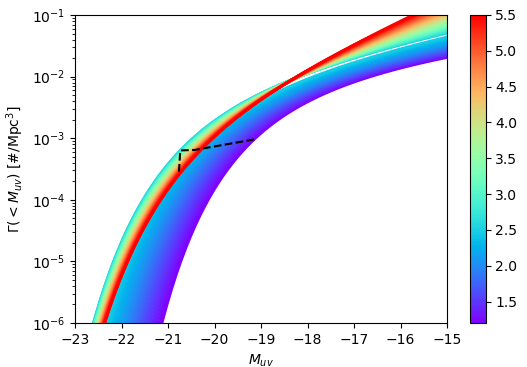}
  \caption{For C15 catalog}
\end{subfigure}
\caption{UV LF-redshift relations separately determined for ASTRODEEP and C15 catalogs. }
\label{CLF_other_cata}
\end{figure}

\begin{figure}
\centering
\begin{subfigure}{.48\textwidth}
  \centering
  \includegraphics[width=\linewidth]{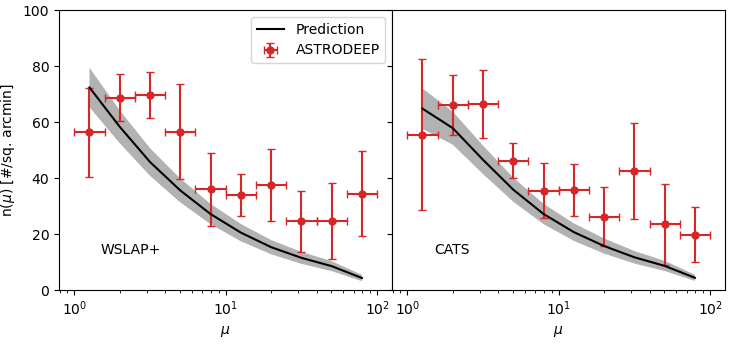}
  \caption{For ASTRODEEP catalog}
\end{subfigure}%
\hfill
\begin{subfigure}{.48\textwidth}
  \centering
  \includegraphics[width=\linewidth]{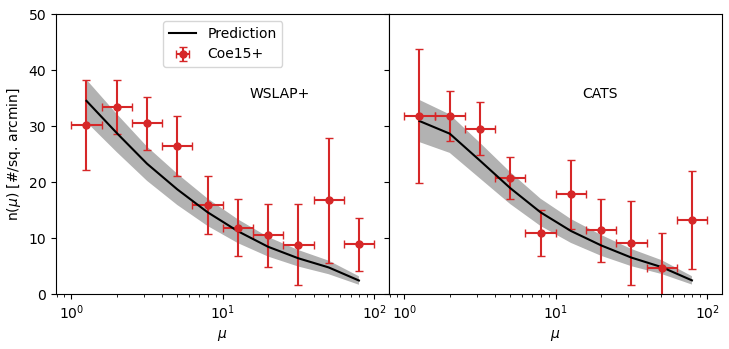}
  \caption{For C15 catalog}
\end{subfigure}
\caption{Magnification bias results over $1.2\leq z\leq 2.4$ for ASTRODEEP and C15 catalog. }
\label{neg_other_cata}
\end{figure}

\begin{figure}
\centering
\begin{subfigure}{.48\textwidth}
  \centering
  \includegraphics[width=\linewidth]{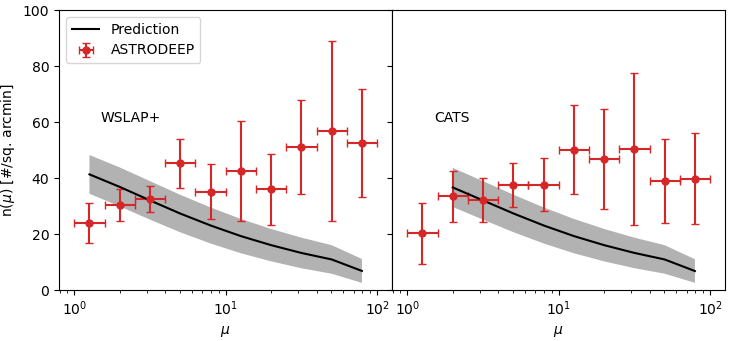}
  \caption{For ASTRODEEP catalog}
\end{subfigure}%
\hfill
\begin{subfigure}{.48\textwidth}
  \centering
  \includegraphics[width=\linewidth]{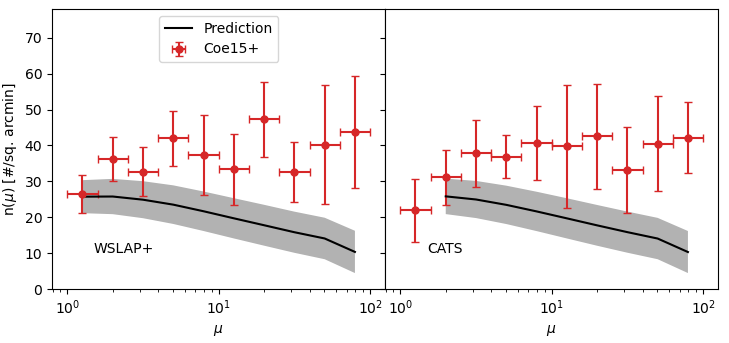}
  \caption{For C15 catalog}
\end{subfigure}
\caption{Magnification bias over $3.5\leq z\leq 5.5$ for ASTRODEEP and C15 catalog. }
\label{pos_other_cata}
\end{figure}

I follow the same magnification bias method to estimate contamination levels as in Sec.\ref{Method}, but using UV LF-redshift relations and data completeness thresholds separately derived for each catalog. Owing to the different photo-z estimates for galaxies, new and separate exclusion region maps were also needed to incorporate the new selection of $z_{phot}>1$ regions, as shown in Fig.\ref{exclusion_ASTRODEEP_C15}. Likewise, when breaking the wide redshift interval of $1.2\leq z\leq 2.4$ and $3.5\leq z\leq 5.5$ into smaller redshift intervals for magnification bias prediction, the redshift interval sizes were separately determined from respective $z_{NMAD}$ for ASTRODEEP and C15 catalogs. I present the separately determined UV LF-redshift relations in Fig.\ref{CLF_other_cata}. The magnification bias result over the redshift range $1.2\leq z\leq 2.4$ is presented in Fig.\ref{neg_other_cata}, and
magnification bias result over redshift range $3.5\leq z\leq 5.5$ is presented in Fig.\ref{pos_other_cata}.

In summary, for both ASTRODEEP and C15 catalogs, I arrived at the same conclusion as with the Shipley catalogs: while distribution of cataloged densities are well reproduced by negative bias over $1.2\leq z\leq 2.4$, a large population of contamination is needed over $3.5\leq z\leq 5.5$ to explain the increasing trend toward higher magnification regions in the derived densities. 

The contamination level for the C15 catalog was estimated to be $52.68 \pm 11.99\%$ when using CATS models, and $46.39 \pm 10.84\%$ when using WSLAP+ models. This is smaller than for the S18 catalog, but still very severe. Moreover, I recall that from Fig.\ref{excess_allcata} there was no sign of any excess for the cluster A1063. I cross-checked the contamination level for this field alone, and it was found to be $13.28 \pm  21.72 \%$ when based on WSLAP+ models and $18.00 \pm 22.37 \%$ when based on CATS models, both of which are consistent with no contamination. The reason for this field having no contamination, however, is unclear.

For the ASTRODEEP catalog, I found a similar contamination level of $52.15 \pm 14.47 \%$ and $44.15 \pm 14.16 \%$ when adopting CATS and WSLAP+ models, respectively. But unique to this catalog, large deviations between the observed and predicted densities were also observed over $1.2\leq z\leq 2.4$, even in low magnification regions. Moreover, the predicted densities in low magnification regions are seen to overwhelm the observed densities over $3.5\leq z\leq 5.5$. This observation contradicts the behavior observed for S18 and C15 catalogs, where the predicted levels are overall in good agreement with the observed densities in the low magnification bins. 

Both of the aforementioned phenomena were later understood to result from the specific way ASTRODEEP catalogs report their photo-zs. In particular, ASTRODEEP catalogs adopt the median from six separate photo-z estimates as the output redshift, with a semi-interquartile range (SIQR) as an indicator of photo-z uncertainty. Such a photo-z reporting method then faces problems with galaxies (e.g., interlopers) with two equally probable photo-z outputs. In such a scenario, three of the photo-z outputs may be in the low-z range, whereas the other three are in the high-z range. As a consequence, the reported median redshift for such a galaxy would be at an intermediate value. 

In Fig.\ref{contaminating_median}, I demonstrate this issue does occur for the ASTRODEEP catalog by simulating potential photo-z and SIQR outputs for interloper-like galaxies. Red data points correspond to simulated galaxies that could either be interpreted as a $z \sim 0.5$ galaxy, or as a $z \sim 4$ galaxy, as indicated by the probability distribution function\footnote{These PDFs are composed of two Gaussian distributions, both with integrated probabilities of 0.5, that correspond to the two equal likelihood photo-z outputs.} in photo-z outputs shown in the insert figure. Blue data points correspond to simulated galaxies at slightly higher redshifts, and may be interpreted as either $z \sim 1$ galaxies, or $z \sim 6$ galaxies. For comparison, the reported redshifts and SIQRs of galaxies in the ASTRODEEP catalogs are also plotted as black points in the background. It could be seen that there are three distinct stripes in the high SIQR regions in both my simulated outputs and in the ASTRODEEP catalog. The middle strip near $z\sim 2$ corresponds exactly to the case where interlopers are inaccurately assigned to an intermediate redshift range. Correspondingly for ASTRODEEP catalogs, the contamination level at $z \sim 4$ is lower than actual, whereas an excess of galaxies (from interlopers) at $z \sim 2$ would be observed, matching exactly the observations from Fig.\ref{neg_other_cata} and Fig.\ref{pos_other_cata}.

\section{Some concluding remarks}
\label{conclusion_section}

To conclude, in this chapter, I have demonstrated that photometric redshift catalogs constructed in HFF cluster fields suffer from very severe contamination. Aided with magnification bias, I was also able to provide quantitative estimates of the contamination levels in S18, ASTRODEEP, and C15 catalogs. This way, I have shown that the contamination levels are generally high $\sim 50\%$, reaching as high as $\sim 60\%$. Through diagnostic checks, contaminants were found to be intrinsically redder than genuine high-z galaxies, and likely have more diffuse light profiles. The dominant source of contaminants is likely dim cluster members, given their clustering trend to the faint-end extension of the cluster Red Sequence, and also towards the center of galaxy clusters. I also demonstrated that the presence of contaminants would lead to profound impacts on the UV LFs, especially at the faint end (unphysical turn-ups, washing out turnovers). Statistical selection rules such as LBG-like criteria were also argued to be non-ideal for mitigating contamination, with the best strategy being to individually mitigate low-z interlopers by acquiring additional measurements (especially measurements extending the wavelength coverage). For the redshift range ($ z\sim4$) that I focused on in this chapter, deep JWST observations covering the rest frame Balmer breaks of $z\sim4$ galaxies will be particularly helpful. In the later chapter (Chap.\ref{chap:contamination_jwst}), I will take advantage of the deep JWST measurements (on the field of M0416) acquired by collaborators from the PEARLS team, to demonstrate that individual mitigation of interlopers is indeed possible. In addition, with the aid of machine learning and an appropriate training set, an 'interloper identifier' with excellent performance could be constructed for individually picking interlopers in the remaining HFF catalogs, alleviating the need for separate time-intensive JWST observations. 


\chapter{Undestanding Contaminants with JWST}
\label{chap:contamination_jwst}

In this chapter, I demonstrate that the combination of deep HST and JWST observations allows for high-z galaxies and low-z contaminants to be individually identified. The stellar properties of identified interlopers will also be investigated. To achieve this goal, I first construct the JWST+HST combined photo-z catalog for the field of M0416 using \texttt{GNUastro} (detailedly introduced in Chap.\ref{chap:gnuastro}) in Sec.\ref{catalog_for_M0416}. Along with the discussion, I also demonstrate that the specific source detection philosophy of \texttt{GNUastro} allowed my constructed catalog to be more data-complete, in particular, reaching $\sim 1$ mag fainter than previously achieved by \cite{Shipley2018}. This constructed catalog is subsequently used to identify low-z interlopers within the previous S18 analysis sample over $3.5\leq z\leq 5.5$ in Sec.\ref{sec_identify_interloper_JWST}, followed by detailed analysis on the stellar properties of identified interlopers in Sec.\ref{sec_interloper_stellar_property}. I will argue interlopers are dominantly dwarf cluster members with quenched star-formation, and also (potentially star-forming) dwarf galaxies in the field with an old stellar population. Combining with the previous analysis on S18 observables (Chap.\ref{chap:contamination}), I conclude in Sec.\ref{sec_interloper_machine_learning} that S18 measurements encode intrinsic differences between genuine high-z galaxies and low-z interlopers. These differences could be learned with machine learning methods, such that an "interloper identifier" with excellent performance (100\% accuracy) could be constructed. The constructed interloper identifier was subsequently used to individually pick out interlopers in the other S18 catalogs (for fields without supplementary JWST observations), and hence, paved the way for a more reliable test on the faint-end UV LFs in all HFF lensing fields in Chap.\ref{chap:UVLFtesting}.

\section{JWST+HST combined photo-z catalog for M0416} 
\label{catalog_for_M0416}


\subsection{Relevant JWST+HST Data} 
\label{Sec_relevant_JWSTHST_data}

M0416 is one of the most extensively studied strong lensing galaxy clusters by the HST and serves as a primary target for two deep lensing surveys: HFF and CLASH. Existing HFF and CLASH observations on this cluster involve 16 broadband filters, covering a wavelength range of 0.3-1.6 $\mu m$. But three of the shortest wavelength filters (F225W, F275W, F336W) from the CLASH program were found to suffer from large background variations across ACS detectors. These filters were thus excluded from catalog construction to avoid inconsistent flux measurements. JWST observations on M0416 were acquired by the PEARLS team \citep{Windhorst2022}, consisting of three epochs of observations in eight broadband filters. JWST observations extended the wavelength coverage to $\sim 4.4\, \mu m$, allowing for the detection of rest-frame Balmer breaks of high-z galaxies (up to $z\leq 10$) and hence better determinations of photo-zs. In total, I gathered images in 21 filters from HST and JWST observations, all of which are with a pixel resolution of 0.03 arcsec and aligned to \textit{Gaia} Data Release 3 (DR3) astrometry\footnote{PEARLS observations were aligned to \textit{Gaia} DR3 during the data reduction stage. For existing HST images before \textit{Gaia} DR3 became available, I aligned them to PEARLS observation using astrometry software \texttt{tweakreg}. The alignment was also checked to be at a sub-pixel level. }. 

When imaging the galaxy cluster, both the PEARLS and HFF programs conducted simultaneous observations of a nearby blank region (parallel field) away from the cluster center. However, I do not create a catalog for these parallel fields, as my analysis primarily focuses on identifying low-redshift interlopers (likely cluster members) and strongly lensed high-redshift galaxies surrounding the galaxy cluster. Additionally, the PEARLS parallel fields do not overlap with the HFF parallel fields. As a result, potential contaminants from lower redshifts, albeit much less populated than in cluster fields, could not be individually mitigated.

To optimize the detection of faint galaxies up to high redshifts ($z\leq 10$), I adopt the inverse-variance weighted stack of four longest-wavelength filters (F277W, F356W, F410M, and F444W) from JWST as the detection image. Before stacking, images from different filters are all PSF-matched to the HST F110W band, which is the shallowest image available from the CLASH program. Following the detailed introduction on \texttt{GNUastro} in Chap.\ref{chap:gnuastro}, for my catalog construction, I carefully fine-tuned \texttt{NoiseChisel} and \texttt{Segment} parameters to (1) best detect faint galaxies as completely as possible from the detection image, and (2) best separate faint galaxies from their bright neighbors. The input parameters were also adjusted such that the unwanted splitting of galaxies into smaller pieces is minimized.  Photometries of the final detected galaxies were extracted from PSF-matched (to CLASH F110W band), smooth background-subtracted images of individual filters. The removal of smooth backgrounds was achieved with \texttt{NoiseChisel}, where I adjusted the input parameters separately for each image for a best practice.

\begin{figure}
     \centering
     \includegraphics[width=0.8\textwidth]{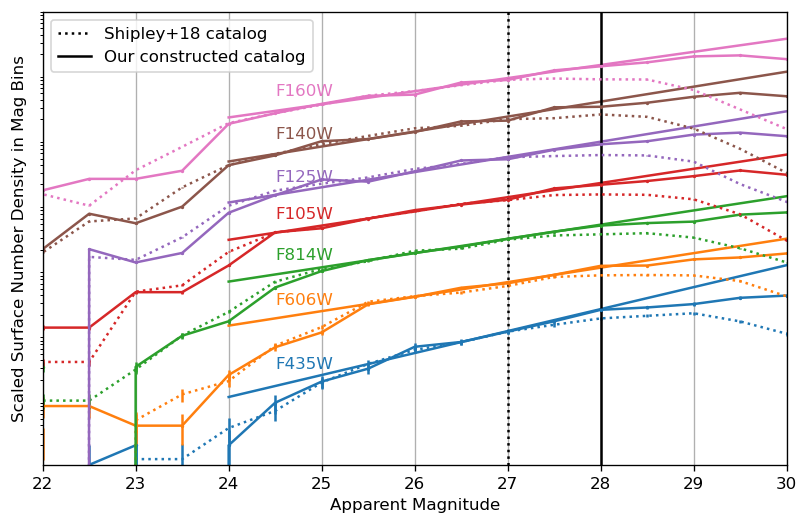}
     \caption{Surface number density of cataloged galaxies in M0416 lensing field as a function of apparent magnitudes for different HFF bands from both my constructed catalog using \texttt{GNUastro} (solid lines) and S18 catalog (dotted lines). These densities are also scaled to be non-overlapping between different HFF bands. At magnitudes brighter than $m<24$, there is a shortage in observed bright galaxies for both my and S18 catalogs. This shortage stems from the exclusion region I imposed, which was necessary to achieve spatial uniformity in the detection thresholds (see the discussion in Sec.\ref{sec_data_complete}). Over apparent magnitudes $25<m<27$, all obtained distributions could be seen well approximated by power laws, with their respective continuation to fainter magnitudes indicated by straight lines of the same colors. Starting from an apparent magnitude of 27 (vertical dotted line), the observed surface number densities from the S18 catalog deviate from power-law behavior. As will be further discussed, such a deviation indicates detection inefficiency and hence data incompleteness. The same phenomenon, however, is delayed to a magnitude of 28 (solid vertical line) for my \texttt{GNUastro} constructed catalog, demonstrating it is more data complete. } 
    \label{compare_threshold}
\end{figure}

\subsection{Data Completeness}
\label{Sec_data_completeness}

To demonstrate my final constructed catalog is more data complete than S18, in Fig.\ref{compare_threshold}, I examine the surface number density of galaxies cataloged as a function of apparent magnitude. For each HFF filter, I plot the obtained distribution as a solid curve for my catalog, and as a dotted curve for the S18 catalog. For a more direct comparison, both of the obtained distributions are scaled to the same level, and could be seen as largely consistent with each other over magnitudes $m<27$. The shortage of bright galaxies observed at magnitudes $m<24$ is the result of the exclusion region imposed, where (see the discussion in Sec.\ref{sec_data_complete}) the exclusion region is needed to ensure spatial uniformity in the detection thresholds. Here, the imposed exclusion region is defined at a brightness level of 24.08 mag/arcsec$^2$ (same as in Sec.\ref{sec_data_complete}), but is defined on the JWST F200W band image instead of on the HFF F160W band. This change in choice was motivated by the observations that the F200W band was more constraining\footnote{This was found by investigating the fraction of FOV area left unremoved, and the F200W band preserved the least FOV area of all other filters. From the Integrated Galaxy Light analysis by \cite{Windhorst2022}, it is also known that lights from galaxies are brightest in the F200W band, further justifying the choice of the F200W band for optimal selection of bright regions.} than the F160W band, and hence better at defining bright regions. 

Over magnitudes $25<m<27$, all obtained distributions could be seen well approximated by power laws as indicated by the solid straight lines. I notice the observed distribution starts to deviate from power law at around $m\sim 27$ (indicated by vertical dotted line) for the S18 catalog, but a rapid drop in densities is not seen until $m > 28.5$. The same trend is also true for surface number densities measured from my constructed catalog, but both features were found to occur at fainter magnitudes. In particular, the deviation from power law starts at around $m\sim 28$ as indicated by the vertical solid line, and the rapid drop in density occurs at $m>29.5$. As I will discuss more in Sec.\ref{sec_injection_recovery}, the observed deviation from power law behavior is the sign of detection inefficiency. Correspondingly, the vertical dotted and solid lines are, in fact, data completeness thresholds for S18 and my catalogs, respectively. As evident, my constructed catalog reached $\sim1$ mag fainter\footnote{The same could also be concluded when adopting the rapid drop in density as the sign of data incompleteness.} than previously achieved by S18, signaling the strength of cataloging with \texttt{GNUastro}.

\subsection{Photo-z Determination}

Given measured photometries, I fit for the photometric redshift of galaxies using template fitting software EAZY \cite{Brammer2008}. Prior to photo-z fitting, measured photometries were corrected for galactic extinction using the python package \texttt{extinction} and \cite{Fitzpatrick1999} extinction function with $R_V = 3.1, A_V = 0.112$ as measured from \cite{SchlaflyFinkbeiner2011}. I also correct for intergalactic medium attenuation using \cite{1402.0677} model with EAZY. For templates, I adopt the ones provided by \cite{Larsons2022}. This set of templates extended upon the existing Flexible Stellar Population Synthesis templates by \citep{Conroy2010FSPS} to better incorporate bluer rest-frame UV color and stronger emission lines anticipated for $z>8$ galaxies. In \cite{Larsons2022}, four new sets of templates generated with BPASS and CLOUDY codes were introduced. Depending on the combination, the additional templates were commented as useful for $4<z<7$ galaxies (set 1+set 3), or $z>8$ galaxies (set 1+set 4). The combination (set 1+set 2) with full Lyman-$\alpha$ emission lines was provided for reference and commented as not likely applicable. For my future analysis, I am interested in both the redshift range facing severe misidentification issues ($3<z<6$) and also a higher redshift range ($6<z<10$) to better probe the effects from $\psi$DM. Hence, I will use the combination set 1+set 3 when picking out interlopers from S18 purported high-z galaxies with my new JWST measurements; whereas I use the combination set 1+set 3+set 4 when testing for any faint-end turnover in $z>6$ UV LFs. 

Shortly after the in-flight operation of JWST, a great deal of attention was spent on cross-checking the precision of JWST's flux calibration \cite{2209.03348, 2209.06547, 2308.10575, 2311.13754}. This is a critical aspect, as any inconsistency between JWST's calibration with existing equipment (e.g., HST) will lead to inconsistent photometries and hence impede the inference of galaxy properties (including photometric redshifts). Large systematic offsets between HST and JWST magnitudes were also found to exist in the earlier stage of JWST operation. What followed was a period of frequent updates on the flux calibration pipeline of JWST. Consequently, PEARLS images that I rely on for catalog construction were also frequently updated.  

\begin{figure}
    \centering
    \includegraphics[width=0.8\textwidth]{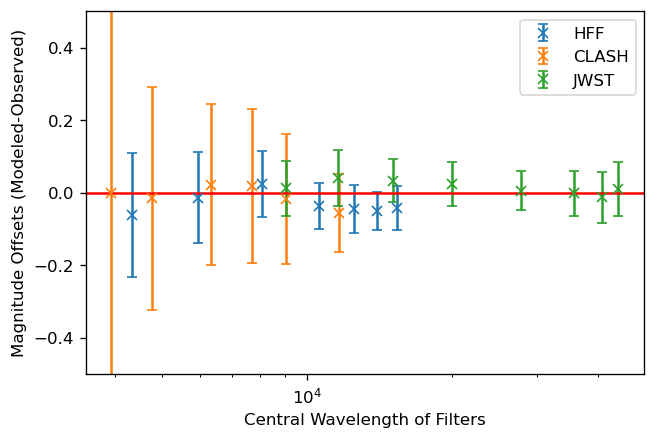}
    \caption{ Median offsets between the modeled (given best-fit spectra fixing to respective spec-zs) and the observed magnitudes for 423 secure spectroscopic redshift galaxies in different HST (blue for HFF and orange for CLASH) and JWST (green) bands. Error bars plotted correspond to the semi-interquartile range (SIQR), reflecting the spread in the distributions of magnitude offsets about the median. The plotted error bars for CLASH bands are seen to be systematically larger than for HFF and JWST bands, reflecting that CLASH images are much shallower than the other images. It is reassuring to observe that all reported median offsets are consistent with zero, given the plotted error bar. At around wavelength 10000\AA, however, there appears to be a small ($\sim 0.086$ mag) but systematic offset in JWST with HFF. This observation then motivated for zero-point correction to achieve better photo-z determinations.}
    \label{zeropoint_correction}
\end{figure}

\begin{figure}
    \centering
    \includegraphics[width=0.95\textwidth]{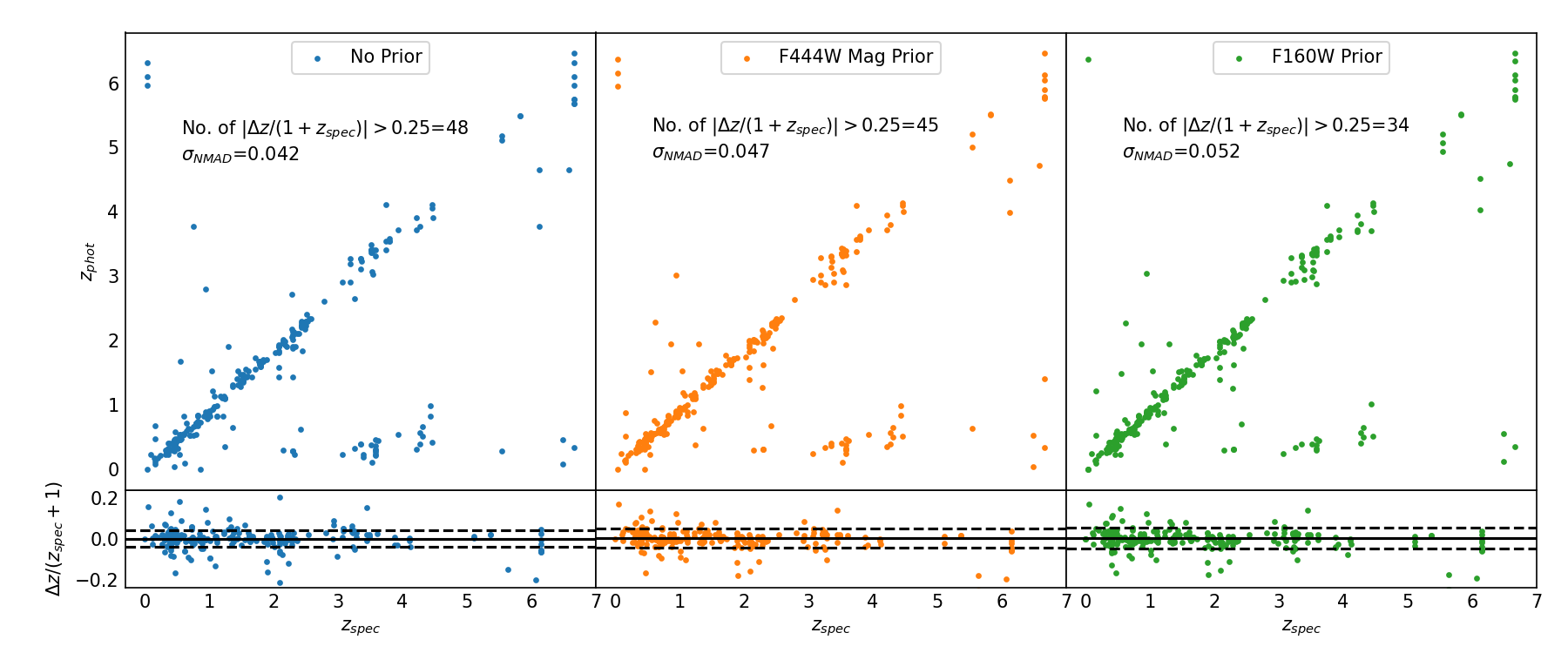}
    \caption{Comparing the photo-z performance against respective spec-z when using no magnitude prior(left panel), magnitude prior on F444W band (middle panel), and magnitude prior on F160W band (right panel). While adopting no magnitude prior leads to the smallest scatter in determined photo-z about respective spec-z (reflected by the smallest $\sigma_{NMAD}$), it also leads to more galaxies wrongly identified to a different redshift range (reflected by the number of galaxies with normalized deviation from spec-z $>0.25$). To achieve more accurate photo-z determinations, I thus compensate for the performance in $\sigma_{NMAD}$, and use a magnitude prior on the F160W band with the least number of galaxies wrongly identified to a different redshift range. }
    \label{compare_prior}
\end{figure}
 
Following the above discussion, I was then motivated to independently test for consistency in the measured magnitudes from HST and JWST PEARLS images. For this, I gathered a total of 423 secure spectroscopic redshift galaxies from \cite{Jauzac2014, 1402.3769, Balestra2016ApJS, Caminha2017, Shipley2018, Vanzella2021} with a one-to-one match in my constructed catalogs. For each of these spec-z galaxies, a best-fit model spectrum was obtained with EAZY\footnote{Here I used set 1+set 3 of \cite{Larsons2022} as template extension, but the same magnitude offsets and hence zero point correction were also obtained when using set 1+set 3+set 4 (relevant for future UV LF faint-end test above $z>6$) as template extension. } at their respective redshift. The median magnitude offsets between modeled and observed magnitudes in different bands are then investigated in Fig.\ref{zeropoint_correction}, with error bars corresponding to semi-interquantile range (SIQR, reflecting the spread of distribution about the median). I used different colors for JWST, HFF, and CLASH filters, and naturally, the spread in magnitude offsets was found to be largest for the shallowest CLASH images. Reassuringly, all of these magnitude deviations are consistent with zero given plotted error bars, but a small ($\sim 0.086$ mag) systematic offset was found to exist between HFF and PEARLS images over filters covering the wavelength range of 1-2$\mu m$. For any subsequent photo-z determination, I thus correct for these magnitude offsets to better determine photometric redshifts. Correspondingly, the plotted error bars of magnitude offset are also added to the uncertainty of measured photometries. 

For my photo-z fitting, I also adopted a prior on the F160W band magnitudes as commonly done in literature, the effectiveness of which is demonstrated in Fig.\ref{compare_prior}. In Fig.\ref{compare_prior}, I compare the scatter of fitted photo-zs about their respective spec-z when applying no magnitude prior (left), a magnitude prior on F444W\footnote{The F444W band magnitude prior was provided by \cite{Giorgio} based on his cosmological simulation.} band (middle) and a magnitude prior on the F160W band (right). The scatter is parameterized by the normalized median absolute deviation $\sigma_{NMAD}$, which is presented explicitly in top panels, as well as in bottom panels by the spacing between dashed lines with solid line (median offsets). It can be seen that the scatter about spec-z is similar for all three choices on prior, with applying no prior performing the best. I also notice, however, that without a magnitude prior (or with a F444W band prior) leads to more $z\sim 4$ galaxies being mistaken as low-z galaxies. This then motivated us to adopt the magnitude prior on F160W band with slightly poorer performance on $\sigma_{NMAD}$, but having least amount of photo-z significantly deviating (i.e., with $|z_{phot}-z_{spec}| > 0.25(1+z_{spec})$) from their respective spec-zs.

\subsection{Source Specific Detection (In-)Efficiency}
\label{sec_injection_recovery}

I now revisit the deviation from power law behavior observed in Fig.\ref{compare_threshold} for the surface number density of galaxies. To examine if this was related to detection inefficiency (or data incompleteness), I injected artificial galaxies on the detection image and investigated the recovery rates at different apparent magnitudes. The source injection was achieved with the \texttt{GNUastro} program \texttt{astmkprof}, allowing us to specify the injection position, light profile (here I adopted Sersic profile), profile parameter (Sersic index, effective radius), axis ratio, position angle, and finally the injected magnitude for each mock galaxy. 


For artificial galaxies to have morphologies as real as possible, I took advantage of three shallower parallel fields (from PEARLS) accompanying cluster field observation to obtain a large unlensed sample. In doing so, I used the same \texttt{GNUastro} input parameters for source detection in the parallel fields as in the cluster field to ensure consistency. The obtained unlensed sample was then used to measure the probability distribution of different parameters\footnote{As brighter galaxies will also visually appear larger, I separately determined the distribution of semi-major axis in different magnitude bins.}. For Sersic profiles, I assumed a simplified log-normal distribution for the Sersic index with mean 0 and spread of 0.45 in $\log(n)$. This distribution was chosen to mimic well the Sersic index distribution reported by \cite{1903.09068, 2210.14713}. And given a Sersic profile (of index $n$), the respective effective radius could be determined analytically from the target semi-major axis by noting semi-major axis is no other than the flux-weighted average radius (in units of effective radius). 

Artificial galaxies generated with aforementioned methods were then injected both onto my detection image to define their respective clump segmentation, as well as onto F150W (chosen as an example, also relevant for future $6\leq z\leq 10$ UV LF analysis) band image for retracting magnitudes. To avoid overlapping with existing detections, I only inject artificial sources in regions away from existing clump segmentation. And I split the whole artificial sample into 15 sub-sets with small enough sample size, such that \texttt{GNUastro} parameters for source detection would not be altered owing to source over-injection\footnote{Over-injecting galaxies onto blank regions may alter the sky-detection separation and hence demand new optimal parameters for \texttt{NoiseChisel}.}. For each sub-sample, I also followed the galaxy number counts as observed by \cite{Windhorst2022} in the F150W band to determine the number of galaxies to be injected in different magnitude bins. Such a practice best mimics real observations and hence is most relevant for understanding the detection inefficiency of real galaxies. 

\begin{figure}
\centering
\begin{subfigure}{.46\textwidth}
  \centering
  \includegraphics[width=\linewidth]{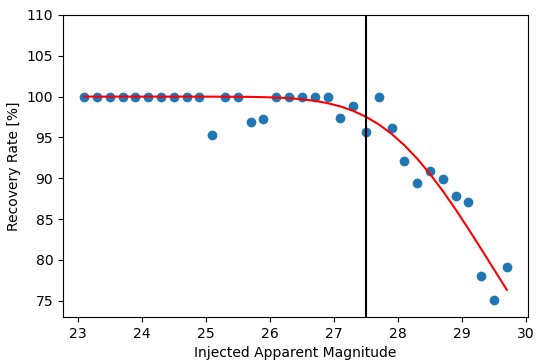}
\end{subfigure}
\hfill
\begin{subfigure}{.49\textwidth}
  \centering
  \includegraphics[width=\linewidth]{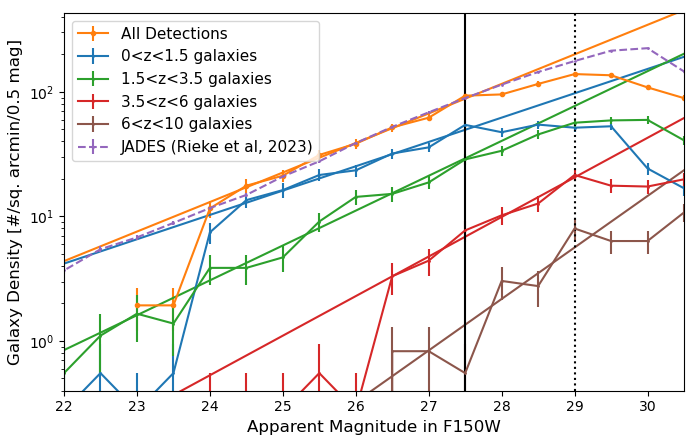}
\end{subfigure}%
\caption{\textit{Left}: Recovery rates (blue data points) of the injected mock galaxies (onto JWST F150W band image) as a function of injected magnitudes. The red solid line indicates the overall trend and is found to decline significantly starting from an apparent magnitude of 27.5 (vertical black line). \textit{Right}: Observed surface number densities of all detected galaxies (orange distribution) in different F150W band magnitude bins. It could be seen that starting from the same magnitude of 27.5 (vertical black line), the orange distribution drops below the power-law behavior (indicated by orange straight line) seen over $24\leq m\leq 27.5$. The observation from the left panel then indicates this deviation from power-law behavior is the result of detection inefficiency, i.e., indicates data incompleteness. In the right panel, I also separately plotted the observed surface number densities of $z<1.5$ (in blue), $1.5<z<3.5$ (in green), $3.5<z<6$ (in red), and $6<z<10$ (in brown) galaxies in different magnitude bins. Interestingly, the blue and green distributions maintain a similar behavior as the orange, whereas the deviation from power-law behavior is delayed to the magnitude of 29 (vertical dotted line) for red and brown distributions. As I explain in the main text, this observation suggests that galaxies above $z>3.5$ are likely to be more compact and hence easier to detect than brighter and more diffuse galaxies at $z<3.5$. This point will be further made clear with Fig.\ref{missed_fraction}. Finally, for reference, I also plotted the observed galaxy counts in the F150W band (purple dashed curve) from the JADES survey for the Hubble Ultra Deep Field. It could be seen that my observed orange distribution largely coincides with theirs over $24\leq m\leq 27.5$. This observation suggests that the galaxies I detected (and their extracted magnitudes) using \texttt{GNUastro} are largely consistent with JADES detections in the deep blank HUDF field, giving confidence to the performance of \texttt{GNUastro}. }
\label{completeness_recover}
\end{figure}

The recovery of injected sources was also separately done for each sub-sample, and F150W band magnitudes were extracted using the respective clump segmentation defined on the detection image. The left panel of Fig.\ref{completeness_recover} shows the final recovery rate obtained as a function of injected apparent magnitudes, combining the results from all 15 sub-samples. The red solid line marks a best fit to the recovery rate, and it could be seen to drop significantly starting from the black vertical line at $m=27.5$, indicating detection inefficiency (hence data incompleteness). This data threshold is also plotted as a solid vertical line on the right panel of Fig.\ref{completeness_recover}, in which the orange solid curve corresponds to the surface number density of galaxies I observed in the M0416 cluster field in different apparent magnitude bins. It could be seen that the observed galaxy surface number density starts to deviate from power law behavior (indicated by straight orange solid line) at exactly the inferred data incompleteness threshold from my source injection-recovery analysis. This justifies my previous assertion that the data incompleteness is reflected by deviation from power-law\footnote{This would not significantly impact my previous determination of data completeness thresholds for S18 catalogs, however, as there I have been conservative and adopted a brighter threshold before the measured density-versus-magnitude distribution peaks. Moreover, very different cluster member populations (also cosmic variance) complicate the situation when analyzing all HFF fields collectively. As a consequence, I verified that the power-law behavior in the observed densities was harder to identify. This observation makes the deviation from power-law behavior not an ideal way of determining the data completeness threshold common to all S18 catalogs. }, and that this occurs at a magnitude brighter than the significant drop in observed surface number densities. 

As another performance check on \texttt{GNUastro}, in the right panel of Fig.\ref{completeness_recover}, I also plotted the observed galaxy counts in the F150W band (purple dashed curve) from the JADES survey for the Hubble Ultra Deep Field (HUDF, deep blank field, \cite{Rieke2023}). Over $24\leq m\leq 27.5$, it is seen that my observed surface number densities in different magnitude bins follow the same distribution as JADES's measurements. This observation suggests that the galaxies I detected (also the extracted magnitudes) using \texttt{GNUastro} are largely consistent with what JADES observed in the deep blank HUDF field, giving confidence to the performance of \texttt{GNUastro}.

\begin{figure}
    \centering
    \includegraphics[width=0.6\textwidth]{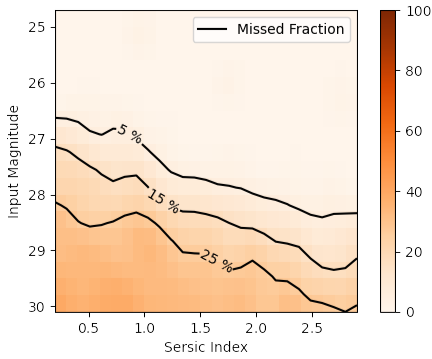}
    \caption{Missed fraction of injected galaxies as a function of injected magnitudes and Sersic index. From the black iso-contours, it could be concluded that galaxies with higher Sersic indices (i.e., light profiles more centrally concentrated) run into detection inefficiency at dimmer magnitudes. This observation explains why $z>3.5$ distributions in Fig.\ref{completeness_recover} (right panel) continue to a fainter magnitude of 29 than $z<3.5$ distributions. }
    \label{missed_fraction}
\end{figure}

From my source injection-recovery analysis, I also noticed that compact galaxies are easier to recover than galaxies with more diffuse light profiles. This is demonstrated in Fig.\ref{missed_fraction}, showing the fraction of injected galaxies unrecovered as a function of both injected magnitude (y-axis) and Sersic index (x-axis). I indicate iso-contours in the missed fraction at levels of 5$\%$,15$\%$, and 25$\%$. Remarkably, all of these contours sit at a dimmer magnitude as the Sersic index gets higher (corresponding to more centrally concentrated light profiles). The same phenomenon is reflected in the right panel of Fig.\ref{completeness_recover}, in which I separately presented the surface number density of galaxies observed at different magnitudes over four different redshift bins. It could be seen that for redshift $z<3.5$, data incompleteness occurs at my previously inferred threshold of 27.5. On the other hand, for redshifts $z>3.5$, deviation from power law behavior is delayed till a fainter magnitude of 29 (indicated by dotted vertical line). This suggests that $z<3.5$ galaxies are more dominated by big and diffuse galaxies, whereas $z>3.5$ galaxies tend to be more compact, in accordance with what one may typically expect. As a consequence of these observations, I will adopt the fainter data completeness threshold of 29 in my future $6\leq z\leq10$ analysis. 

\begin{figure*}
    \centering
    \includegraphics[width=\linewidth]{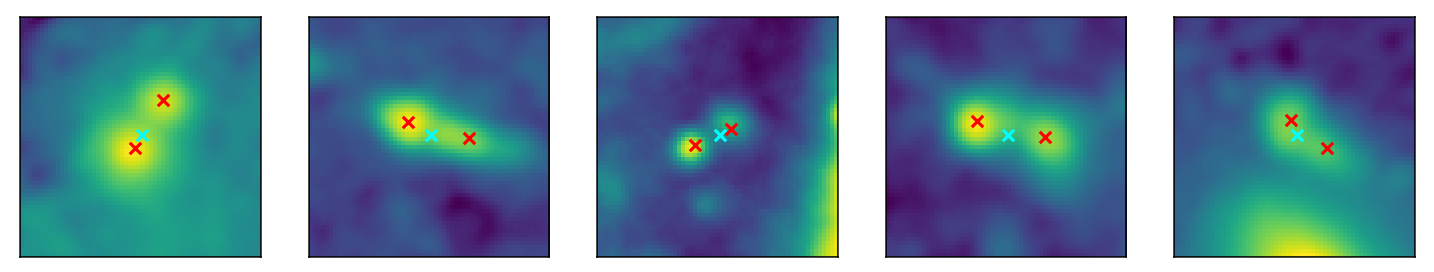}
    \caption{Five sources (indicated by cyan cross) in the previous S18 analysis sample were matched to two objects (red crosses) in the newly constructed catalog using \texttt{GNUastro}. For easier future analysis, these presented galaxies were removed to focus on one-to-one matched galaxies only. The background image shown is from a weighted stack of F090W, F115W, F150W, and F200W JWST images.}
    \label{multiply_matched_objects}
\end{figure*}

\section{Identifying Interlopers}
\label{sec_identify_interloper_JWST}

Equipped now with a catalog incorporating both deep JWST and HST measurements, I now examine whether contaminants among the S18 $z\sim4$ analysis sample could be individually identified. For this, I cross-matched the constructed catalog with the previous S18 analysis sample\footnote{Recall the S18 analysis sample composed of \texttt{use\_phot}=1 galaxies from the S18 catalog, and the selected galaxies have full coverage of HFF and sit outside of the \textit{exclusion region} imposed.} over redshift $3.5 \leq z\leq 5.5$ in the field of M0416 (containing 124 sources). By conditioning a maximum separation of 0.3 arcsec between the matched sources, 106 sources from the S18 analysis sample were matched to 111 sources in my catalog. Among these, five S18 sources were matched to two objects owing to the finer details captured by JWST images and \texttt{GNUastro}'s ability at deblending neighboring galaxies. These five galaxies are collectively shown in Fig.\ref{multiply_matched_objects} and will be removed from subsequent investigation for convenience. For the remaining 101 one-to-one cross-matched sources, 66 were found to be low redshift ($z<3$) interlopers according to the new JWST+HST combined photo-z fits, and only 35 are high-z galaxies ($z>3$).  

\begin{figure}
    \centering
    \includegraphics[width=\linewidth]{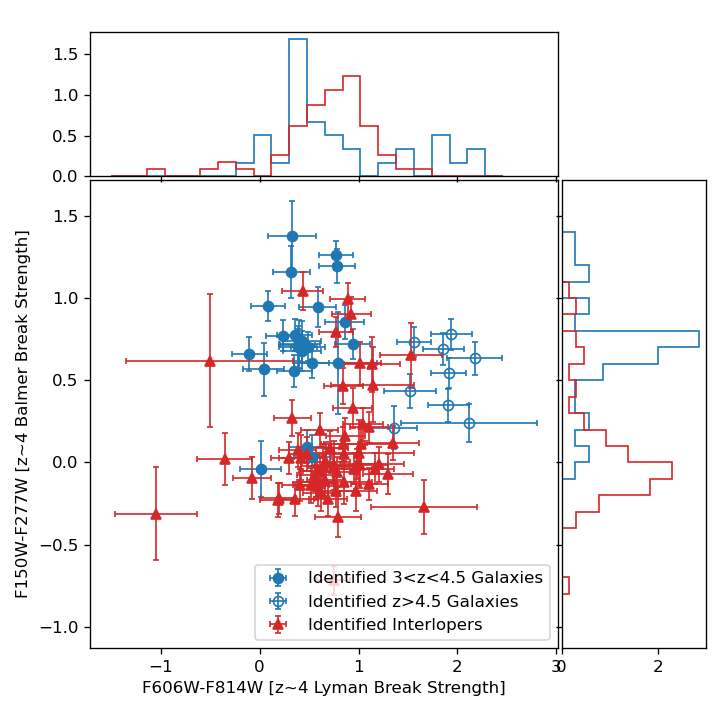}
    \caption{Color-color diagram for both identified interlopers (red data points) and high-z galaxies (blue data points), where color F606W-F814W captures $z\sim4$ Lyman breaks, and color F150W-F277W captures $z\sim4$ Balmer breaks. A bimodality is observed in color F150W-F277W, demonstrating Balmer breaks of $z\sim4$ galaxies are mostly strong enough to be distinguished from low-z interlopers. Interestingly, there appear to be two subgroups of high-z galaxies, corresponding to $z<4.5$ (filled markers) and $z>4.5$ (unfilled markers) galaxies, respectively.}
\label{bimodality_color}
\end{figure}

Previously, I argued that low-z contaminants could be distinguished from high-$z$ galaxies by JWST, mainly as NIRCam images cover the rest-frame optical range (and in particular the Balmer break) of $z\sim 4$ galaxies. To support this argument, in Fig.\ref{bimodality_color} I plot color F150W-F277W (capturing Balmer break of $z\sim4$ galaxies) against color F606W-F814W (capturing Lyman break at $z\sim4$ and Balmer break at $z\sim 0.4$) for both identified interlopers in red and high-z galaxies in blue. I also plot the histogram of respective colors to the right and top of the scatter plot. As evident, a strong difference in F150W-F277W is observed between the two populations, demonstrating Balmer breaks of $z\sim4$ galaxies are mostly strong enough to be distinguished from low-z interlopers. For F606W-F814W, no strong difference between the red and blue histograms could be claimed, reflecting again the cause of misidentification. For the high-z galaxies, it is also seen that they are split into two subgroups, with one being slightly bluer in F606W-F814W than interlopers, whereas the other is slightly redder than interlopers. These two sub-populations were found to be galaxies residing in two different redshift ranges. In particular, the first subgroup corresponds to $z\sim 4$ galaxies (filled blue markers in Fig.\ref{bimodality_color}), and the second subgroup corresponds to $z\sim5$ galaxies (unfilled blue markers) that are naturally fainter in the F606W band.


\section{Stellar Properties}
\label{sec_interloper_stellar_property}

To better understand the type of galaxies more relevant for the misidentification issue, I next investigate the stellar properties of the identified interlopers. For this, I will rely on the Python package \texttt{Bagpipes} \citep{1712.04452}, allowing us to infer the stellar mass, age, metallicity, and amount of dust of galaxies given an assumed star formation history (SFH). 

As I will demonstrate in subsequent sections, however, some of the JWST+HST combined photo-zs are considered to have low confidence. Inclusion of the low-confidence galaxies may impede the inference of stellar properties of true interlopers, and correspondingly, it is desired to have a more robust sample (with interloper/high-z identities more confidently established) for analysis. In sec.\ref{sec_high_confidence}, I detail how such a sample was achieved, and present \texttt{Bagpipes} fitting results in sec.\ref{sec_bagpipes_results}. For completeness, I revisit low photo-z confidence systems excluded from analysis in Sec.\ref{sec_why_confusion}, and demonstrate their relatively low confidence is the result of confusion between dust continuum at low-z with Balmer continuum at high-z. This confusion adds to the known confusion between Lyman breaks with Balmer breaks (in fact, also Balmer jump if with strong nebular emission lines), thereby impeding the mitigating power of JWST. 

\subsection{High Confidence Sample}
\label{sec_high_confidence} 

\begin{figure}
\centering
\begin{subfigure}{.49\textwidth}
  \centering
  \includegraphics[width=\linewidth]{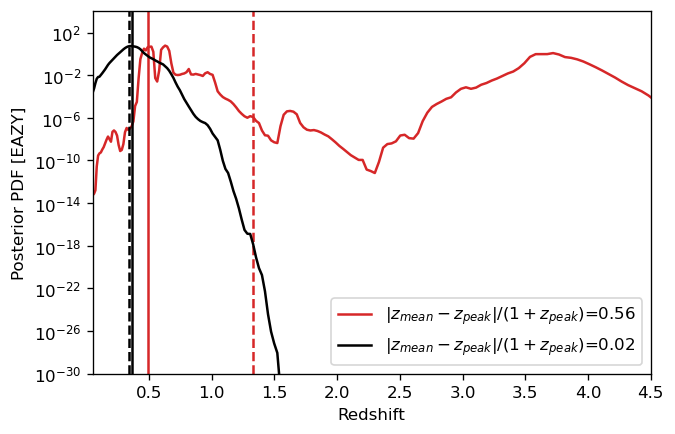}
\end{subfigure}%
\hfill
\begin{subfigure}{.49\textwidth}
  \centering
  \includegraphics[width=\linewidth]{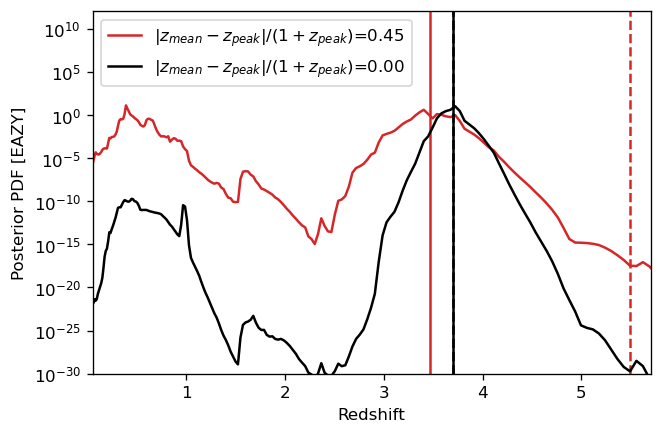}
\end{subfigure}
\caption{Comparison of the posterior PDFs in redshift between galaxies with a low confidence (in red) and a high confidence (in black) in their respective output photo-zs. The left panel corresponds to two galaxies with output photo-z $\sim 0.4$, whereas the right panel compares two $z\sim 3.5$ galaxies. For each of the presented posterior PDFs, confidence in the respective photo-z output is measured by the normalized deviation between the average over PDF ($z_{mean}$, vertical dashed line) and the output redshift ($z_{peak}$, vertical solid line) associated with maximum posterior likelihood. It could be seen that a higher normalized deviation (i.e., lower confidence in output photo-z) corresponds to the existence of comparable peaks in the posterior PDF distribution, and this is true for both galaxies outputted with a high or low $z_{peak}$ value. }
\label{photoz_confidence}
\end{figure}

\begin{figure}
    \centering
    \includegraphics[width=\linewidth]{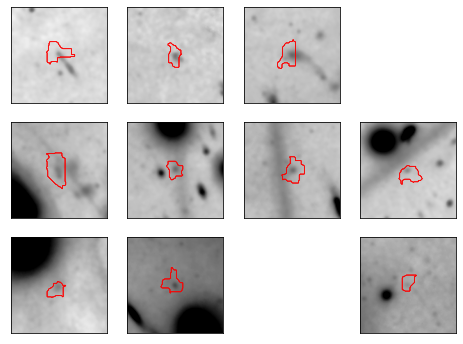}
    \caption{Galaxies removed from robust sample through visual inspection. \textit{Top} \& \textit{Middle} panels are galaxies removed owing to segmentation running across bright regions: through galaxy in the top panel, through diffraction spike in the middle panel. The first two galaxies in the \textit{Bottom} panel are very faint galaxies with relatively large segmentation and sitting close to a much brighter cluster member. All of the aforementioned galaxies were removed as their photometry might have been affected by the reported issues. The last galaxy in the bottom panel was found to be a very faint galaxy with only 5 (whereas the rest of the galaxies had at least 9) $>3\sigma$ flux measurements. Correspondingly, the photo-z of this source was dominantly driven by the less significant flux measurements, and considered less confident.  }
\label{removed_objects}
\end{figure}

If the photo-z of a galaxy could be very precisely determined with high confidence, the posterior probability distribution in redshift from EAZY would be strongly peaked at the output solution. On the other hand, if the SEDs could be equally reproduced by spectrum at different redshifts, there would be corresponding comparable peaks in the posterior PDF distribution as illustrated by Fig.\ref{photoz_confidence}. In Fig.\ref{photoz_confidence}, the posterior PDF of redshift is plotted in black if the PDF has a single strong peak at the output redshift, and in red if the PDF has comparable likelihoods for both low-z and high-z solutions. Given a posterior distribution from EAZY, a good measure for confidence in photo-z could then be the (normalized) deviation between the mean redshift $z_{mean}$ (averaged over posterior PDF), and the maximum likelihood redshift $z_{peak}$ (i.e., the output redshift). For each PDF presented in Fig.\ref{photoz_confidence}, I indicate the respective $z_{mean}$ as a vertical dashed line, and report their normalized deviations from $z_{peak}$ (vertical solid line) in the legend. It could be concluded that for both low (left panel) and high (right panel) $z_{peak}$ output range, the presence of comparable peaks (i.e., ambiguity in redshift solutions) in PDF leads to $z_{mean}$ significantly deviating from $z_{peak}$. To achieve a more robust sample for my subsequent analysis, I therefore remove galaxies with $|z_{mean}-z_{peak}|/(1+z_{peak})>5\sigma_{NMAD}$, corresponding to the galaxies with very low confidence in their reported photo-zs. In this condition, the $\sigma_{NMAD}$ was measured from spec-z systems, and was determined to be 0.052 for my constructed photo-z catalog from Fig.\ref{compare_prior}. For reference, such a selection condition removes 7 out of the 66 low-z interlopers identified, and one high-z candidate. Out of these 8 galaxies, 3 were removed owing to their faintness, and 5 were removed owing to the confusion to be discussed in more detail in Sec.\ref{sec_why_confusion}. 

For the remaining sample with higher confidence in photo-z, I further performed visual inspection and removed 10 additional galaxies presented in Fig.\ref{removed_objects}. The three galaxies on the top panel are seen to have \texttt{GNUastro} segmentation (red contour) running across the relatively bright region of the galaxy, an issue already discussed in Sec.\ref{sec_pro_cons_gnuastro}. In such a case, the local sky background may be overestimated, hence leading to unreliable photometries. The four galaxies in the middle panel are removed owing to a similar reason: their \texttt{GNUastro} segmentations were found to accidentally lie on/close to a diffraction spike, also leading to a potential overestimation of local background. The first two galaxies shown on the bottom panel are instead removed as they are compact galaxies with relatively spacious \texttt{GNUastro} segmentation and are sitting close to a much brighter neighbor. Such a configuration may result in the bright neighbor strongly influencing the measured photometries, consequently lowering the confidence in their new photo-zs. The last galaxy on the bottom panel was removed as it has only 5 $>3\sigma$ measurements, whereas the rest of the galaxies had at least 9 $>3\sigma$ flux measurements. Having only 5 prominent detections was considered not enough to capture prominent spectral features in the SED, and consequently, I regarded the corresponding photo-z as less confident.

\subsection{\texttt{Bagpipes} fitting results} 
\label{sec_bagpipes_results}


\begin{figure}
\centering
\begin{subfigure}{.49\textwidth}
  \centering
  \includegraphics[width=\linewidth]{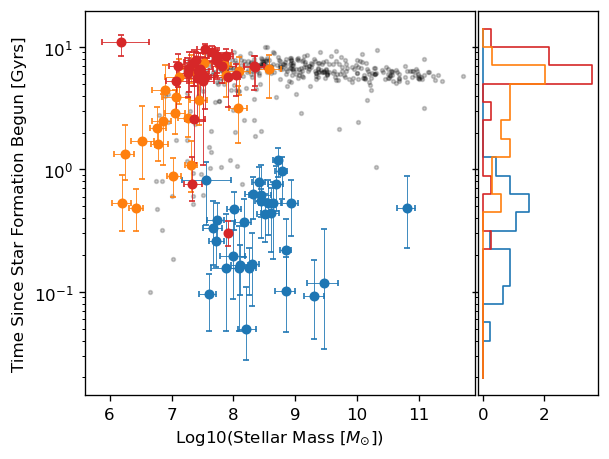}
  \caption{}
\end{subfigure}%
\hfill
\begin{subfigure}{.47\textwidth}
  \centering
  \includegraphics[width=\linewidth]{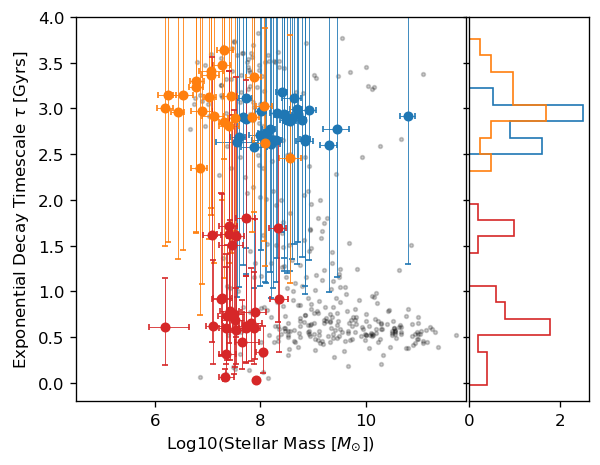}
  \caption{}
\end{subfigure}
\vfill
\begin{subfigure}{.48\textwidth}
  \centering
  \includegraphics[width=\linewidth]{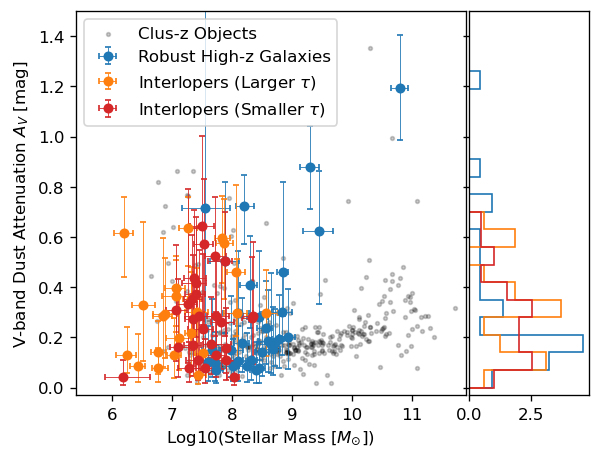}
  \caption{}
\end{subfigure}
\hfill
\begin{subfigure}{.49\textwidth}
  \centering
  \includegraphics[width=\linewidth]{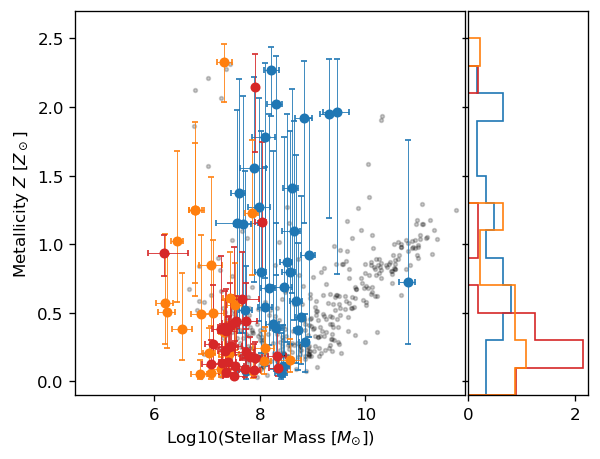}
  \caption{}
\end{subfigure}
\caption{Fitted stellar ages (time since onset of star formation, panel a), SFH exponential decay time scale $\tau$ (panel b), V-band dust attenuation (panel c), and metallicity $Z$ (panel b) plotted against fitted stellar masses. To the right of each panel, I also plot the density histogram of the corresponding parameter. In these figures, robust high-z galaxies identified are associated with a blue color, whereas interlopers are divided into two subgroups with a red (with $\tau < 2$Gyrs on panel b) and orange (with $\tau > 2$Gyrs) color, respectively. For comparison, I also plot the distribution of cluster-redshift galaxies (galaxies having redshift within $z_{clus}\pm 0.15$ both in S18 and my constructed JWST+HST combined catalog) as black dots. It can be seen that the low-z interlopers (red and orange) follow nicely the distribution (as low mass extension) of cluster-z galaxies in panels a, c, and d. In panel b, orange data points (also high-z galaxies) sit above most of the cluster-z galaxies and also red data points, demonstrating there are two types of galaxies among the interlopers. As I further explain in the main text, orange data points correspond to interlopers in the field, and their large $\tau$ values suggest they may still be star-forming. Red data points (those with $\tau\sim1.5$Gyrs may also be green valley galaxies), owing to their close resemblance with the cluster-z galaxies, are likely dwarf cluster members with quenched star formation.  }
\label{Stellar_properties}
\end{figure}

In Fig.\ref{Stellar_properties}, I present \texttt{Bagpipes} fitting results for the remaining (high confidence) high-z galaxies as blue data points, and low-z galaxies as red and orange data points (depending on the SFH decay rate fitted, see later). In obtaining these results, I assumed a simple exponential decay model\footnote{I also performed \texttt{Bagpipes} fitting assuming burst SFH, and obtained similar results except for stellar ages, which were systematically higher than the values obtained with the exponential decay SFH model. } for the star formation history, which is known to be well-suited to a broad class of galaxies. For the \texttt{Bagpipes} fits, I allowed for a relatively wide redshift fitting range of $\pm 0.3z$ about the reported JWST+HST photo-z. For galaxies at redshifts above 0.6 (corresponding to the end of the cluster member peak on the photo-z histogram), I also corrected for the effect of gravitational lensing based on the average of lensing magnification factors predicted by v4 CATS model \citep{Caminha2017}, v4.1 WSLAP+ model \citep{Diego2015a}, internal glafic model from collaborator, and a more recent JWST based free-form model by \cite{Diego2024}.

The fitted stellar ages (time since the onset of star formation) are plotted against the fitted stellar masses in Fig.\ref{Stellar_properties}(a). It is seen that galaxies tend to have stellar mass of the order $\sim 10^{8.44 \pm 0.64} M_{\odot}$, whereas low-z interlopers are mostly with $\sim 10^{7.39\pm 0.52} M_{\odot}$. In contrast, interlopers have old stellar populations with age $\sim5.14\pm 2.71$ Gyrs, whereas high-z galaxies are tentatively still star forming with stellar ages $\sim0.41\pm 0.28$ Gyrs. In Fig.\ref{Stellar_properties}(a), I also indicated the distribution of cluster-redshift galaxies using black data points. These cluster-redshift galaxies were selected by having photo-zs within $z_{clu}\pm0.15$ both in S18 and my catalog, and are expected to be dominated by cluster members. It could be seen that interlopers (especially the red data points) fall on the low-mass extension of the cluster sequence, suggesting they are likely dwarf (given the measured stellar masses) cluster members. 

In Fig.\ref{Stellar_properties}(a), red data points are found more clustered about the cluster sequence than the orange data points, and they correspond to interlopers with SFH exponential decay timescale $\tau$ less than 2 Gyrs as indicated by Fig.\ref{Stellar_properties}(b). The characteristic stellar age for these red data points could also be read off from Fig.\ref{Stellar_properties}(a) to be $\gtrsim5$ Gyrs, which are larger than their characteristic $\tau$, hence I suspect they are dwarf cluster members with quenched star formation. 

\begin{figure}
\centering
\begin{subfigure}{.375\textwidth}
  \centering
  \includegraphics[width=\linewidth]{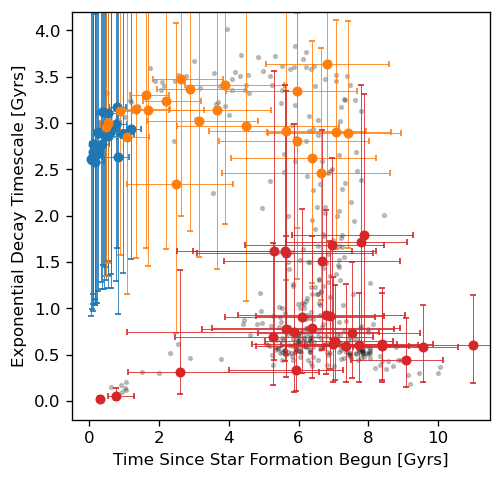}
\end{subfigure}%
\hfill
\begin{subfigure}{.6\textwidth}
  \centering
  \includegraphics[width=\linewidth]{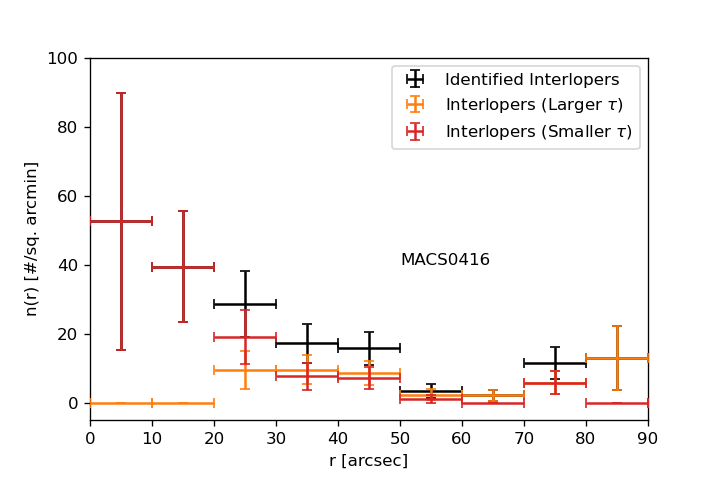}
\end{subfigure}
\caption{\textit{Left}: Time since star formation plotted against characteristic time scale ($\tau$) of exponential decay in SFH. As in Fig.\ref{Stellar_properties}, blue data points correspond to robust high-z galaxies, red and orange are the interlopers with distinct $\tau$ values, and black data points are cluster-redshift galaxies. An 'evolution' trend is made clear, where the star-forming galaxies in the top-right corner are anticipated to gradually evolve with time into the region occupied mostly by red data points. From this trend, I argue that the orange data point interlopers may still be star-forming. \textit{Right}: Radial distribution of $\tau>2$Gyrs (in orange) and $\tau<2$Gyrs (in red) interlopers, with $r$ the radius from FOV center. The radial distribution of the combined interloper population is also presented in black. It could be seen that the red distribution is strongly concentrated towards the cluster center, indicating interlopers with $\tau<2$Gyrs are dwarf cluster members. Orange distribution, on the other hand, is uniform across the FOV, indicating $\tau>2$Gyrs interlopers are likely field galaxies. }
\label{different_population}
\end{figure}

Orange data points, on the other hand, have exponential decay timescales $>2$Gyrs and a broader range of stellar age (reaching as low as 0.5 Gyrs). It is also seen that they distribute closer to the high-z galaxies (blue points) in Fig.\ref{Stellar_properties}(b), suggesting they may be star-forming just as the high-z galaxies. In the left panel of Fig.\ref{different_population}, the orange data points are also seen as located closer to high-z galaxies on the 'evolutionary track' than the red data points and the cluster sequence. These observations suggest they may be dwarf cluster members in the relatively outskirts of the galaxy cluster (suffering less ram pressure stripping) or field galaxies with ongoing star formation.

In the right panel Fig.\ref{different_population}, I test the above hypothesis by plotting the radial distributions of red data point interlopers and orange data point interlopers with the same respective color. Black data points reflect the combined (red + orange) interloper population. Indeed, red data point interlopers could be seen to strongly cluster toward the cluster center, suggesting their star formation is likely quenched owing to ram pressure stripping. Whereas orange data point interlopers have a more uniform radial distribution with a slight tentative elevation in the last radial bin. This is consistent with them being field galaxies or cluster members at the galaxy cluster outskirts.

It could also be seen that orange data points extend to lower stellar masses than most of the red data points from Fig.\ref{Stellar_properties}(a), and the less massive ones have a younger stellar age (i.e., in an earlier stage of star formation). The same trend is also observed for high-z galaxies, and I suspect this is the result of detection bias. I.e., given the same detection threshold, lower mass galaxies need to be intrinsically brighter for them to be detected. Finally, it is also seen that some of the red data points are elevated from the rest in Fig.\ref{Stellar_properties}(b) with an exponential decay timescale $\tau\sim 1.8$ Gyrs. These may be the 'green valley' galaxies.

\begin{figure}
\centering
\begin{subfigure}{.49\textwidth}
  \centering
  \includegraphics[width=\linewidth]{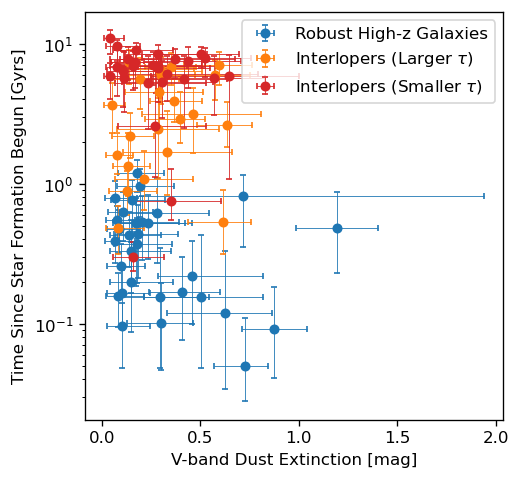}
\end{subfigure}%
\hfill
\begin{subfigure}{.49\textwidth}
  \centering
  \includegraphics[width=\linewidth]{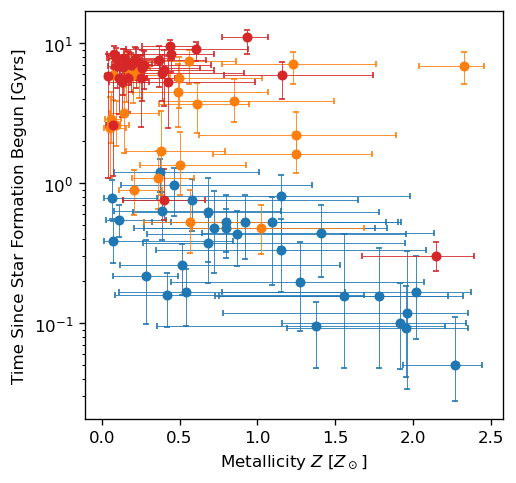}
\end{subfigure}
\caption{Testing for fitting degeneracy. \textit{Left}: Fitted stellar ages plotted against the V-band dust attenuation. \textit{Right}: Fitted stellar ages plotted against fitted metallicities. From the right panel, there is clear evidence for age-metallicity degeneracy for galaxies younger than $<1$Gyr, i.e., a lower stellar age fitted is associated with a higher metallicity. From the left panel, age-dust degeneracy (i.e., younger galaxies are fitted to be more dust extincted) could also be seen from the lower left corner, relevant dominantly for galaxies younger than 0.3 Gyrs and $A_V<1$ mag.  }
\label{testing_fitting_degenarcy}
\end{figure}

From Fig.\ref{Stellar_properties}(c), no significant statistical difference is seen in the V-band dust attenuation $A_V$ of quenched (red) and star-forming (orange) interlopers. This may contradict my ram pressure stripping conjecture for the cluster members, but could be explained if (a) there were gas re-accretion from the interstellar medium (ISM) and hence the slightly higher stellar mass, or simply (b) dust within the ISM also obscures the measured fluxes. In Fig.\ref{Stellar_properties}(c), it is also seen that some high-z galaxies are more dust extincted (with $A_V>0.7$) than all interlopers, whereas most high-z galaxies are with $A_V<0.2$. As I demonstrate with the left panel of Fig.\ref{testing_fitting_degenarcy}, such a high amount of dust is likely due to the age-dust degeneracy (i.e., younger galaxies are fitted to be more dust extincted) encountered for stellar ages younger than 0.3 Gyrs. The extremely dust-extincted galaxy with $A_V\sim 1.2$, however, appears to be located away from the age-degeneracy trend in the left panel of Fig.\ref{testing_fitting_degenarcy}. This galaxy may be a genuine strongly dust-extincted high-z galaxy, and following my previous discussion and with evidence from \cite{2023arXiv230315431A}, such galaxies may be the source of misidentification to a very high photo-z range $\sim16$ when working with JWST measurements only. 

Finally, from Fig.\ref{Stellar_properties}(d), I notice that field interlopers (orange) can have higher metallicity than cluster member interlopers (red). Additionally, those field interlopers that are more metal-rich tend to be less massive. Combining with the observation that lower mass field interlopers also have a younger stellar age from Fig.\ref{Stellar_properties}(a), I conclude that the resulting higher metallicities are likely the result of age-metallicity degeneracy. This degeneracy could be more clearly seen from the right panel of Fig.\ref{testing_fitting_degenarcy}, where galaxies (including a few field interlopers) are more metal-rich if fitted with a younger stellar population $\lesssim 1$Gyr. In Fig.\ref{Stellar_properties}(d), it is also seen that most high-z galaxies are fitted with very high metallicity values, but with very large uncertainties. These high metallicity values are evidently also the result of age-metallicity degeneracy, and hence should not be over-interpreted. 

\begin{figure}
\centering
\begin{subfigure}{.49\textwidth}
  \centering
  \includegraphics[width=\linewidth]{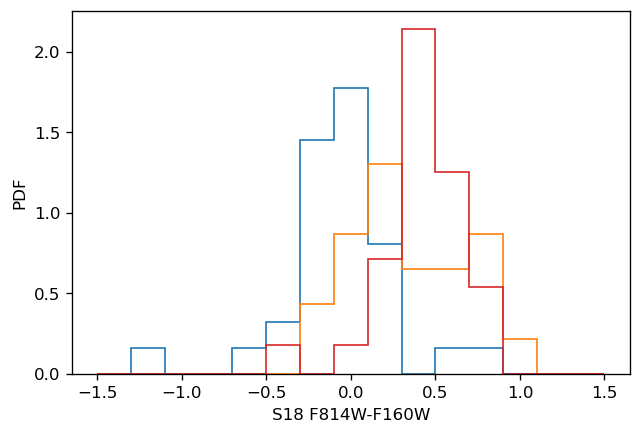}
\end{subfigure}%
\hfill
\begin{subfigure}{.49\textwidth}
  \centering
  \includegraphics[width=\linewidth]{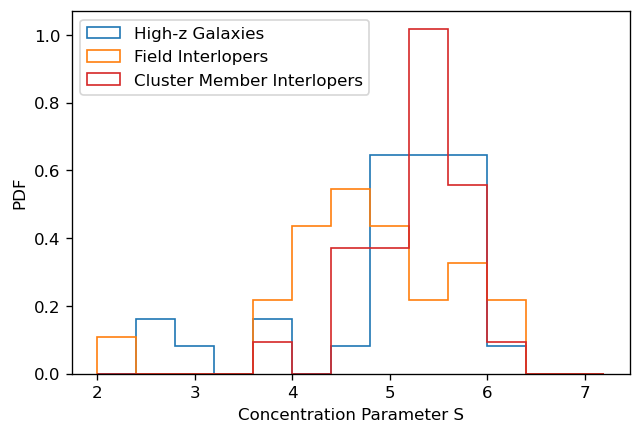}
\end{subfigure}
\caption{\textit{Left}: Histogram of F814W-F160W (based on S18 measurements) for the identified high-z galaxies (blue), field interlopers (orange), and cluster member interlopers (red). Both of the interloper populations are seen to have redder F814W-F160W colors than high-z galaxies, which is the result of interlopers having older stellar populations. \textit{Right}: Histogram of concentration parameter $S$ for the identified high-z galaxies (blue), field interlopers (orange), and cluster member interlopers (red). Cluster member interlopers are seen to have a similar distribution as the high-z galaxies, and have centrally concentrated light profiles. Field interlopers, on the other hand, were found to have less centrally concentrated light profiles. These observations were understood as the outer stellar wings of cluster members were removed from the ram pressure stripping effect, whereas field interlopers do not suffer from ram pressure stripping. }
\label{inspect_redorangeblue}
\end{figure}

Combining all aforementioned observations, I thus conclude that the dominant source of low-z contamination to the S18 $z\sim4$ sample is dwarf galaxies with stellar mass $10^{7.39\pm 0.52} M_{\odot}$. Over this mass range, cluster members with quenched star formation (owing to ram pressure stripping) and field galaxies possibly with ongoing star formation contribute equally to the misidentification problem. This is reflected by similar sample sizes observed for red (28 galaxies) and orange (24 galaxies) data points, with cluster members contributing more dominantly near the galaxy cluster center. Overall, both field and cluster member interlopers have an older stellar population than high-z galaxies of the same mass, which then explains the redder S18 F814W-F160W color observed in Sec.\ref{section_implications} and the re-illustrated version in the left panel of Fig.\ref{inspect_redorangeblue}. Cluster member interlopers are naturally anticipated to have experienced ram pressure stripping. As a result, from Fig.\ref{inspect_redorangeblue} it is also inferred that a large fraction of cluster member interlopers lost their outer stellar wings and hence have more centrally concentrated light profiles (concentration parameter $S$ was introduced in Sec.\ref{sec_contamination_csbcon}). Field galaxies, on the other hand, do not suffer from ram pressure stripping and hence correspond to more diffuse galaxies previously identified in Sec.\ref{sec_contamin_visualinspect}. As I will discuss further in Sec.\ref{sec_ML_results}, however, past star formation may also drive out the outer wings of galaxies, and result in more compact morphologies for the quiescent field interlopers.  


\subsection{Understanding confusion}
\label{sec_why_confusion}

\begin{figure}
\centering
\begin{subfigure}{.9\textwidth}
  \centering
  \includegraphics[width=\linewidth]{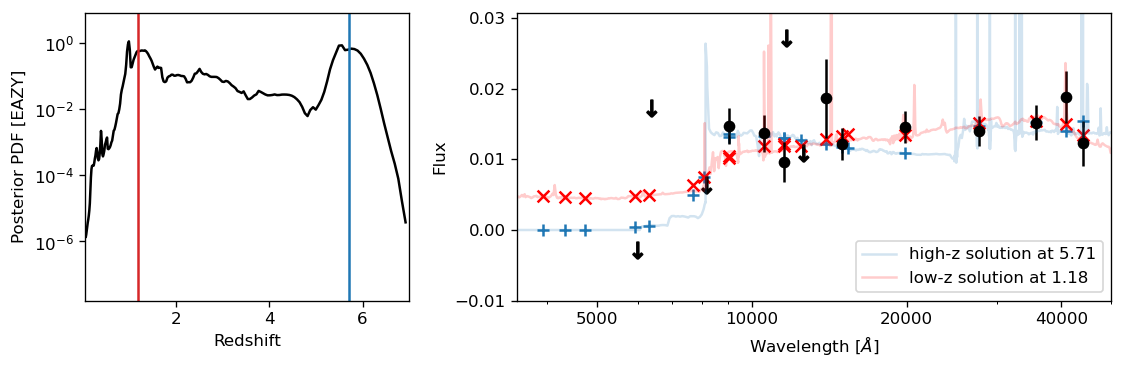}
\end{subfigure}%
\vfill
\begin{subfigure}{.9\textwidth}
  \centering
  \includegraphics[width=\linewidth]{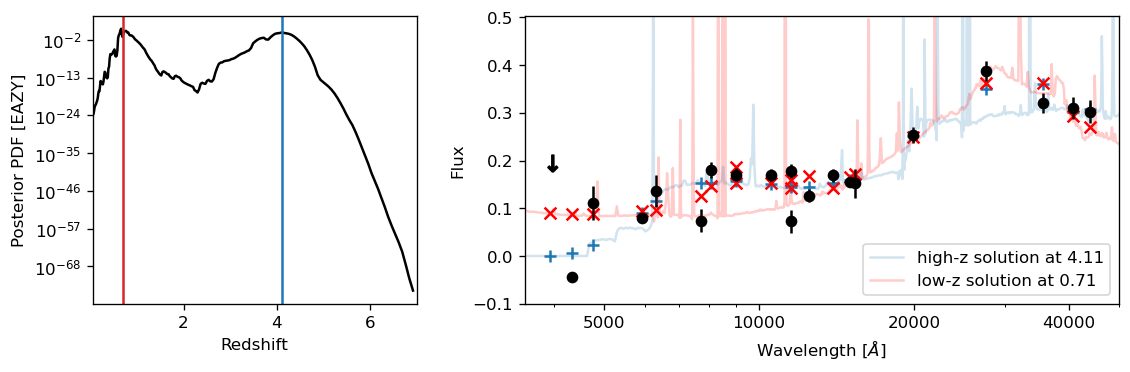}
\end{subfigure}
\caption{To understand why the confusion between high-z and low-z solutions persists with the combination of HST and JWST data, here I examine two example galaxies with low confidence in photo-z. \textit{Left}: Posterior PDF in redshifts for these two galaxies, with red and blue vertical lines marking low-z and high-z solutions (respectively) with comparable likelihood. \textit{Right}: Measured SEDs (black data points) of the considered galaxies, in comparison with modeled SEDs and best-fit spectrum (red for low-z solution and blue for high-z solution) corresponding to the two comparable likelihood photo-z solutions. It is seen that the modeled SEDs over rest-frame Balmer wavelengths of the high-z solution are similar to the modeled SEDs from the dust continuum of the low-z solution. This observed similarity, combined with the known degeneracy between high-z Lyman break and low-z Balmer break, results in high-z and low-z solutions having comparable likelihood.}
\label{why_confusion}
\end{figure}

In this subsection, I revisit the lower photo-z confidence systems and further investigate why confusion between low-z and high-z solutions persists despite the inclusion of JWST measurements. This was ultimately understood as owing to either (a) galaxies being very faint, such that photo-zs are in general hard to constrain; or (b) potential confusion between low-z dust continuum and high-z Balmer continuum, in addition to the established confusion between low-z Balmer and high-z Lyman breaks. The latter point is demonstrated in Fig.\ref{why_confusion}, where I plot the posterior PDF of redshift in the left panel, and the observed SED\footnote{Measurements with SNR$<3$ are plotted as upper limits (downward-pointing arrows).} (right panel) for two representative low photo-z confidence objects. On the left panel, I also indicate the position of two comparable likelihood solutions as vertical lines (red for low-z and blue for high-z solutions). Best fit spectra fixing to these respective redshifts are presented on the right panel, with the corresponding modeled SEDs also indicated as crosses of the same color. For both objects, the dust continuum of the low-z solution leads to a modeled SED at wavelengths $>2 \mu m$ remarkably similar to that from the high-z Balmer continuum. This similarity, combined with similar modeled SEDs over $0.8 \mu m<\lambda<1.6\mu m$, then explains the equivalent likelihoods of the two redshift solutions.  

For the SED shown on the bottom panel, the similarity in modeled SEDs over $0.8 \mu m<\lambda<1.6\mu m$ actually arises from strong emission lines combined with a Balmer jump at $z=0.71$, versus the Lyman continuum at $z=4.11$. The fitted Balmer jump and strong emission lines suggest the object is potentially undergoing bursts of star formation, resulting in highly ionized gas and prominent nebular continuum emission from free-bound transitions (free electrons captured by nuclei) \cite{2408.03189}. Among the 66 (now including galaxies with both high confidence and low confidence in photo-z) low-z interlopers I previously identified from the S18 $3.5\leq z\leq 5.5$ sample, seven galaxies exhibited such a Balmer jump in their best-fit spectra. In selecting the robust sample presented in Sec.\ref{sec_high_confidence}, five of these seven Balmer jump interlopers were removed as low photo-z confidence galaxies (owing to the confusion with high-z solutions discussed previously). One of these seven objects was instead removed owing to the segmentation issue. The last Balmer jump interloper was included in the robust analysis sample (ID 5201), and it was fitted with a younger stellar population (0.48 Gyrs since onset of star formation, with SFH decay time scale 2.961 Gyrs), higher metallicity (1.023 $Z_\odot$), less dust extinction ($A_V = 0.085$ mag), and a lower stellar mass ($10^{6.43} M_\odot$) than most of the other low-z interlopers. These fitted values are consistent with it being a field galaxy (given its association with the $\tau>2$Gyrs sample) that has experienced multiple star formation bursts. 



\section{Machine learning interloper identifier}
\label{sec_interloper_machine_learning}

As evident from the extensive analysis on S18 observables in Chap.\ref{chap:contamination} and also Fig.\ref{inspect_redorangeblue}, subtle differences between genuine high-z and low-z galaxies are already encoded in the S18 measurements. While these differences can be visualized statistically, they are too subtle and intricate to enable accurate identification of individual high-z and low-z galaxies, thereby necessitating the need for supplementary JWST data. Conversely, if these intricate differences could be exploited, time-intensive supplementary JWST observations may be avoided. One promising approach to achieve this goal is with machine learning, which has been demonstrated to perform exceptionally (even outperforming humans) in identifying intricate patterns. Consequently, machine learning methods have been widely applied in astronomy for object detection \citep{2403.19912, 2504.00054}, classification \citep{2403.12120, 2412.02112}, and estimation of redshifts and stellar properties  \citep{2411.07305, 2501.14408}.

In this section, I demonstrate that a supervised machine learning approach, leveraging the known (inferred with high confidence) identities of 83 objects in the M0416 field, an "interloper identifier" with excellent performance (reaching 100\% accuracy) can be constructed. Applying the constructed identifier to other S18 catalogs (without supplementary JWST data) reveals the contamination rate to be $57.25\%$, in great agreement with my previous magnification bias estimates ($56.87 \pm 11.84 \%$). In what follows, I first review key concepts of deep learning and introduce my adopted neural network architecture in Sec.\ref{sec_neural_network_intro}. Sec.\ref{sec_ML_data} details the training sample needed for the interloper identifier model. Finally, Sec.\ref{sec_ML_results} presents my results when the constructed model is applied to the other S18 catalogs, and provides diagnostic cross-checks demonstrating the robustness of my interloper-mitigated sample.

\subsection{Deep Learning Neural Network}
\label{sec_neural_network_intro} 

\begin{figure}
    \centering
    \includegraphics[width=0.8\linewidth]{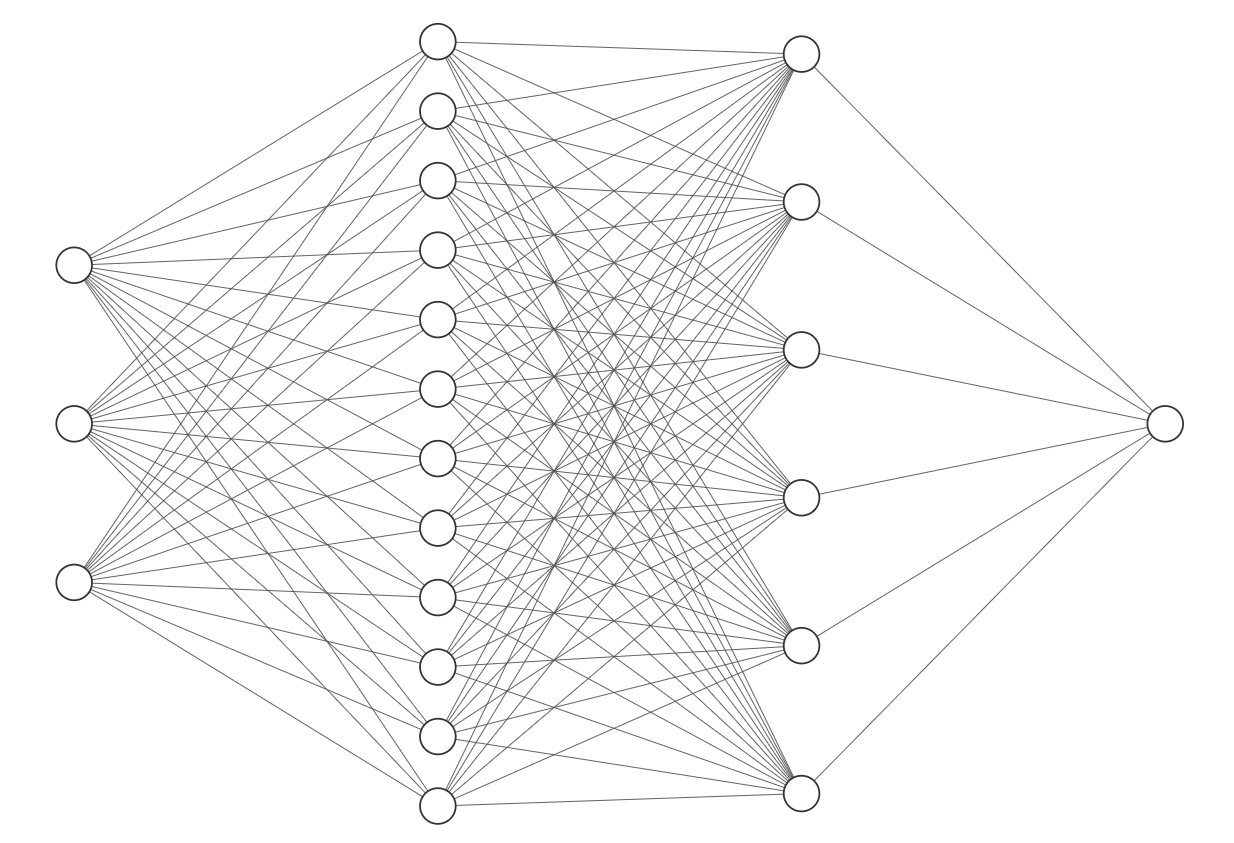}
    \caption{An illustrative example of a deep neural network. Here, from left to right, the layers correspond to the input layer taking 3 input values, 2 hidden layers containing 12 and 6 neurons respectively, and 1 output layer. Each neuron is depicted as a node, and the links between neurons demonstrate that output from the previous layer is used as input for the next layer.}
    \label{neural_network_example}
\end{figure}

In typical machine learning settings, such as predicting continuous variables using polynomial functions\footnote{Predicting continuous variables is called regression, while predicting discrete labels is classification.}, the number of adjustable model parameters must be smaller than the training sample size to avoid overfitting\footnote{Overfitting occurs when a model memorizes training data instead of learning general patterns.}. This is often achieved with regularization terms (parameterizing model complexity) to be minimized alongside the loss function (e.g., $\chi^2$ in $\chi^2$ minimization). Owing to this desired simplicity, constructed (shallow) machine learning models identify only strong patterns in the data, with the advantage that identified patterns are easy to interpret. 

Unlike these shallow approaches, deep learning models are often opaque. Rather than establishing direct input-output links, they iteratively reprocess identified patterns through multiple layers to uncover intricate relationships. Fig.\ref{neural_network_example} illustrates this with an example multi-layer neural network. Here, the leftmost layer is the input layer (taking in 3 input variables), and the rightmost layer is the output layer (predicting the value for a single variable). The middle two \textit{hidden} layers, consisting of 12 and 6 \textit{neurons} respectively, iteratively process identified patterns before output. In analogy to my previous contamination analysis (Chap.\ref{chap:contamination}), the 12-neuron layer generates colors from three magnitude measurements, and the 6-neuron layer then inspects the color-color distribution. As such, deep learning can capture intricate patterns typically missed by shallow methods, but their complex, multi-layered structure suggests they are often over-parametrized. Despite being over-parameterized, various applications of the deep learning methods have demonstrated that deep learning models maintain a good predictive power on unseen data \footnote{The good predictive power of deep learning models remains to be better understood, but it has been demonstrated by \cite{Canatar2021NatCo} that deep learning models preferentially learn the simplest feature functions (i.e., relation among inputs and also between inputs and outputs) among those that align with the machine learning objective (in my case, to distinguish high-z and low-z galaxies). }. For my study, I then leverage this established predictive power of deep learning models to examine whether S18 measurements alone can separate interlopers from genuine high-$z$ galaxies. My final adopted model structure is akin to the illustrated case, but with more input variables (10, see Sec.\ref{sec_ML_data} for details) and 4 hidden layers consisting of 128, 64, 32, 16 neurons respectively to vastly search for intricate relations among inputs. 

Just as minimum $\chi^2$ fitting, training a deep neural network means to optimize model parameters such that a \textit{loss function} is minimized. The common loss functions include Mean Squared Error (MSE) when predicting a continuous variable, and cross-entropy ($L = - \sum_i y_i \log \hat{y}_i,  \; \; y_i = \text{data}, \; \hat{y}_i = \text{predicted}$) for classification problems. In my case, I design my model to output 0 (more technically, a value $<0.5$) for a galaxy considered to be genuinely high-z, or 1 ($>0.5$) if considered to be an interloper. Hence, I adopt binary cross-entropy as my loss function, best suited for such a binary classification problem. Since I know the true identities of my training sample (established with JWST), this constitutes a \textit{supervised} learning scheme where the knowledge of true identities allows for model performance monitoring.


The parameters of the model that are to be trained for a deep learning model are, in effect, neuron parameters. For each neuron, inputs ($v_i$) are combined in a linear and weighted (with weights $w_i$) fashion to determine the neuron output, $z = f(\sum_i w_i v_i + b)$. In the last expression, $b$ is the bias term (analogous to the interception point in the $y=ax+b$ relation), and $f$ is the \textit{activation function} determining whether a neuron is activated and used for subsequent layers. Through shutting off unwanted nodes (or measured relations), activation functions effectively facilitate the learning of complex patterns, especially patterns that cannot be parametrized with linear functions. This power of activation functions is granted by their non-linear functional form, with an example being the rectified linear unit (ReLU), which introduces a break\footnote{This break may also be thought of as rejecting weak (e.g., noise-induced) or negative correlations among the input variables. } in the output by replacing negative values with zero. For my model construction, I follow the common practice and use ReLU as the activation function in all hidden layers. For the last output layer, I use sigmoid\footnote{Also called logistic function, given by $f(x) = \frac{1}{1+e^{-x}}$. } activation function (outputting a value in range [0,1]) instead to best suit my anticipated output.

The training of a deep learning model typically proceeds as follows. Model parameters (weights and biases of all neurons) are initially randomly sampled and updated iteratively to minimize the loss function. To optimize both computation time (total number of times for parameter updates) and memory usage (number of data points involved per update), it is common practice to split (randomly) the training data into smaller \textit{batches}. Subsequently, the gradients of the loss function are computed separately for each batch (averaged across samples) to determine the parameter update direction. One full pass through all batches constitutes an \textit{epoch}, a hyperparameter users need to adjust via trial and error for optimal performance. While a higher number of epochs generally improves results, excessive epochs may cause overfitting and reduced accuracy. Similarly, batch size affects how faithfully gradient estimates reflect the global geometry of the loss function, hence also needs to be optimized\footnote{Too large a batch size reduces the total number of times for parameter updates, whereas too small a batch size leads to unfaithful estimates on the gradient of the loss function. } given the training sample size. In my case, I chose to train models over 200 epochs, with a batch size of 64 (twice the commonly adopted value) for more robust determinations of the gradients to the loss function. 



\subsection{Training Data and Data Augmentation}
\label{sec_ML_data}

Currently, my sample comprises 83 galaxies with confirmed (inferred with high photo-z confidence) identities: 31 genuine high-z sources, 52 low-z interlopers. This dataset is inadequate for machine learning, however, due to (1) limited sample size, (2) too discrete sampling in the redshift and also variables to be used as input parameters, and (3) the observed imbalance in the size of the two populations, which would bias the model to preferentially output interloper labels. To obtain a sufficiently large (and with more continuous parameter sampling) and balanced training sample, I therefore perform data augmentation, i.e., synthesizing artificial galaxies with predefined labels (high-z or interloper).

\begin{figure}
    \centering
    \includegraphics[width=0.8\linewidth]{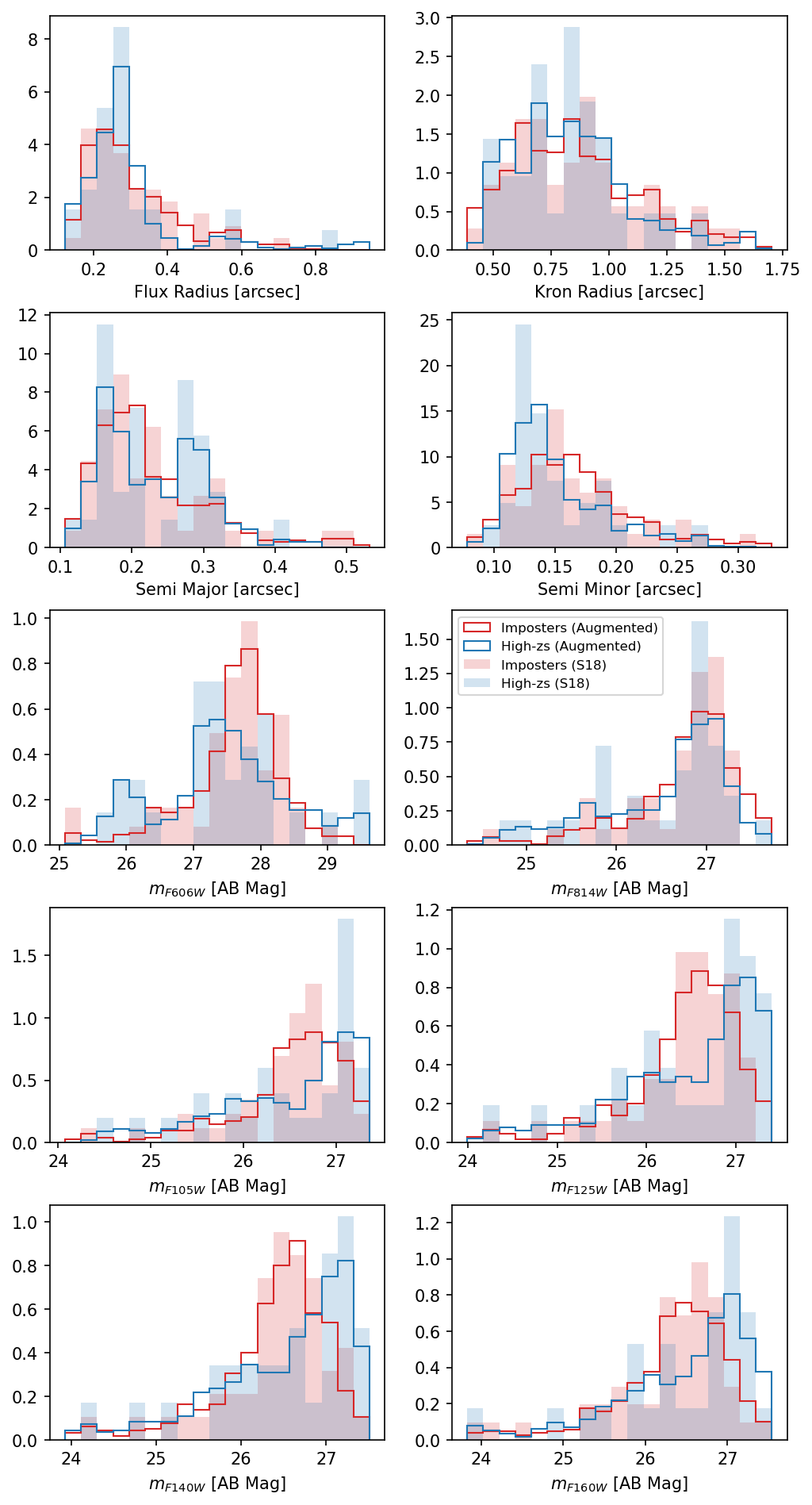}
    \caption{Distributions of ten S18 measurements used as machine learning input variables before (shaded histograms) and after (step histograms) data augmentation. These distributions are plotted separately for interlopers (red) and high-z galaxies (blue), and data augmentation could be seen to preserve the original parameter distributions for both populations.}
    \label{inspect_augment}
\end{figure}


I first define relevant measurements from S18 catalogs, encoding intricate differences between high-z and low-z galaxies, to use as input variables. These include (1) flux measurements (allowing for negative values, normalized to AB zero-point of 25) in six HFF band excluding F435W band, where many galaxies lack observations; and (2) four shape measurements: half light radius (in S18, called flux radius), flux weighted radius (Kron radius), and semi-major and semi-minor axes. Fig.\ref{inspect_augment} shows the distributions of these parameters (flux measurements converted to magnitudes for easier visualization) as shaded histograms, where blue is for high-z and red for low-z sample.

High-z galaxies and low-z interlopers arguably exhibit similar distributions in flux radius, Kron radius, semi-major axis (despite a slight excess at $\sim$0.3 arcsec for high-z), and also brightnesses in F606W and F814W bands. But high-z galaxies appear to favor smaller semi-minor axes than low-z interlopers, suggesting more elongated morphologies. More prominent differences between the two samples are in longer wavelength bands (F105W to F160W) magnitudes, echoing the previously identified F814W-F160W difference. Moreover, the right panel of Fig.\ref{inspect_redorangeblue} reveals distinct light-profile concentrations (defined from the ratio between flux radius and Kron radius) in field interlopers relative to both high-redshift galaxies and cluster members. These statistical differences may grant machine learning models the ability to distinguish between high-z and low-z galaxies. It is therefore vital that these statistical differences are preserved when augmenting the sample. More specifically, newly generated samples must exhibit the same statistical properties as the 31 high-z galaxies and 52 low-z interlopers identified, such that the augmented sample remains representative of S18 measurements.

\begin{figure}
    \centering
    \includegraphics[width=0.8\linewidth]{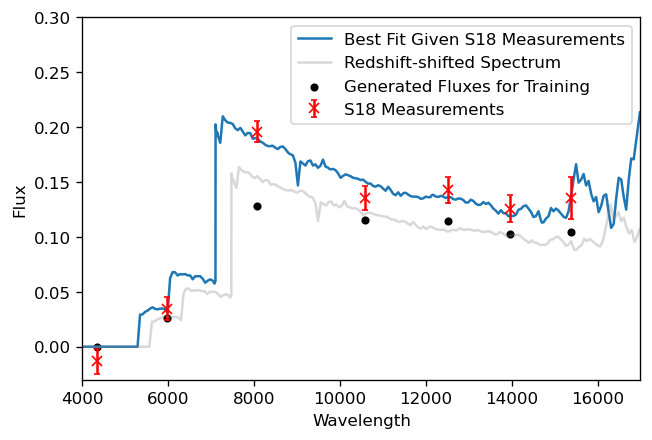}
    \caption{Illustration of how to synthesize a new galaxy sample with a predefined high-z identity, given a confidently identified high-z galaxy at $z = 4.86$. Best-fit spectrum to S18 flux measurements (red data points, normalized to AB magnitude of 25) is shown in blue, and is perturbed slightly along redshift to produce the gray spectrum for a new $z=5.16$ sample. This gray spectrum has been corrected for the cosmological dimming effect, and incorporates a random scaling on the overall brightness to reflect variations in the intrinsic galaxy luminosity. Black data points represent generated SEDs given the gray spectrum to use as machine learning inputs, with additional small noise added to mimic the scatter of red data points about the original blue spectrum. Shape parameters are adjusted from the $z=4.86$ sample based on the random scaling on the overall brightness, with additional small perturbations on each parameter to produce slightly different intrinsic morphology for the synthesized sample. 
    }
    \label{generated_example}
\end{figure}

For this purpose, I refit the spectrum of 83 high photo-z confidence galaxies using EAZY, but now restricted to only S18 flux measurements and a redshift fitting range of $<3$ for low-z interlopers and $>3$ for high-z galaxies. Fig.\ref{generated_example} shows the spectrum (blue) thus obtained for one of the high-z galaxies at $z=4.86$, where red data points are S18 flux measurements. To generate new high-z samples similar to this galaxy, I perturb the best-fit spectra slightly along redshift (of a step maximally $\pm z_{NMAD}=(1+z)\sigma_{NMAD}$) with the gray spectrum being an example. In doing so, I incorporate cosmological dimming/brightening effect due to the change in redshift, and also apply a random scaling (maximally $\pm0.3$mag) on the overall brightness of the synthesized galaxy. This additional scaling may either be interpreted as a reflection of different intrinsic luminosities (possibly also different intrinsic sizes) for the generated galaxies, or due to slightly different lensing magnification factors of high-z galaxies. Correspondingly, all four shape parameters are scaled simultaneously to meet this random scaling on overall brightness. And I further introduce small perturbations (maximally $\pm 0.05$\%) on each of the shape parameters to allow for slightly different morphologies and light profiles. Synthetic fluxes (black data points in Fig.\ref{generated_example} for the shown example) as machine learning inputs are computed by convolving the redshift-shifted, brightness-scaled (gray) spectra with HFF filter transmission curves. And to be more realistic, here and likewise for other generated samples, I also generate small random noise on the generated fluxes. The maximum size of added noise was determined from the root-mean-square spread of S18 magnitudes about the best-fit blue spectrum, i.e., preserving the properties of the original data.  

For each of the 31 high-z galaxies, I repeated the above exercise 19 times to generate a final high-z training sample of size 620. The final low-z training sample was also obtained with the same method, and comprises 624 $z<3$ interlopers.

In Fig.\ref{inspect_augment}, I present distributions of the ten input variables for the augmented high-z sample as blue steps and the low-z sample as red steps. These distributions are largely consistent (effectively the smoothed version owing to random scalings applied) with original shaded histograms, justifying that the augmented sample remains representative of S18 measurements and suitable for model training.

When training the model, I use only 80\% of the augmented sample (randomly selected, hereon referred to as the training set) for parameter optimization, reserving the remaining 20\% (validation set) for performance evaluation on unseen data. To prevent bias, both high-z and low-z galaxies are checked to contribute $\sim 50\%$ of the training and validation sets\footnote{For my identifier model, interlopers constitute 49.9\% of the training set and 51.2\% of the validation set.}. To reduce model training complexity and focus on relative statistical differences of objects to the full population, it is also a common practice in machine learning to \textit{standardize} input variables. For each input variable $v$, this means to replace absolute values by the normalized deviations from the mean, $(v-mean(v))/std(v)$. To ensure consistency, the mean and standard deviation of each input variable are measured from the full augmented sample and subsequently applied to standardize the input variables of both training and validation sets. The same standardization also applies to S18 inputs from other HFF catalogs for more robust predictions. 

I end this subsection by commenting that in papers such as \cite{2501.01942}, cutout images in different bands were also incorporated as inputs for a better determination of galaxy redshifts (via separate Convolutional Neural Networks). While desirable, it was unable to make use of similar cutout images. This is because HFF cluster fields are crowded fields, hence it is generally difficult to obtain large enough cutouts containing only my interested galaxies. 


\begin{figure}
\centering
\begin{subfigure}{.49\textwidth}
  \centering
  \includegraphics[width=\linewidth]{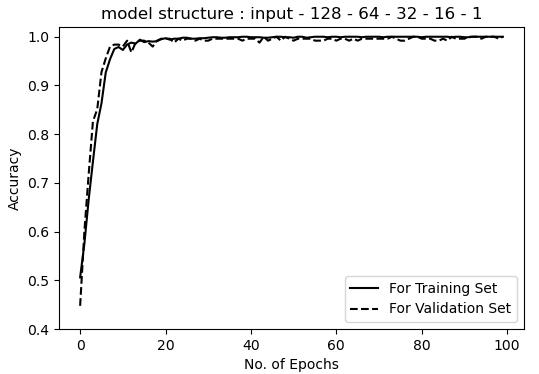}
\end{subfigure}%
\hfill
\begin{subfigure}{.49\textwidth}
  \centering
  \includegraphics[width=\linewidth]{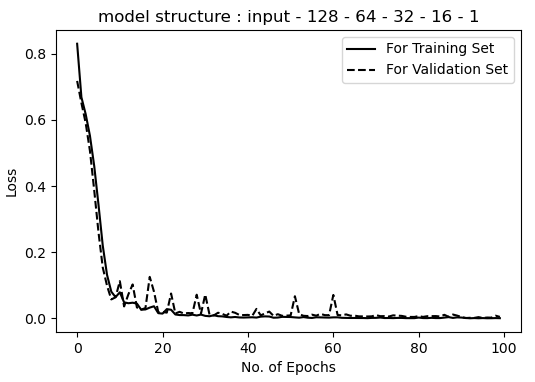}
\end{subfigure}
\caption{Performance of my final adopted model after different training epochs. \textit{Left}: prediction accuracy when the trained model is applied on the training set (solid curve) and validation set (dashed curve). \textit{Right}: loss function values after different training epochs. My adopted model structure is summarized at the top of both panels. After 10 training epochs, the trained model is seen to have stable performance and gradually improves with more training epochs. The near-overlap of solid and dashed curves also reflects that the trained model handles unseen data equally well as the training set. This model achieved a final prediction accuracy of 100\%. }
\label{model_performance_history}
\end{figure}

\subsection{Identified Interlopers in remaining S18 HFF catalogs}
\label{sec_ML_results}

Fig.\ref{model_performance_history} presents the performance of my final adopted model, whose structure is summarized on top of both left and right panels: it consists of an input layer accepting ten S18 measurements, four hidden layers with 128, 64, 32, 16 neurons respectively, and finally an output layer predicting a single label ($>0.5$ for interloper and $<0.5$ for high-z). Black solid curves represent prediction accuracy (left panel) and loss function values (right panel) when the trained model is applied to the training set after different training epochs, while dashed curves correspond to prediction accuracy and loss function values when applied to the unseen validation set. My adopted model demonstrates stable and excellent performance after 10 training epochs, and separates unseen interlopers from unseen high-z galaxies as accurately as the galaxies used for training, as evidenced by the near-overlap of dashed and solid curves. In fact, when applying the final trained model to the validation set, \textit{all} interlopers and high-z galaxies were found accurately identified, achieving a final prediction accuracy of 100\%. I also applied this model to the S18 analysis sample (over $3.5\leq z_{S18} \leq 5.5$, see Sec.\ref{subsec_uv_bright_sample}) from the other five HFF cluster fields. The contamination level was found to be $57.25\%$, in excellent agreement with my previous magnification bias estimate of $56.87 \pm 11.84 \%$.

\begin{figure}
    \centering
    \includegraphics[width=0.95\linewidth]{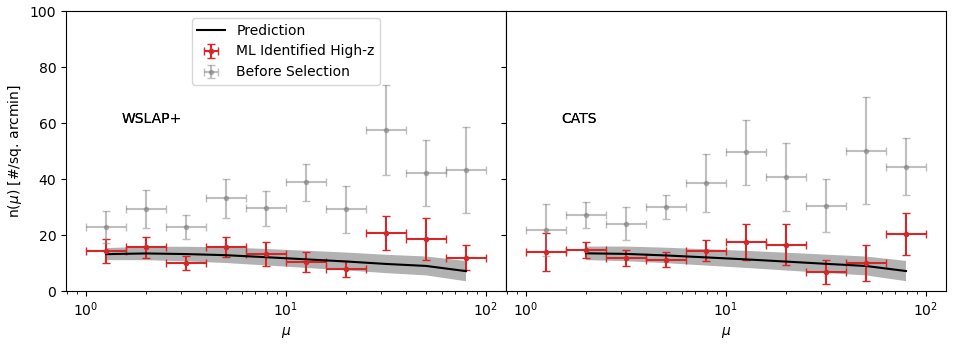}
    \caption{Magnification bias analysis over $3.5\leq z_{S18} \leq 5.5$ after machine learning mitigation. Here, I used only CATS and WSLAP+ models for defining lensing magnification bins, as these are the only two sets commonly available to the remaining five HFF cluster fields. For reference, the surface number densities in different lensing magnification bins prior to any machine learning mitigation are presented as gray data points, and they increase towards higher magnification regions. This behavior contradicts the negative bias previously predicted (Sec.\ref{bias}) and also reflected by the black solid curves. In obtaining the black solid curves, I corrected for the true abundance of high-z galaxies in parallel fields following the contamination levels identified in Fig.\ref{ML_contamination_para}. The observed surface number densities of machine learning selected high-z galaxies are presented as red data points, and they are seen to match excellently with the updated magnification bias prediction. This agreement observed thus demonstrates the robustness of the constructed machine learning model in mitigating interlopers. }
    \label{ML_updated_magbias}
\end{figure}

\begin{figure}
\centering
\begin{subfigure}{.45\textwidth}
  \centering
  \includegraphics[width=\linewidth]{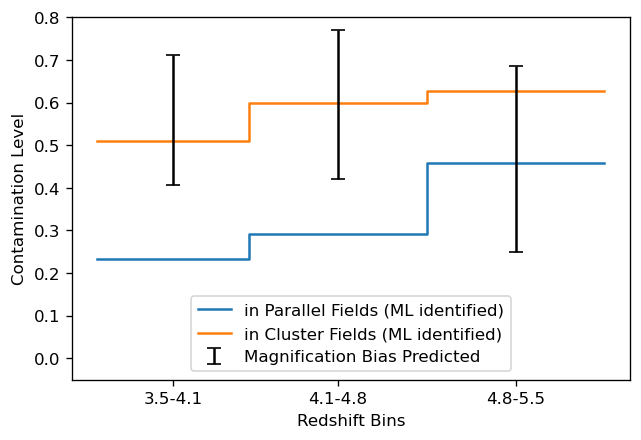}
\end{subfigure}%
\hfill
\begin{subfigure}{.51\textwidth}
  \centering
  \includegraphics[width=\linewidth]{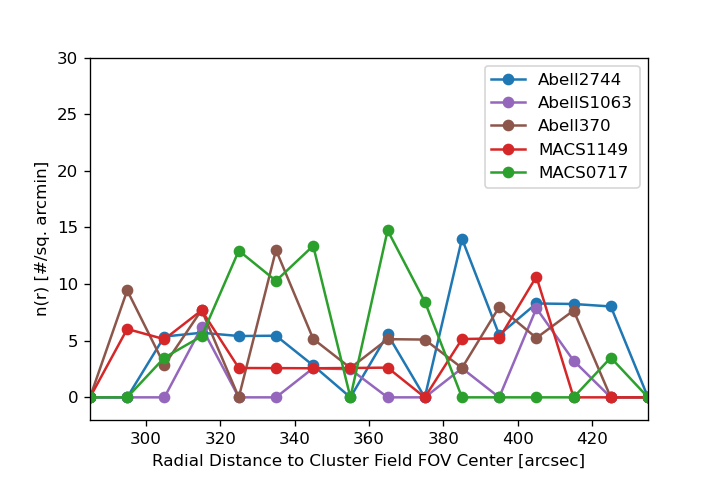}
\end{subfigure}
\caption{\textit{Left}: Contamination levels identified with machine learning method in both cluster (orange step) and parallel (blue) fields in different redshift bins involved in magnification bias analysis over $3.5\leq z\leq 5.5$. Magnification bias predicted contamination levels are also plotted as black error bars, which are seen as largely consistent with the machine learning results. \textit{Right}: Spatial distribution of machine learning identified interlopers in the parallel fields, as measured relative to the respective cluster field FOV center. No clear trend could be claimed in these spatial distributions, suggesting interlopers are randomly distributed across parallel field FOVs. }
\label{ML_contamination_para}
\end{figure}

\begin{figure}
    \centering
    \includegraphics[width=0.95\linewidth]{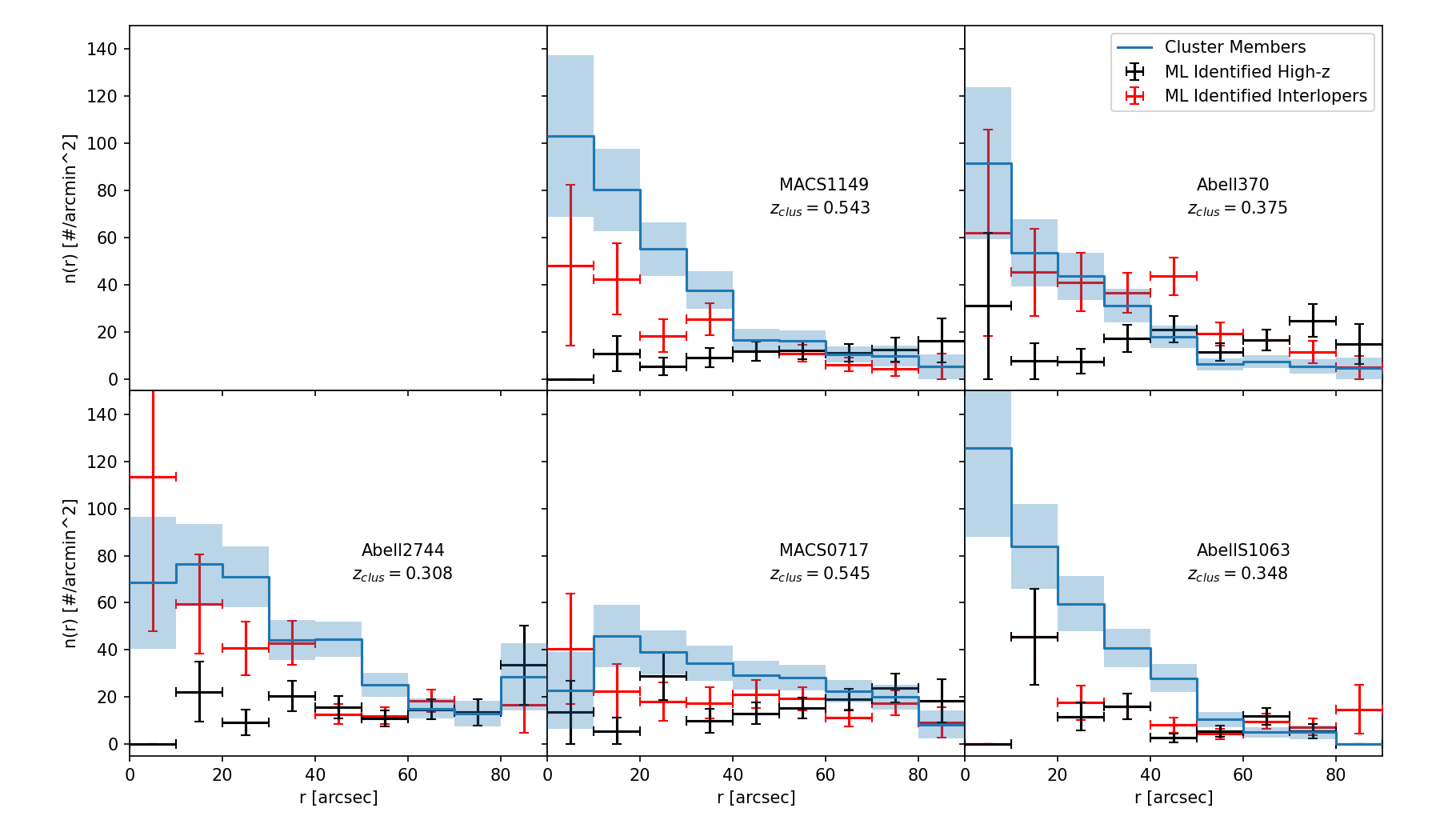}
    \caption{Spatial distribution of machine learning identified interlopers (red data points) and high-z galaxies (black data points) in the remaining five HFF cluster fields without supplementary JWST observations. Identified interlopers in M1149, A370, and A2744 are seen to follow a similar centrally concentrated distribution as cluster redshift galaxies (dominated by cluster members), demonstrating the power of my constructed machine learning model at identifying cluster member interlopers. This centrally clustering trend is less obvious for the other two cluster fields, which may be due to limited statistics or field interlopers being more dominant. }
    \label{ML_updated_radial}
\end{figure}

As a further cross-check, I also examine whether machine learning selected high-z galaxies are magnification-bias-wise clean from contamination (see Chap.\ref{chap:contamination} for detailed methodology). In Fig.\ref{ML_updated_magbias}, the surface number densities of machine learning selected high-z galaxies are shown as red data points in different lensing magnification bins. Here, I focused only on WSLAP+ and CATS models, which were the only sets commonly available to all five HFF cluster fields besides M0416. Observed surface number densities without any machine learning mitigation are also shown as gray data points for reference. It could be seen that the previously identified increasing trend (of gray data points) to higher lensing magnification regions is now replaced with a decreasing trend, consistent with the previously identified negative bias and also magnification bias prediction (black solid curve). In Fig.\ref{ML_updated_magbias}, the magnification bias prediction shown is updated compared to Fig.\ref{pos_bias_magbin_diego_collective}. The update was based on the observation that parallel fields were also contaminated by low-z interlopers, as indicated by the blue step in the left panel of Fig.\ref{ML_contamination_para}. From the right panel of Fig.\ref{ML_contamination_para}\footnote{Here, I used radial distance to the respective cluster field FOV center to test for any distributional signature from distant cluster members. }, it could be argued that identified interlopers are mostly randomly distributed across parallel field FOVs. Correspondingly, the predicted abundance of high-z galaxies was corrected through a uniform scaling across the analysis FOVs to account for the true abundance of high-z galaxies in parallel fields. Such a scaling was also separately applied in each of the small redshift bins involved (see Sec.\ref{sec_mag_bias_equation}), following the different contamination level identified in each bin from the left panel of Fig.\ref{ML_contamination_para}. With this correction, total predicted densities (black curves presented) in cluster fields could be seen in excellent agreement with observed densities (red data points) after machine learning mitigation, justifying the robustness of the constructed interloper identifier model. 

In Fig.\ref{ML_contamination_para}, I have also indicated the contamination level anticipated from the original magnification bias analysis (i.e., without correction on the actual abundance of high-z galaxies) in different redshift bins as black error bars. Owing to the contamination in parallel fields, I expect the contamination level in cluster fields to be underestimated, which is indeed observed in the highest redshift bin of 4.8-5.5. For the other two redshift bins, however, magnification bias analysis yields contamination levels that are comparable to, or even exceed, those determined by the machine learning method. This suggests that cluster members are likely a more dominant source of contamination in those redshift bins. Overall, I find that the contamination levels predicted by the magnification bias method agree with those identified by machine learning within the error bars. This alignment may be largely due to limited statistics (i.e., large error bars) in each smaller bin, but it also demonstrates the effectiveness of magnification bias for this type of analysis.

Another check on the robustness of the constructed model is to examine the radial distribution of identified interlopers in the remaining five HFF cluster fields. This check is performed in Fig.\ref{ML_updated_radial}, where I compare the radial distribution of identified interlopers (red data points) and high-z galaxies (black data points) with cluster-redshift galaxies ($z_{S18} \in z_{clu}\pm 0.15$, blue step histograms). For three of the cluster fields (M1149, A370, and A2744), identified interlopers could be seen to agree well with the distribution of cluster members. This agreement is less obvious for the other two cluster fields, which may be due to limited statistics or field interlopers being more dominant.

\begin{figure}
\centering
\begin{subfigure}{.49\textwidth}
  \centering
  \includegraphics[width=\linewidth]{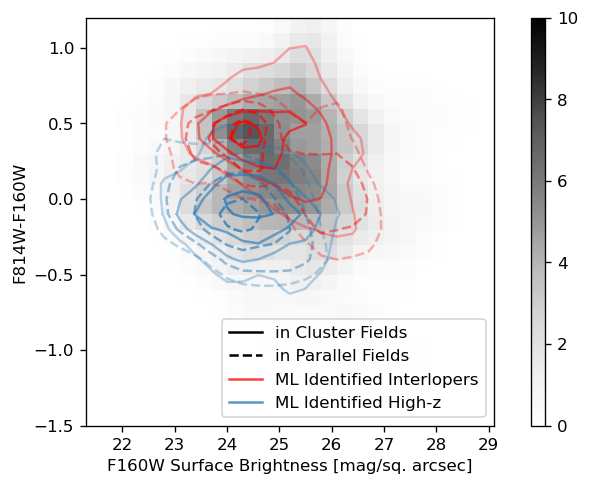}
\end{subfigure}%
\hfill
\begin{subfigure}{.46\textwidth}
  \centering
  \includegraphics[width=\linewidth]{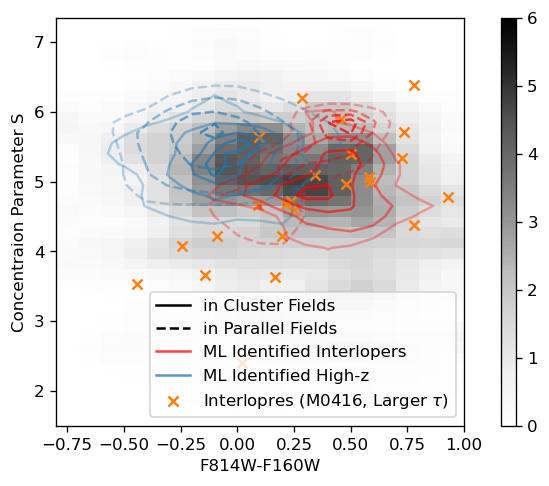}
\end{subfigure}
\caption{\textit{Left}: color-surface brightness distribution of machine learning identified interlopers (red) and high-z galaxies (blue) in both cluster (solid contours) and parallel (dashed contours) fields. The underlying background histogram is from the full S18 analysis sample, combining all HFF cluster fields. \textit{Right}: color-concentration parameter $S$ distributions of machine learning identified interlopers and high-z galaxies, presented in the same manner as in the left panel. Orange crosses correspond to field interlopers identified in the field of M0416 with \texttt{bagpipes} analysis, and they are shown to demonstrate redder field interlopers tend to have more centrally concentrated light profiles (likely due to past cycles of star formation driving out the outmost stellar contents).  }
\label{ML_check_csbcon}
\end{figure}

In Fig.\ref{ML_check_csbcon}, I also examine the color-surface brightness (left panel) and color-concentration parameter $S$ (right panel) distribution of machine learning identified interlopers (red contours) and high-z (blue contours) galaxies in both cluster (solid lines) and parallel (dashed lines) fields. In these figures, background 2d histograms are again measured from all high-z candidates in cluster fields as claimed by S18 catalogs. It could be seen that high-redshift galaxies and interlopers in parallel fields are largely consistent with those found in cluster fields in both figures. This observation is not surprising, given that the identifier model would preferentially assign the same label to galaxies with similar statistical properties. 

As a continuation of the discussion in Sec.\ref{sec_bagpipes_results}, it is interesting to notice from the right panel of Fig.\ref{ML_check_csbcon} that identified interlopers in parallel fields (also field interlopers in M0416, shown as orange crosses) tend to have less concentrated light profiles when they exhibit a bluer color (possibly owing to be more star-forming). In contrast, interlopers with redder colors, which are likely quiescent, tend to have more concentrated light profiles. This observation could be understood if past star formation has driven out the outer stellar wings of quiescent galaxies to a fuller extent, thereby leaving only the central core.

Finally, from the left panel of Fig.\ref{ML_contamination_para}, I notice that the machine learning identified contamination levels in both cluster (orange step) and parallel (blue step) fields are, in fact, higher in the higher redshift bins. This can be understood as galaxies are naturally less abundant at higher redshifts. In contrast, many low-redshift (low-z) galaxies may potentially contaminate the sample. This observation highlights the importance of accurately identifying low-z interlopers for constructing any high-z UV LFs. And in this section, I have demonstrated that machine learning can be particularly effective for such a task.

\chapter{Magnification Bias to Constrain Faint-end UV LF}
\label{chap:UVLFtesting}

Following the mitigating power of combined HST and JWST, in this chapter, I aim to perform a more reliable test on the faint-end of UV LF, free from contamination by low-z interlopers. In particular, I am interested in testing for the faint-end turnover predicted by wave-like $\psi$DM models, previously already hinted by the analysis of \cite{Leung2018} (favoring a $\psi$DM mass of $1.6\times 10^{-22}$eV). For such a test, I again (as in Chap.\ref{chap:contamination} and also by \cite{Leung2018}) rely on the concept of magnification bias for a simpler and more direct analysis, as the complicated task of identifying multiply lensed pairs can be avoided (see Chap.\ref{chap:intro}). With magnification bias, the specific signature of faint-end turnovers is to result in a sharper drop (relative to the $p$CDM paradigm) in the predicted surface number density of high-z galaxies in higher magnification bins. By observing, or not observing, such a sharper decrease, I can correspondingly claim the evidence for $\psi$DM or provide a bound on the respective $\psi$DM mass. 

\section{Motivation for $\psi$DM}

I begin this chapter by briefly motivating the consideration of $\psi$DM models. As introduced in Chap.\ref{chap:intro}, $\psi$DM, with its astronomical-scale wave nature, is promising at resolving small-scale problems faced by conventional heavy particle CDM (see \cite{2005.03254} for an extensive review). More recently, $\psi$DM has also received further phenomenological support. For instance, the distribution of stars measured around the inner cores of dwarf galaxies supports the presence of solitons associated with $\psi$DM of mass $\sim 1.5-1.8 \times 10^{-22}$eV \citep{Pozo2024PhRvD, Pozo2020}. In addition, the velocity dispersions measured for different dwarf galaxies collectively display an inverse scaling relation with the respective soliton core radius, matching the anticipated behavior based on the uncertainty principle, i.e., the product of (uncertainty in) position and momentum of wave packets to be a constant.

The wave interference pattern distinct to $\psi$DM of mass scale $\sim 10^{-22}$eV has also been demonstrated to be capable of explaining gravitational lensing anomalies typically seen in smooth lens models (often associated with $p$CDM) \citep{Amruth}. Such an interference pattern also provides an explanation for the observed skewness of micro-lensed stars to the negative parity side of the lensing critical curve, whereas adding sub-halos to the $p$CDM lens model leads to opposite behavior \citep{TomKeith2024}. It has also been demonstrated that the wave interference pattern is promising at explaining the oblate morphology of dwarf spheroidals, by inducing random walks on stars, gradually kicking them away from the stellar disk \citep{1809.04744, 1809.07673, 2308.14664}. 

One of the predictions by $\psi$DM models is that the abundance of low mass halos will be suppressed relative to the heavy particle CDM paradigm ($p$CDM). Following this suppression, UV LFs, especially at high redshifts, would display a faint end turnover owing to the suppressed abundance of intrinsically faint galaxies associated with low mass halos. As previously discussed (also in Sec.\ref{psiDM_UVLF} below), such $\psi$DM-induced turnovers occur at $M_{UV}\gtrsim -15$, a magnitude scale that remains hard to access with deep blank field surveys. While strong gravitational lensing by massive galaxy clusters allows the fainter magnitudes to be probed, dim cluster members (also field galaxies with old stellar population) may severely contaminate the high redshift galaxy sample, as I have extensively demonstrated in Chap.\ref{chap:contamination}. As I will further discuss in Sec.\ref{sec_comparions_existing_massbound}, the potential severe contamination then casts doubts on previous lensing field analysis of the UV LFs at the faint end. 

The promise for a more reliable lensing field analysis was presented in Chap.\ref{chap:contamination_jwst}, in which I demonstrated that the combination of deep JWST observations with deep HST observations allows for individual mitigation of low-z interlopers. Moreover, the deep learning method was also shown to be able to capture intricate differences encoded in existing S18 measurements, thereby extending the individual mitigating power of interlopers to the remaining five HFF catalogs without supplementary JWST observations. In this chapter, I thus perform a more reliable test of faint-end turnover in UV LF, focusing only on machine learning identified high-z galaxies among S18 analysis samples (over $3.5\leq z\leq 5.5$), free from any contamination. For such an analysis, magnification bias will be shown particularly helpful given it avoids the need for individually identifying multiply lensed pairs (see the discussion in Chap.\ref{chap:intro}), and also its sensitivity to the faint-end of UV LFs. With subsequent magnification bias analysis, I demonstrate that I found no hint of faint-end turnover in UV LFs, thereby leading to a bound on the mass of $\psi$DM particles of $>2.97\times 10^{-22}$eV at the 95\% confidence level. 

In addition, the mitigating power of combined JWST and HST observation also extends to higher redshifts. This fact, combined with the construction of a more data-complete catalog for the field of M0416 (reaching $\sim$1 mag fainter than \cite{Shipley2018}), enabled me to reliably test for any faint-end turnovers in UV LFs at $z>6$. In particular, I will focus on the redshift range of $6\leq z\leq 10$, chosen to receive stronger impacts from $\psi$DM, and also having JWST data for a more robust determination of galaxy redshifts (JWST covers the rest frame Balmer wavelength for galaxies up to $z\sim10$). Similar to $3.5\leq z\leq 5.5$ range, I found no evidence of faint end turnover, which then translated to a slightly weaker (as only for one field) of $>2.53\times 10^{-22}$eV at 95\% confidence level.

The rest of this chapter is organized as follows. In Sec.\ref{psiDM_UVLF}, I first introduce how the $\psi$DM-induced faint-end turnover in UV LFs is parameterized. Then in Sec.\ref{sec_UVLF_faintend_magbias_method}, I recap on the magnification bias methodology, including also the selection of reliable $6\leq z\leq 10$ candidates from constructed JWST+HST-combined catalogs. Results will be presented in Sec.\ref{Sec_UVLF_results}, followed by some comparison with existing mass bounds on $\psi$DM particles. In particular, I will mention how the existing stringent mass bounds on $\psi$DM have separate issues, and hence may not be reliable compared to my new bounds. Finally, I provide some concluding remarks and discussions in Sec.\ref{sec_UVLF_conclu}.


\section{$\psi$DM induced UV LF faint end turnover}
\label{psiDM_UVLF}

The relative suppression on UV LF at the faint-end has been investigated by various teams, but for my analysis, I will follow the specific parametrization given by \cite{Schive2016} (S16) as I further motivate below. It is also worth mentioning that the same parametrization by S18 was also used by \cite{Leung2018} (L18) and subsequently found tentative evidence for the faint-end turnover. To check if the same faint-end turnover signature could be recovered with my more reliable test, then serves as another motivation for adopting S16's parametrization. 

\begin{figure}
    \centering
    \includegraphics[width=0.9\linewidth]{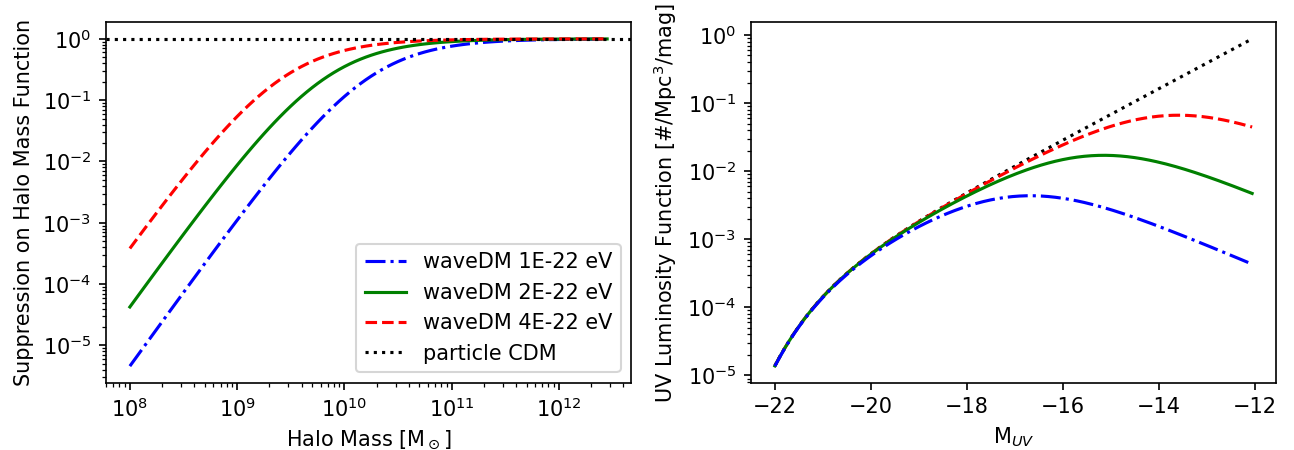}
    \caption{Suppression on HMF (left) and on UV LFs (right) induced by $\psi$DM of different masses (colored lines) relative to $p$CDM paradigm (black dotted lines) as determined by \cite{Schive2016}. A heavier $\psi$DM copy is seen to result in a lower mass scale for suppression on HMF, and hence a fainter faint-end magnitude turnover scale. }
    \label{waveDM_suppresion}
\end{figure}

In S16, the relative suppression induced by $\psi$DM on the faint-end of UV LFs was determined through N-body simulation with appropriate initial conditions (i.e., small-scale fluctuations are suppressed on initial density perturbations). Through their N-body simulations, S16 first determined the halo mass function (HMF) in the $\psi$DM context, finding them suppressed below a halo mass of $\lesssim 10^{10} m_{22}^{-4/3} M_\odot$, where $m_{22}$ is $\psi$DM mass in units of $10^{-22}$eV. More specifically, the HMF $dn/dM_h$ determined by S16 is 
\begin{equation}
\frac{dn(M_h,z)}{dM_h}\bigg{|}_{\psi DM} = \frac{dn(M_h,z)}{dM_h}\bigg{|}_{pCDM} \bigg{[} 1+ \bigg{(}\frac{M_h}{1.6\times 10^{10} m_{22}^{-4/3} M_\odot} \bigg{)}^{-1.1} \bigg{]}^{-2.2},  
\end{equation}
\noindent and the suppression factor relative to $p$CDM HMF is presented on the left panel of Fig.\ref{waveDM_suppresion} for $m_{22}=1,2,4$. Naturally, a heavier $\psi$DM copy allows for the formation of smaller halos, and hence associates with a lower mass scale for the suppression. 

To translate the relative suppression obtained on HMFs to a faint-end turnover in UV LFs, S16 adopted the conditional LF method \citep{Cooray_2005}. This method fits for the conditional probability for halos of mass $M$ to have luminosity $L$, by matching the converted UV LFs to the non-lensing field UV LFs (for S16, the measurements were from \cite{Bouwens2015ApJ}, B15, reaching $M_{UV}< -16$). To acknowledge that the matching condition determined in the $p$CDM context does not necessarily apply to a $\psi$DM-dominated Universe, fitting to non-lensing field UV LFs was also separately performed for $p$CDM and $\psi$DM models of different masses. Relative to $\psi$DM UV LFs, the faint-end turnover induced by $\psi$DM was found to be well parametrized by the following functional form
\begin{equation}
\phi_{\psi DM}(L) = \phi_{pCDM}(L) \bigg{[} 1 + \bigg{(} \frac{L}{L_\phi} \bigg{)}^\gamma \bigg{]}^{\beta/\gamma}
\label{Truncated_Schechter}
\end{equation}
\noindent where
\begin{align}
     M_{\phi} &= -17.44 + 5.19\log_{10}\bigg{(}\frac{m_{22}}{0.8}\bigg{)}- 2.71 \log_{10}\bigg{(}\frac{1+z}{7}\bigg{)} \label{M_psiDM_cutoff_formula}\\
     \beta & = 1.69+0.03(z-6), \; \; \; \gamma = -1.10.
\end{align}
\noindent In Equ.\ref{Truncated_Schechter}, $\phi_{pCDM}$ corresponds to UV LF without any faint-end turnover, and is conventionally parametrized as a Schechter function
\begin{equation}
    \phi_{pCDM}(L) = \frac{\phi_*}{L_*} \bigg{(} \frac{L}{L_*}\bigg{)}^\alpha e^{-L/L_*},
\label{Schechter_UVLF}
\end{equation}
\noindent where $\phi_*$ is the normalization factor, $L_*$ is characteristic luminosity for exponential suppression, and $\alpha$ is the faint end slope. In the right panel of Fig.\ref{waveDM_suppresion}, I present the more recently determined Schechter UV LF (effectively $\phi_{pCDM}$) at $z\sim 6$ from deep blank fields by \cite{Bouwens2021} (B21, largely consistent with B15) as black dotted line, and present the corresponding $\phi_{\psi DM}$ (for $\psi$DM masses of $1,2,4 \times 10^{-22}$eV) determined from Equ.\ref{Truncated_Schechter} as colored curves. Rather than exhibiting a sharp cut-off below a faint-end turnover magnitude scale, $\phi_{\psi DM}$s are seen to asymptote to a power-law behavior at magnitudes $M \gg M_\psi$. Consequently, Equ.\ref{Truncated_Schechter} is also referred to as \textit{truncated} Schechter function. And following the smaller halo mass scale for HMF suppression, the truncation (or faint-end turnover) scale is seen fainter for a heavier $\psi$DM mass.


Note that relative suppression of $\phi_{\psi DM}$ t $\phi_{pCDM}$ may be (slightly) altered if a different prescription is used for translating HMF to UV LFs. For example, \cite{Winch2024ApJ} (also with conditional LF method) and \cite{Corasaniti2017} (with halo abundance matching method) calibrated the star formation rates (SFRs) for halos of different masses, instead of fitting for a direct map between luminosity and halo mass as S16 did. The more convoluted link established with star formation rate then likely produces slightly different UV LFs than Equ.\ref{Truncated_Schechter}, but the resulting UV LFs were not provided for a direct comparison, unfortunately. The difference in the resulting $\phi_{\psi DM}$, however, is not anticipated to be huge as all methods were implemented to reproduce largely consistent UV LFs from deep blank fields. Extrapolating the compatibility of different methods in reproducing similar UV LFs over bright magnitudes to fainter magnitudes, I thus anticipate different prescriptions to not significantly change the faint-end suppression induced by $\psi$DM. 

Instead, I comment that the more determining influence is from potential differences in HMFs, directly characterizing the abundance of lower mass halos available. For instance, HMFs determined with analytical extended Press-Schechter formalism \citep{KulkarniOstriker2022}, or semi-analytic methods \citep{1608.02575, Winch2024ApJ} tend to have a sharper cut-off at the lower mass end than what S16 determined. Correspondingly, a much stronger faint-end turnover on UV LFs would be anticipated for these HMFs. Nonetheless, HMFs determined by S16 better agree with the results from full wave simulation of $\psi$DM \citep{2209.14886}, which is considered to more realistically reflect the halo abundance in $\psi$DM context. The last point thus gives credibility for adopting the $\psi$DM HMF parametrization, and hence the suppression on UV LFs, as determined by S16.



Before proceeding to discuss magnification bias methodology, I still need to specify the choice of $\phi_{pCDM}$, UV LFs without faint-end turnover, at different redshifts. In S16 (also L18), $\phi_{pCDM}$ was determined from the measurements of B15, but these deep blank field UV LFs were subsequently updated by B21. The consistency between B21 and B15 measurements suggests that B21 could also be used for $\phi_{pCDM}$, despite the precise form of Equ.\ref{Truncated_Schechter} was determined with fitting to B15 measurements.  

From my previous analysis in Chap.\ref{chap:contamination}, I also separately determined the UV LF-redshift relation over $1.2\leq z\leq 5.5$, and the resulting relations were commented to be largely consistent with B21 determinations (i.e., may also be used as $\phi_{pCDM}$). Relative to B15 or B21 measurements, my previously determined UV LF-redshift relation has the advantage of being derived from S18 catalogs for HFF parallel fields, which were imaged to the same observational depth as in the cluster fields. I.e., I consider my previously derived UV LF-redshift relation to best suit the S18 catalogs in HFF cluster fields over the same observed magnitudes. As such, for my subsequent $3.5\leq z\leq 5.5$ analysis (relying on S18 catalogs), I will also use the UV LF-redshift relation previously derived in Chap.\ref{chap:contamination}. 

\begin{figure}
\centering
\begin{subfigure}{.49\textwidth}
  \centering
  \includegraphics[width=\linewidth]{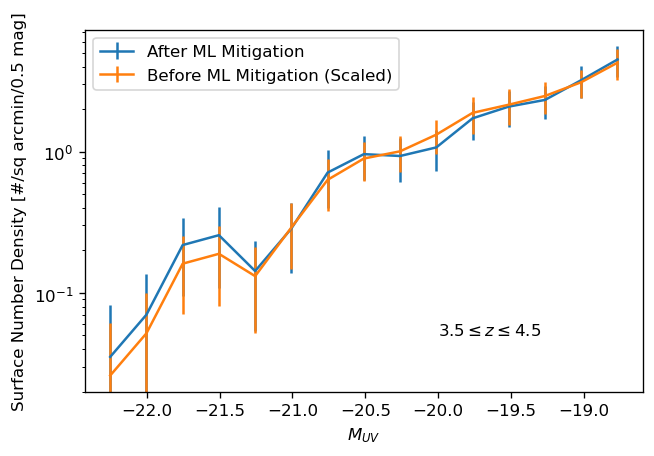}
\end{subfigure}%
\hfill
\begin{subfigure}{.49\textwidth}
  \centering
  \includegraphics[width=\linewidth]{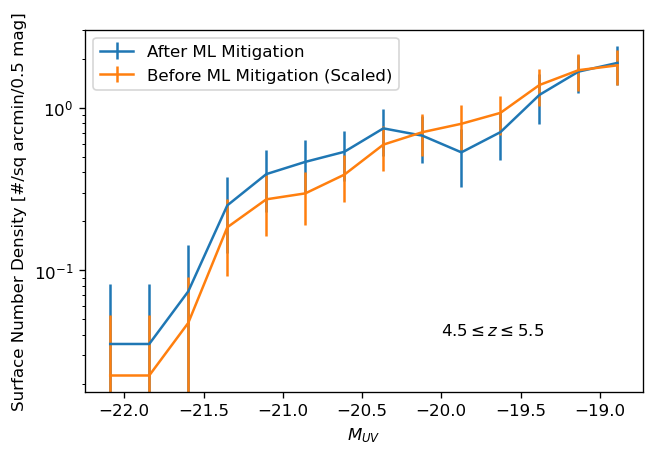}
\end{subfigure}
\caption{Observed surface number density of galaxies in different $M_{UV}$ bins before (orange) and after (blue) machine learning mitigation over redshifts $3.5\leq z\leq 4.5$ (left) and $4.5\leq z\leq 5.5$ (right). The orange distributions shown are also downscaled to match the number of high-z galaxies identified by machine learning for a more direct visual comparison in the trends. The resulting distributions could be seen to match well over $3.5\leq z\leq 4.5$. Over $4.5\leq z\leq 5.5$, the smaller sample size left by machine learning leads to larger scatters in observed surface number densities, but the resulting blue distribution displays a similar overall trend as the orange distribution. These observations justify the usage of previously determined UV LF-redshift relations before machine learning mitigation.}
\label{ML_Mitigated_UVLF}
\end{figure}

It is also important to note that my previously determined UV LFs were based on S18 catalogs before machine learning mitigation, but their suitability to the machine learning mitigated S18 analysis sample can be motivated from Fig.\ref{ML_Mitigated_UVLF}. In Fig.\ref{ML_Mitigated_UVLF}, I compare the observed surface number density of galaxies in different magnitude bins (directly reflecting the underlying UV LFs) before (in orange) and after (in blue) machine learning mitigation. It could be seen that the distribution of machine learning-identified high-z galaxies is largely consistent (albeit with larger scatter) with the orange distribution before machine learning mitigation across both $3.5\leq z\leq 4.5$ (left panel) and $4.5\leq z\leq 5.5$ (right panel). Consequently, I anticipate UV LFs measured from the orange distributions as suitable for the machine learning selected sample, an argument that could also be extended to lower redshifts (facing less contamination). Moreover, previously in Fig.\ref{ML_updated_magbias} of Sec.\ref{sec_ML_results}, the updated magnification bias prediction (relying on UV LF-redshift relations presented in Chap.\ref{chap:contamination}) was seen to reproduce excellently the observed surface number densities of machine learning-identified high-z galaxies. This agreement thus further justifies the usage of previously determined UV LF-redshift relations.

My previously determined UV LF-redshift relations, however, do not extend to the higher redshift range of $6\leq z\leq 10$. A separate determination of UV LF-redshift relation from parallel fields was also not possible given (1) the three epochs of PEARLS observations covered three separate, non-overlapping parallel fields, preventing measurements of UV LFs to the same observational depth as achieved in the M0416 cluster field; and (2) JWST parallel fields do not overlap with the HFF parallel fields, hindering more robust determination of photo-zs and also hence the individual mitigation of potential low-$z$ contaminants (an issue more severe at higher redshifts). For this redshift range, I will instead adopt the UV LF-redshift relations as measured by B21.

Finally, I briefly comment on how the faint-end turnovers induced by baryonic physics (see Fig.\ref{UVLF_turnover} of Chap.\ref{chap:intro}) can be neglected from my subsequent analyses. From Fig.\ref{ML_Mitigated_UVLF}, the faintest $M_{UV}$ reachable, assuming a maximum lensing magnification factor of 100, is seen $\sim -14.5$ over $3.5\leq z\leq 5.5$. A similar check for the $6\leq z\leq 10$ redshift range reveals a similar magnitude limit of $\sim -14$. Such a magnitude limit is relevant for testing $\psi$DM-induced turnovers, but the baryonic physics-induced turnover on $p$CDM UV LFs occurs at even a fainter magnitude of $M_{UV}\sim -12$. Consequently, in the magnitude range relevant, $p$CDM UV LFs ($\phi_{pCDM}$) can be safely assumed to continuously and steadily increase to the fainter end as shown in the right panel of Fig.\ref{waveDM_suppresion} (black dotted line), with the effect from baryonic physics (which is also uncertain) negligible.





%
\subsection{Magnification Bias Methodology}
\label{sec_UVLF_faintend_magbias_method}

I now quickly recap the magnification bias methodology. At any given redshift $z$, magnification bias extrapolates the underlying UV LF ($\phi_{\psi DM}$ or $\phi_{pCDM}$) to magnitudes fainter than the detection limit, and predicts the total surface number density of gravitationally lensed high-z galaxies at different levels of lensing magnification. Owing to the suppression at the faint-end, $\phi_{\psi DM}$ UV LFs lead to a significant drop in predicted number density in comparison with $\phi_{pCDM}$ at higher lensing magnifications. Such an anticipated behavior can then be compared with the observed surface number density of galaxies at the same level of lensing magnifications, to test for evidence in favor of or against $\psi$DM models. And as magnification bias is a local effect on the image plane, multiply lensed images of the same background galaxy could be treated independently, furnishing a simpler and more direct test for the faint-end behavior of UV LFs.  

The implementation of magnification bias over redshift $3.5\leq z\leq 5.5$ largely follows the previously introduced method in Sec.\ref{sec_mag_bias_equation}, and the essential details could be summarized as follows. To measure the surface number density of machine learning-identified high-z galaxies, I divide HFF cluster field FOVs into ten magnification bins logarithmically equally spaced in lensing magnifications up to a factor of 100. To mitigate the dependence of results on the lens modeling technique, I will adopt two sets of lens models (commonly available for all cluster fields) for defining the lensing magnifications: v4 CATS \textit{lenstool} models representative of the parametric modeling method, and v4.1 WSLAP+ models representative of the free-form lens modeling method. For each magnification bin, I predict the total surface number density of galaxies (at a given redshift $z$ and an underlying UV LF) by collecting the total number of galaxies predicted on relevant image plane pixels, and subsequently dividing this number by the total area. In contrast to previous implementation, HFF parallel fields were also found to suffer from contamination by low-z field interlopers (Sec.\ref{sec_ML_results}). As a consequence, the true abundance of high-z galaxies in HFF parallel fields needs to be corrected via a uniform scaling across the lensing field FOV when implementing magnification bias prediction.   

Within the same considered magnification bin, the observed surface number density could be obtained with a similar method as the predicted level: by counting the total number of galaxies within $z\pm z_{NMAD}$ from the machine learning selected sample. Here, $z_{NMAD}$ characterizes the typical scatter of photo-zs against respective spec-zs for secure sources confirmed with spectroscopy, and was determined to be $z_{NMAD} \sim 0.062(1+z)$ for S18 catalogs. Upon obtaining the predicted and observed surface number densities at a particular redshift $z$, the same procedure will be repeated at a slightly higher redshift, such that the new redshift range for counting galaxies continues from the previous selection range. The whole process is also repeated several times\footnote{In fact, from the discussion in Sec.\ref{sec_ML_results}, $3.5\leq z\leq 5.5$ is in total split into three smaller redshift bins of 3.5-4.1, 4.1-4.8, and 4.8-5.5. Here, I am summarizing the methodology on a more general basis. }, until the full redshift range of $3.5\leq z\leq 5.5$ is covered. Subsequently, the predicted and observed surface number densities at different central redshifts will be summed up to yield the total predicted and observed density over the full redshift range. 

The implementation of magnification bias over $6\leq z\leq 10$ is largely consistent with the above method, except for the following key differences. First, the analysis over $6\leq z\leq 10$ focuses only on the M0416 cluster field. Correspondingly, more recent lens models available for this cluster will also be incorporated. The two new lens models I considered include a state-of-the-art glafic model (carefully examined for its predictive power) provided by collaborator, and an updated free-form lens model by \cite{Diego2024}, incorporating the latest JWST lensing constraints. Secondly, there are much fewer galaxies available over $6\leq z\leq 10$. I therefore adopted a wider magnification bin width of 0.3 (in log-scale, 0.2 over $3.5\leq z\leq 5.5$) for better density measurements. In addition, I define magnification bins up to a magnification factor of 32 (instead of 100), as there were no robust $6\leq z\leq 10$ candidates magnified above this level. Finally, the $z_{NMAD}$ relevant is different from the $3.5\leq z\leq 5.5$ range, and was determined from my JWST+HST combined photo-z catalog to be $z_{NMAD}\sim 0.052(1+z)$ (see Sec.\ref{catalog_for_M0416}).

\subsection{Analysis Sample}

One last thing to recap for magnification bias methodology is that not all galaxies passing the redshift selection criteria will be considered. For counting and hence measuring the observed surface number density, I restrict to galaxies that are brighter than the data completeness threshold in the band capturing their respective rest-frame 1600\AA. For the $3.5\leq z\leq 5.5$ analysis, the corresponding bands relevant are F814W (with data completeness threshold of 27.5) for redshifts $z\leq 4.9$, and F105W (27.75) above $z>4.9$. For M0416 only analysis over $6\leq z\leq 10$, the relevant bands are instead the F115W band below $z<7.3$, and F150W band above $z>7.3$, both with a data completeness threshold of 29 (see the detailed discussion in Sec.\ref{sec_injection_recovery} for F150W). 

\begin{figure}
    \centering
    \includegraphics[width=0.9\linewidth]{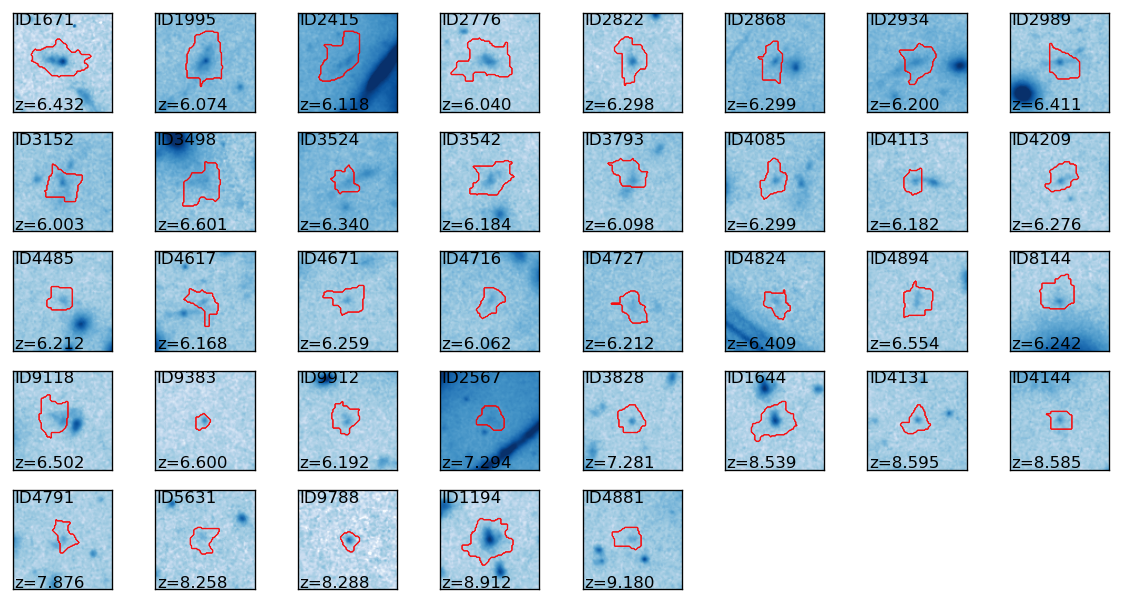}
    \caption{Selected sources in the robust $6\leq z\leq 10$ sample, with red contours indicating their respective \texttt{GNUastro} segmentation. All these sources have high confidence in their respective photo-z, and suffer no segmentation nor bright neighbor issues, as could be visually verified. These sources are also the UV bright ones of direct relevance to magnification bias analysis, i.e., they are brighter than the data completeness threshold in the band capturing their respective rest-frame 1600\AA. }
    \label{z610_galaxies}
\end{figure}

Furthermore, galaxies sitting in very bright regions (exclusion regions) are also not included, as the local brighter background lowers the confidence in photometries, and also alters the effective detection threshold by impeding the detectability of intrinsically fainter galaxies. Following my previous contamination level analysis, I implement the same exclusion regions over $3.5\leq z\leq 5.5$, defined with a surface brightness threshold of 24.08 mag/arcsec$^2$ on HFF F160W band. The exclusion region for $6\leq z\leq 10$ analysis was also defined with the same surface brightness threshold for consistency, but on JWST F200W band instead, as F200W band was found to be more constraining (Integrated Galaxy Light also peaks in F200W, \cite{Windhorst2022}). It should also be noted that the data completeness thresholds quoted previously were all determined after implementing exclusion regions. The removal of bright regions ensures spatial uniformity in the detection thresholds, a critical condition for a consistent implementation of magnification bias across the entire lensing field FOV.

Over $3.5\leq z\leq 5.5$, the analysis sample thus selected is same as previously introduced in Sec.\ref{subsec_uv_bright_sample}, with the additional condition that galaxies are not flagged as low-z galaxies by the constructed machine learning model. Over the redshift range $6\leq z\leq 10$, I obtained a total of 37 UV-bright sources following the aforementioned selection procedure. Motivated from my previous analysis on selecting robust interlopers (Chap.\ref{chap:contamination_jwst}), these 37 sources are also restricted to be the ones that have (1) high confidence in respective photo-z (posterior PDFs in redshift are conditioned to have a single strong likelihood peak), and suffers no \texttt{GNUastro} segmentation issues (running across bright region) nor very bright neighbor issues to impede their photometry measurements. 

\begin{figure}
\centering
\begin{subfigure}{.9\textwidth}
  \centering
  \includegraphics[width=\linewidth]{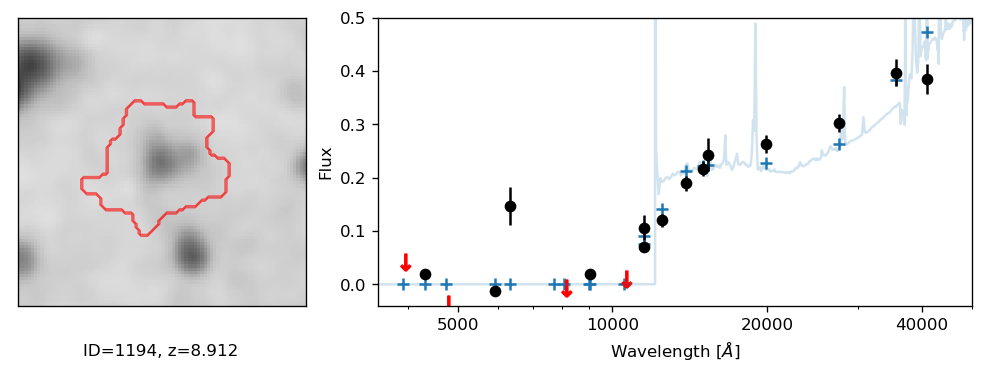}
\end{subfigure}%
\vfill
\begin{subfigure}{.9\textwidth}
  \centering
  \includegraphics[width=\linewidth]{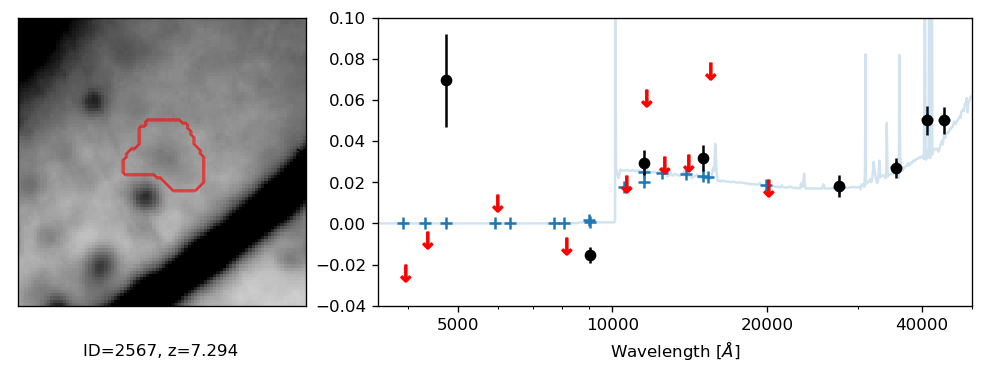}
\end{subfigure}
\caption{Examples of a very bright (top panel) and very faint galaxy selected from the high confidence $6\leq z\leq 10$ sample, with their cut-out image shown on the left and corresponding SEDs shown on the right. For both of these galaxies, it could be verified that the best-fit spectrum (light blue curve) and modeled SEDs (blue pluses) reflect well spectral features captured by respective SEDs (black data points, red arrows correspond to $3\sigma$ upper limits). This demonstrates the robustness of our photo-z determination. }
\label{z610_spec_fitting}
\end{figure}

All of 37 selected sources are presented in Fig.\ref{z610_galaxies} with a cutout size of 2.7 arcsec$^2$, and red contours indicate their respective \texttt{GNUastro} segmentation. For all of these selected sources, I also examined their respective best-fit spectra to confirm the robustness of the photo-z fitting. Two examples are provided in Fig.\ref{z610_spec_fitting}, where the top panel features one of the brightest galaxies among the sample, and the bottom panel features one of the faintest galaxies. Irrespective of their very different brightness, their respective SEDs (black data points, with red down-pointing arrows reflecting upper limits for $<3\sigma$ measurements) could be seen to nicely capture the Lyman break and the rise associated with the Balmer break at the corresponding redshifts. Both of these spectral features are also well captured by the best-fit spectrum (in blue) and modeled SEDs (blue pluses), giving credit for their determined photo-zs. 

\begin{figure*}
    \centering
    \includegraphics[width=0.9\textwidth]{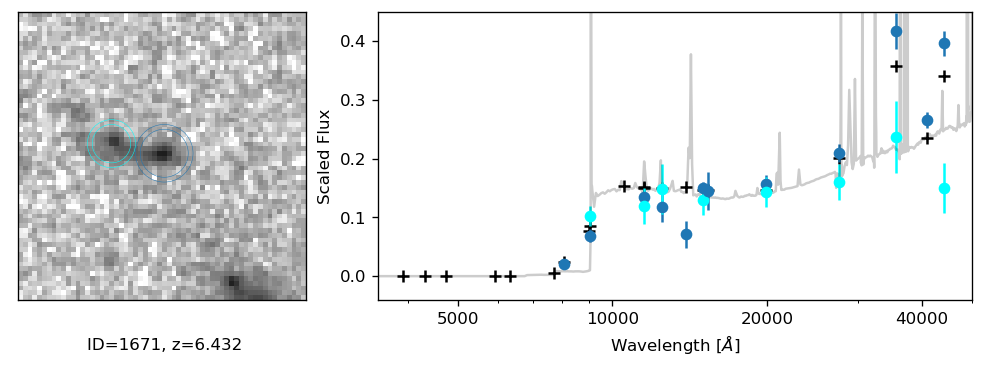}
    \caption{ \textit{Left}: Cutout image for \texttt{GNUastro} identified source (ID 1671) from the stacked image of four JWST short wavelength bands. Blue and cyan circular apertures indicate the two galaxies within the same \texttt{GNUastro} segmentation associated with source ID 1671. \textit{Right}: Respective SEDs (blue and cyan data points) measured using the circular apertures shown on the left hand side for those two galaxies. For easier visual inspection, here, I plotted only the fluxes with $\geq 3\sigma$ significance. These presented fluxes are also scaled to match the best-fit spectrum (black) to the original \texttt{GNUastro} SEDs (at $z=6.432$), which enclosed the flux from both of these galaxies. It could be seen that the two neighboring galaxies both have similar spectral features as the best-fit spectrum, indicating they are both $z\sim 6.4$ galaxies. }
    \label{1671_twogalaxy}
\end{figure*}

One last but very important comment regarding this sample is that the \texttt{GNUastro} segmentation associated with ID 1671 actually encloses two neighboring compact galaxies (see Fig.\ref{z610_galaxies}). These two galaxies could be more clearly seen in Fig.\ref{1671_twogalaxy}, as indicated by blue and cyan circular apertures in the left panel. In the right panel, I also plot their respective SEDs (above $>3\sigma$) measured from the circular apertures using the same respective colors. These flux measurements are scaled to match the level of the best-fit template (shown in black) for direct comparisons, but keep in mind that the best-fit template was obtained through fitting to the original \texttt{GNUastro} SEDs (capturing the fluxes from both galaxies). After the scaling, it could be seen that both cyan and blue SEDs share similar spectral features as the black best-fit spectrum. This observation suggests that the two galaxies shown in the left panel of Fig.\ref{1671_twogalaxy} are in fact both $z\sim6.4$ galaxies. Correspondingly, both of these galaxies need to be included when performing magnification bias analysis, which could be easily done by double-counting the source with ID 1671.


\begin{figure*}
    \centering
    \includegraphics[width=0.95\textwidth]{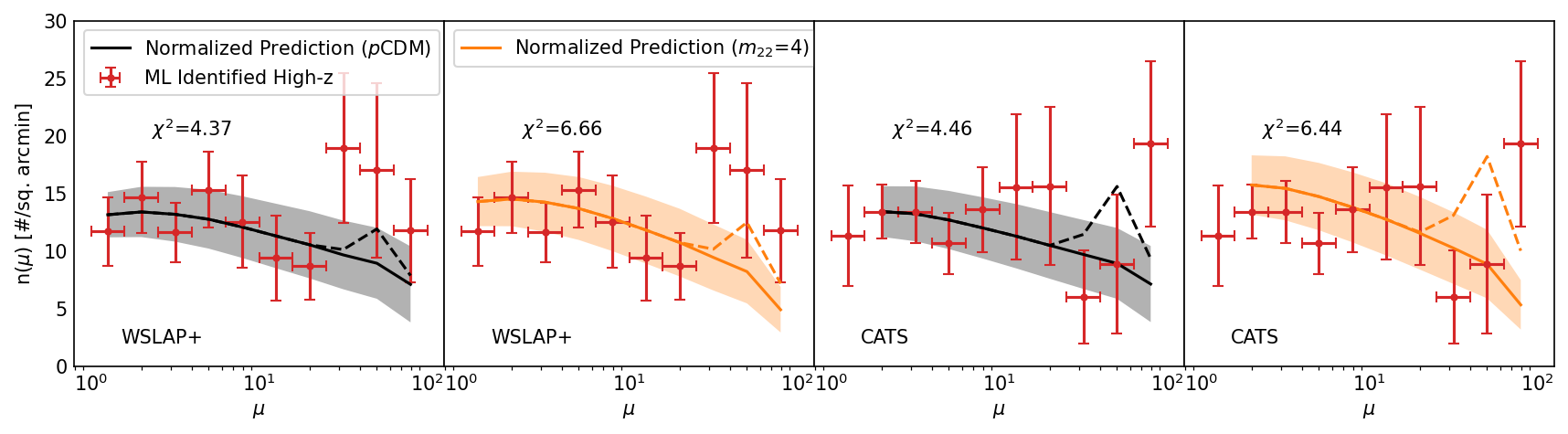}
    \caption{Comparison of the predicted surface number densities with observed densities (red data points) from machine learning selected $3.5\leq z\leq 5.5$ sample in different lensing magnification bins. In the first (using WSLAP+ model) and third (using CATS model) panels from left, I present the magnification bias prediction as a black solid line assuming no faint end turnover in UV LFs. The shaded region corresponds to how the fitting uncertainties of UV LF-redshift relations propagate into my magnification bias prediction. The same comparison with the orange distribution, based on UV LFs with a faint-end turnover induced by $\psi$DM of mass $4\times 10^{-22}$eV ($m_{22}=4$), is presented in the second (for WSLAP+) and last (for CATS) panels from the left. In all panels, magnification bias predictions are normalized to match the total number of machine learning selected galaxies for a more direct comparison in the trends of surface number densities. Such a normalization procedure is seen to shift the orange prediction slightly above the observed surface number densities in low magnification bins, in compensation for the faint-end suppression induced by $\psi$DM. As a consequence, this resulted deviation in lower magnification bins contributes to a higher $\chi^2$ value than the black curves based on $p$CDM UV LFs. Finally, in the first and second panels from left, I also noticed that the observed densities in high magnification bins lie systematically above the predicted trends. This behavior is likely the result of very small area involved in those bins, as reflected from the deviation between dashed curves from the solid curves: the dashed curves are obtained by conditioning that at least one galaxy is observed in each magnification bin (and for all smaller redshift intervals, see the earlier discussion on magnification bias methodology), and hence directly reflects small-area encountered if leading to very large surface number densities. }
    \label{35_ML_mitigated_mag_bias_results}
\end{figure*}

\begin{figure*}
    \centering
    \includegraphics[width=0.95\textwidth]{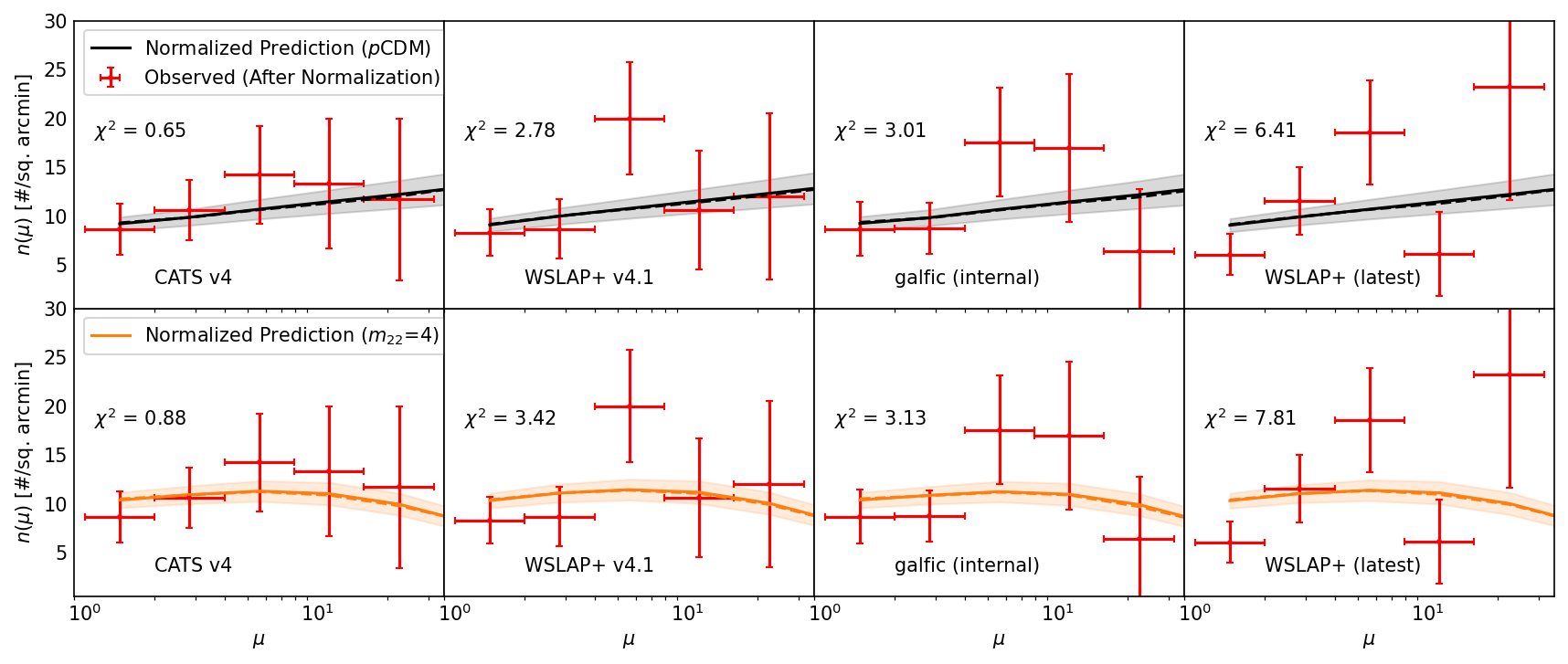}
    \caption{Comparison of the magnification bias predicted surface number density with observed (red data point) behind the M0416 cluster field over redshift $6\leq z\leq 10$ in different lensing magnification bins. In the top panel, I compare the observed densities with the prediction (black solid line) assuming no faint end turnover in UV LF, whereas in the bottom panel, observed densities are compared with the prediction (orange solid line) assuming a faint end turnover induced by $\psi$DM of mass $4\times 10^{-22}$eV. The comparison is presented separately for all four sets of lens models adopted. As in Fig.\ref{35_ML_mitigated_mag_bias_results}, magnification bias predictions are also normalized to match the total number of $6\leq z\leq 10$ galaxies involved for more direct trend comparison. Unlike in Fig.\ref{35_ML_mitigated_mag_bias_results}, however, $6\leq z\leq 10$ analysis does not encounter small area effect in high magnification bins, as reflected by dashed curves overlapping with solid curves. In all panels, I also present the $\chi^2$ value to reflect how well the predicted (after normalization) trends match with the observed densities. It could be seen that the UV LF without faint-end turnover is slightly favored over the $\phi_{\psi DM}$ UV LF of mass $4\times 10^{-22}$eV, irrespective of the lens model involved.  }
    \label{zabove6_mag_bias_results}
\end{figure*}

\section{Results}
\label{Sec_UVLF_results}
\subsection{Observed vs Predicted Densities}

I now present key results from my magnification bias analysis. The observed surface number densities of machine learning selected $3.5\leq z\le 5.5$ galaxies at different levels of lensing magnification are presented as red data points in Fig.\ref{35_ML_mitigated_mag_bias_results}. It is evident that the observed densities have an overall tendency to decrease towards higher magnification regions, i.e., suggestive of negative magnification bias\footnote{To quickly recap, negative magnification bias occurs when the newly accessed (owing to flux magnified above detection thresholds) intrinsically faint galaxies cannot compensate for the loss in sky area probed given the same telescope FOV. }. In contrast, the observed surface number densities over $6\leq z\leq 10$ (red data points in Fig.\ref{zabove6_mag_bias_results}) appear to increase towards higher magnification regions, indicating positive magnification bias. 

In both Fig.\ref{35_ML_mitigated_mag_bias_results} and Fig.\ref{zabove6_mag_bias_results}, I compare these observed surface number densities against the magnification bias prediction based on $\phi_{pCDM}$ without faint-end turnovers (black solid curves), and the prediction based on $\phi_{\psi DM}$ with faint-end turnover induced by an example $\psi$DM mass of $4\times 10^{-22}$eV (orange solid curves). For a more direct comparison of the trends, I have also normalized the magnification bias predictions shown to match the total number of galaxies observed. Owing to its faint-end suppression, magnification bias prediction based on $\phi_{\psi DM}$ manifests a sharper decrease towards higher magnification bins than the prediction based on $\phi_{pCDM}$. Such a behavior is more clearly seen from Fig.\ref{zabove6_mag_bias_results}, in which the black solid curves in the top panel nicely reproduce the observed positive bias trend, whereas orange curves (based on $\phi_{\psi DM}$) in the bottom panel first increases and then decreases to higher magnification regions. 

In all panels of Fig.\ref{35_ML_mitigated_mag_bias_results} and Fig.\ref{zabove6_mag_bias_results}, I test for the statistical resemblance of respective predicted trends ($n_{pred}$s) with the observed densities ($n_{obs}$s) by measuring the $\chi^2$ value:
\begin{equation}
    \chi^2 \equiv \sum_i \frac{(n_{pred,i} - n_{obs,i})^2}{\sigma_{obs,i}^2 + \sigma_{pred,i}^2},
\end{equation}
\noindent where the sum is over different magnification bins. It could be seen that the predicted trend based on $\phi_{pCDM}$ is (slightly) favored over the trend with a faster decrease to higher magnification bins, irrespective of the lens model involved. Such a preference could also be inferred by direct visual inspection. In particular, the presented black curves (after normalization) are seen to match amazingly well with the observed surface number densities in lower magnification bins, especially over $3.5\leq z\leq 5.5$. On the other hand, the presented orange curves are shifted slightly above the observed levels in those bins by the same normalization procedure, in compensation for the faster decrease to higher magnification regions. Overall, I am led to conclude that the observed surface number densities over both $3.5\leq z\le 5.5$ and $6\leq z\leq $ do not provide hints of faint-end turnovers (at least not by $\psi$DM of mass $4\times10^{-22}$eV). Given that the faint-end suppression is stronger for a lighter $\psi$DM mass, I anticipate $\phi_{pCDM}$ UV LF to be even strongly favored when compared with magnification bias prediction based on $\phi_{\psi DM}$ of lighter masses. In the next subsection, I will then follow along this logic, and derive a mass bound for the $\psi$DM models.  

To complete the discussion, I also quickly comment on the systematically high observed surface number densities in high lensing magnification bins, as seen from the first and second panel of Fig.\ref{35_ML_mitigated_mag_bias_results}. Such a behavior is likely the result of small magnification bin areas encountered when measuring surface number densities. To illustrate this point, in all panels of Fig.\ref{35_ML_mitigated_mag_bias_results}, I also plotted the magnification bias prediction (dashed curves) by requiring that at least one galaxy is observed in each magnification bin (also per redshift intervals involved, see Sec.\ref{sec_UVLF_faintend_magbias_method}). The observed deviation in dashed curves with the solid curve in those high magnification bins thus reflects very small areas encountered when measuring surface number densities. For consistency, I also performed the same check over $6\leq z\leq 10$ in Fig.\ref{zabove6_mag_bias_results}. There, the resulting dashed curves are seen nicely overlapping with the respective solid curves, i.e., no small area effect. 

\subsection{Mass Bound on $\psi$DM Models}

\begin{figure}
\centering
\begin{subfigure}{.49\textwidth}
  \centering
  \includegraphics[width=\linewidth]{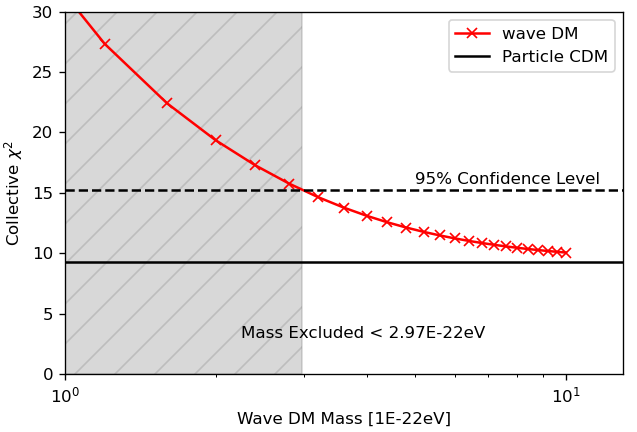}
  \caption{Over $3.5\leq z\leq 5.5$}
\end{subfigure}%
\hfill
\begin{subfigure}{.49\textwidth}
  \centering
  \includegraphics[width=\linewidth]{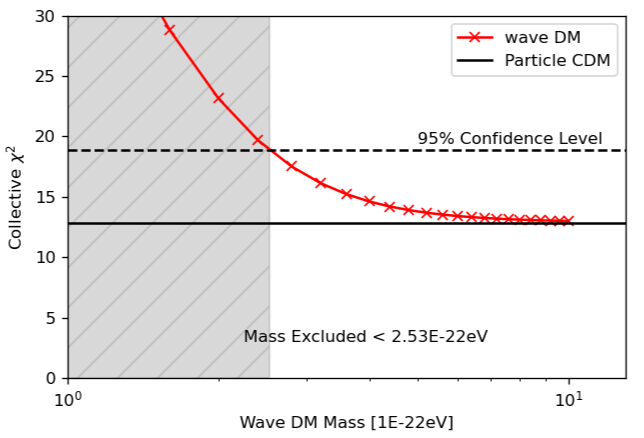}
  \caption{Over $6\leq z\leq 10$}
\end{subfigure}
\caption{The collective $\chi^2_{collective}$ values obtained when collectively comparing observed surface number densities with the magnification bias predictions based on different lens models over $3.5\leq z\leq 5.5$ (left) and $6\leq z\leq 10$ (right). Black horizontal lines correspond to using $\phi_{pCDM}$ without any faint-end turnovers in generating magnification bias prediction, whereas red crosses correspond to using $\phi_{\psi DM}$ with different $\psi$DM masses ($0.4-10\times10^{-22}$eV). Red solid lines are obtained through spline fitting to the red crosses. The shaded region corresponds to $\psi$DM masses with $\chi^2_{collective}(\psi DM) - \chi^2_{collective}(pCDM) \geq 5.991$, and is the excluded $\psi$DM mass range at $\geq 95\%$ confidence level.}
\label{mass_bound}
\end{figure}

Following my previous discussion, the preference in magnification bias results on $\phi_{pCDM}$ relative to $\phi_{\psi DM}$ can be extended to a quantitative mass bound on the $\psi$DM mass. To derive this bound, also to do so independently of the lens model involved, I examine a collective chi-squared value ($\chi_{collective}^2$) defined as 
\begin{equation}
    \chi_{collective}^2 \equiv \sum_{lens}\sum_{bin} \frac{(n_{pred,bin,lens} - n_{obs,bin,lens})^2}{\sigma_{pred,bin}^2+\sigma_{obs,bin,lens}^2},
\end{equation}
\noindent where the sum is over all observed and predicted (after normalization) densities in different magnification bins ($bin$), and over different lens models ($len$). 

For both $3.5\leq z\leq 5.5$ and $6\leq z\leq 10$ redshift ranges, I compute such an collective $\chi^2$ when (1) using $\phi_{pCDM}$ to generate magnification bias predictions, and (2) using $\phi_{\psi DM}$ with different $\psi$DM masses ($0.4 - 10 \times 10^{-22}$eV). For the second exercise, the resulting $\chi^2_{collective}$ distribution as a function of $\psi$DM mass is presented as red crosses in Fig.\ref{mass_bound}. The red solid curve joining the shown red crosses was obtained through spline fitting. In Fig.\ref{mass_bound}, I also indicate the $\chi^2_{collective}$ value obtained when using $\phi_{pCDM}$ as black horizontal line. As would be anticipated, the red crosses (and hence red solid line) get closer to the black horizontal line as the mass of the $\psi$DM gets larger, i.e., with a more similar behavior to $\phi_{pCDM}$ at the faint-end. To define a bound on the $\psi$DM mass, I notice that the difference between $\chi_{collective}^2(\psi DM) - \chi^2_{collective}(pCDM)$ follows the distribution of a second degree $\chi^2_{(2)}$ variable. With this method, I found that $\psi$DM of mass lighter than $2.97 \times 10^{-22}$eV is not favored relative to the $p$CDM paradigm at 95\% confidence level given my $3.5\leq z\leq 5.5$ magnification bias analysis. Over the redshift range of $6\leq z\leq 10$, a slightly weaker bound was obtained, that $\psi$DM lighter than $2.53 \times 10^{-22}$eV is not favored at 95$\%$ confidence level.



\subsection{Comparison with existing Bounds}
\label{sec_comparions_existing_massbound}

In S16, the predicted UV LF faint-end turnover was compared with non-lensing field UV LF measurements by B15, and they obtained a mass bound of $\gtrsim 1.2\times 10^{-22}$eV at $2\sigma$ confidence. More recently, relying on more recent non-lensing field UV LFs measurements by B21, \cite{Winch2024ApJ} derived a stronger mass bound of $>2.51\times 10^{-22}$eV at 95$\%$ confidence. Their more stringent bound, however, may be driven by the semi-analytical approach they adopted in generating HMFs. As I commented earlier in Sec.\ref{psiDM_UVLF}, adopting such a semi-analytic strategy tends to generate a sharper cut-off at the lower mass end on HMFs, hence leads to stronger faint-end turnover at the faint-end of UV LFs (thereby also more constraining on the effects from $\psi$DM). 

In comparison with these constraints based on blank field UV LFs, my $3.5\leq z\leq 5.5$ analysis led to a more stringent mass constraint owing to the better sensitivity on the fainter end of UV LFs enabled by gravitational lensing. Over redshift $6\leq z\leq 10$, the mass constraint derived was also seen comparable to that of \cite{Winch2024ApJ}, despite with a single strong lensing field (M0416) available for analysis, and assumed a more modest cut-off at the low mass end of HMF. 

Example of existing $\psi$DM constraints from HFF lensing fields include the bound from \cite{Menci2017}: $1.2\times 10^{-21}$eV at $2\sigma$ level, much more stringent than existing constraints from deep blank fields and also my constraints. The analysis of \cite{Menci2017}, however, relied on the lensing field UV LFs measured by \cite{Livermore2017}, which I previously commented as likely contaminated by low-z interlopers (see Chap.\ref{chap:intro}). More specifically, an excess at the faint end was observed by \cite{Livermore2017} relative to existing UV LF measurements from deep blank fields, matching exactly the anticipated behavior if the high-z sample is severely contaminated following my previous discussion in Sec.\ref{contaminated_LF_sec}. The presence of contamination then likely has led to an inaccurately stringent bound. 

In fact, the derived mass bound by \cite{Menci2017} was weakened considerably to $>5\times 10^{-22}$eV at $2\sigma$ level, when relying on lensing field UV LFs measurements by \cite{Bouwens2017} (now displaying a slight faint-end turnover, see Chap.\ref{chap:intro}) instead. Nevertheless, \cite{Bouwens2017} constructed their UV LFs following the Lyman Break Galaxies-like (LBG-like) selection criteria, which I previously discussed in Sec.\ref{sec_LGB_failure} as (very likely) incapable of mitigating low-z interlopers. Furthermore, LBG-like selection criteria would also reject genuine high-z galaxies, and hence may not best suit the analysis of UV LFs at the very faint-end (e.g., intrinsically faint high-z galaxies may also be intrinsically redder than allowed by LBG-like criteria). Combining these discussions, I therefore consider my derived mass bounds (albeit they appear less stringent) as more robust. 

Finally, I quickly comment on the most noticeably stringent constraint on $\psi$DM mass from Lyman Alpha Forest: excluding $\psi$DM mass lighter than $\sim 3 \times 10^{-21}$eV by \cite{2017PhRvL.119c1302I, Kobayashi2017, Armengaud2017}, and more recently, $\gtrsim 2\times 10^{-20}$eV by \cite{RogersPeiris2020}\footnote{For the interested reader, the more stringent constraint by \cite{RogersPeiris2020} is likely the result of both (1) adopting more recent observations reaching smaller scales, and (2) assigning a higher prior probability to higher $\psi$DM masses. For the second point, I noticed that previous analysis assumed a uniform prior in $m_{\psi DM}^{-1}$, whereas \cite{RogersPeiris2020} assumed a uniform prior in $\log(m_{\psi DM})$, effectively giving more weights to higher $\psi$DM masses than previous approach.}. These analyses investigate the spatial correlation in the total transmitted flux of Lyman Alpha Forrest, to test for suppressed (as anticipated from $\psi$DM) small-scale density fluctuations in the intergalactic neutral hydrogen. It is assumed that such a small-scale fluctuation in the intergalactic medium (IGM) traces directly the underlying dark matter perturbations, but there may be additional complications in assessing the implications on the dark matter power spectrum. For instance, IGM may be more patchy than anticipated from the $\psi$DM model, owing to fluctuations in ionizing backgrounds and temperature, or induced by galactic winds (see \citep{2101.11735} and references therein). The corresponding details, however, remain elusive, and hence the constraining power of Lyman Alpha Forest analysis on $\psi$DM models may be lower than claimed. 

\section{Some Concluding Remarks and Discussion}
\label{sec_UVLF_conclu}
 
In this chapter, I have demonstrated that $\psi$DM with mass lighter than $2.97\times 10^{-22}$eV is not favored at 95\% confidence level, by investigating the surface number density of gravitationally lensed $3.5\leq z\leq 5.5$ and $6\leq z\leq 10$ galaxies behind massive galaxy cluster M0416. This mass bound is more constraining than the existing $\psi$DM mass bound based on UV LFs measured from deep blank fields. My constraint is also considered more reliable than previous lensing field analysis, given the individual mitigating power on low-z interlopers I (but previous analysis likely fails to) achieved with the combination of deep HST and JWST observations. The lack of supplementary JWST observation for the HFF cluster fields besides M0416 was also compensated by a carefully constructed machine learning model, exploiting the subtle differences encoded in existing S18 measurements between genuine high-z and low-z galaxies. The stringent mass bounds from Lyman Alpha Forest were also quickly discussed at the end of Sec.\ref{sec_comparions_existing_massbound}, and I mentioned several caveats that need to be properly dealt with before evaluating the robustness of corresponding constraints. 

One aspect that I have neglected, and could have improved on, however, is to incorporate the effect of wave interferences (as have done by \cite{Amruth}) when implementing magnification bias. In particular, the wave interference pattern of $\psi$DM was shown to cause a large perturbation on the lensing critical curve, and hence may considerably alter the definition of magnification bins. Nonetheless, relative to the $\sim 10^{12}M_\odot$ lensing galaxy considered in \cite{Amruth}, HFF galaxy clusters are generally much heavier ($\sim 10^{14}-10^{15}M_\odot$) and hence the velocity of $\psi$DM particles would be much higher. As a consequence, $\psi$DM particles would have much smaller de Broglie wavelengths, potentially leading to less noticeable (even negligible) modulation on lensing magnifications. As such, I consider my derived mass bound not to be significantly altered when including the effects of wave interference patterns.

\begin{figure}
    \centering
    \includegraphics[width=0.8\linewidth]{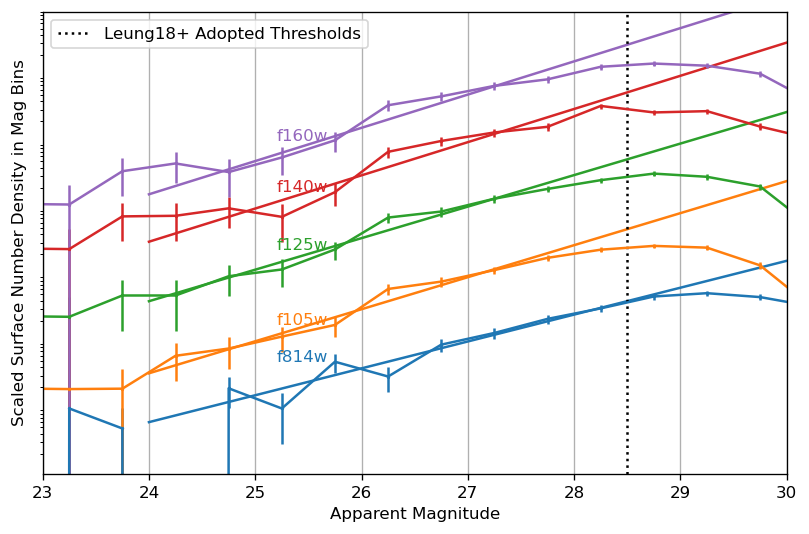}
    \caption{Surface number densities of $z>4.75$ galaxies in different magnitude bins measured from the C15 catalog as used by L18 for their magnification bias analysis. The distributions measured for different HFF bands (F814W-F160W, relevant for $z>4.75$ candidates selection of L18) are scaled to be non-overlapping. The vertical dotted line marks the data completeness threshold adopted by L18, whereas it can be seen that the deviation in the observed distribution from power-law behavior (indicated by straight lines) occurs at magnitudes brighter than 28.5 for F105W-F160W bands (relevant for $z>5$ candidates selection of L18). As previously discussed in Sec.\ref{sec_injection_recovery}, such a deviation from power-law behavior signals data incompleteness. As a consequence of earlier (than reaching 28.5) deviation observed here, I suspect the faint-end turnover observed by L18 was instead driven by data incompleteness. }
    \label{Enoch_completeness_threshold}
\end{figure}

I also need to comment that our derived mass bound is inconsistent with the faint end turnover previously identified by L18 (favoring a mass of $\sim 1.6\times 10^{-22}$eV), despite the fact that they also applied a similar magnification bias analysis (relying on C15 catalogs instead). Previously in Sec.\ref{sec_LGB_failure}, I checked that their selected sample above $z>4.75$ is magnification bias-wise clean from contamination, the inconsistency between my mass bound and their detected faint-end turnover is thus puzzling. Here, I argue that their observed faint-end turnover is likely not the signature of $\psi$DM but instead due to data incompleteness. In particular, they adopted a data completeness threshold of 28.5. On the other hand, by focusing on the $z>4.75$ sample\footnote{Here I removed my previous \textit{odds}$>0.5$ selection condition, as such a condition was not applied by L18. I have, however, applied the exclusion region defined in Sec.\ref{sec_data_complete}, which I commented was less constraining than theirs. On the other hand, when determining the data completeness threshold of 28.5, L18 did not adopt any exclusion region as I have done here. As a result, I suspect the data completeness threshold implied by Fig.\ref{Enoch_completeness_threshold} to be more reflective of what they may actually expect. } of direct relevance for their analysis, I determined in Fig.\ref{Enoch_completeness_threshold} that data incompleteness occurs brighter than 28.5 (shown as dotted vertical line) in F105W-F160W bands relevant for their $z>5$ candidate selection. As previously discussed in detail in Sec.\ref{sec_injection_recovery}, this data incompleteness is reflected by the deviation of observed surface number density (in different magnitude bins) from the power-law behavior. Such a deviation, i.e., data incompleteness, might have been confused in L18 as a faint-end turnover induced by $\psi$DM. 

Finally, I comment that there remains one key underlying assumption that I could not overlook: that S16 (and hence my previous analysis) assumed that the $\psi$DM is composed of a single particle copy, with a unique mass. From the perspective of String Axiverse \citep{Arvanitaki2010}, however, having multiple copies of ultralight particles is more theoretically motivated than having a single $\psi$DM copy. Observationally, the presence of more than one $\psi$DM copies has also been hinted at by dwarf galaxies when interpreting their respective inner cores as solitons \citep{1811.03771, Pozo2024PhRvD}. In particular, dwarf spheroidals and ultra-faint dwarfs were found to reflect two distinct galaxy classes, with the solitons hinted by their stellar distributions favoring $\psi$DM masses of $\sim 1.9\times 10^{-22}$eV and $2.3\times 10^{-21}$eV, respectively. The existence of such distinct galaxy classes in the multi-copy $\psi$DM scenario has also been justified recently with the simulation by \cite{2506.16915}. In such a multi-copy $\psi$DM scenario, it is evident that the mass bound derived above could not be interpreted easily, given that the lighter $\psi$DM copy suggested by dwarf spheroidals is clearly incompatible. In the next chapter, I therefore attempt to understand how my derived mass bound can be interpreted in the multi-copy $\psi$DM scenario. In particular, by considering the growth of density perturbations, I will demonstrate that different $\psi$DM copies reach a "gravitational equilibrium". As a consequence, different $\psi$DM copies will behave similarly on large scales, and my mass bound could instead be interpreted by an effective mass scale, characterizing the same large-scale behavior of all $\psi$DM copies.



\chapter{Multi-copy $\psi$DM and Astronomical Implications}
\label{chap:multi_axion}

\section{String Axiverse and Motivation for Multi-copy $\psi$DM}

String Theory is arguably the most celebrated candidate for the Theory of Everything so far, in the sense that it provides natural candidates for force carriers of all four fundamental forces, along with many matter fields that could align with the observed particle content. In relevance of $\psi$DM models, ultralight axion-like particles (ALPs\footnote{For the rest of this chapter, I will use the phrase ALPs and axions interchangeably. These axions are not to be confused with QCD axions first proposed to solve the \textit{Strong CP Problem}, as QCD axions occupy a different mass range $\sim 10^{-10}-10^{-3}$eV and have distinct origins from String Axiverse ALPs. }) may copiously exist in the framework of String Theory (leading to String Axiverse, \cite{Arvanitaki2010}), depending on the details of how large extra dimensions (needed for the self-consistency of String Theory) are compactified to be non-observable. Broadly speaking, such ultralight ALPs may arise either owing to the integration of background fluxes along closed surfaces/cycles in compact extra dimensions, or as the phases of matter fields in non-compact dimensions. In both cases, the presence of multiple copies of ultralight ALPs is generic, with the precise number of copies depending on the details of the model: e.g., the number of independent closed cycles for integration in the first case, and the number of matter fields in the second case. And owing to their String Theory origin, non-perturbative effects generating the mass of these ALPs are naturally associated with very large values for their decay constants $\sim 10^{16}$GeV, which in turn lead to highly suppressed mass scale (e.g. $\sim 10^{-28}-10^{-18}$eV) allowing these ALPs to serve as $\psi$DM candidates. 

While the impact of single copy $\psi$DM on structure formation has been extensively considered/studied\footnote{In fact, also the scenario where there is a mixture of $\psi DM$ with heavy particle CDMs, e.g., see \cite{Winch2024ApJ}. }, there exist only a few works that investigated the structure formation in multi-copy $\psi$DM scenario. \cite{Luu2024} serves as one of the works in this direction, by providing the first simulation with two copies of axions differing by a mass factor of 5. The same set of simulations was also demonstrated to produce distinct galaxy classes \citep{2506.16915}, potentially explaining distinct inner soliton cores hinted by dwarf spheroidals and ultra-faint dwarf galaxies \citep{Pozo2024PhRvD}. For a better understanding of structure formation in the multi-copy $\psi$DM scenario, such a simulation is expected to be generalized to higher mass ratios and with more copies, but the relevant initial conditions for each species may be hard to generate. 

In this chapter, I attempt to understand the anticipated structure formation in such a multi-copy $\psi$DM Universe using linear perturbation theory, and present a simplified parametrization for the respective transfer functions. The latter, I emphasize, would be helpful for future simulations to more easily implement the appropriate initial condition (density perturbation) for each copy of ALPs. In this chapter, I will also argue that the large-scale structure formation in a multi-copy $\psi$DM Universe may resemble the Universe with a single "equivalent" copy of effective mass $m_{eff}$, whereas the effect from individual copies would be apparent in the non-linear regime (e.g., at the galactic level). This conjecture then allows me to better interpret various existing mass bounds on $\psi$DM from observations (especially my derived mass bound in the previous chapter). In particular, existing mass bounds are likely on the effective mass of the "equivalent" copy instead of individual copies. In what follows, I will also provide an analytical expression for the effective mass, demonstrating it is determined by the relative contribution to the total dark matter density of individual copies. 

In Sec.\ref{sec_reproduce_Schrodinger}, I start by reproducing the Schrodinger-Poisson equation in an expanding Universe for each copy of ALPs, based on which I perform linear perturbation analysis in Sec.\ref{sec_multicopy_transfer} and provide a simplified parametrization for multi-copy ALPs transfer functions. I will then conclude in Sec.\ref{sec_multicopy_interp}, and comment on potential astronomical implications of the multi-copy $\psi$DM scenario and the effective mass identified in my derivation. 

\section{Deriving Schrodinger Poisson equation in Expanding Universe}
\label{sec_reproduce_Schrodinger}

To reproduce the Schrodinger-Poisson equation for axion fields, I shall denote each axion field as $\phi_I$, and write the full particle Lagrangian as 
\begin{equation}
    S = \int \sqrt{-g} \mathcal{L} = \int \sqrt{-g} \bigg{[} \sum_I \bigg{(} - \frac{1}{2}\partial_\mu \phi_I \partial^\mu \phi_I -  \frac{1}{2} m_I^2 \phi_I^2 \bigg{)} -  U(\phi_1, \phi_2, \cdots) \bigg{]}
\end{equation}
\noindent where the $U(\phi_1, \phi_2, \cdots)$ describe any interactions among axion fields. In what follows, I will focus on the simplest case of $U=0$ as a starting point, and leave $U\neq0$ scenarios for future investigation. 

For an expanding Universe with gravitational perturbation, the metric (in Newtonian gauge) could be written as
\begin{equation}
    ds^2 =  -(1+2\Phi) dt^2 + (1-2\Phi)a(t)^2 \delta_{ij} dx^i dx^j 
\end{equation}
\noindent where $\Phi \ll 1$ being the gravitational potential. The Euler-Lagrange equation for each copy of axion in such a spacetime could be derived to be
\begin{equation}
    \ddot{\phi_I} + 3\mathcal{H}\dot{\phi_I} - \frac{1}{a^2} \nabla^2 \phi_I - \frac{1+2\Phi + 2\Phi^2}{1-4\Phi^2} 4\dot{\phi_I} \dot{\Phi} + \frac{4 \Phi \partial_i \phi_I \partial^i \Phi }{a^2(1-2\Phi)^2}+ (1+2\Phi) m_I^2 \phi_I = 0.
\end{equation}
\noindent where I have defined the Hubble expansion rate to be $\mathcal{H} \equiv \dot{a}/a$, and $\dot{ } \equiv d/dt$. To first order in perturbative quantities $\Phi$ and $\partial_i\phi_I$ (recall zeroth order quantities are cosmologically homogeneous), I could simplify the above equation to 
\begin{equation}
    \ddot{\phi_I} + 3\mathcal{H}\dot{\phi_I} - \frac{1}{a^2} \nabla^2 \phi_I - 4\dot{\phi_I} \dot{\Phi} + (1+2\Phi) m_I^2 \phi_I  = 0. 
\end{equation}

Owing to their ultralight mass, axion fields could be split into a rapidly oscillating part $\sim e^{-im_I t}$ with a slowly time-varying envelope $\varphi_I$ as $\phi_I = (\varphi_I e^{-im_It}+c.c.)/\sqrt{2}$. Substituting this expansion into the previous equation, one could obtain the following equation for $\varphi_I$:
\begin{equation}
    -2im_I \dot{\varphi_I} - m_I^2 \varphi_I + \ddot{\varphi_I} + 3\mathcal{H} (\dot{\varphi_I} - im_I \varphi_I) -\frac{\nabla^2 \varphi_I}{a^2} - 4\dot{\Phi} (\dot{\varphi_I} - im_I \varphi_I) + (1+2\Phi) m_I^2 \varphi_I = 0,
\end{equation}
\noindent within which, it is fair to assume $\ddot{\varphi_I} \ll m_I \dot{\varphi_I} \ll m_I^2 \varphi_I$ owing to the very slow time variation of $\varphi_I$. Similarly, as the gravitational potential/collapse is sourced by the perturbation to axion fields encoded in $\varphi_I$, I could further assume time evolution of $\Phi$ to be negligible (i.e. $\dot{\Phi} \ll \Phi$). These simplifications then allow me to reproduce the Schrodinger equation for each axion copy as
\begin{equation}
    i(\partial_t + \frac{3}{2}\mathcal{H} )\varphi_I = -\frac{\nabla^2 \varphi_I}{2m_Ia^2} + \Phi m_I \varphi_I. 
\end{equation}

The Poisson equation could be derived by combining $tt-$ and $ti-$Einstein equations, leading to
\begin{equation}
    \frac{2}{a^2} \nabla^2 \Phi  + 3\mathcal{H}^2(1+2\Phi) = 8\pi G (1+2\Phi) \rho_{tot} 
\end{equation}
\noindent where the full functional form of density could be read off from the stress energy tensor as  
\begin{equation}
    \rho_{tot} =\sum_I \bigg{[} \frac{\dot{\phi_I}^2}{2(1+2\Phi)} + \frac{(\nabla \phi_I)^2}{2a^2(1-2\Phi)}  + \frac{1}{2} m_I^2 \phi_I^2\bigg{]}. 
\end{equation}

Using $\varphi_I$ notation and keeping to first order in perturbative quantities $\Phi$, $\partial_i \varphi_I$, I find 
\begin{equation}
    \rho_{tot} \approx \sum_I \bigg{[} \frac{1+\Phi}{1+2\Phi} m_I^2|\varphi_I|^2 + \frac{|\nabla \varphi_I|^2}{2a^2(1-2\Phi)} \bigg{]} \approx \sum_I m_I^2 |\varphi_I|^2 \equiv \sum_I \rho_I
\end{equation}
\noindent where in the last equality, I have defined the density field $\rho_I \equiv m_I^2 |\varphi_I|^2$ for each axion copy. With the above expression, one notices that the Friedmann equation and Poisson equation correspond to the zeroth and first orders of the Einstein equation, respectively, and are 
\begin{align}
    3\mathcal{H}^2 = 8\pi G \overline{\rho_{tot}}& = \sum_I 8\pi G \overline{\rho_I},\\
    \nabla^2 \Phi = 4\pi G a^2 \delta \rho_{tot} &\equiv 4\pi G a^2 \sum_I \delta\rho_I.
\end{align}

\section{Multi-copy axion Perturbation equations}
\label{sec_multicopy_transfer}

In the early Universe, the growth of gravitational inhomogeneities (which are seeds to future cosmological structures) is controlled by the perturbations to the mean dark matter densities. To derive the relevant perturbation equations, I perform Madelung decomposition on each axion copy via $\varphi_I = \sqrt{\rho_I} e^{iS_I}/m_I$, and rearrange Schrodinger-Poisson equations into 
\begin{align}
\dot{\rho_I} + 3\mathcal{H} \rho_I + \frac{1}{a} \nabla\cdot (\rho_I v_I) = 0, \\
\dot{v_I} + \mathcal{H} v_I+ \frac{1}{a^3} \nabla Q_I + \frac{1 }{ a}(v_I \cdot \nabla)v_I + \frac{1}{a} \nabla \Phi = 0,
\end{align}
\noindent which corresponds to the real and imaginary part, respectively. In above equations, velocity field is defined as $v_I \equiv \nabla S_I/(m_I a)$, and $Q_I$ denotes the quantum pressure for each axion copy given by 
\begin{equation}
    Q_I \equiv -\frac{\nabla^2(\sqrt{\rho_I}) }{2m_I^2 \sqrt{\rho_I}} = -\frac{1}{2m_I^2}\bigg{(} \frac{\nabla^2 \rho_I}{2\rho_I} + \frac{(\nabla \rho_I)^2}{4\rho_I^2} \bigg{)}.
\end{equation} 

By considering the density contrasts $\delta_I \equiv \delta \rho_I/\overline{\rho_I}$, I may combine the above two equations into a single linear order time evolution equation for $\delta_I$ as 
\begin{equation}
    \ddot{\delta_I} + 2\mathcal{H} \dot{\delta_I}  + \frac{\nabla^4 \delta_I}{4 m_I^2 a^4} - \sum_I 4\pi G \delta_I \overline{\rho_I} =0. 
\label{Matrix_constrast_equation}
\end{equation} 
\noindent Notice the gravitational term receives the contribution from all axion copies. 

In analyzing the properties of density contrasts, it will be seen useful to define a 'total' perturbation contrast as $\overline{\rho_{tot}}\delta_{tot} = \sum_I\overline{\rho_I} \delta_I $, whose time derivative could be verified (owing to the zeroth order equation in density $\dot{\overline{\rho_I}} + 3\mathcal{H}\overline{\rho_I}=0$) to satisfy

\begin{equation}
    \dot{\delta_{tot}} = \frac{\sum_I\overline{\rho_I} \dot{\delta_I} }{\overline{\rho_{tot}}}, \;\;  \ddot{\delta_{tot}} = \frac{\sum_I\overline{\rho_I} \ddot{\delta_I} }{\overline{\rho_{tot}}}. 
\label{total_contrast_def}
\end{equation} 
\noindent And the evolution equation for $\delta_{tot}$ could be obtained by summing over density contrast equations for all axion copies, giving
\begin{equation}
    \ddot{\delta_{tot}} + 2 \mathcal{H} \dot{\delta_{tot}} - 4\pi G \delta_{tot}\overline{\rho_{tot}}+ \frac{k^4}{4m_{eff}^2a^4}\delta_{tot} = 0
\label{limiting_case_total}
\end{equation}  
\noindent where I have defined an effective mass scale as

\begin{equation}
    \frac{1}{m_{eff}^2} \delta_{tot} \overline{\rho_{tot}} \equiv \sum_I \frac{\overline{\rho_I} \delta_I}{m_I^2 }.
\label{effective_mass_defin}
\end{equation} 

Note Equ.\ref{limiting_case_total} is essentially the same as the single copy axion case, hence the corresponding transfer function for $\delta_{tot}$ could be well-approximated with the solution by \citep{Huetal2000}

\begin{equation}
    T_{tot} \approx \bigg{|}\frac{\cos x^3}{1+x^8}\bigg{|}T_{pCDM}, \; \; x = \frac{1.61}{9}\frac{k}{m^{4/9}_{eff}}
\label{transfer_1}
\end{equation} 
\noindent in which the mass $m_{eff}$ is measured in units of $10^{-22}$eV, $k$ is wavenumber measured in Mpc$^{-1}$, and $T_{pCDM}$ the transfer function expected from heavy particle CDM. In \cite{2201.10238,2412.15192}, an alternative fitting form was provided:

\begin{equation}
    T_{tot} \approx \bigg{|}\frac{\sin x^{5/2}}{x^{5/2}(1+Bx^{7/2})}\bigg{|}T_{pCDM}, \; \; x = \frac{A}{9}\frac{k}{m^{1/2}_{eff}}
\label{transfer_2}
\end{equation} 
\noindent where $A = 2.22 m_{eff}^{1/25 - \ln(m_{eff})/1000}$, $B = 0.16 m_{eff}^{-1/20}$. Likewise, the transfer function determined from Boltzman codes (e.g., AxionCAMB, \cite{axioncamb}) for the single copy axion case could also be used.

\subsection{Limiting cases consideration}

To better understand the behavior of each axion copy, I note there are two separate limiting cases for Equ.\ref{Matrix_constrast_equation}.

Near the frequency $k$ where quantum pressure-induced suppression from $\psi$DM just begins, the quartic contribution $k^4$ term remains relatively small compared to the gravity term and hence could be neglected. As a consequence, the last term in Equ.\ref{Matrix_constrast_equation} and also Equ.\ref{limiting_case_total} could be dropped, leading to an equivalency between all copies of axions with the total density contrast $\delta_{tot}$. In \cite{Luu2024}, this feature was demonstrated to happen for the double axion case, i.e., lighter and heavier copies were observed to both follow the same transfer function near the onset of quantum pressure-induced suppression. Here, I demonstrated that this equivalence is generic and anticipated for an arbitrary number of axion copies. This 'equivalence' among all copies and with the total density contrast (which I may also term as an effective copy with mass $m_{eff}$) stems from the fact that all axion copies are coupled to the same mutual gravitational potential, and hence establish a gravitational equilibrium. Correspondingly, the onset of small-scale structure suppression is also governed by the effective mass scale $m_{eff}$. 

The identified 'equivalence' between $\delta_I$s with $\delta_{tot}$ also allows for a simpler expression for the effective mass, by canceling density contrasts on both side of Equ.\ref{effective_mass_defin}: $\overline{\rho_{tot}}/m_{eff}^2 = \sum_I \overline{\rho_I}/m_I^2$. This suggests that the effective mass is solely determined by the relative contribution of each copy to the total matter density\footnote{To be more precise, this ratio will be fixed at the time for the lightest copy to start oscillate}, as long as there are no mixing mechanism beside gravity among different axion copies (i.e. $U=0$), and that there are no decay channels for axions.

In \cite{Luu2024}, it was also observed that at much larger $k$, the lighter and heavier copy approximates back to their respective single copy axion transfer functions. From Equ.\ref{Matrix_constrast_equation}, this behavior could also be inferred to hold true in general, as in the limit of very large $k$, the gravity term is overwhelmed by the quantum pressure term and could be dropped off. 

\subsection{Parameterizing Multi-copy Transfer Function}

The two limiting cases discussed above could be recast into the following \textit{approximate} equations for each axion copy: 
\begin{align}
    \ddot{\hat{\delta}}_I + 2 \mathcal{H} \dot{\hat{\delta}}_I - 4\pi G \overline{\rho_{tot}} \hat{\delta}_I + \frac{k^4 \hat{\delta}_I }{4a^4m_{I}^2} = 0 
\label{limiting_cases_1} \\
    \ddot{\hat{\delta}}_{tot} + 2 \mathcal{H} \dot{\hat{\delta}}_{tot} - 4\pi G \overline{\rho_{tot}} \hat{\delta}_{tot} + \frac{k^4 \hat{\delta}_{tot} }{4a^4m_{eff}^2} = 0
\label{limiting_cases_2}
\end{align}  
\noindent where $\hat{\delta}_{tot}$ is the solution in single-copy Universe with mass $m_{eff}$ and transfer function $T_{tot}$, and $\hat{\delta}_{tot}$ is the single-copy Universe solution with mass $m_I$ and transfer function $T_{I}$. Here, the hatted notation for these solutions was adopted to avoid confusion with density contrasts $\delta_{tot}, \delta_I$s in the multi-copy Universe that I aim to solve. In the $\hat{\delta}_I$ equation, the gravity term should really be $4 \pi G \overline{\rho_I} \hat{\delta}_I$ instead of  $4 \pi G \overline{\rho_{tot}} \hat{\delta}_I$. But following the equivalence identified between $\delta_I$ with $\delta_{tot}$ at low-$k$ limit, and that $\hat{\delta}_I$ is the solution to high-$k$ limiting case for which the gravity term could be neglected, it is reasonable to replace $\overline{\rho_I}$ with $\overline{\rho_{tot}}$ as done here.


Following my discussion, it is evident that $\delta_I$ in the multi-copy Universe will transit from being $\hat{\delta}_{tot}$ like to being $\hat{\delta}_I$ like as $k$ increases, and I can parameterize this transition as follows:
\begin{equation}
    \delta_I = A(k) \hat{\delta}_I + (1-A(k))\hat{\delta}_{tot}
    \label{density_decomposition}
\end{equation}
\noindent where the frequency dependent coefficient $A(k)$ limits to 0 at low $k$ and $1$ at high $k$. By subbing this assumed decomposition into Equ.\ref{Matrix_constrast_equation} and using limiting case equations (also note $\hat{\delta}_{tot} = \delta_{tot}$ as defined by Equ.\ref{total_contrast_def}), I can get 
\begin{equation}
4\pi G \overline{\rho_{tot}} A(\hat{\delta}_I - \hat{\delta}_{tot}) = (1-A)\frac{k^4}{4a^4}\bigg{(} \frac{1}{m_{eff}^2} - \frac{1}{m_I^2} \bigg{)} \hat{\delta}_{tot}.
\end{equation}
\noindent Hence the solution for $A(k)$ is given by 
\begin{equation}
    A = \frac{\frac{k^4}{4a^4}\bigg{(} \frac{1}{m_{eff}^2} - \frac{1}{m_I^2} \bigg{)} \hat{\delta}_{tot}}{4\pi G \overline{\rho_{tot}}(\hat{\delta}_I - \hat{\delta}_{tot})+\frac{k^4}{4a^4}\bigg{(} \frac{1}{m_{eff}^2} - \frac{1}{m_I^2} \bigg{)} \hat{\delta}_{tot}},
\end{equation}
\noindent, which could be further simplified to 
\begin{equation}
    A = \frac{\frac{k^4}{\tilde{k}^4}\bigg{(} 1 - \frac{m_{eff}^2}{m_I^2} \bigg{)}}{\frac{T_I}{T_{tot}} -1+\frac{k^4}{\tilde{k}^4}\bigg{(} 1 - \frac{m_{eff}^2}{m_I^2} \bigg{)}}
\label{Ak_solution_simplified}
\end{equation}
\noindent where I have used the fact $\hat{\delta}_{tot}/\hat{\delta}_I = T_{tot}/T_I$, and defined the quantity 
\begin{equation}
    \tilde{k} \equiv a^{3/4}k_J \approx \frac{44.7}{\text{Mpc}}\,\bigg{(}6E(a)\frac{\Omega_{DM}}{0.3}\bigg{)}^{1/4} \bigg{(}\frac{H_0}{70 \,\frac{\text{km/s}}{\text{Mpc}}}\frac{m_{eff}}{10^{-22} \, \text{eV}}\bigg{)}^{1/2}
\end{equation}
\noindent in which $E(a) \equiv a^4(\Omega_\Lambda+\Omega_m/a^3)$, $k_J$ is the comoving quantum Jeans scale for $\hat{\delta}_{tot}$ given $\overline{\rho_{tot}}$:
\begin{equation}
4\pi G\overline{\rho_{tot}}\cdot4a^4 \equiv \frac{a^3 k_J^4}{m_{eff}^2} = 6 \mathcal{H}^2 \Omega_{DM}.
\label{jeans_scale}
\end{equation}

In the transfer function given by Equ.\ref{transfer_1}, recall there is also the highly oscillating $\cos x^3$ term (or $\sin x^{5/2}$ for Equ.\ref{transfer_2}). In the determination of $A(k)$, such an oscillatory term may be considered as a higher-order effect and less important. Consequently, in a practical implementation of Equ.\ref{Ak_solution_simplified}, one may approximate the ratios between any two transfer functions as the ratio between envelop functions, i.e., $T_a /T_b = (1+x_b^8)/(1+x_a^8)$ (likewise for transfer functions given by Equ.\ref{transfer_2}). For transfer functions determined by Boltzmann codes, smoothing could be applied to find the corresponding envelop functions. 

\begin{figure}
    \centering
    \includegraphics[width=1\linewidth]{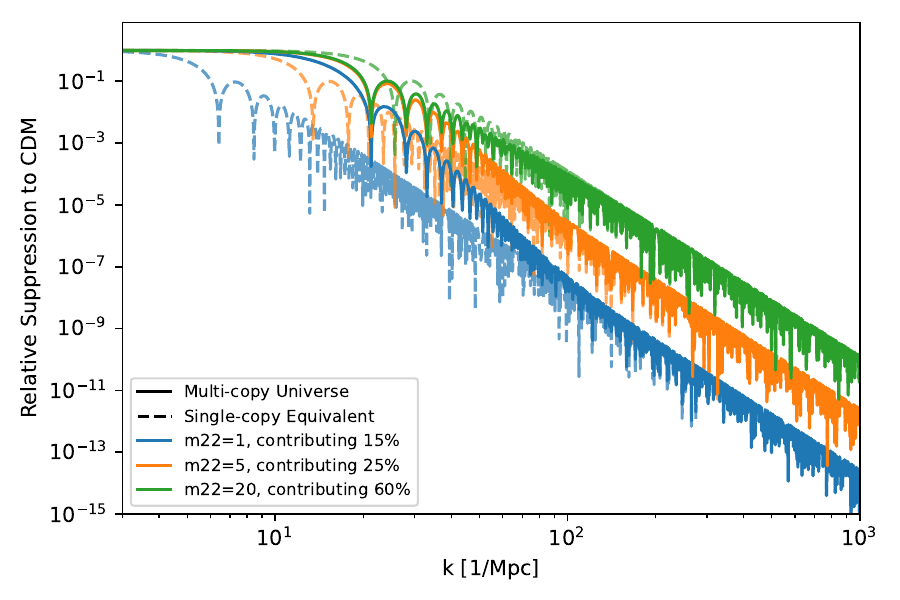}
    \caption{Example transfer functions for each copy of axion for a Universe at redshift 127. The three copies have mass $1, 5, 20\times10^{-22}$eV respectively and contribute $15\%, 25\%, 60\%$ of the total matter density $\overline{\rho_{tot}}$. Transfer functions $T_I$s, $T_{tot}$ in single copy axion Universe are computed using Equ.\ref{transfer_2}.}
    \label{example_3_copy}
\end{figure}

With my assumed decomposition in Equ.\ref{density_decomposition}, the transfer function for each axion copy $\delta_I$ could also be decomposed as an interpolation between two limiting cases:
\begin{equation}
    T = A T_I + (1-A) T_{tot}.
\end{equation}
\noindent This then allows me to determine easily the transfer functions of axions in the multi-copy scenario from known transfer function solutions in a single-copy $\psi$DM Universe. An example is given in Fig.\ref{example_3_copy} (at redshift 127) for a Universe with three copies of axion with masses $1, 5, 20\times10^{-22}$eV, and contributing $15\%, 25\%, 60\%$ of the total matter density $\overline{\rho_{tot}}$ respectively. 

\section{Conclusion and Discussion}
\label{sec_multicopy_interp}

In this chapter, I considered the transfer function for the multi-copy axion Universe and provided a simplified, analytic expression that may be useful for future simulations to determine the initial conditions for each axion copy. In particular, I show that the transfer functions in a multi-copy axion could be parametrized as an interpolating sum of single-copy axion transfer functions that are easier to compute (e.g., via axionCAMB). 

I also identified an 'equivalence' among all copies of axion and with total density contrast $\delta_{tot}$ at small wavenumber, owing to their mutual coupling to the gravitational potential (potentially reflecting a gravitational equilibrium). This 'equivalence' was also identified by \citet{2303.00999}, but here I removed their assumption of heavy particle dark matter being the dominant contribution to CDM and considered a solely $\psi$DM Universe with an arbitrary number of axion copies. This identified 'equivalence' suggests that the large-scale structure formed in such a multi-copy $\psi$DM Universe may resemble an equivalent single-copy Universe where the mass of the equivalent copy is given by a mass $m_{eff}$ determined by 
\begin{equation}
    \frac{1}{m_{eff}^2}  = \sum_I \frac{\overline{\rho_I}}{\overline{\rho_{tot}} } \frac{1}{m_I^2}.
\end{equation}
\noindent Correspondingly, the onset of faint-end turnover in galactic luminosity functions would also be solely determined by this mass scale. Whereas on a smaller (galactic) scale, effects from independent copies of axions will be apparent owing to the 'decoupling' seen at larger wavenumbers. This may then collectively explain the preference for ultralight $\psi$DM with mass $10^{-22}$eV at low redshifts (in dwarfs) while also being able to accommodate the stringent constraints on $\psi$DM mass (e.g., $>2.97\times10^{-22}$eV as I obtained in Chap.\ref{chap:UVLFtesting}). The precise determination of Halo Mass Function and galactic luminosity function, however, remains best investigated with future cosmological simulations.

The same 'equivalence'/gravitational equilibrium may also occur in the late Universe on scales larger than the de Broglie wavelength of the lightest copy. This argument, for instance, may apply to galaxy clusters located on the junction of cosmic filaments, such that the net interference pattern may be well described by the effective de Broglie wavelength scale set by $m_{eff}$. One potential complication for the late Universe is that the corresponding effective mass scale may vary from galaxy cluster to galaxy cluster owing to cosmic variance. But $m_{eff}$ could still be expected to be set by relative contributions of each $\psi$DM to the full galaxy cluster density. Following this logic, the effective mass/de Broglie wavelength scale may be more relevant for lensing interpretation of galaxy clusters, for instance, the apparent skewness of transients detected by \cite{TomKeith2024}. The similar argument (i.e., interference determined by $m_{eff}$) may also apply to individual galaxies, but there the $\psi$DM composition would be more uncertain (galaxies can be dominantly contributed by one particular copy as demonstrated by simulation of \cite{2506.16915}). 

Interpreting mass constraints from Lyman Alpha Forest in the multi-copy $\psi$DM scenario, however, would be more challenging. More specifically, such an interpretation is complicated by the facts that: (1) Lyman Alpha Forest deals more with small-scale density perturbations that cannot be fully described by the linear matter power spectrum; and (2) the heavier $\psi$DM copies lead to additional small-scale density perturbations than the equivalent copy (of mass $m_{eff}$) determining the large-scale structure, and the impact of having these additional perturbations on non-linear matter power spectrum is unclear. As a consequence, mass bounds obtained from Lyman Alpha Forest (e.g., $>2\times10^{-20}$eV by \cite{RogersPeiris2020}) could not be directly interpreted as on the effective mass scale as for UV LF faint-end turnovers. How to better understand these constraints in the multi-copy $\psi$DM scenario is a future direction worthy of pursuing.

\chapter{Summary And Future Prospects}
\label{chap:summary}

In this thesis, I focused on achieving a more reliable test of faint-end UV LF in strong lensing fields. In particular, I was particularly interested in testing whether UV LF displays any turnover at the faint end owing to dark matter being ultralight ($m\sim10^{-22}$eV) and behaving like waves on macroscopic scales. 

In strong lensing fields, one critical challenge for a reliable test of faint-end UV LF at high redshift is contamination by low-z galaxies sharing similar SEDs. As the first key message to the community, in this thesis, I demonstrated that such a contamination issue cannot be overlooked, particularly with the contamination level reaching $\sim 50\%$ over the redshift range $3.5\leq z\leq 5.5$ in existing HFF photometric redshift catalogs. This finding calls for effective ways to mitigate interlopers, and the commonly adopted selection criteria (such as Lyman Break Galaxy-like criteria) were criticized as likely not suitable for such a task. The (potential) presence of contamination could explain the unnatural turn-up in the faint-end UV LF observed by \cite{Livermore2017} (L17), and also casts doubts on existing mass bounds for $\psi$DM particles building on those UV LF constructions (by L17, also by \cite{Bouwens2017} based on Lyman Break Galaxy-like criteria).

In this thesis, I achieved individual mitigation of low-z interlopers by combining deep HST and deep JWST observations on one of the massive galaxy clusters, M0416. It was found that low-z interlopers are dominantly dwarf galaxies with old stellar populations, with half being galaxy cluster members (likely with quenched star formation) and the other half being field galaxies (potentially still star-forming). Quiescent interlopers were also found to have a more compact morphology than star-forming interlopers, possibly owing to ram pressure stripping for cluster members and past star formation for field interlopers driving out their outer stellar contents. 

Through machine learning methods, leveraging the known identities of identified high-z galaxies and low-z interlopers, I also demonstrated that subtle differences between high-z and low-z galaxies encoded in existing measurements can be learned and used for individual mitigation. In particular, with a suitably constructed training set, I demonstrated that an interloper identifier with excellent performance (reaching prediction accuracy of 100\%) can be constructed. The construction of such an interloper identifier alleviates the need for time-intensive supplementary JWST observations on the remaining HFF cluster fields, and paved the way for constructing an interloper-free sample for a more reliable test of faint-end UV LFs. 

When applying the constructed interloper identifier to neighboring blank fields (HFF parallel fields), I also noticed that parallel fields are equally suffering from contamination by low-z interlopers, and the contamination level is higher at higher redshifts. This observation further necessitates the need for additional observations to individually mitigate low-z interlopers, even for robust high-z UV LF faint-end constructions in deep blank fields. 

By inspecting the observed surface number densities (interpreted with magnification bias) of machine learning-selected $3.5\leq z\leq 5.5$ galaxies in all HFF cluster fields, I found no evidence for faint-end turnover in UV LFs, leading me to place a bound on the mass of $\psi$DM particles of $>2.97\times10^{-22}$eV. 

Thanks to the individual mitigating power on interlopers by combining HST and JWST data on M0416, I could similarly perform a reliable test for UV LF faint-end turnovers at higher redshifts. For this analysis, I took advantage of a newly developed and novel source detection algorithm, \texttt{GNUastro}, to construct a catalog that is data complete to 29 mag (in JWST F115W and F150W bands) over $6\leq z\leq 10$. With a similar methodology of comparing the observed surface number densities over $6\leq z\leq 10$ with magnification bias predictions (in both $\psi$DM and $p$CDM paradigms), I derived a slightly weaker bound of $>2.53\times10^{-22}$eV at 95\% confidence level on the $\psi$DM mass. 

I also commented that my derived mass bounds are inconsistent with the faint-end turnover previously hinted by \cite{Leung2018}, favoring a $\psi$DM mass of $1.6\times 10^{-22}$eV. I commented that they have likely suffered from data incompleteness, where, as discussed in detail with my source injection and retraction analysis, data incompleteness leads first to a deviation in galaxy counts against observed apparent magnitudes from power-law behavior, and hence may mimic faint-end turnover induced by $\psi$DM. 

Following the longstanding theoretical motivation from String Axiverse, and also recent observational motivation from dwarf galaxies \citep{Pozo2024PhRvD}, I also considered the scenario where the entire $\psi$DM budget is composed of multiple particle copies with distinct masses. In such a Universe, my linear perturbation analysis suggests different $\psi$DM copies to behave similarly on macroscopic scales, owing to the gravitational equilibrium reached. As a consequence, the suppression effect of the entire $\psi$DM budget is likely determined by an effective de Broglie scale characterizing the macroscopic behavior of all copies. Correspondingly, my derived mass bounds are to be interpreted on an effective mass, determined by the relative contribution of different $\psi$DM copies to the full dark matter budget. On small scales, I argued that different copies to retain their characteristic behavior set by respective mass scale, and hence may explain the diverse dwarf galaxy classes observed at late time \citep{Pozo2024PhRvD, 2506.16915}. 

Along with the analysis, I have also provided a simplified determination of the transfer functions in the multi-copy $\psi$DM paradigm, building on transfer functions in the single-copy $\psi$DM paradigm that could be more easily constructed. My provided parametrization works for an arbitrary number of $\psi$DM copies, and also arbitrary relative contributions, and hence may be particularly helpful for future cosmological simulations in determining the corresponding initial conditions of each $\psi$DM copy.

Finally, I mention several possible future continuations of my analyses in this thesis. The first would be to consider the implications of Lyman Alpha Forest constraints on the multi-copy $\psi$DM scenario, as mentioned in Chap.\ref{chap:multi_axion}. The second is to investigate the effect of incorporating distortions on lensing magnification regions induced by wave interference patterns, to justify that my magnification bias analysis (and hence mass constraint) will not be significantly altered. I have also not considered other dark matter models (such as warm dark matter or self-interacting dark matter) that also lead to a suppression on small scales and hence faint-end turnovers in UV LFs. Interpreting the null detection of faint-end turnovers in those contexts will also be interesting. Finally, $\psi$DM models also predict the onset of structure formation to be delayed relative to the $p$CDM paradigm, which could be tested by counting the number of highest redshift galaxies observed once a reliable sample free from contamination could be constructed. In this thesis, I demonstrated how such a sample could be achieved. It will then be interesting to subsequently apply the methods I applied in this thesis to testing the delayed onset of structure formation.






\printbibliography[heading=bibintoc]

\end{document}